\documentclass[fleqn,usenatbib]{mnras}

\usepackage[T1]{fontenc}

\DeclareRobustCommand{\VAN}[3]{#2}
\let\VANthebibliography\thebibliography
\def\thebibliography{\DeclareRobustCommand{\VAN}[3]{##3}\VANthebibliography}

\usepackage{graphicx}	% Including figure files
\usepackage{amsmath}	% Advanced maths commands

\usepackage{amssymb}	% Extra maths symbols

\usepackage{verbatim}
\usepackage{lscape}     % e.g. for vertical tables \begin{landscape}
\usepackage[dvipsnames]{xcolor} % colours for comments

\usepackage{newtxtext,newtxmath}

\newcommand{\Msol}{M_{\odot}}

\title[Nebular Phase KNe]{On the Validity of Steady-State for Nebular Phase Kilonovae}

\author[Q. Pognan et al.]{
Quentin Pognan,$^{1}$\thanks{E-mail: quentin.pognan@astro.su.se}
Anders Jerkstrand,$^{1}$
Jon Grumer$^{2}$
\\
$^{1}$The Oskar Klein Centre, Department of Astronomy, Stockholm University, AlbaNova, SE-10691 Stockholm, Sweden\\
$^{2}$Theoretical Astrophysics, Department of Physics and Astronomy, Uppsala University, Box 516, SE-751 20 Uppsala, Sweden \
}

\date{Accepted XXX. Received YYY; in original form ZZZ}

\pubyear{2021}

\begin{document}
\label{firstpage}
\pagerange{\pageref{firstpage}--\pageref{lastpage}}
\maketitle

\begin{abstract}
The radioactively powered transient following a binary neutron star (BNS) merger, known as a kilonova (KN), is expected to enter the steady-state nebular phase a few days after merger. Steady-state holds until thermal reprocessing time-scales become long, at which point the temperature and ionisation states need to be evolved time-dependently. We study the onset and significance of time-dependent effects using the non-local thermodynamic equilibrium (NLTE) spectral synthesis code \texttt{SUMO}. We employ a simple single-zone model with an elemental composition of Te, Ce, Pt and Th, scaled to their respective solar abundances. The atomic data are generated using the Flexible Atomic Code (\texttt{FAC}), and consist of energy levels and radiative transitions, including highly forbidden lines. We explore the KN evolution from 5 to 100 days after merger, varying ejecta mass and velocity. We also consider variations in the degree of electron magnetic field trapping, as well as radioactive power generation for alpha and beta decay (but omitting fission products). We find that the transition time, and magnitude of steady-state deviations are highly sensitive to these parameters. For typical KN ejecta, the deviations are minor within the time-frame studied. However, low density ejecta with low energy deposition show significant differences from $\sim 10$ days. Important deviation of the ionisation structure solution impacts the temperature by altering the overall line cooling. Adiabatic cooling becomes important at $t \geq 60$ days which, in addition to the temperature and ionisation effects, lead to the bolometric light curve deviating from the instantaneous radioactive power deposited.
\end{abstract}


%%%%%%%%%%%%%%%%%%%%%%%%%%%%%%%%%%%%%%%%%%%%%%%%%%

%%%%%%%%%%%%%%%%% BODY OF PAPER %%%%%%%%%%%%%%%%%%

\section{Introduction}
\label{sec:intro}
The merging of a binary neutron star (BNS) system has long been a promising site for the creation of heavy elements via the r-process \citep{Symbalisty.Shramm:82,Eichler.etal:89,Freiburghaus.etal:99,Rosswog.etal:99}, due to the production of neutron rich ejecta. Radioactive decay of newly created unstable isotopes power an associated thermal electromagnetic transient known as a kilonova (KN) \citep{Li.Paczynski:98,Metzger.etal:10,Metzger:19}. The discovery of the first BNS gravitational wave signal GW170817, and subsequent electromagnetic follow up of the KN AT2017gfo \citep[see]{Abbott.etal:17} has propelled the study of KNe from the purely theoretical into the observational, with properties generally agreeing with those predicted by theory \citep[e.g.][]{Arcavi.etal:17,Coulter.etal:17,Cowperthwaite.etal:17,Drout.etal:17,Kasen.etal:17,Smartt.etal:17,Tanvir.etal:17}. The amount of r-process elements produced, combined with rate estimates, indicate that these events contribute strongly to the cosmic abundance of heavy elements and may even be the dominating contributors \citep{Rosswog.etal:18,Metzger:19}, though the picture is far from clear \citep[see e.g.][]{Cote.etal:19}.

Much work covering a wide range of aspects has been conducted on KN modelling, with the aim of providing a theoretical framework for the prediction and interpretation of current and future observations. Energy deposition via radioactive decay has been studied extensively both in terms of energy generation from nuclear network calculations \citep[e.g.][]{Metzger.etal:10,Rosswog.etal:18,Barnes.etal:21,Zhu.etal:21}, as well as how decay products effectively thermalise, or deposit their energy to the KN, over time \citep{Barnes.etal:16,Hotokezaka.etal:16,Kasen.Barnes:19,Waxman.etal:19,Hotokezaka.etal:20}. Thorough research on KN lightcurves has also been conducted, with an emphasis on how the evolving, composition dependent opacity influences the observed intensity and photometric colour \citep[e.g.,][]{Barnes.Kasen:13,Tanaka.Hotokezaka:13,Kasen.etal:17,Tanaka.etal:18,tanaka:opacities:2020,Wollaeger.etal:21}. Combined with hydrodynamic studies of the various ejecta types associated to KNe \citep[e.g.][]{Freiburghaus.etal:99,Rosswog.etal:99,Wanajo.etal:14,Perego.etal:14,Perego.etal:17,Nedora.etal:21}, a more complete understanding of KN is starting to form.

Element identification from the emergent spectrum remains difficult. Several factors contribute to this difficulty: many r-process elements are extremely line-rich, and high expansion velocities lead to severe line blending. Furthermore, the wavelengths and strengths of these r-process transitions are not as well known as for lighter elements. Thus far, only strontium (Sr) has been tentatively identified \citep{Watson.etal:2019,Domoto.etal:21}, while other elements expected to be present from nuclear network calculations, such as platinum (Pt) and gold (Au), remain elusive in their identification \citep{Gillanders.etal:21}.

Due to the relatively high velocities ($\sim 0.1\mathrm{c}$) and low ejecta masses ($\sim 0.05~\mathrm{\Msol}$) of the BNS ejecta, the KN is expected to rapidly evolve to the 'nebular' phase after a matter of days to weeks. The term 'nebular phase' is applied when the spectra appear dominated by emission lines, rather than P-Cygni like scattering lines superimposed on a quasi-black body continuum, as during the 'photospheric phase'. For this to happen, the optical depth of the nebula needs to drop enough such that continuum absorption becomes unimportant. Associated with that is a reduced radiation transport time-scale that ends the diffusion-phase of the transient, giving instead a light curve that starts to follow the instantaneous energy deposition by radioactivity. For AT2017gfo this transition happened after 2-3d \citep[see e.g.][]{Smartt.etal:17}. 

Associated with decreasing density is also stronger deviation from LTE, and as for supernovae we can expect that modelling nebular-phase KNe requires an NLTE approach. Predicting emergent NLTE spectra requires determination of temperature, ionisation, excitation, and radiation field, all interconnectedly. The rewards can, however, be significant, as element identification is typically easier from emission-line spectra. Not only this; an improved understanding of the photometric post-peak behaviour of the KN will allow to constrain global parameters such as ejecta mass and radioactive decay rate. 

Apart from \citet{Hotokezaka.etal:21}, current models for KN evolution use an LTE approach. These predict a reddening of the SED as lanthanide elements continue to provide sustained infra-red (IR) emission \citep[see e.g.][]{Kasen.etal:17}. Observations from \textit{Spitzer} have detected such IR emission at late times from AT2017gfo in the $4.5\mu m$ band at 43 and 74 days after merger \citep{Villar.etal:18,Kasliwal.etal:19}. As such, observations of nebular phase emission are feasible to very late times, and the modelling of this phase relevant to current and future research. 

% Steady-state vs time-dep.
The nebular phase is initially in a steady-state regime, where short thermal microphysical reprocessing times lead to the bolometric light curve following the instantaneous radioactive energy thermalization \citep[see e.g.][]{Hotokezaka.etal:21}. A transition to the time-dependent nebular phase is eventually reached when these reprocessing time-scales become long, and processes such as adiabatic cooling become important \citep{Fransson.Kozma:93,Jerkstrand:11}. At this point, the bolometric light curve no longer tracks the energy deposition. The deviations from steady-state lead to different temperature and ionisation structure solutions, which directly affect the atomic line emission. It is thus important to understand when this transition occurs, and how significantly it affects the ejecta conditions. The time-dependent effects studied here are for the thermal particles, different from the time-dependent energy degradation of non-thermal particles \citep{Barnes.etal:16,Hotokezaka.etal:16,Kasen.Barnes:19,Waxman.etal:19,Hotokezaka.etal:20}; both effects together will lead to the light curve deviating from the radioactive decay rate. 

Much of the physics of nebular phase SNe have close analogues in the KN environment \citep{Jerkstrand:17}, and NLTE codes used for SN spectral synthesis may be adapted to model KNe. Here, we describe adaptions of the \texttt{SUMO} code \citep{Jerkstrand:11,Jerkstrand.etal:11} for KN simulations, and present some first applications. \texttt{SUMO} solves for Compton electron degradation \citep{Kozma.Fransson:92}, temperature, NLTE ionisation/excitation coupled to detailed radiative transfer, including fluorescence, in the Sobolev approximation. It has been used to model a wide variety of supernovae in the nebular phase, and is a well-suited tool to tackle KN NLTE modelling. 

In particular, we use the KN-modified version of \texttt{SUMO} to investigate the significance of time-dependent terms in the ionisation and energy equations, with the goal to delineate when the steady-state approximation is sufficient and when a full time-dependent evolution is necessary. We consider the evolution up to 100 days after the merger, corresponding roughly to the time at which the last electromagnetic emission was detected for AT2017gfo (in the infra-red by \textit{Spitzer}, $\sim 74$ days). These conditions and associated time-scales, as well as the treatment for radioactive heating, are described in Section \ref{sec:input_equations}. We then present our atomic data and treatment of relevant physical processes in Section \ref{sec:atomic_data}. In Section \ref{sec:results} we show the resulting evolution of gas state quantities in both steady-state and time-dependent modes. Finally, we discuss the significance of these results and present our conclusions in Section \ref{sec:conclusions}.

\section{Radioactive Heating, and Nebular Phase Equations}
\label{sec:input_equations}

In this section we present the temperature, ionisation structure, and state population equations solved by \texttt{SUMO} for KN ejecta. We begin by considering the energy input via radioactive decay, and continue to address the equations relevant to temperature, ionisation and excitation. 

\subsection{Radioactive Energy Deposition}
\label{subsec:energy_dep}

Ongoing energy deposition to the KN generally comes from the radioactive decay of unstable r-process isotopes, though can also have contributions from a central engine \citep{Yu.etal:13,Wollaeger.etal:19}. The neutron rich isotopes that compose the ejecta are expected to primarily decay via $\beta$ decay, with $\alpha$ decay and spontaneous fission also contributing if translead nuclei are created \citep{Rosswog.etal:17,Kasen.Barnes:19}. Alongside the energy release from the radioactive decays, the thermalisation efficiency of the various decay products, which depends on the geometry and density of the ejecta, must also be taken into account. The thermalisation at a given time depends on the previous decay and deposition history, but may be written in terms of an efficiency $f_i$(t) of the current decay rate \citep{Barnes.etal:16}:

\begin{equation}
    \dot{q}_{\mathrm{tot}}(t) = \sum_{i} \dot{Q}_i(t) \; f_i(t) \: \mathrm{erg \; s^{-1} \; g^{-1}}
    \label{eq:tot_dep}
\end{equation}

\noindent where $\dot{Q}_i(t)$ is the energy released in the decay via a particular decay product, and $f_i(t)$ accounts for the time-delayed, partial thermalisation of earlier decays. Since r-process elements are expected to decay mostly by $\beta$ decay, we first consider the energy input from this channel. Nuclear network calculations generally agree on early time $\beta$ decay energy generation evolving as \citep{Metzger.etal:10,Barnes.etal:16,Rosswog.etal:18,Kasen.Barnes:19}: 

\begin{equation}
    \dot{Q_{\beta}}(t) \approx 10^{10} \: t_{\mathrm{d}}^{-1.3} \: \mathrm{erg \: s^{-1}\: g^{-1}}
    \label{eq:Q_Beta}
\end{equation}

\noindent where $t_d$ is the time in days after the merger. The above energy generation rate includes electrons/positrons, gamma rays and neutrinos. Equation \ref{eq:Q_Beta} shall be taken as indicative, as recent studies have shown that the decay power may vary significantly depending on nuclear network inputs, with subsequent associated quantities such as effective heating rate varying by over an order of magnitude at 1 day after merger \citep[see e.g.][]{Barnes.etal:21,Zhu.etal:21}.

Thermalisation efficiencies $f_i(t)$ depend on the decay product and ejecta properties, but are all expected to be near unity at early times (i.e. all products thermalise efficiently on time-scales shorter than any evolutionary time-scale), and then transition to a late time behaviour after an associated thermalisation time-scale. We follow the treatment of thermalisation efficiency described in \citet{Kasen.Barnes:19}, with the fraction of energy associated to each $\beta$ decay product (electrons/positrons, gamma rays and neutrinos) taken to be $(p_e,p_\gamma,p_{\nu}) = (0.2,0.5,0.3)$ respectively \citep{Barnes.etal:16,Hotokezaka.etal:16}. Neutrinos are assumed to provide negligible thermalisation at the time-scales studied here, instead freely streaming out of the KN. For the time evolution of the electron/positron thermalisation efficiency, we include the added consideration from \citet{Waxman.etal:19,Hotokezaka.etal:20} regarding the energy losses by ionisation of the plasma, thus yielding a steeper power law ($f_{\beta,e}(t) \propto t^{-1.5}$) compared to the default one in \citet{Kasen.Barnes:19} ($f_{\beta,e}(t) \propto t^{-1}$).

 We currently ignore fission products, but do consider $\alpha$ decays with an initial energy input corresponding to 10 per cent of the $\beta$ decay (e.g. $\dot{Q}_{\alpha} = 0.1 \dot{Q}_{\beta}$), well within the broadly estimated contribution range of $5 - 40$ per cent \citep{Barnes.etal:16,Kasen.Barnes:19} for low electron fraction outflows ($Y_e \lesssim 0.15$) \citep[see e.g.][]{Mendoza.etal:15}. Higher $Y_e$ outflows are expected to synthesise less trans-lead nuclei, and thus $\alpha$ decay will play a less important role. Our chosen $\alpha$ decay contribution corresponds adequately to our chosen elemental composition which includes heavier elements like Ce and Th, and thus implies a relatively low electron fraction (see Section \ref{sec:atomic_data}).

The inclusion an actinide in our model suggests that fission may play a role in the ejecta's composition, as well as for the energy deposition, since this process is found to affect the abundance of $\alpha$ decaying nuclei in the regime of $222 \leq A \leq 225$ \citep[][]{Giuliani.etal:20}. Direct thermalisation of fission products from isotopes such as $^{254}\mathrm{Cf}$ may also provide significant late time energy deposition to the ejecta directly, potentially even becoming the dominating energy source due to longer effective thermalisation times \citep[see e.g.][]{Wanajo.Shinya:18,Zhu.etal:18,Wu.etal:19}. As such, it is possible that our late time energy deposition is underestimated due to the exclusion of fission products, which may affect the time-dependent effects seen at those times. However, our models are mostly composed of $^{127}\mathrm{Te}$, and have an average atomic mass of $<A> = 142$. In this mass range, $\beta$ decay is expected to be the dominating channel \citep[see e.g.][]{Barnes.etal:16,Wanajo.Shinya:18,Kasen.Barnes:19}, and so we believe excluding fission products in the context of this study remains acceptable.

Therefore, the overall energy deposition per mass follows the form of Equation \ref{eq:tot_dep}, with the thermalisation efficiencies for $\beta$ and $\alpha$ decay given by \citep{Kasen.Barnes:19}:

\begin{align}
    \label{eq:beta_therm}
    f_{\beta}(t) &= 0.2 \left(1 + \frac{t}{t_e} \right)^{-1.5} + 0.5 \left( 1 - e^{-t_{\gamma}^2/t^2} \right) \\
    \label{eq:alpha_therm}
    f_{\alpha}(t) &= \left(1 + \frac{t}{3t_e}\right)^{-1.5}
\end{align}

\noindent The thermalisation time-scales for $\beta$ decay electrons ($t_e$) and gamma rays ($t_{\gamma}$) are given by: 

\begin{align}
    \label{eq:t_e} 
    t_e &\approx 150 \: \left( \frac{M_{\mathrm{ej}}}{0.05~\mathrm{\Msol}}\right)^{2/3} \: \left( \frac{v_{\mathrm{ej}}}{0.1 \mathrm{c}} \right)^{-2} \: \mathrm{days} \\
    \label{eq:t_y}
    t_{\gamma} &\approx 1.3 \: \left( \frac{M_{\mathrm{ej}}}{0.05~\mathrm{\Msol}}\right)^{1/2} \: \left( \frac{v_{\mathrm{ej}}}{0.1 \mathrm{c}}\right)^{-1} \: \mathrm{days}
\end{align}

\noindent These time-scales correspond to when thermalisation becomes inefficient for the decay product, presently in the case of a uniform density spherically symmetric ejecta undergoing homologous expansion with maximum velocity $v_{\mathrm{ej}}$. The electron time-scale expression assumes a time dependent electron decay energy of the form $E(t) = E_{0}t^{-1/3}$ \citep{Kasen.Barnes:19}. 

The heating rate per volume is then found from the energy deposition as follows:

\begin{equation}
\begin{split}
    h(t) &= M_{\mathrm{ej}} \dot{q}_{\mathrm{tot}}(t)/(\frac{4\pi}{3} v_{\mathrm{ej}}^3 t^3)  \\
   & \propto \left(\frac{M_{\mathrm{ej}}}{0.05~\mathrm{\Msol}} \right) \left(\frac{v_{\mathrm{ej}}}{0.1 \mbox{c}} \right)^{-3} t_d^{-4.3} f(t) \mathrm{ \; erg \: s^{-1} \: cm^{-3}}
    \label{eq:heating_pervolume}
\end{split}
\end{equation}

\noindent Where $f(t)$ is the summation of efficiencies of the various decay products i.e. $\alpha, \beta, \gamma$. Assuming $\beta$ and $\alpha$ decay to dominate during the nebular phase, and thus taking the thermalisation efficiency to evolve as $f(t) \propto t^{-1.5}$, we find that the heating rate evolves as $h(t) \propto t^{-5.8}$ at $t \gg t_e$.

\subsection{Energy Equation and Temperature}
\label{subsec:temperature}

The energy deposited to the KN from radioactive decay particles is split into heating, excitation and ionisation according to the Spencer-Fano equation \citep{Spencer.Fano:54,Kozma.Fransson:92}. Although heating can generally occur from a variety of other sources such as photoionisation, free-free absorption, and collisional de-excitation, non-thermal collisional heating is normally dominant in nebular phase SNe, and expected to be even more so in KNe where the ionisation state is higher, which increases the Spencer-Fano heating fraction. Cooling occurs by line emission following thermal collisional excitation, recombination, free-free emission, and adiabatic expansion. 

The first law of thermodynamics for the particles in a thermal distribution can be written as \citep[see e.g.][]{Jerkstrand:11}:

\begin{equation}
    \frac{\rm{d}T(t)}{\rm{d}t} = \frac{h(t) - c(t)}{\frac{3}{2}k_B n(t)} - \frac{2T(t)}{t} - \frac{T(t)}{1 + x_{e}(t)} \frac{\rm{d}x_{e}(t)}{\rm{d}t}
    \label{eq:temperature_equation}
\end{equation}

\noindent where $h(t)$ and $c(t)$ are the heating and cooling rates per unit volume respectively, $n(t)$ is the total thermal particle number density, $k_B$ the Boltzmann constant, and $x_e(t)$ is the electron fraction which equals the number of free electrons divided by the number of atoms and ions. The $2T(t)/t$ term corresponds to adiabatic cooling, and we call the final term the ``ionisation cooling'' term. 

When assuming steady-state conditions, Equation \ref{eq:temperature_equation} reduces to the condition $h(t) = c(t)$. Cooling in the nebular phase is expected to be dominated by line cooling, which for a single radiative transition in ion $i$, from upper level $k'$ to lower level $k$, can be written as: 

\begin{equation}
        c_{\mathrm{line}}^{i,kk'} = \Delta E_{kk'}n_e q_{k'k}(T) [\frac{g_{k'}}{g_k} n_k e^{-\Delta E_{kk'}/k_B T} - n_{k'}]
        \label{eq:cline_specific}
\end{equation}

\noindent where $E_{kk'}$ is the transition energy, $q_{k'k}$ is the rate for downwards collisions, $n_k$ and $n_{k'}$ are the level populations of the lower and upper levels respectively, and $g_k$ and $g_{k'}$ are the statistical weights for the lower and upper levels respectively. The total line cooling from a particular ion is the summation over all transitions. 

Since $c_{\mathrm{line}}$ in many regimes is proportional to $n_e$ and $n_i$, it is convenient to define the line cooling for an ion in terms of an intrinsic 'line cooling function' $\Lambda^i(T,n_e,n_i)$, such that we have:

\begin{equation}
    c_{\mathrm{line}}^i (t) = \Lambda^i(T,n_e,n_i)n_e n_i \: \mathrm{erg \; s^{-1} \; cm^{-3}}
\end{equation}
    
\noindent where $n_e$ and $n_i$ are the electron and ion number densities respectively. The dependence of $\Lambda^i$ on electron and ion density is complex and non-linear, with both quantities affecting level populations, directly by collisions and additionally via line-self absorption for ion density\footnote{In principle there may be dependencies also on $n_i+$ through recombinations and on other ions through charge transfer reactions, but we omit these dependencies here as they are typically unimportant.}. In the case of spherically symmetric, homologously expanding media, the ion and electron density are:

\begin{align}
    \label{eq:ion_density}
    n_i &= \frac{M_i}{A_i m_p} (\frac{4\pi}{3} v_{\mathrm{ej}}^3 t^3)^{-1} \\
    \label{eq:elec_density}
    n_e &= \frac{M}{<A> m_p} (\frac{4\pi}{3} v_{\mathrm{ej}}^3 t^3)^{-1} x_e
\end{align}

\noindent where $M_i$ is the mass of element $i$, $M$ is the total ejecta mass, $<A>$ the average atomic weight of the ejecta, and $m_p$ the mass of a proton. The average atomic weight here is taken to be $<A>=140$, corresponding to the chosen elemental composition detailed in Section \ref{sec:atomic_data}. The total number density used in Equation \ref{eq:temperature_equation} is found simply by adding the ion and electron densities, $n(t) = n_e + \sum_i n_i(t)$. The line cooling per volume can then be written as:

\begin{equation}
\begin{split}
    c_{\mathrm{line}}(t) =  0.34 \left(  \sum_i x_i(t) \frac{ \Lambda^i(T,n_e,n_i)}{10^{-20}~\mathrm{erg \; cm^3 \; s^{-1}}} \right) \left(\frac{M_{\mathrm{ej}}}{0.05~\mathrm{\Msol}} \right)^2 \left( \frac{v_{\mathrm{ej}}}{0.1\mathrm{c}} \right)^{-6} \\
    \times \; x_e(t) \; t_d^{-6} \; \mathrm{ \: erg \: s^{-1} \: cm^{-3}}
    \label{eq:line_cooling}
\end{split}    
\end{equation}

\noindent where $x_i(t)$ is the abundance of ion $i$. The total line cooling function, $\Lambda(T,n_e,n_1,..,n_l)=\sum_i x_i \Lambda^i(T,n_e,n_i)$, is expected to be solely a function of temperature when densities are low enough, and generally increases with temperature. Although the exact evolution remains sensitive to ejecta composition, $\Lambda$ can still be used for rough estimates of how steady-state temperature solutions will evolve with time. Using the assumptions for steady-state and setting Equations \ref{eq:heating_pervolume} and \ref{eq:line_cooling} to be equal, we solve for $\Lambda(T,n_e,n_1,...,n_l)$ and find the following evolution: 

\begin{align}
    \Lambda \propto \left(\frac{M_{\mathrm{ej}}}{0.05~\mathrm{\Msol}} \right)^{-1} \left(\frac{v_{\mathrm{ej}}}{0.1\mathrm{c}} \right)^{3} x_e(t)^{-1} \; t_d^{1.7} \; \mathrm{ \: erg \: cm^{3} \: s^{-1}}~~t \lesssim t_e 
\label{eq:lambda_evolution_early}\\
    \Lambda \propto \left(\frac{M_{\mathrm{ej}}}{0.05~\mathrm{\Msol}} \right)^{-1} \left(\frac{v_{\mathrm{ej}}}{0.1\mathrm{c}} \right)^{3} x_e(t)^{-1} \; t_d^{0.2} \; \mathrm{ \: erg \: cm^{3} \: s^{-1}}~~t \gtrsim t_e
\label{eq:lambda_evolution}
\end{align}

\noindent From the above equation, if $x_e(t)$ is approximately constant, then we have an initial evolution of $\Lambda \propto t^{1.7}$ which flattens to $\Lambda \propto t^{0.2}$ at late times when thermalisation efficiency evolves as $t^{-1.5}$. Since $\Lambda$ increases with T, temperature is also expected to rise as time goes on, rapidly at first, then more slowly at later times. 
 
There will come a point however, when time-dependent effects will cause a deviation from steady-state solutions, with processes like adiabatic cooling preventing temperature from rising indefinitely. These effects arise if the other terms in Equation \ref{eq:temperature_equation} become important relative to the heating and line cooling terms. The ``ionisation cooling'' term is expected to remain quite small, as the change in electron fraction usually obeys $dx_e/dt \ll (1 + x_e)/t$, making the last term much smaller than the adiabatic term. The adiabatic term, given by $c_{\mathrm{adia}} = 3nk_B T/t$, will on the other hand, eventually become comparable to, and then exceed the line cooling, due to decreasing densities as time goes on; $c_{\mathrm{line}}(t) \propto x_e(t) t^{-6}$, $c_{\mathrm{adia}}(t) = 3 k_B n(t) T(t)/t \propto (1+x_e(t))T(t) t^{-4}$, with $x_e \gtrsim 1$ and $T(t) \propto t^{0.2}$, yielding $c_{\mathrm{adia}}(t)/c_{\mathrm{line}}(t) \propto t^{2.2}$, where we have assumed $T \propto \Lambda$. The time-scale on which this will occur can be found in detail from setting the ratio

\begin{equation}
\begin{split}
    \frac{c_{\mathrm{adia}}(t)}{c_{\mathrm{line}}(t)} = 1.4\times 10^{-6} \left( \frac{M_{\mathrm{ej}}}{0.05~\mathrm{\Msol}} \right)^{-1} \left(\frac{v_{\mathrm{ej}}}{0.1\mathrm{c}} \right)^3 \left(\frac{T(t)}{10^4 K} \right) \\ 
    \times \left( \frac{(1+ x_e(t))/x_e(t)}{1.6} \right) \left( \sum_i x_i \frac{\Lambda^i(T,n_e,n_i) }{10^{-20} \: \mathrm{erg \; cm^3 \; s^{-1}}} \right)^{-1} t_d^{2}
    \label{eq:adia_to_line}
\end{split}
\end{equation}

\noindent equal to unity and solving for the time. Taking ejecta parameters of $M=0.05~M_\odot$ and $v_{ej}=0.1\mathrm{c}$, and fixing $\left(1+x_e\right)/x_e=1.6$ and $T=10^4$K, yields a value $845$ days, implying that adiabatic cooling is not significant for these parameter choices within a relevant period. However, choosing ejecta parameters representing low mass, high velocity dynamical ejecta, e.g. $M_{\mathrm{ej}} = 0.01~\mathrm{\Msol}$ and $v_{\mathrm{ej}} = 0.2 \mathrm{c}$, leads to a time-scale of only $\sim 133$ days. Depending on the temperature and ionisation structure solutions, as well as the line cooling function for those solutions, it is thus possible that adiabatic cooling may be significant within $t < 100 $ days, at least for the high velocity components of the KN ejecta. 

\subsection{Ionisation Structure}
\label{subsec:ionisation}

Calculations of ionisation structure are done in NLTE, and thus use rate equations. For an element \textit{j} in ionisation state \textit{i}, the rate of change of ion abundance $x_{j,i}$ is given by:

\begin{equation}
    \frac{dx_{j,i}}{dt} = \Gamma_{j,i-1}x_{j,i-1} - (\Psi_{j,i} + \Gamma_{j,i})x_{j,i} + \Psi_{j,i+1}x_{j,i+1}
    \label{eq:ion_balance}
\end{equation}

\noindent where $\Gamma$ is the ionisation rate per particle and $\Psi = \alpha(T) n_e $ is the total recombination rate per particle. \texttt{SUMO} allows ionisation by both non-thermal electron collisions and photoionisation. Recombination can occur by radiative or dielectronic recombination such that the total temperature dependent recombination coefficient for a particular ion is given by $\alpha = \alpha_{rr} + \alpha_{dr}$. The exact treatment of the various ionisation and recombination processes are detailed in Section \ref{sec:atomic_data}. 

In the steady-state approximation, the time derivative term is set to zero and ionisation structure is given simply by balancing the ionisation and recombination terms. As for the temperature evolution, this approximation is only valid if the time derivative is small relative to the other terms, which will cease to be the case if either the ionisation or recombination time-scale is long. The recombination time-scale is $t_{\mathrm{rec}} = 1/\left(\alpha n_e\right)$, where $\alpha$ is the total recombination rate for that particular ion. A critical recombination time $t_{\mathrm{crit,rec}}$ can be identified where the typical recombination time-scale equals the evolutionary time-scale, i.e. $t_{\mathrm{rec}} = t$, giving:

\begin{equation}
    t_{\mathrm{crit,rec}} = 71\mathrm{d}  \left(\frac{M}{0.05~\mathrm{\Msol}} \right)^{1/2} \left(\frac{v}{0.1\mathrm{c}} \right)^{-3/2} x_e^{1/2} 
    \left( \frac{\alpha_{\mathrm{typical}}}{10^{-11} \mathrm{cm^3 \; s^{-1}}} \right)^{1/2}   
    \label{eq:trec_crit}
\end{equation}

The critical recombination time depends on the ejecta properties and the value of the recombination coefficient (see also Section \ref{sec:atomic_data}). From the above expression, it is quite possible that time-dependent effects in the ionisation structure will be relevant within 100 days after merge, especially for lower density models ($M_{\mathrm{ej}}=0.01~\mathrm{\Msol}$ and $v_{\mathrm{ej}}=0.2c$ yields $t_{crit} \sim 13$d). 

A similar analysis can be carried out for the ionisation time-scale considering ionisation by non-thermal electrons, which is expected to dominate for most ions. The non-thermal ionisation rate for a particular ion is given by $\Gamma_{nt,i} = \dot{q}_{\mathrm{tot}}(t) \; SF_{\mathrm{ion},i}(t) / ( \chi_i \; N_i)$, where $SF_{\mathrm{ion},i}(t)$ is the amount of radioactive energy going to ionisation of ion $i$ from the Spencer-Fano equation, and $\chi_i$ is the ionisation potential of the ion. This becomes, averaged over the ions:

\begin{equation}
\begin{split}
    \Gamma_{\mathrm{nt}} &= 10^{10} \; t_d^{-1.3} \; f(t) \; SF_{\mathrm{ion}}(t) \; \frac{<A> \; m_p}{\bar{\chi}} s^{-1}\\
    & = 7.31\times 10^{-2} \; t_d^{-1.3} \; f(t) \; SF_{\mathrm{ion}}(t) \left(\frac{\bar{\chi}}{20~eV}\right)^{-1} \; s^{-1}
\end{split}
\end{equation}

\noindent In the last step we take $<A> ~ = 140$. The ionisation time-scale is then found by taking the reciprocal of the ionisation rate such that $t_{\mathrm{ion}} = 1/\Gamma_{\mathrm{nt}}$. As for recombination, a critical time can be identified at which the ionisation time-scale is equal to the evolutionary time:

\begin{equation}
t_{\mathrm{crit,ion}} \approx 10^3 \mbox{d}  \left(\frac{f(t)}{0.25}\right)^{10/3} \left(\frac{SF_{\mathrm{ion}}(t)}{0.01}\right)^{10/3}\left(\frac{\chi}{20~eV}\right)^{-10/3}.
\label{eq:tion_crit}
\end{equation}

\noindent Here we have estimated typical values for thermalisation efficiency $f(t)$ and the amount of radioactive energy going to ionisation $SF_{\mathrm{ion}}(t)$. From Equation \ref{eq:tion_crit}, it appears that the critical ionisation time will not be reached within 100 days, however this critical time is highly sensitive to every parameter. Generally, models with less efficient thermalisation, lower initial energy input (i.e. $\dot{q}_{\mathrm{tot}}(1 \mathrm{day}) < 10^{10} \; \mathrm{ erg \; s^{-1} \;~g^{-1}}$, and/or higher ionisation levels (reduces $SF_{\mathrm{ion}}(t)$ and increases $\chi$), will have earlier critical times. For example, for $f(t)=0.1$ and $\chi=30$ eV (as for Te II and Pt III) the time-scale is $t_{\mathrm{crit,ion}} = 12$ days.

As will be demonstrated later, time-dependent ionisation structure effects may also begin to be noticeable well before these critical times, as long as the conditions $t_{\mathrm{ion}}/t,t_{\mathrm{rec}}/t \ll 1$ are no longer satisfied (e.g. if the ratios are in the range 0.1-1). Since \texttt{SUMO} solves ionisation structure, excitation, and temperature self consistently, a change in ionisation structure can also affect the temperature solution, as ions are in general not expected to have identical cooling capacities.

\subsection{Excitation and Radiation Field}
\label{subsec:excitation}

Each ion within the model will have level populations that depend not only on internal transitions, but also recombinations and ionisations into excited states. As such, the fraction of an ion \textit{i} in excitation state $k \neq k'$ is given by:

\begin{equation}
\begin{split}
    \frac{\rm{d}x_{i,k}}{\rm{d}t} &= \sum_{k'}x_{i-1,k'}\Gamma_{i-1,k',i,k} + \sum_{k'} x_{i+1}\Psi_{1+1,k',i,k}  \\
    &+ \sum_{k' \neq k} x_{i,k'}\xi_{k'k} \\
    &- x_{i,k} \left(\sum_{k' \neq k} \xi_{k,k'} + \sum_{k'}\Gamma_{i,k,i+1,k'} + \sum_{k'} \Psi_{i,k,i-1,k'}  \right)
    \label{eq:exc_balance}
    \end{split}
\end{equation}

\noindent where $\Gamma$ denotes ionisation, $\Psi$ recombination, and $\xi$ internal transition rates respectively. The first three terms represent processes into the particular excitation state, while the last three are processes out of the particular excitation state. The recombination and ionisation processes are the same as described above in Section \ref{subsec:ionisation}. In general, the internal transition rates $\xi$ encompass spontaneous emission, stimulated emission, photoabsorptions, and both non-thermal and thermal collisions. \texttt{SUMO} considers every internal transition process when calculating the excitation structure, though here we ignore non-thermal excitations due to the lack of r-process cross sections for these.

The steady-state approximation for excitation corresponds to setting the time derivative on the left hand side of Equation \ref{eq:exc_balance} to zero. Similarly to temperature and ionisation, this condition is satisfied if the de-excitation time-scale is short relative to the evolutionary time. Although some excited states can only radiatively decay by slow forbidden transitions, the total de-excitation time taking into account other radiative transitions and collisional de-excitations will never be long enough to warrant a consideration of the full time-dependent equation (see Appendix \ref{app:ssexc}). As such, the steady-state approximation for excitation is used here even when the full time-dependent temperature and ionisation equations are employed.

\texttt{SUMO} does not treat the radiation-field time-dependently but assumes $c=\infty$. While the decreasing optical depths post-peak push the radiation transport times towards the free-streaming limit, $v_{ej}/c \; \times \; t$, remaining line opacity lengthens this to $N_{\lambda} v_{ej}/c \; \times \; t$, where $N_{\lambda}$ is the number of optically thick lines per unit radius \citep{Jerkstrand.etal:16}. As this can remain $\gg 1$ for an extended period in certain wavelength ranges, there may be residual time-dependent effects for the radiation field. Although a full investigation of the scale of such effects remains to be done, the relatively high probability of fluorescence to occur at longer, more optically thin wavelengths, limits the impact. The radiation field is also of secondary importance for setting physical conditions in SN and KN ejecta, with non-thermal and thermal electrons playing the most important roles.

\begin{figure}
    \centering
    \includegraphics[trim={0.5cm 1.0cm 0.5cm 0.0cm},width=\linewidth]{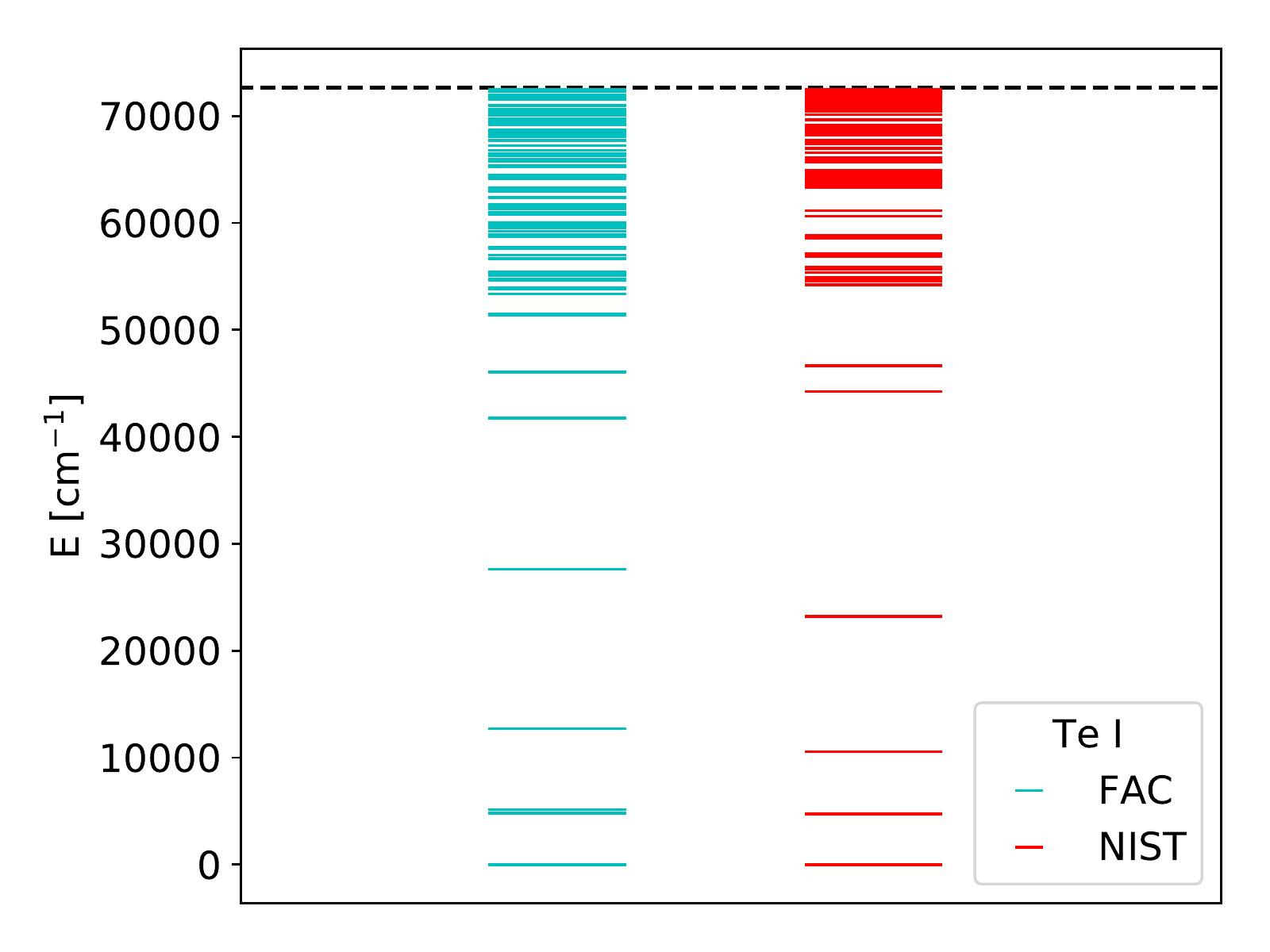}
    \caption{Theoretical energy levels of Te I from this work compared to those compiled in the NIST atomic spectra database \citep{NIST_ASD}. The dashed line indicates the ionisation energy as given by NIST. The same kind of plots for the other ions can be found in Appendix \ref{app:atomicdata}.}
    \label{fig:TeI_levels}
\end{figure}

\section{Atomic Data and Methodology}
\label{sec:atomic_data}

In order to investigate the importance of time-dependent effects on the thermodynamic evolution of the KN during the nebular phase, we construct a simple constant-density, single-zone ejecta model expanding homologously in spherical symmetry. We choose a grid of ejecta masses of $M_{\mathrm{ej}} = 0.01, 0.05, 0.1 \mathrm{\Msol}$, and maximum ejecta velocities of $v_{\mathrm{ej}} = 0.05, 0.1, 0.2 c$. These parameters cover the broad range of KN ejecta, from faster dynamical ejecta to relatively slower disc wind ejecta \citep[see e.g.][]{Metzger:19}. 

We choose four elements to be included in our model composition: tellurium (Te, Z=52), cerium (Ce, Z=58), platinum (Pt, Z=78) and thorium (Th, Z=90). Te and Pt correspond to relatively abundant elements at the second and third r-process abundance peaks respectively, while Ce and Th are chosen as representatives for the lanthanides and actinides respectively. Since Ce and Th are included while no elements from the first r-process abundance peak are present, this composition represents a typically redder, lanthanide-rich KN, as opposed a blue lanthanide-poor KN. Each element is allowed to be triply ionised at most, resulting in a total of 16 unique ions in the model. The mass fractions are scaled to each element's respective solar r-process abundance \citep[see e.g.][]{Freiburghaus.etal:99,Rosswog.etal:18,Prantzos.etal:20}, such that the mass fractions are given by $(\mathrm{Te,Ce,Pt,Th}) = (0.65,0.08,0.25,0.02)$. These mass fractions are broadly consistent with nuclear network calculations for a KN with an electron fraction within the reasonable range \citep{Rosswog.etal:18,Nedora.etal:21,Zhu.etal:21}.

\subsection{Energy Levels and Lines}
\label{subsec:levelsandlines}

Experimental atomic data for heavy, r-process, elements complete enough to be applicable to spectral modelling remains largely unavailable, and must therefore be generated from numerical calculations \citep[see e.g.][]{Kasen.etal:13,Gaigalas.etal:19,Gamrath.etal:19,Fontes.etal:20,Radziute.etal:20,tanaka:opacities:2020}. In this work, we determine fundamental atomic data with the Flexible Atomic Code (\texttt{FAC}) \citep{Gu:08}. FAC is a generally applicable atomic structure package suitable for the computation of atomic data such as energy levels and bound-bound radiative transition rates. \texttt{FAC} employs a fully relativistic approach suitable for heavy atoms and ions, and as such the fundamental electron basis set is constructed from Dirac orbitals which are optimised to self-consistency in a Dirac-Fock-Slater procedure on a non-physical, mean configuration defined with fractional occupation numbers in order to account for the screening of more than one configuration. With this orbital basis at hand, a many-electron basis of $jj$-coupled configuration state functions (CSFs) is constructed from a set of target configurations. Final atomic eigenstates and energies are then obtained through a standard configuration-interaction (CI) approach. Radiative transition rates between the obtained eigenstates are calculated in the single multipole approximation. The spin-angular integration required in the evaluation of various one- and two-particle operators throughout the code is carried out using the techniques of \cite{Gaigalas:1997, Gaigalas:2001}. See the FAC documentation for further details\footnote{\url{https://github.com/flexible-atomic-code/fac/}}.

As this work is focused on the gas state evolution of the ejecta, we target an as complete set of atomic energy levels and processes as possible, rather than e.g. a more limited set of spectral lines sufficiently accurate to infer the presence or absence of specific elements in observational data.  As such, for each ion, we limit the many-electron basis to the CSFs that correspond to those configurations which are estimated to cover the energy range up to the first ionisation limit; i.e. the smallest possible CI space where all eigenstates are of physical interest. The radial orbitals are optimised on the ground configuration of each ion, except for Pt I where a weighted mean of [Xe]4f$^{14}$\{5d$^9$6s + 5d$^{10}$ + 5d$^9$6p\} was found to better represent the energy spectrum. A complete list of configurations included in the present calculations is given in Tab. \ref{tab:configs} of Appendix \ref{app:atomicdata}.
Based on these atomic state functions, both allowed and forbidden transitions with transition rates down to the order of $\mathrm{A} \sim 10^{-12} \: \mathrm{s^{-1}}$ are calculated. We exclude those transitions with an upper level above the experimental ionisation limit. This approach reproduces energy levels to an accuracy as shown in Figure \ref{fig:TeI_levels} (see also Appendix \ref{app:atomicdata} for other ions). The level of accuracy obtained with this approach is likely sufficient for a reasonable prediction of the evolution of thermodynamic quantities such as temperature and ionisation.

\subsection{Ionisation and Recombination Treatment}
\label{subsec:ionandrec}

Ionisation is treated in \texttt{SUMO} by considering non-thermal collisions and photoionisations. The non-thermal collisions for r-process elements are treated following \citet{Lotz:67}

\begin{equation}
    \sigma(E) = \psi \; a \; \frac{1}{\chi^2} \frac{\ln{E/\chi}}{E/\chi}
\end{equation}

\noindent where $a=4\times 10^{-14}$ cm$^2$, $\chi$ is the ionisation potential in eV, $E$ is the energy of the electron in eV, and $\psi$ is a parameter of order unity. We set a constant $\psi=1$, which was determined by comparing the formula against measured cross-sections for various lighter elements.  We apply this formula to the valence shell, giving an ionisation cross section of order $10^{-17} \: \mathrm{cm^{2}}$ at 1~keV for $\chi=10$ eV. It should be noted that this remains an approximative treatment, in that ideally such a cross-section formula should be applied to every shell. However, treatment of inner shells requires the inclusion of additional physical processes such as the Auger process and X-ray fluorescence, which are not presently included in \texttt{SUMO}\footnote{Some inner shells are considered for lighter elements, but with ionisation potentials set to the valence one to avoid treating these processes.}.  Considering that inner shells have higher ionisation potentials, as well as typically smaller cross-sections, ionisation from the valence shell is expected to dominate. However, the sum of inner shells may make a non-negligible contribution, so it is possible that our non-thermal ionisation rates are somewhat underestimated. 

Photoionisation is expected to be important for the neutral atoms due to their low ionisation thresholds (5.5-9 eV for the four elements considered here) being reachable for a significant part of the radiation field, and may also be important for the ions, especially those with low ionisation potentials (e.g. Ce II and Th II with 10-12 eV thresholds). Currently, due to lack of photoionisation cross sections for r-process elements, hydrogenic cross sections are used. 

As mentioned in Section \ref{subsec:ionisation}, recombinations in \texttt{SUMO} occur by radiative and dielectronic processes. These come into effect by combining into a total recombination coefficient $\alpha = \alpha_{rr} + \alpha_{dr}$ (unit $\mathrm{cm^3 s^{-1}}$). In the case of many lighter elements considered for SNe, only radiative recombination is relevant as dielectronic recombination becomes significant only at high temperatures ($T > 10^4 \mathrm{K}$) not reached in nebular-phase SNe. However, we expect KN temperatures to be potentially higher than this threshold and dielectronic recombination may become important. As for many aspects of r-process atomic data however, the recombination rates are largely unknown. As such, we consider how the recombination coefficients of some lighter elements (C, N, O, Ne, Mg, Si, S, Ar, Ca, Fe, Ni)  evolve with temperature (see Appendix \ref{app:recombination}, Figure \ref{fig:recrates}), as empirical formulas and fits for the temperature dependence of these rates can be readily found \citep[see e.g.][]{Shull.Steenberg:82,Nussbaumer.Story:83,Arnaud.Rothenflug:85,Nahar:95,Nahar.etal:97}. We find that the general evolution of recombination with temperature is complex and highly dependent on the shell into which an electron is captured.

Using the atomic data code \texttt{HULLAC} \citep{Bar-Shalom.etal:01}, \citet{Hotokezaka.etal:21} find that for the lanthanide element neodymium (Nd, Z=60), dielectronic recombination from  a few autoionising states dominates recombination. The dielectronic recombination rates are found to be over an order of magnitude greater than the radiative recombination rates, with values $\alpha_{dr} \sim 10^{-10} - 10^{-9} \: \mathrm{cm^3 \; s^{-1}}$ for recombination to Nd II and Nd IV respectively. However, the behaviour of the recombination coefficient depends heavily on the electron shell properties of the ion, and can change immensely between ions when recombining to a different shell (see Figure \ref{fig:recrates}). As such, it is difficult to directly compare recombination coefficients between elements with different shell structures. Nevertheless, this calculation highlights the importance of dielectronic recombination for heavy elements, such as the lanthanides, from which we include Ce. 

Considering both the range of values from lighter elements and the aforementioned work on Nd, we elect to pick a constant $\alpha = 10^{-11} \mathrm{\: cm^3 \; s^{-1}}$ for the model calculations here. Although this is of course a highly simplistic treatment of recombination, it does appear to represent quite well the average value when both heavier and lighter elements are considered. The choice of a temperature-independent value is further motivated by the fact that recombination coefficients do not systematically evolve the same way with temperature. For example, we find that for the lighter elements, the coefficients evolve in complex manners that are heavily shell dependent, whereas \citet{Hotokezaka.etal:21} find decreasing rates with increasing temperature for every Nd ion. It is possible that our fiducial choice under or overestimates recombination in certain temperature regimes, however we believe the overall behaviour to be a reasonable approximation.

\subsection{Collision Strengths and Radiative Processes}
\label{subsec:collstrenths}

Collisional excitations and de-excitations play key roles during the post-peak evolution of the KN. Since cooling by emission lines following collisional excitations is expected to dominate the nebular phase until adiabatic cooling becomes important, the thermal collision strengths have significant impact on the temperature evolution of the KN.

\begin{figure*}
\center
\includegraphics[trim={0.1cm 0.2cm 0.4cm 0.3cm},clip,width = 0.32\textwidth]{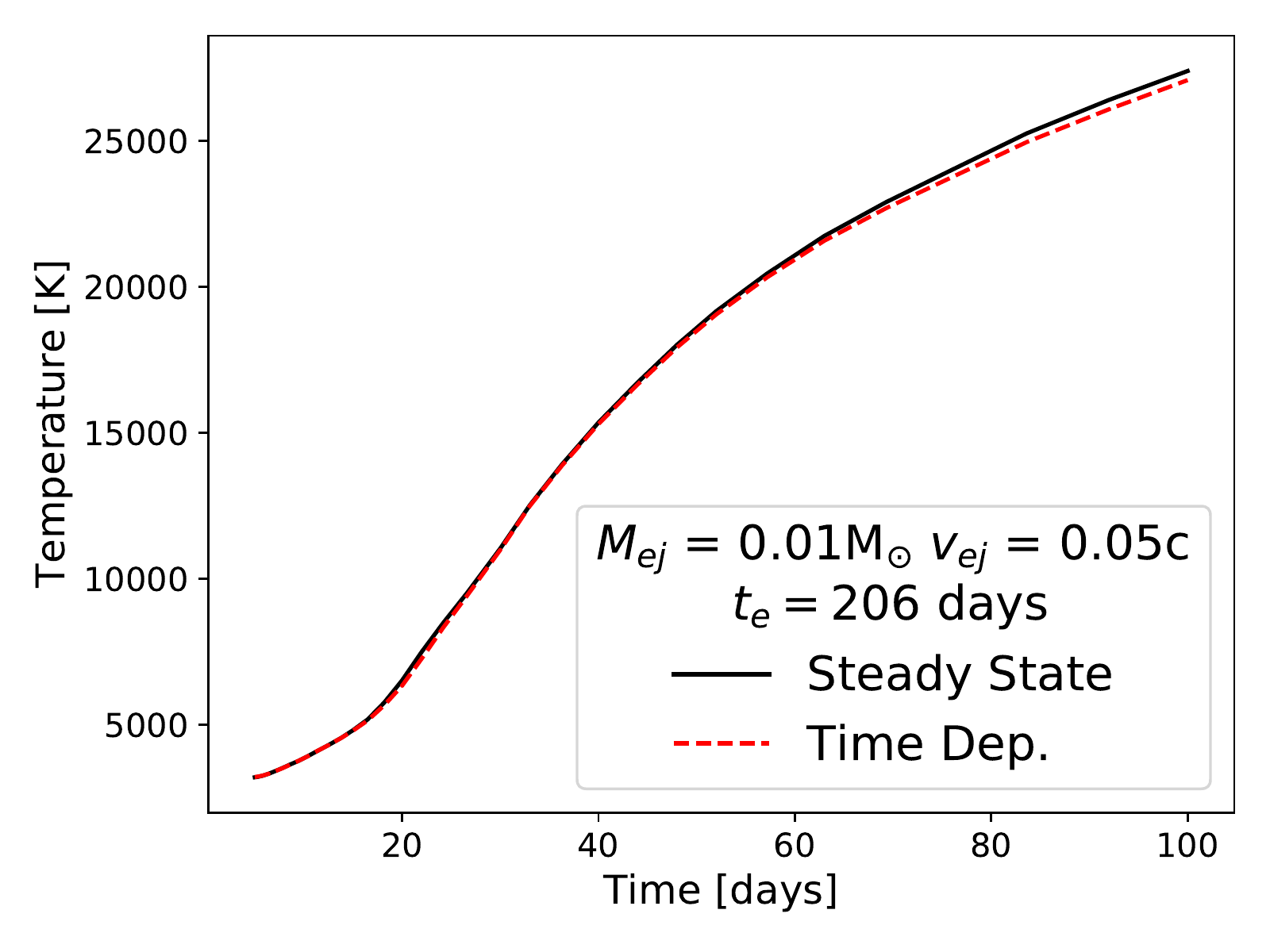}
\includegraphics[trim={0.1cm 0.2cm 0.4cm 0.3cm},clip,width = 0.32\textwidth]{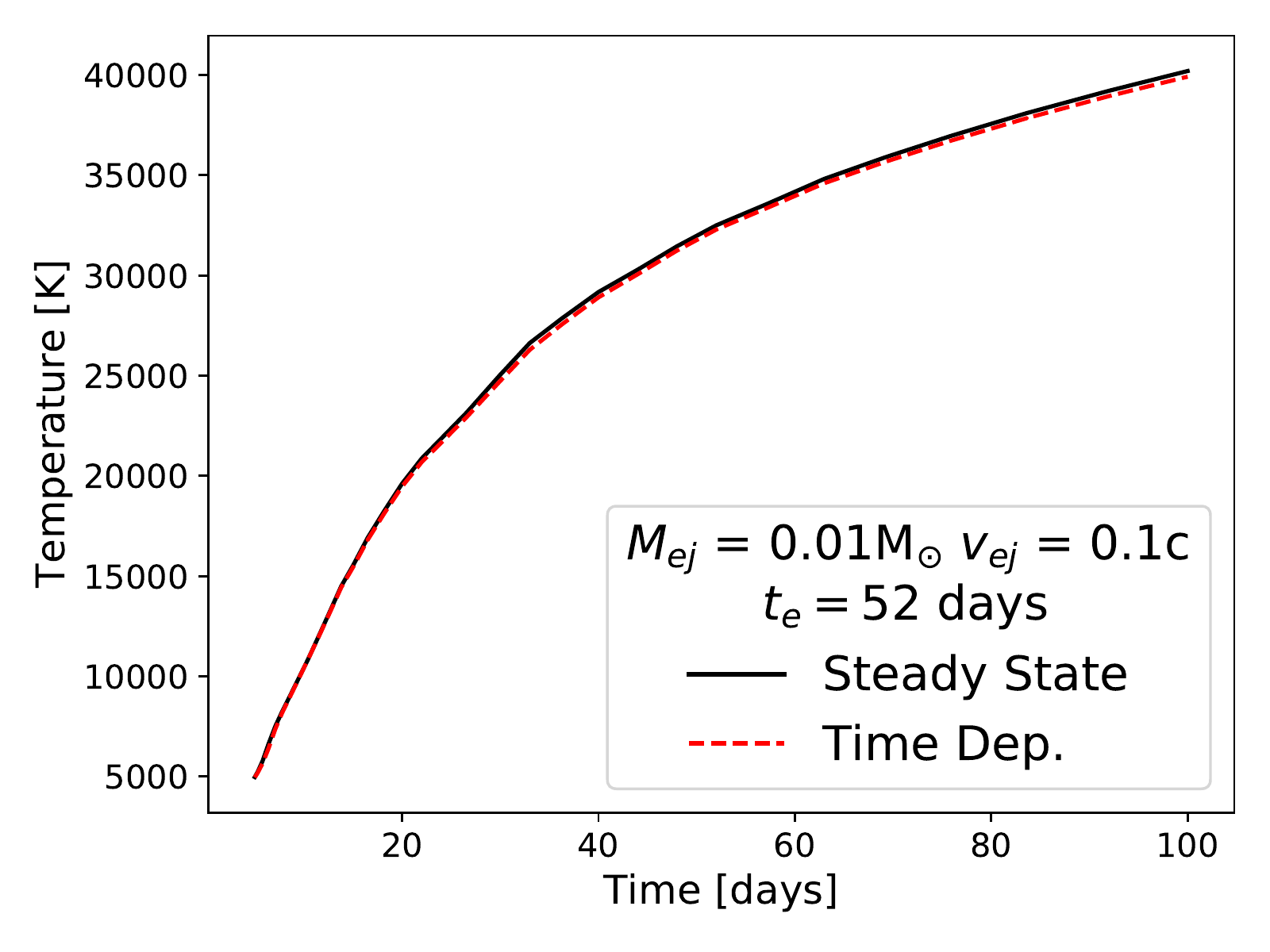}
\includegraphics[trim={0.1cm 0.2cm 0.4cm 0.3cm},clip,width = 0.32\textwidth]{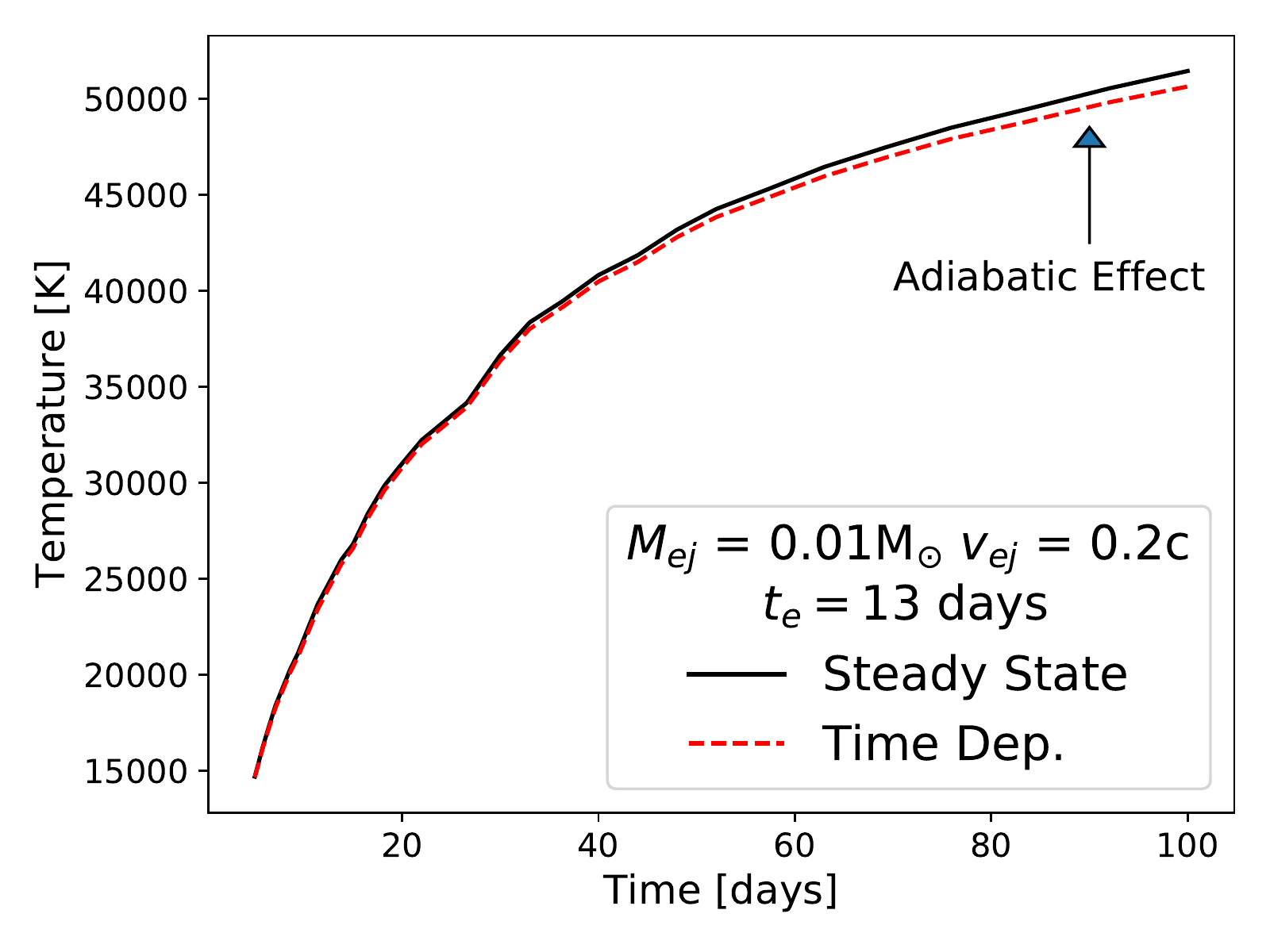}
\includegraphics[trim={0.1cm 0.2cm 0.4cm 0.3cm},clip,width = 0.32\textwidth]{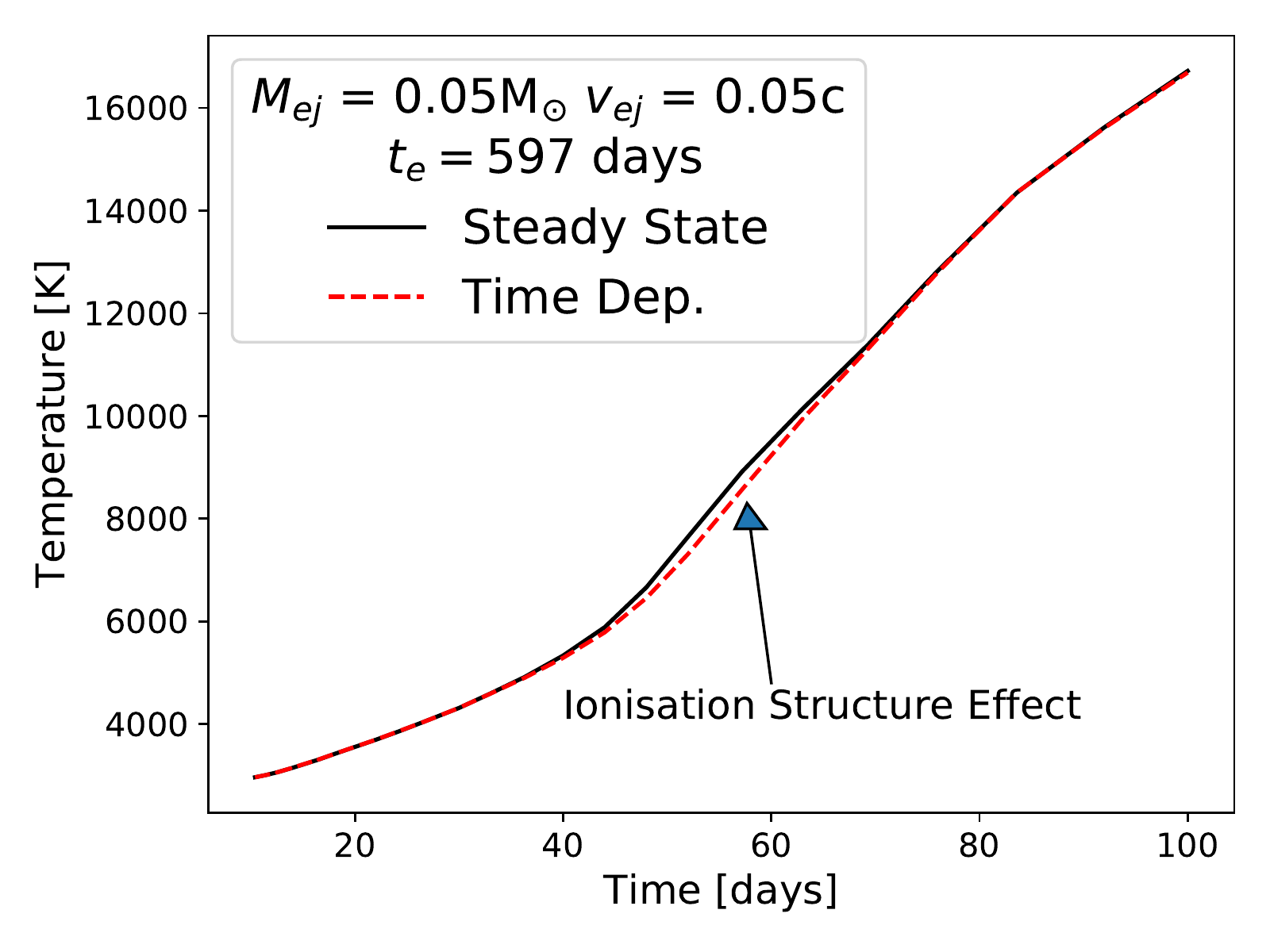}
\includegraphics[trim={0.1cm 0.2cm 0.4cm 0.3cm},clip,width = 0.32\textwidth]{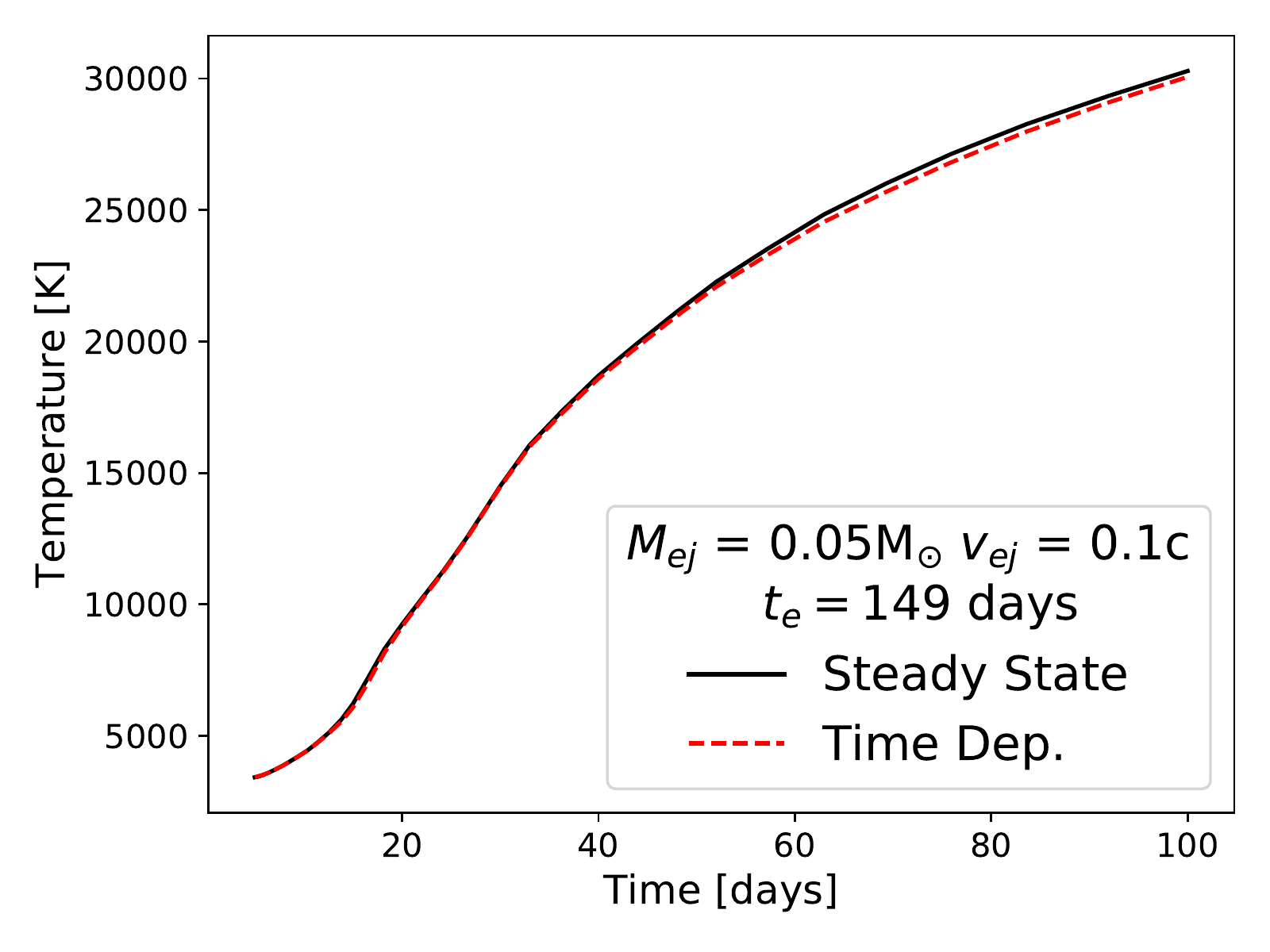}
\includegraphics[trim={0.1cm 0.2cm 0.4cm 0.3cm},clip,width = 0.32\textwidth]{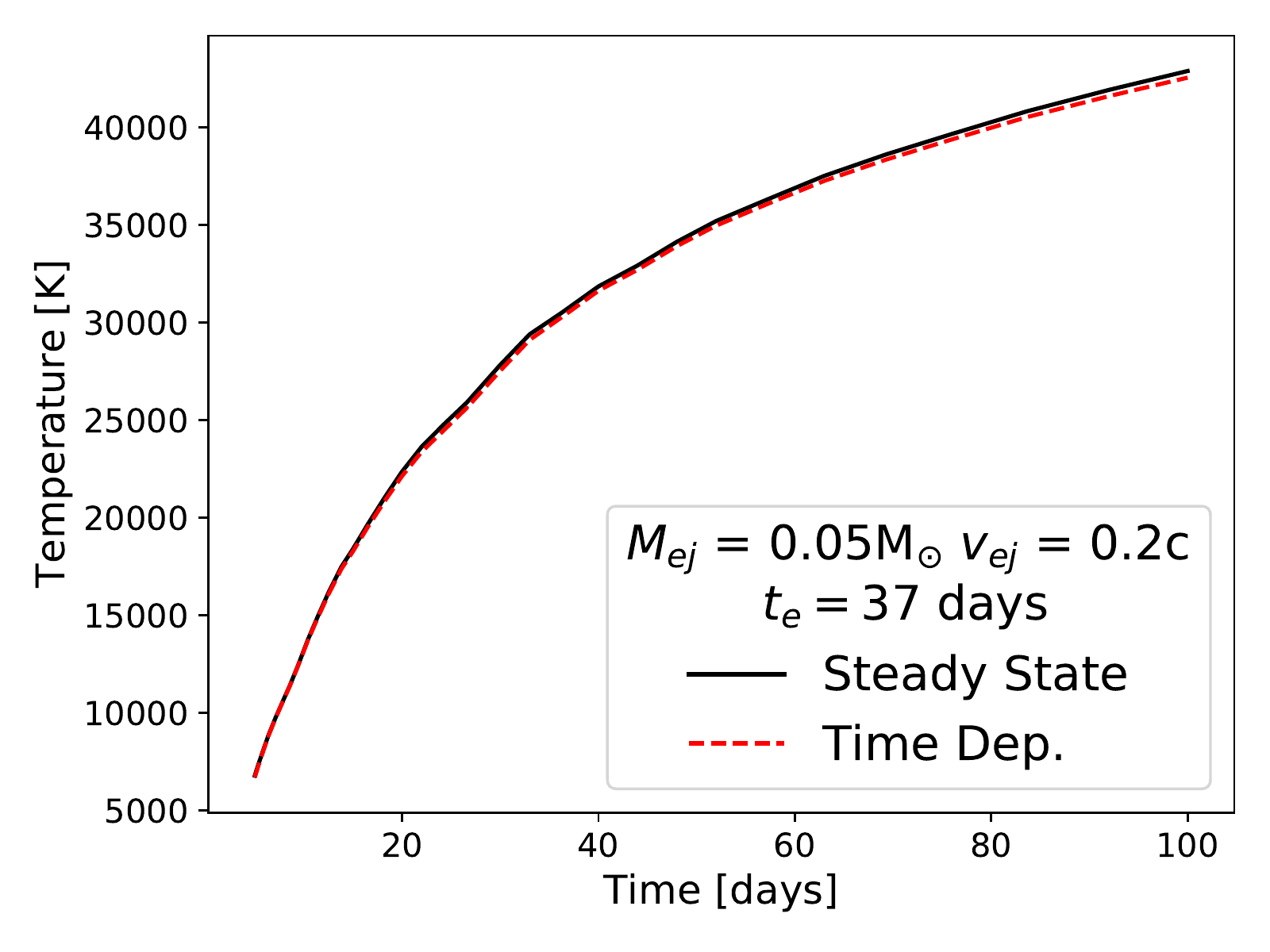}
\includegraphics[trim={0.1cm 0.2cm 0.4cm 0.3cm},clip,width = 0.32\textwidth]{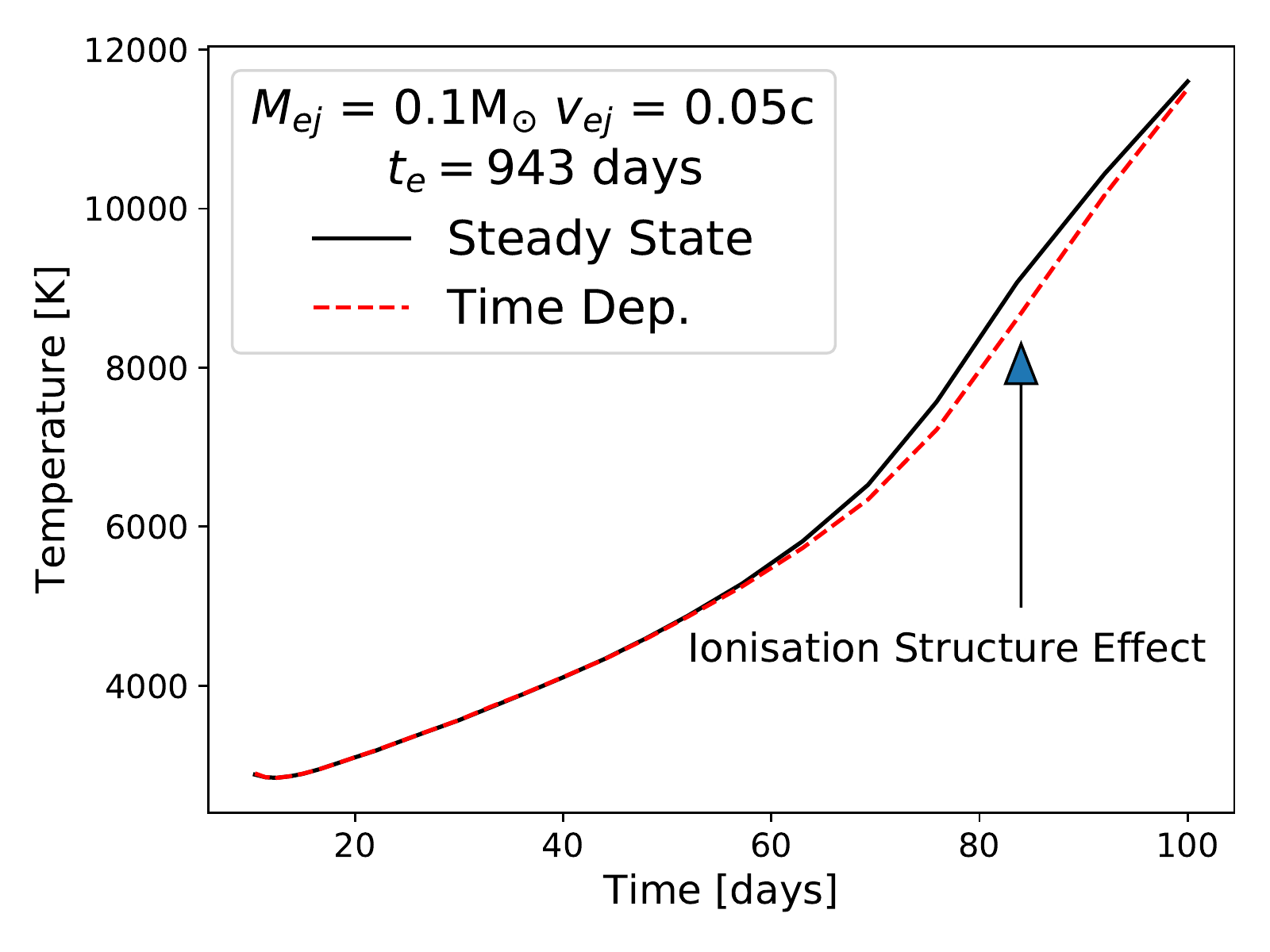}
\includegraphics[trim={0.1cm 0.2cm 0.4cm 0.3cm},clip,width = 0.32\textwidth]{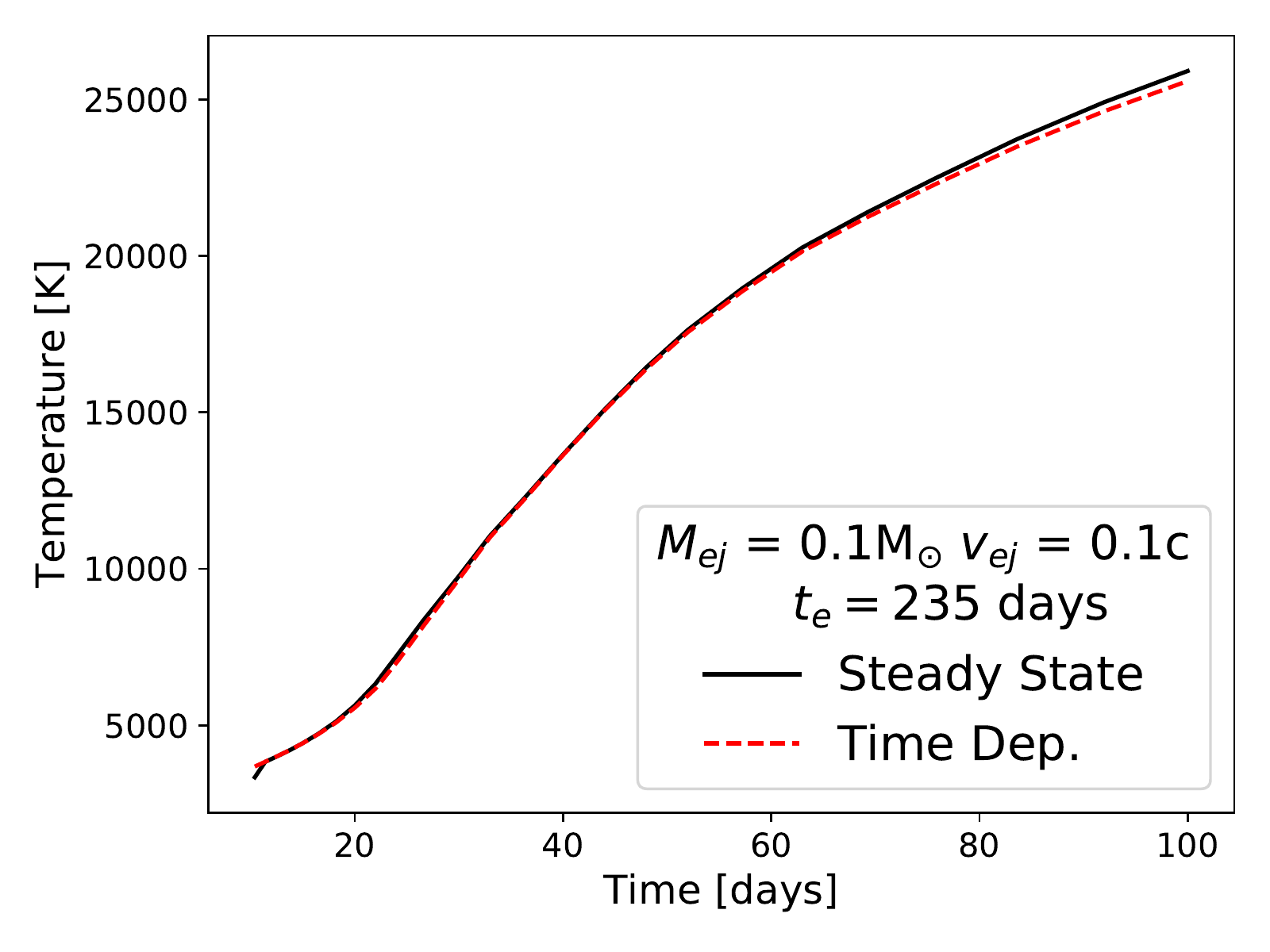}
\includegraphics[trim={0.1cm 0.2cm 0.4cm 0.3cm},clip,width = 0.32\textwidth]{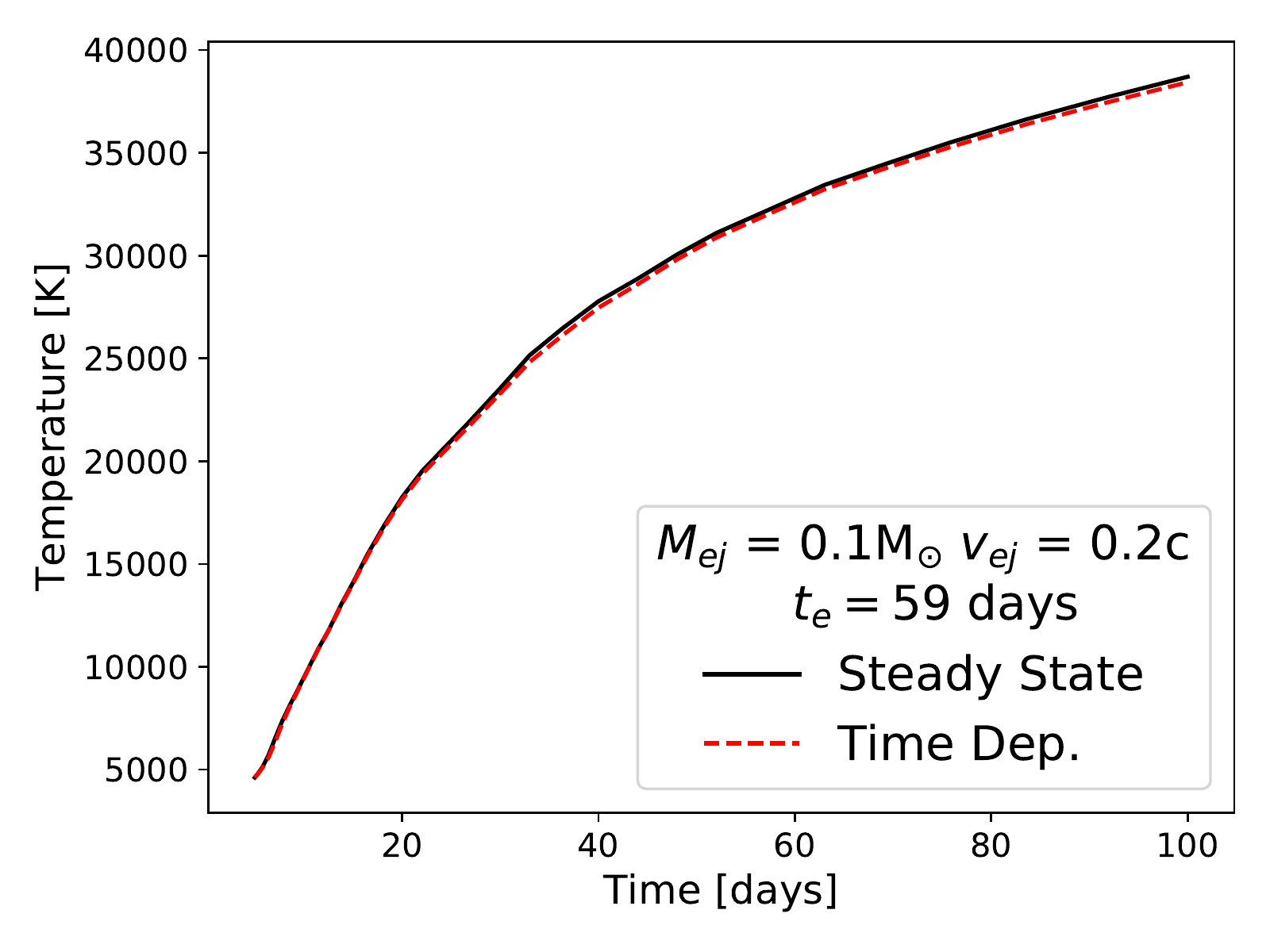}
\caption{Temperature evolution of models with 1 day radioactive decay power of  $\dot{Q_{\alpha}} + \dot{Q_{\beta}} = 10^{10}~\mathrm{erg~s^{-1} g^{-1}}$, with mass increasing downwards and velocity increasing to the right (i.e. the upper leftmost panel shows the lowest mass and velocity model), and non-thermal electron thermalisation time-scale written out (see Equation \ref{eq:t_e}). For the majority of models, time dependence makes almost no difference to the temperature evolution, though deviations are seen for three of the models. These are marked out by arrows indicating the nature of the effect behind the different temperature solution.}
\label{fig:temperature_grid}
\end{figure*}

The thermal collisional de-excitation rate from an upper level \textit{k'} to a lower level \textit{k} is given by \citet{Osterbrock:06}: 

\begin{equation}
    \xi_{k',k} = \frac{8.63\times 10^{-6} \; \Upsilon_{k,k'}(T_e)}{g_{k'} \; T_e^{1/2}} \: \mathrm{cm^3 \: s^{-1}}
    \label{eq:colldeexc}
\end{equation}

\noindent and is related to the thermal collisional excitation rate by $\xi_{k,k'} = \frac{g_{k'}}{g_k} \xi_{k',k} e^{-E_{k',k}/k_B T_e}$. Here, $T_e$ is the electron temperature, $g_k$ and $g_{k'}$ are the statistical weights of the levels, and $\Upsilon_{k',k}=\Upsilon_{k,k'}$ is the effective collision strength (dimensionless). Unless known for the specific transition, \texttt{SUMO} estimates $\Upsilon$ differently depending on whether the transition is forbidden or allowed. The distinction is made according to the oscillator strength of the transition (e.g. equation 2.68 of \citet{Rutten:03}),

\begin{equation}
    f_{\mathrm{osc}} = 1.49 \frac{g_{k'}}{g_k} \lambda^{2} A_{k',k} 
    \label{eq:oscstrength}
\end{equation}

\noindent where $A_{k',k}$ is the Einstein coefficient for spontaneous emission ($s^{-1}$), and $\lambda$ is the transition wavelength in cm. 
For allowed transitions, which we define as $ f_{osc} \geq 10^{-3}$, the effective collision strength is estimated using the formula derived by \citet{Regemorter:62}:

\begin{equation}
    \Upsilon_{k',k} = 2.39 \times 10^{6} P(y) \lambda^{3} A_{k',k} g_{k}
    \label{eq:omegaallowed}
\end{equation}

\noindent where $P(y)$ is the Gaunt factor integrated over the electron velocity distribution, typically in the range of $0.02 \sim 1.2$. Forbidden transitions with an oscillator strength less than $10^{-3}$ are instead calculated using the formula from \citet{Axelrod:80}:

\begin{equation}
    \Upsilon_{k',k} = 0.004 g_k g_k'
    \label{eq:omegaforb}
\end{equation}

\noindent which is the standard method used in many NLTE codes \citep[e.g.][]{Botyanski.Kasen:17,Shingles.etal:20}. This formula can yield values between $\sim 0.01 - 1$ depending on the transition. 

\texttt{SUMO} has the capability to calculate collisional excitation from non-thermal decay products, however the non-thermal excitation cross sections are not readily available in the literature for r-process elements. Since most non-thermal collisions are generally expected to produce either heating or ionisation  \citep{Kozma.Fransson:92,Jerkstrand:11}, this omission is not expected to lead to any significant errors for the physical gas state.
 
\begin{figure*}
\center
\includegraphics[trim={0.2cm 0.cm 1.6cm 0.8cm},clip,width = 0.49\textwidth]{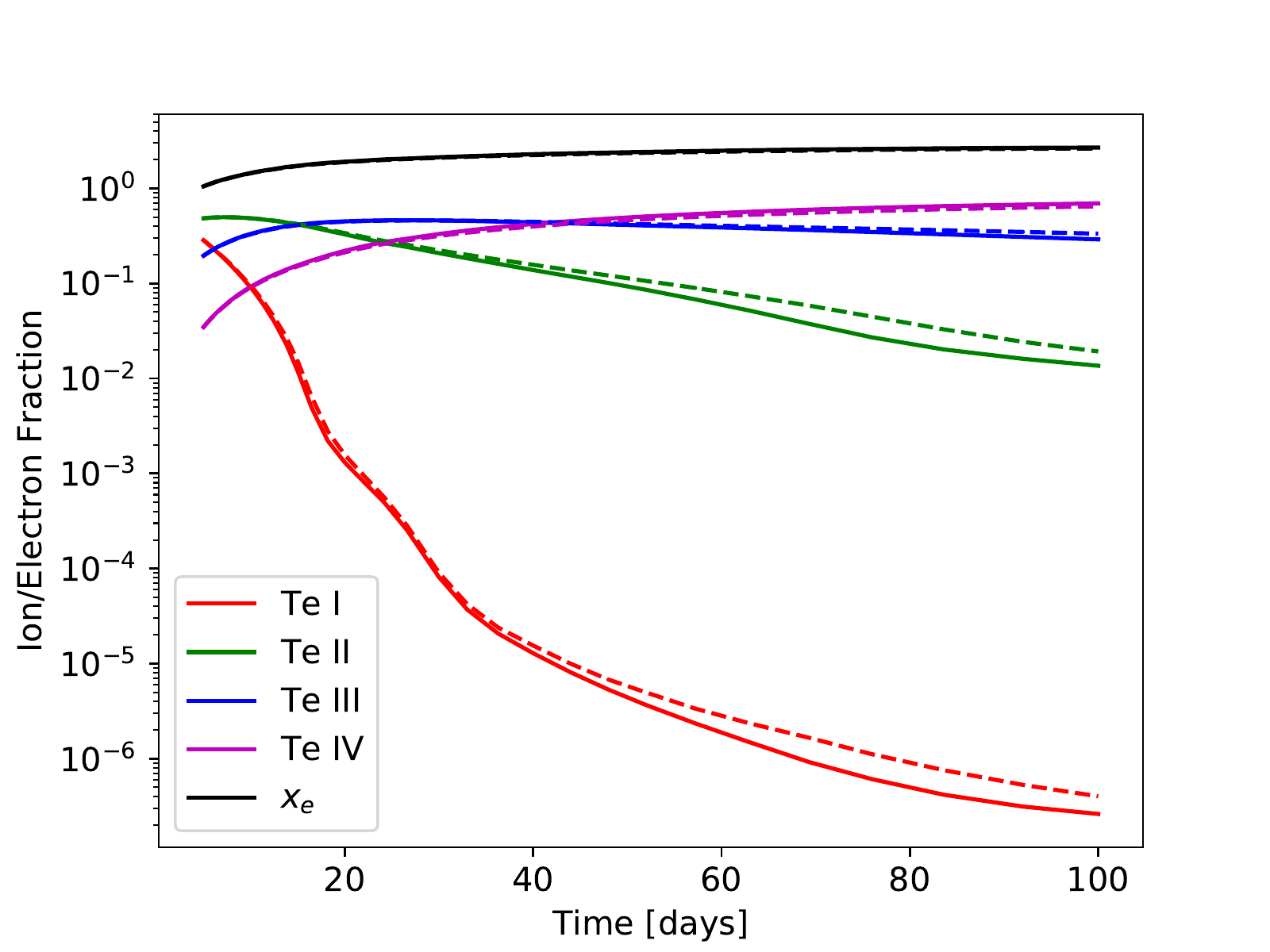}
\includegraphics[trim={0.2cm 0.cm 1.6cm 0.8cm},clip,width = 0.49\textwidth]{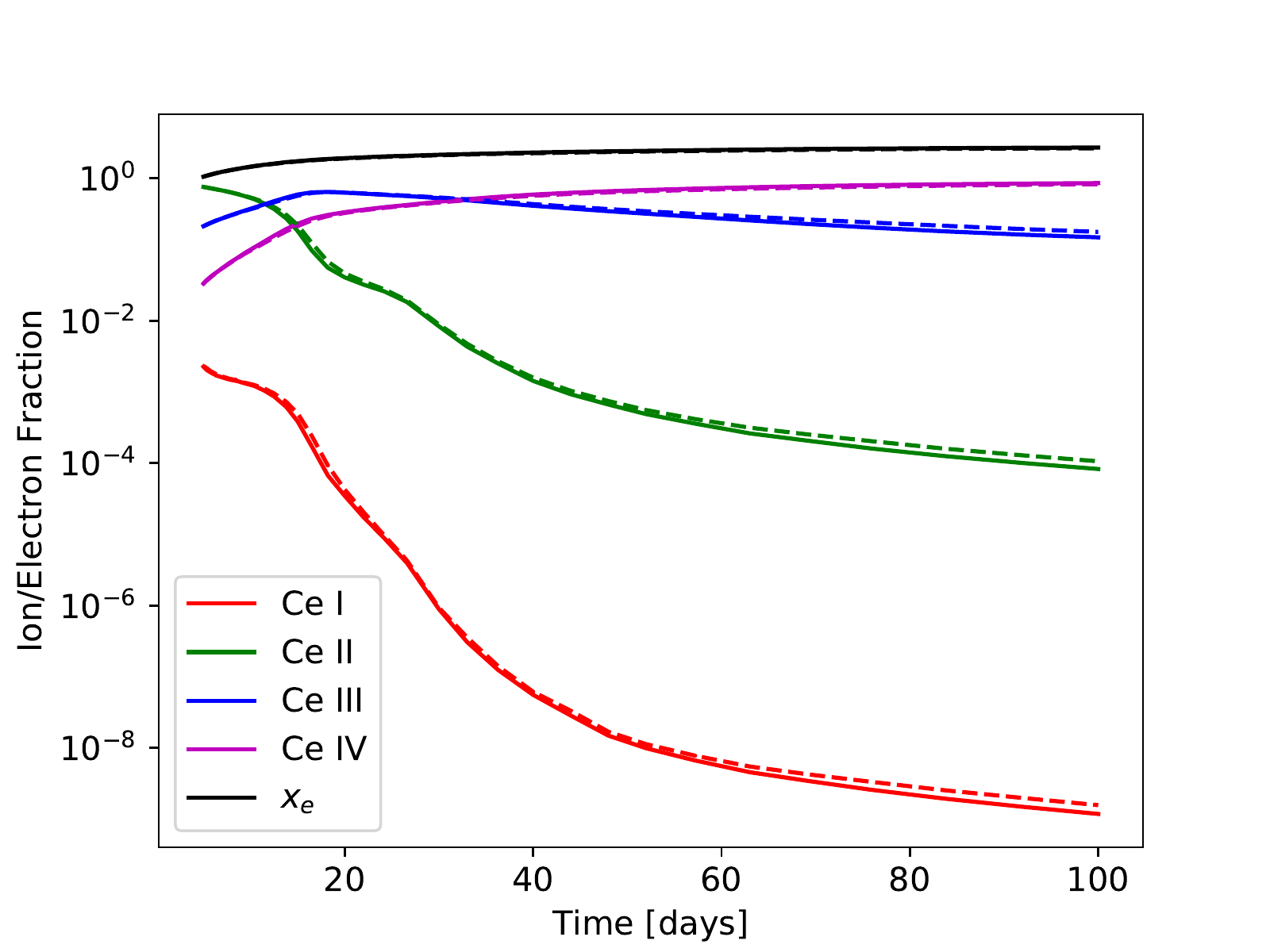}
\includegraphics[trim={0.2cm 0.cm 1.6cm 0.8cm},clip,width = 0.49\textwidth]{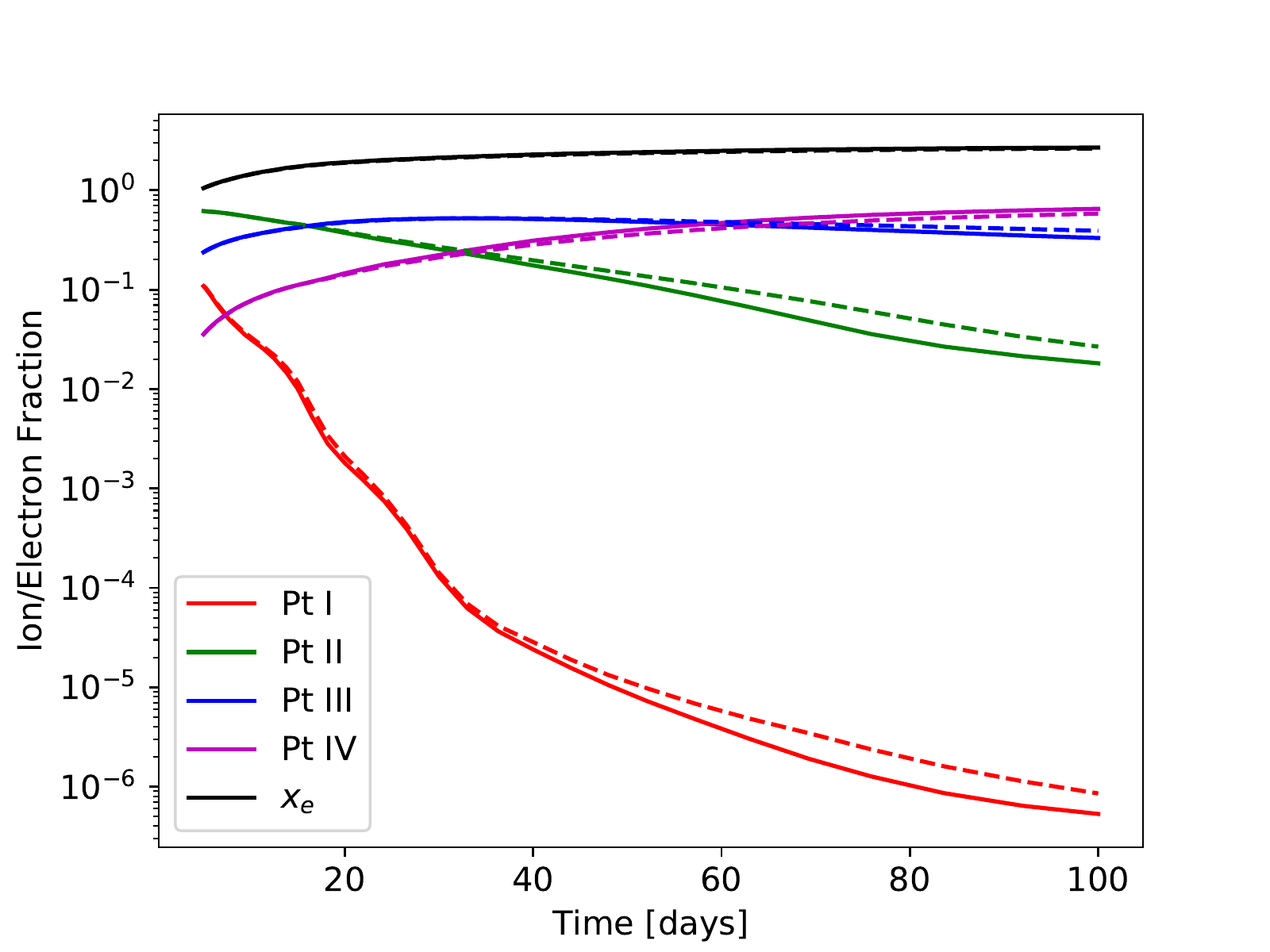}
\includegraphics[trim={0.2cm 0.cm 1.6cm 0.8cm},clip,width = 0.49\textwidth]{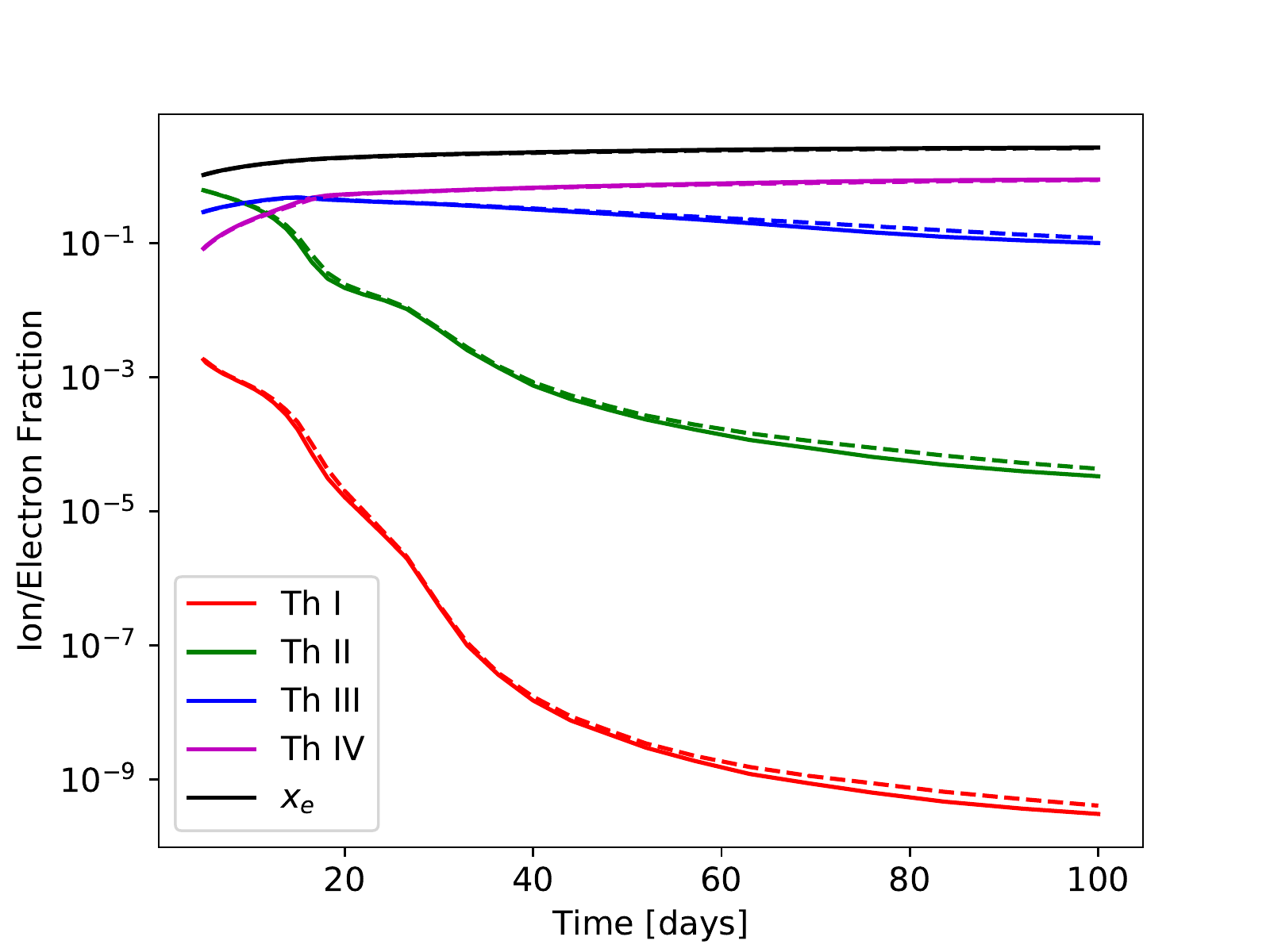}
\caption{Ionisation structure evolution for the model with $M_{\mathrm{ej}} = 0.05\mathrm{\Msol}$, $v_{\mathrm{ej}} = 0.1$c. The solid lines are the steady-state results, while the dashed lines are the time-dependent results.}
\label{fig:005M_01v_ionfrac}
\end{figure*}

Since \texttt{SUMO} is primarily a nebular phase code, line emission from both allowed and forbidden lines is important. Radiative transition rates are included as spontaneous and stimulated emission as well as radiative absorption. These are treated under the Sobolev approximation, appropriate for modelling of nebular phase SNe and KNe where the velocity gradients are high.

\subsection{Methodology for time-dependent calculations}

The nebular phase differential equations for the time evolution of temperature and ionisation, Equations \ref{eq:temperature_equation} and \ref{eq:ion_balance},  are discretised such that they can be solved by implicit time differentiation. This, as well as the steady-state equations for excitation, results in a set of non-linear algebraic equations that are solved using standard Newton-Raphson methods \citep[see e.g.][]{Axelrod:80,Press.etal:92}. This method is sensitive to the accuracy of the initial guess, which we take to be either the LTE solution at that epoch, or the NLTE solution at the previous epoch. In time-dependent mode, these equations require knowledge of the solutions at the previous epoch, and so the previous NLTE solution is included in the equations to be solved, and not only as a starting guess. 

Having too large time-steps may impact the accuracy of the solution, and thus must be chosen carefully. A time-step given by some fraction $\Delta t$ of the current epoch is implemented. We have tested the speed and accuracy of convergence with time-steps of $\Delta t = 0.1, 0.2 \: t_{\mathrm{d}}$, where $t_d$ is the current epoch in days, finding that a 10 per cent time-step provides a good balance of accurate solutions without excessively long runtimes. Appendix \ref{app:time-step} shows that the solutions with 20 per cent time-step reproduce the same results as the 10 per cent time-step, thus implying the smaller time-step to be reliably accurate. 

\section{Temperature and Ionisation Structure Evolution}
\label{sec:results}

Using \texttt{SUMO}, we use a simple single-zone KN ejecta model to explore the evolution of the ejecta's gas state. The parameters of the ejecta in the model grid are presented in Section \ref{sec:atomic_data}. For each model, the ejecta is initially evolved under the assumption of steady-state, then once again using the full time-dependent form of Equations \ref{eq:temperature_equation} and \ref{eq:ion_balance}. We allow the ejecta to evolve from 5 days after merge up to 100 days. While the upper limit was justified previously in Section \ref{sec:intro} as a good combination of late time IR observations and potential importance of time-dependent effects, we choose a lower limit of 5 days because this is early enough for steady-state conditions to hold throughout parameter space (see Equations \ref{eq:adia_to_line} and \ref{eq:trec_crit}), while late enough for the diffusion phase to be over such that time-dependent radiation transport is not needed. 

\subsection{Temperature Evolution}
\label{subsec:standard_temperature}

The temperature evolution in the main model grid, with initial 1 day radioactive decay power of $\dot{Q}_{\alpha} + \dot{Q_{\beta}}(t = 1 ~ \mathrm{day}) =  10^{10} \; \mathrm{erg \; s^{-1} \; g^{-1}}$ (see Equation \ref{eq:tot_dep}), is shown in Figure \ref{fig:temperature_grid}. The densest models with $M_{\mathrm{ej}} = 0.1~\mathrm{\Msol}$ and $v_{\mathrm{ej}} = 0.05, 0.1 \mathrm{c}$ are only plotted from 10 days onwards, as convergence issues were encountered prior to this epoch. We see here that time-dependent temperature effects are minor for the majority of this parameter space, and thus the temperature evolution follows the steady-state solutions to good approximation. At all times $t_d > 10$ days, the temperature is increasing, in accordance with the indications from Equations \ref{eq:lambda_evolution_early} and \ref{eq:lambda_evolution}. The flattening of the temperature at late times is driven by the on-set of thermalisation inefficiency, and appears earlier for models with smaller electron thermalisation break times ($t_e$ in Equations \ref{eq:beta_therm} and \ref{eq:alpha_therm}). This time-scale is longer for denser models, and so the lowest density model in the top right panel experiences this effect the earliest due to a shorter efficient thermalisation time.

\begin{figure*}
\center
\includegraphics[trim={0.2cm 0.cm 1.6cm 0.8cm},clip,width = 0.49\textwidth]{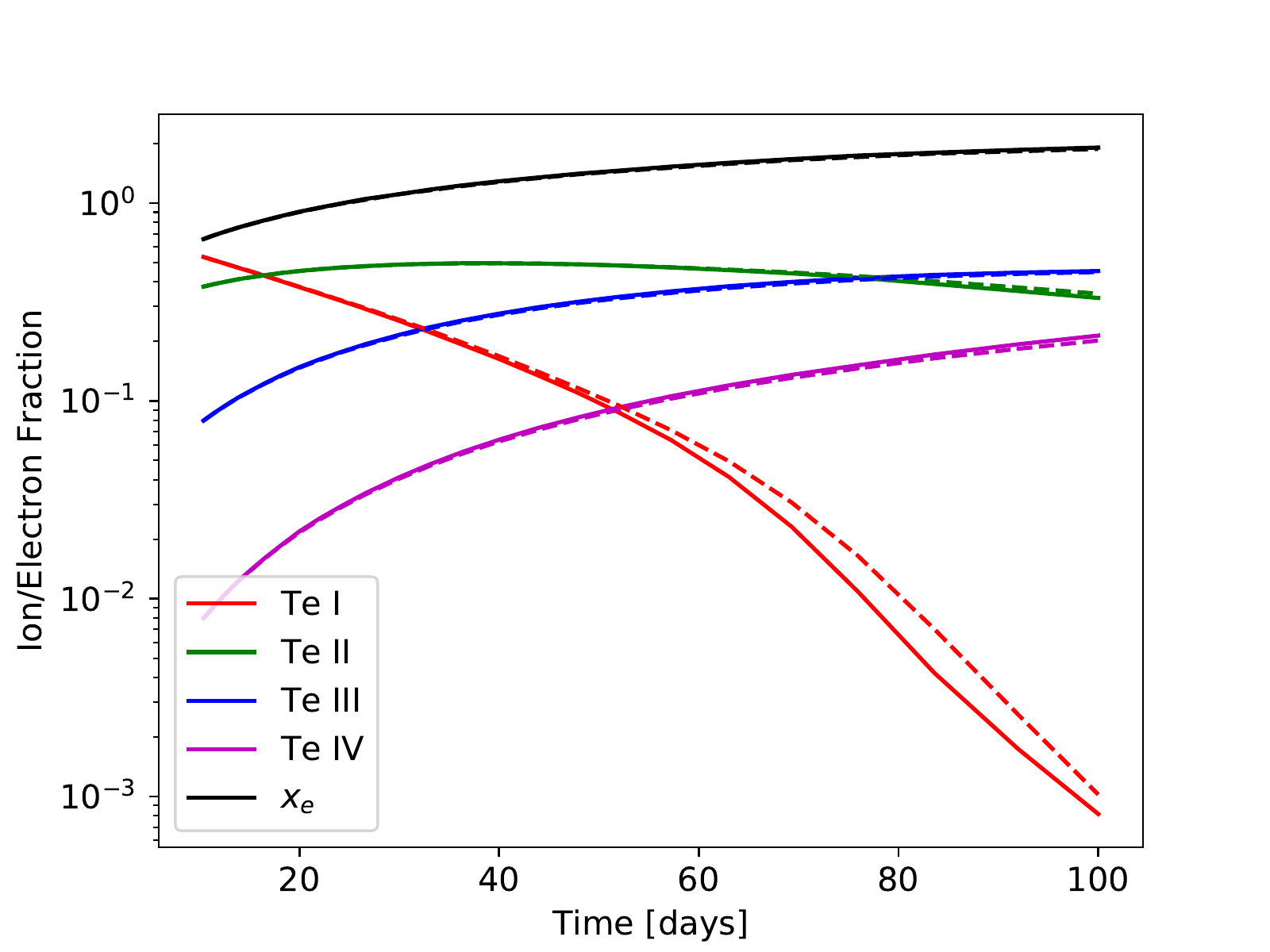}
\includegraphics[trim={0.2cm 0.cm 1.6cm 0.8cm},clip,width = 0.49\textwidth]{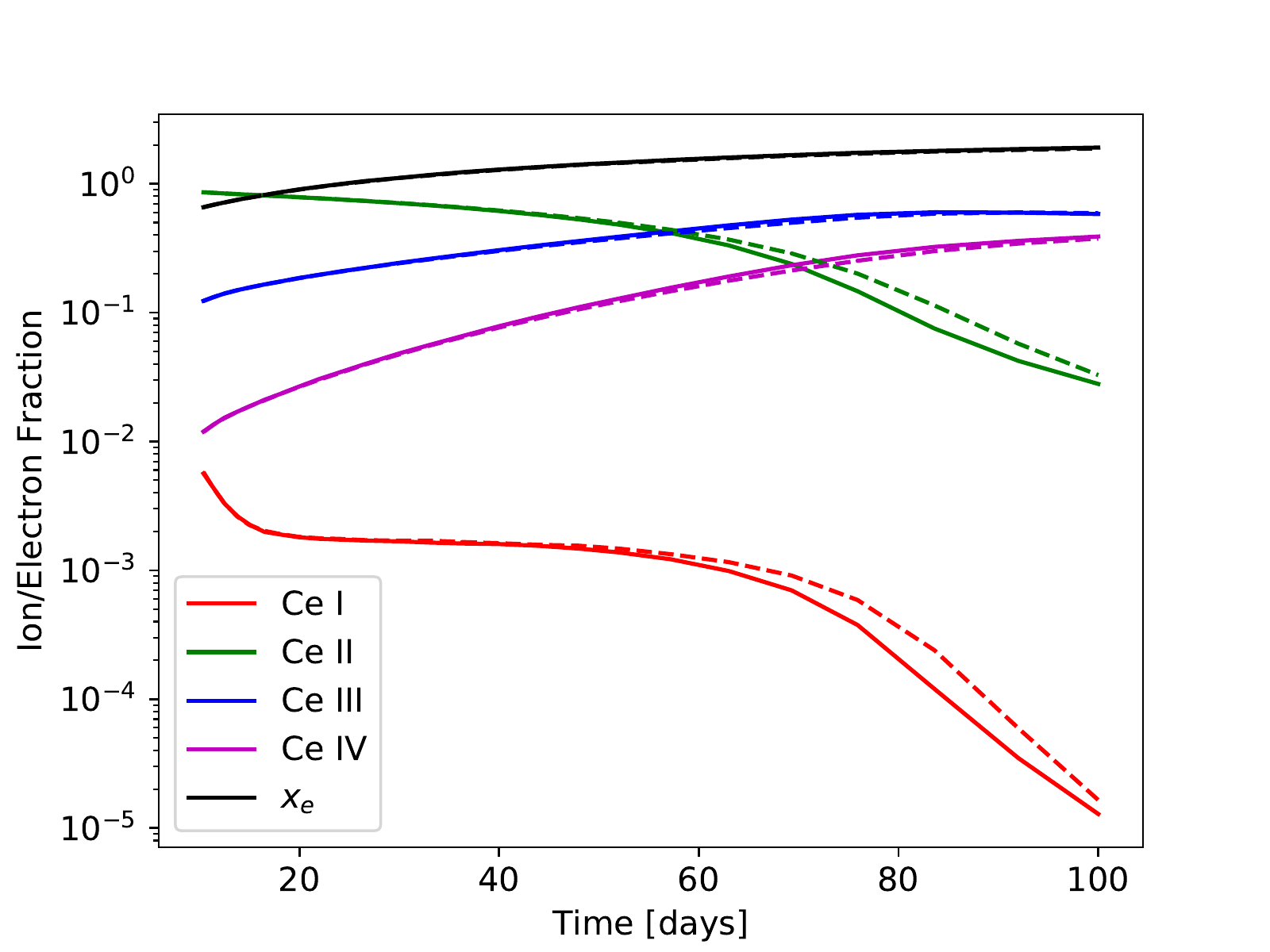}
\includegraphics[trim={0.2cm 0.cm 1.6cm 0.8cm},clip,width = 0.49\textwidth]{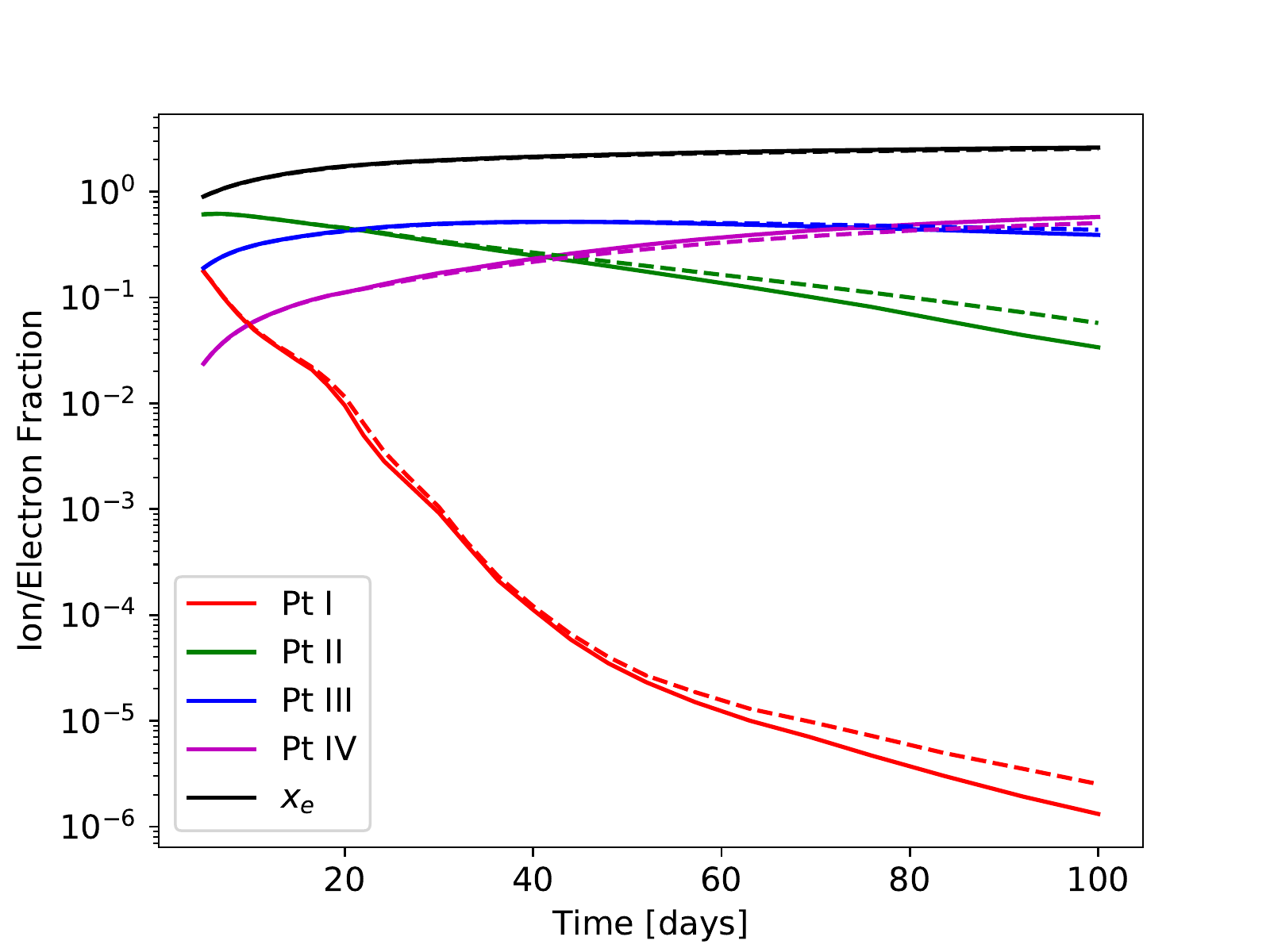}
\includegraphics[trim={0.2cm 0.cm 1.6cm 0.8cm},clip,width = 0.49\textwidth]{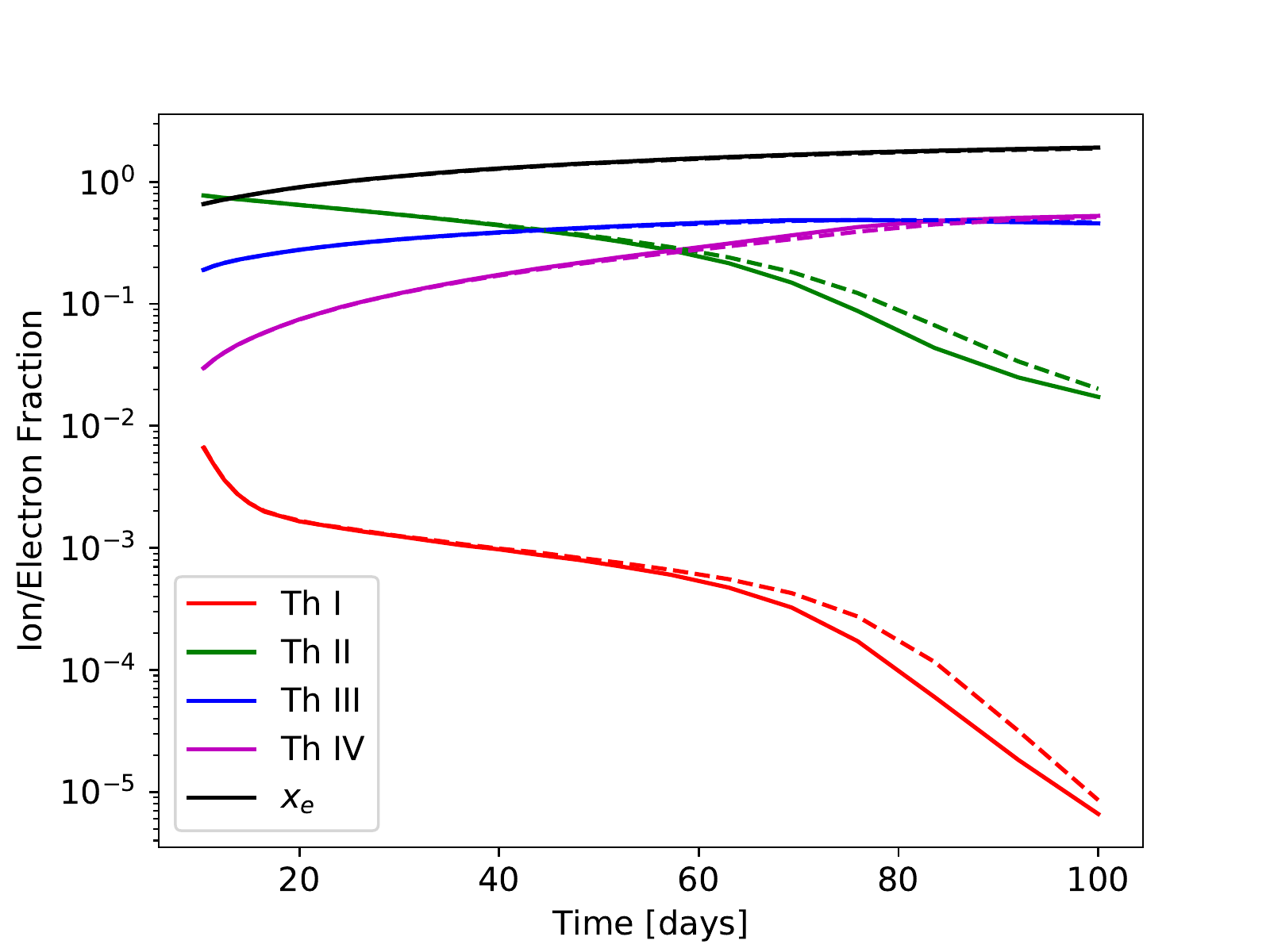}
\caption{Ionisation structure evolution for the highest-density model with $M_{\mathrm{ej}} = 0.1\mathrm{\Msol}$, $v_{\mathrm{ej}} = 0.05$c. The solid lines are the steady-state results, while the dashed lines are the time-dependent results.}
\label{fig:01M_005v_ionfrac}
\end{figure*}

Since the time-dependent models mostly follow the steady-state solutions, adiabatic cooling mostly has little effect, as one could anticipate if r-process atomic line cooling functions are $\lesssim 10^{-20}$ erg cm$^3$ s$^{-1}$ (see the discussion in Sec. \ref{subsec:temperature}). The adiabatic to line cooling ratio reaches a maximum value of 8 per cent at 100 days in the model grid, for the lowest density model, $M_{\mathrm{ej}}=0.01~\mathrm{\Msol}$, $v_{\mathrm{ej}} = 0.2$c, and a slight effect can be seen in this model, as shown in the top right panel of Figure \ref{fig:temperature_grid}. 

We also see a slight effect earlier than 100 days in the lowest velocity models, with this being most visible for the densest model $M_{\mathrm{ej}}=0.1~\mathrm{\Msol}$, $v_{\mathrm{ej}} = 0.05$c in the range of $\sim 60 - 100 $ days, while the model with $M_{\mathrm{ej}}=0.05~ \mathrm{\Msol}$, $v_{\mathrm{ej}} = 0.05$c shows a similar, but less significant effect in the range $\sim 40 - 70 $ days. Again here, the time-dependent temperature solution is slightly cooler than the steady-state solution. However, adiabatic cooling is not responsible for this deviation, having a contribution of $< 0.01$ per cent at 80 days for the densest model. Similarly, the 'ionisation cooling' term in Equation \ref{eq:temperature_equation},  also has a contribution of $< 0.01$ per cent at this time, and as mentioned in Section \ref{subsec:temperature}, is generally expected to be small at all times (see Figure \ref{fig:ioncool} in Appendix \ref{app:results}). Instead, this temperature difference arises from a different ionisation solution, the details of which are investigated in the next section.

\subsection{Ionisation Structure Evolution}
\label{subsec:standard_ionisation}

The ionisation evolution of the $M_{\mathrm{ej}} = 0.05~\mathrm{\Msol}$, $v_{\mathrm{ej}} = 0.1$c model, taken here to be our fiducial model with average ejecta parameters, is shown in Figure \ref{fig:005M_01v_ionfrac}, and the same plots for the rest of the models are shown in Appendix \ref{app:results}. The time-dependent ionisation structure appears to generally be less ionised than the steady-state solution, with slightly higher fractions of neutral, singly and doubly ionised states, at the expense of the triply ionised ion. For every element, in both steady-state and time-dependent modes, the neutral fraction drops rapidly after the initial 5 day solution, and already starts below 1 per cent for Ce and Th, which both have relatively low ionisation potentials of 5.5~eV and 6.3~eV respectively. The singly ionised ions follow a similar, albeit shallower evolution, while the doubly ionised ions initially increase, before starting to decrease in favour of the triply ionised state. It is interesting to note how little neutral atoms, as well as Ce II and Th II, remain later on, being reduced to essentially trace amounts from 40 days onwards. As such, while there does generally appear to be more neutral ions and singly ionised Ce II and Th II at late times, this usually remains a negligible quantity. 

The ionisation structure for the densest model with $M_{\mathrm{ej}} = 0.1~\mathrm{\Msol}$, $v_{\mathrm{ej}} = 0.05$c is shown in Figure \ref{fig:01M_005v_ionfrac}. The general evolution is similar to that of the fiducial model seen in Figure \ref{fig:005M_01v_ionfrac}, though it should be noted that the neutral and singly ionised ion abundances are much higher here, with neutral Te and Pt remaining above 1 per cent up to 80 days. The time-dependent solutions here are also less ionised, with the changes to those ion fractions being significant, e.g. at 75 days, the Ce II fraction goes from $\sim 14$ per cent in steady-state to 20 per cent in time-dependent mode. This difference in ionisation structure peaks around 75 days, and the solutions appear to converge again around 100 days, though the time-dependent solution does remain slightly less ionised ($x_e = 1.91$ in steady-state, and $x_e = 1.88$ in time-dependent mode). Checking the recombination and ionisation timescales (see Section \ref{subsec:ionandrec}) at 80 days where the ion structure deviation is maximised, we find that $t_{\mathrm{rec}}/t_d \approx t_{\mathrm{ion}}/t_d = 0.05$, showing that even ratios of $< 0.1$ may produce minor effects.

Looking again at the bottom left panel of Figure \ref{fig:temperature_grid}, which corresponds to this densest model, we see that the time-dependent temperature solution is cooler than the steady-state solution over the same timespan as the ionisation structure solution deviation, implying a direct connection. Since line cooling entirely dominates at this epoch, it is probable that the different ions' line cooling capacities are responsible for the temperature difference arising from a different ionisation structure solution. In this case, it is not only the relative difference of ion fraction between the steady-state and time-dependent solutions that matters, but also the overall fraction of that ion relative to the total species' abundance. For example, changing the ion fraction of Ce II in Figure \ref{fig:01M_005v_ionfrac} on the order of $~10^{-1}$ will be more significant than that of Figure \ref{fig:005M_01v_ionfrac} on the order of $10^{-3}$. This is explored in detail in Section \ref{subsec:suppressed_linecool}, with the analysis of line cooling functions appearing to support this hypothesis.

\subsection{Reduced Energy Deposition Models}

The results presented in Sections \ref{subsec:standard_temperature} and \ref{subsec:standard_ionisation} show that time-dependence appears to be relatively minor over the range of model parameters for a 'standard' energy deposition. However, as mentioned in Section \ref{subsec:energy_dep}, the exact form and evolution of the energy generation rate and thermalisation efficiency are both uncertain. Notably, there is a compositional dependency for the initial day 1 radioactive decay power, with values varying around the canonical $10^{10} \: \mathrm{erg \; s^{-1} \; g^{-1}}$ (see e.g. \citet{Barnes.etal:21,Zhu.etal:21}). The thermalisation efficiencies $f(t)$ given in Equations \ref{eq:beta_therm}, and \ref{eq:alpha_therm} are also uncertain, notably with regards to geometry dependent time-scales, as well as the degree of electron trapping by magnetic fields \citep{Kasen.Barnes:19,Waxman.etal:19,Hotokezaka.etal:20}. As such, the amount of radioactive energy deposited to the ejecta at any given time is uncertain both with regards to the initial energy generation as well as the time-dependent thermalisation process. It is therefore possible that lower thermalisation and/or energy generation rate may lead to an overall less energetic KN.

\begin{figure*}
\center
\includegraphics[trim={0.0cm 0.2cm 0.4cm 0.2cm},clip,width = 0.48\linewidth]{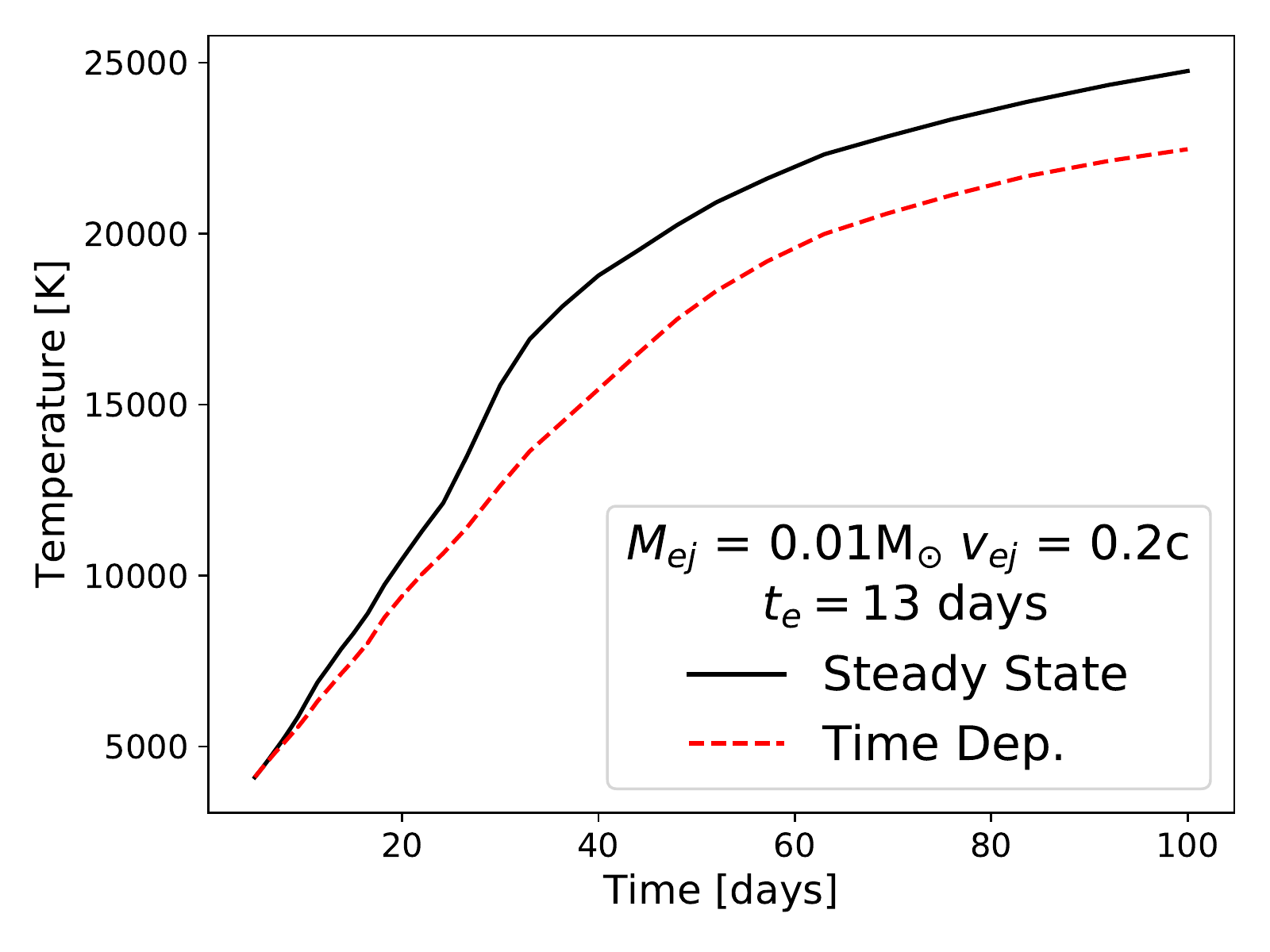}
\includegraphics[trim={0.0cm 0.2cm 0.4cm 0.2cm},clip,width = 0.48\linewidth]{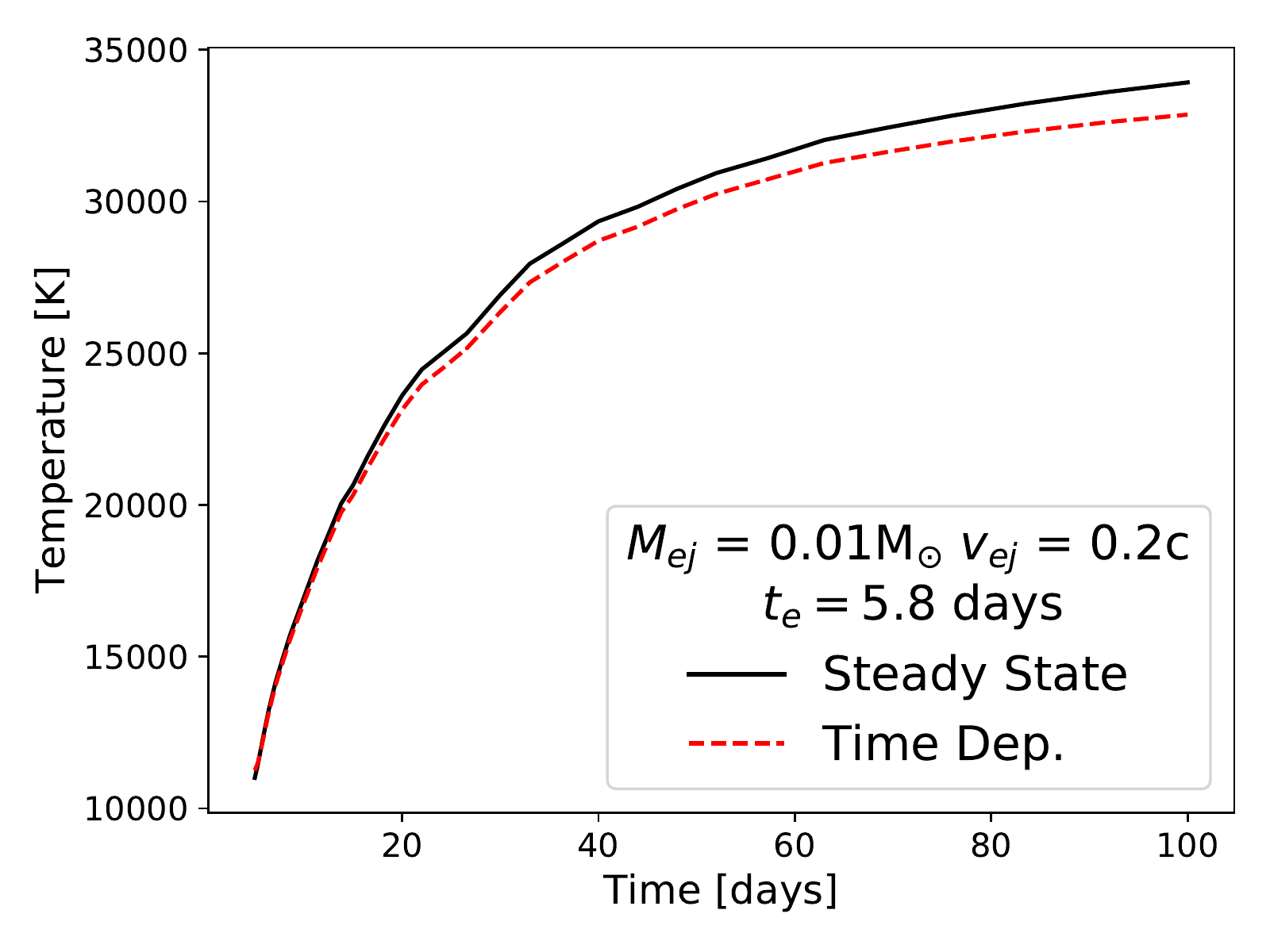}
\caption{Left panel: Temperature evolution of the suppressed $M_{\mathrm{ej}} = 0.01~\mathrm{\Msol}$, $v_{\mathrm{ej}} = 0.2$c model with day 1 radioactive decay power set to $\dot{Q_{\alpha}} + \dot{Q_{\beta}} = 10^9~\mathrm{erg \: s^{-1} \: g^{-1}}$. Right panel: Temperature evolution of the steep thermalisation $M_{\mathrm{ej}} = 0.01~\mathrm{\Msol}$, $v_{\mathrm{ej}} = 0.2$c model with steeper $\beta$ decay electron thermalisation. }
\label{fig:reduced_temperature}
\end{figure*}

Following this line of reasoning, we study two reduced energy deposition models, aimed at representing the uncertainty in energy generation rate from nuclear network calculations, and the possibility of a less efficient $\beta$ decay electron thermalisation from reduced magnetic field trapping. The first model has a reduced initial day 1 radioactive decay power by an order of magnitude to $\dot{Q_{\alpha}} + \dot{Q_{\beta}} = 10^{9} \; \mathrm{erg \: s^{-1} \: g^{-1}}$, which we call the 'suppressed' energy deposition model. The second model has a non-thermal electron thermalisation decreasing as $f_{\beta,e} \propto t^{-2.0}$, with the thermalisation time-scale set to $t'_{e} = t_{e} \sqrt{\frac{v_{\mathrm{ej}}}{c}}$, where $t_{e}$ is the original thermalisation time-scale shown in Equation \ref{eq:t_e}. This corresponds to electron free streaming as opposed to trapping by magnetic fields \citep{Waxman.etal:18,Waxman.etal:19} (see Appendix \ref{app:electron_freestream} for a full derivation), and we call this model the 'steep thermalisation' model. We test these reduced energy depositions on the lowest density model ($M_{\mathrm{ej}} = 0.01~\mathrm{\Msol}$, $v_{\mathrm{ej}} = 0.2$c), and the highest density model ($M_{\mathrm{ej}} = 0.1~\mathrm{\Msol}$, $v_{\mathrm{ej}} = 0.05$c). The densest model does not show particularly strong time-dependent effects and those results are included at the end of Appendix \ref{app:results}. The low density model on the other hand, shows significant time-dependent effects which are discussed further below.

\subsubsection{Temperature Evolution}
\label{subsec:suppressed_temperature}

The effect of time dependence is immediately visible for both reduced energy models (Figure \ref{fig:reduced_temperature}), with temperature deviation starting as early on as $\sim 10 $ days. In both cases, the time-dependent temperature is cooler, with the suppressed deposition model having a more significant deviation from steady-state. As seen in the left panel of Figure \ref{fig:reduced_temperature}, a temperature gap of $\sim 2500$ K appears around 30 days, and remains more or less constant onwards. The steady-state temperature appears to undergo a steeper increase from around 20 - 30 days, which is not seen in the time-dependent temperature. Both curves then follow a similar evolution, albeit along different tracks. 

\begin{figure}
    \center
    \includegraphics[trim={0.0cm 0.0cm 1.6cm 0.6cm},clip,width = \linewidth]{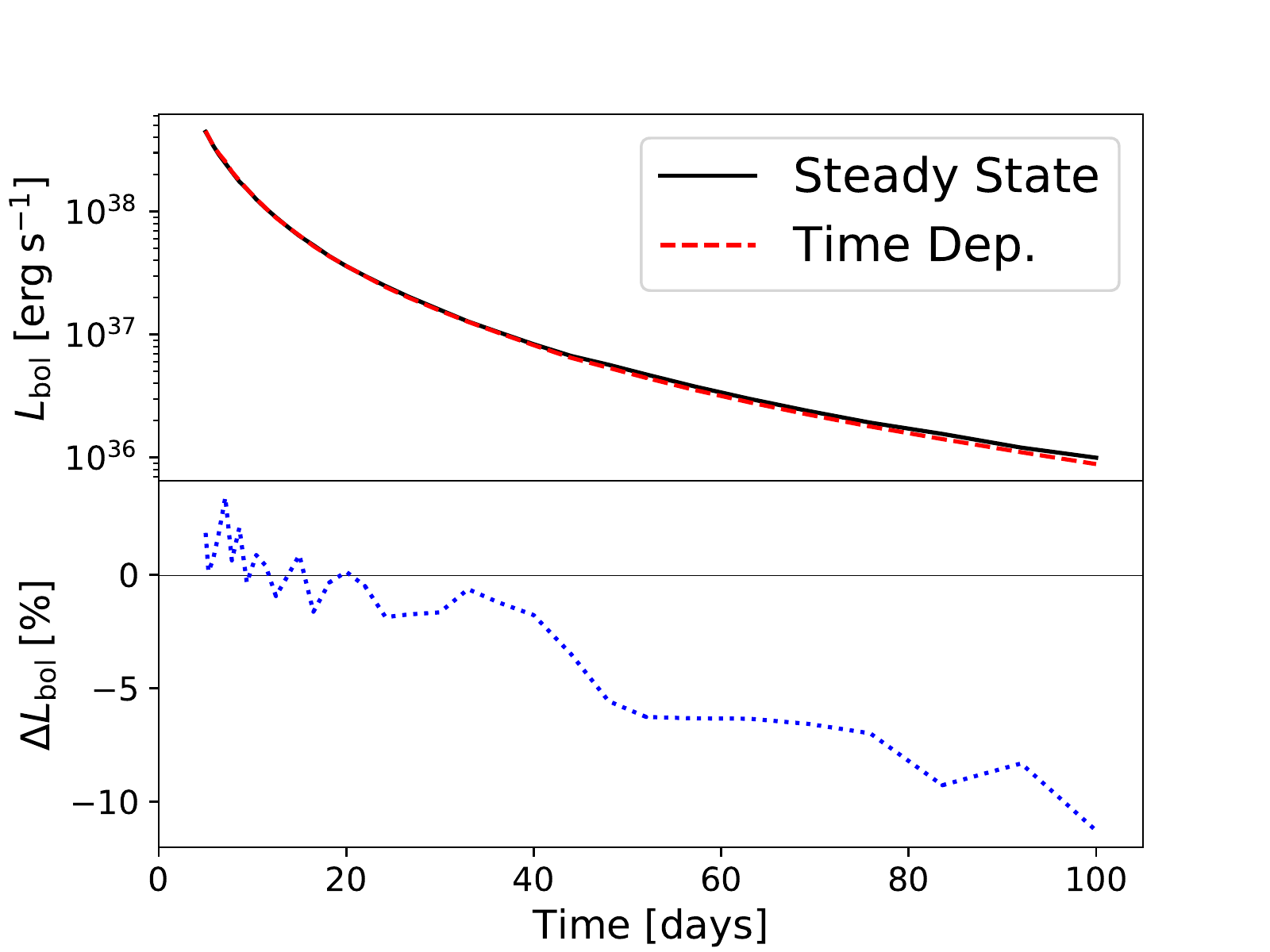}
    \caption{Bolometric luminosity of the suppressed energy deposition model with $M_{\mathrm{ej}} = 0.01~ \mathrm{\Msol}$, $v_{\mathrm{ej}} = 0.2$c (top panel) and the percentage difference given by $(L_\mathrm{bol}^{\mathrm{time}} - L_\mathrm{bol}^{\mathrm{steady}} )/L_\mathrm{bol}^{\mathrm{steady}}$ (bottom panel). The 1-2\% fluctuations are due to Monte Carlo noise.}   \label{fig:luminosity}
\end{figure}

The evolution of the steep thermalisation model, shown in the right panel of Figure \ref{fig:reduced_temperature}, is similar but with certain key differences. The time-dependent solution appears to follow the steady-state temperature more closely, with a slowly diverging solution as time goes on. The temperature in this model is slightly higher than the suppressed model, though broadly in a similar range. The late time behaviour appears to show the time-dependent solution flattening more so than the steady-state model, signalling the onset of the adiabatic dominated cooling regime, consistent with the reduced electron thermalisation time-scale for this model. 

The deviation of the time-dependent temperature from that of the steady-state solution can be attributed to various time-dependent terms, either directly related to temperature (Equation \ref{eq:temperature_equation}) or indirectly by ionisation structure (Equation \ref{eq:ion_balance}). Considering first the temperature equation, this effect is unlikely to arise from the last term (the 'ionisation cooling' term), for the same reasons as in the main model grid (see again Figure \ref{fig:ioncool} in Appendix \ref{app:results}). 

For adiabatic cooling, we use Equation \ref{eq:adia_to_line} to estimate its significance compared to line cooling. Plugging in the ejecta parameters for the suppressed energy deposition model with $M_{\mathrm{ej}} = 0.01 \mathrm{\Msol}$, $v_{\mathrm{ej}} = 0.2 \mathrm{c}$, $T \sim 22500 \mathrm{K}$, $x_e = 2.0$ , $t_d = 100$, and assuming the line cooling functions to be on average of order $10^{-20}  \; \mathrm{erg \: s^{-1} \: cm^{-3}}$, we find a ratio of adiabatic to line cooling of $\sim 1.2$, implying that adiabatic cooling is highly significant at 100 days for this model. Naturally, the exact value of the adiabatic cooling contribution will also depend on the line cooling functions at this temperature and density (see Section \ref{subsec:suppressed_linecool}). Using \texttt{SUMO}'s internal cooling calculation, we find the actual ratio of adiabatic to line cooling at 100 days to be 0.30, implying that the line cooling function at that epoch is slightly higher than the estimated value of  $\sim 10^{-20}  \: \mathrm{erg \; cm^{3} \; s^{-1}}$ for this particular solution. The model with the steeper thermalisation evolution has a lower adiabatic cooling fraction at 100 days, of $\sim 0.18$. 

\begin{figure*}
\center
\includegraphics[trim={0.2cm 0.2cm 1.6cm 0.8cm},clip,width = 0.49\textwidth]{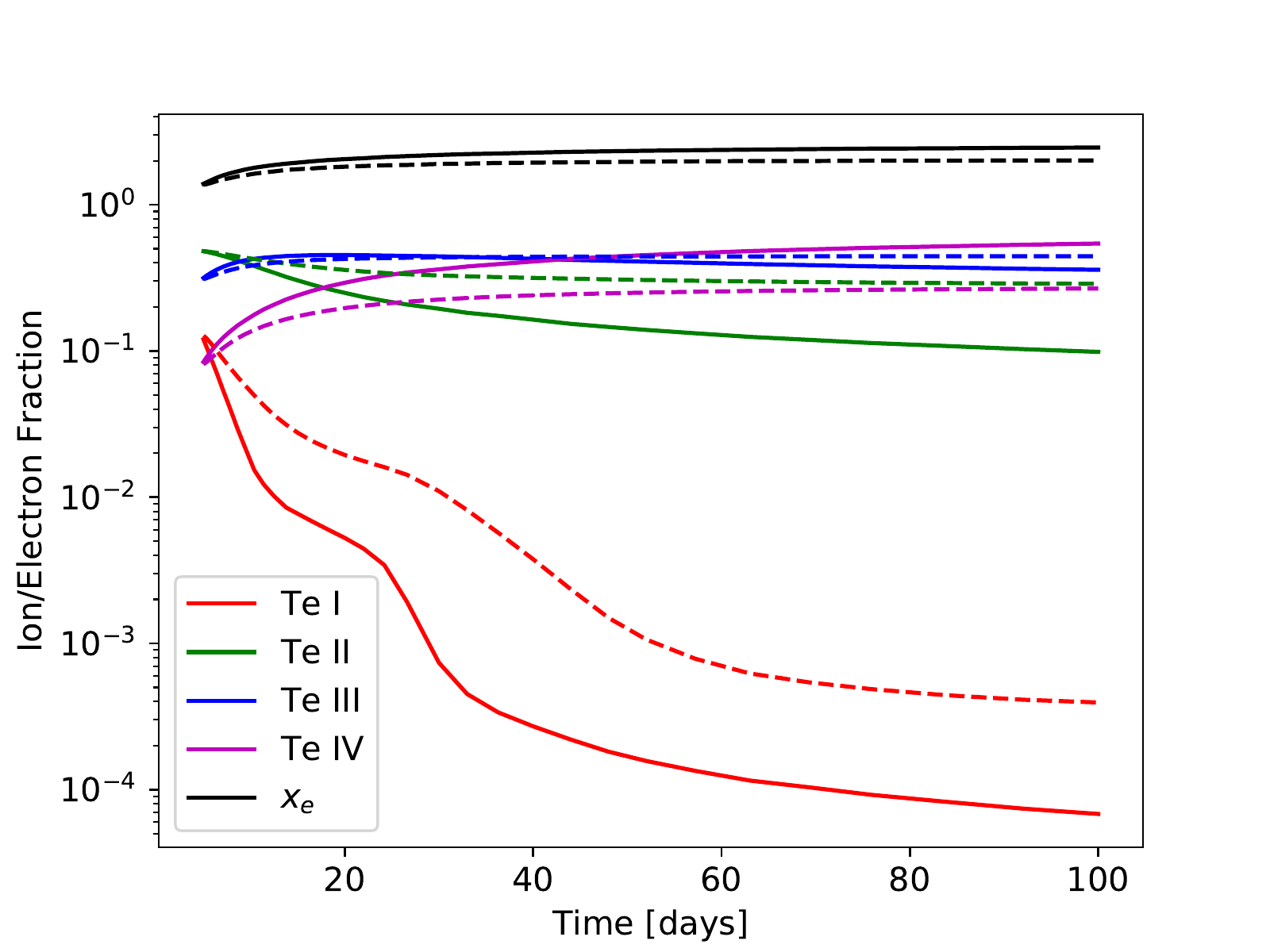}
\includegraphics[trim={0.2cm 0.2cm 1.6cm 0.8cm},clip,width = 0.49\textwidth]{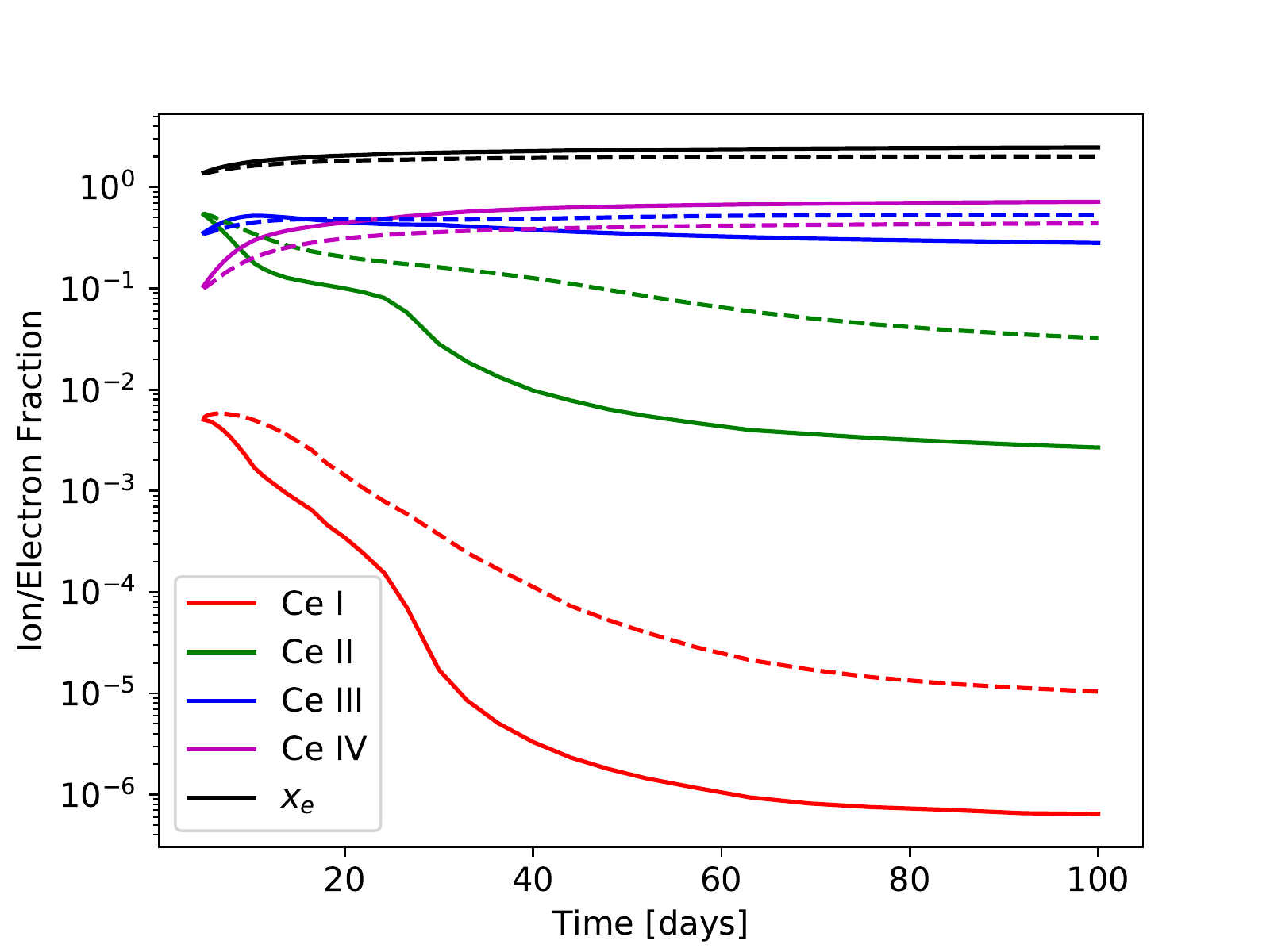}
\includegraphics[trim={0.2cm 0.2cm 1.6cm 0.8cm},clip,width = 0.49\textwidth]{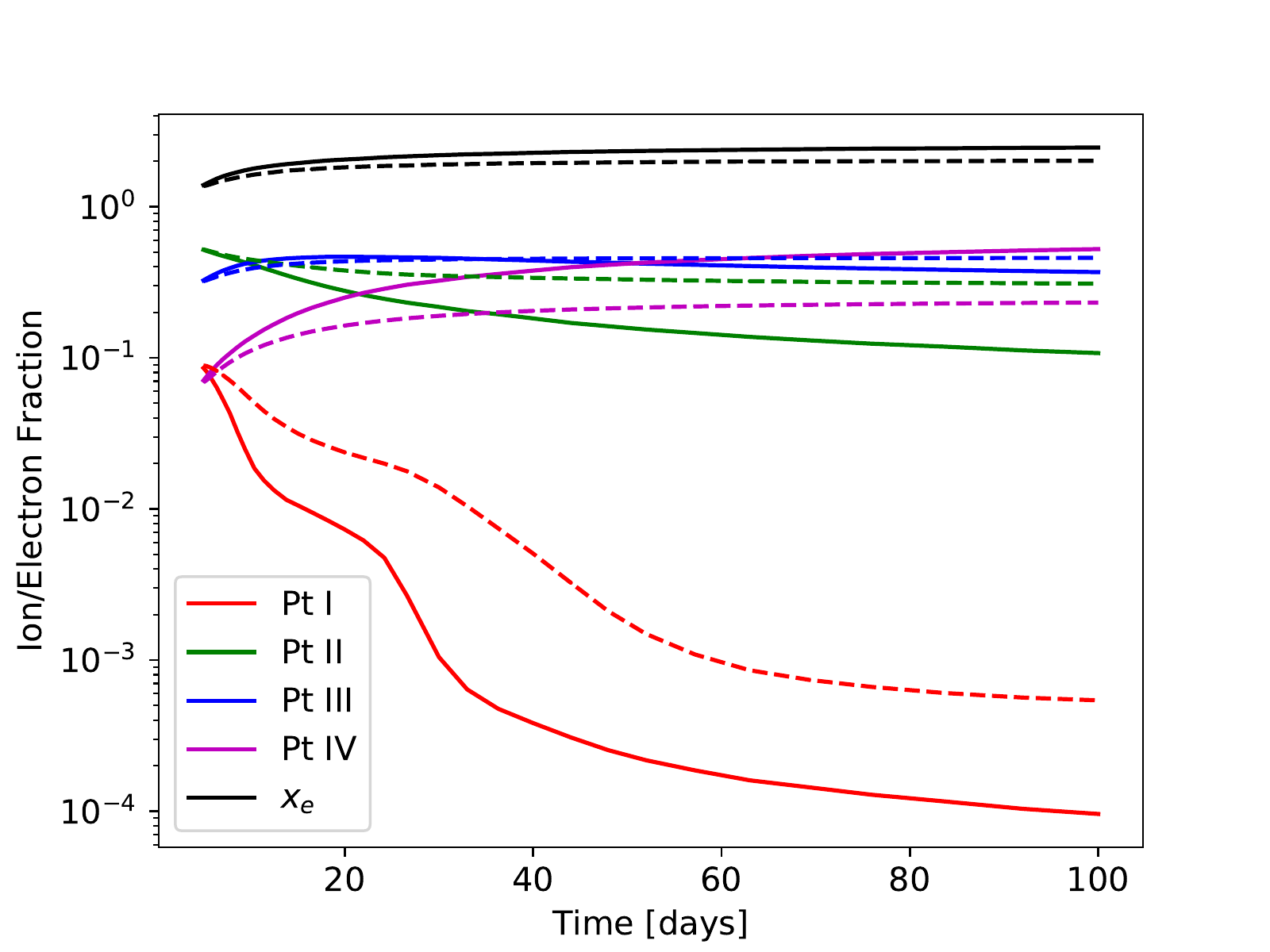}
\includegraphics[trim={0.2cm 0.2cm 1.6cm 0.8cm},clip,width = 0.49\textwidth]{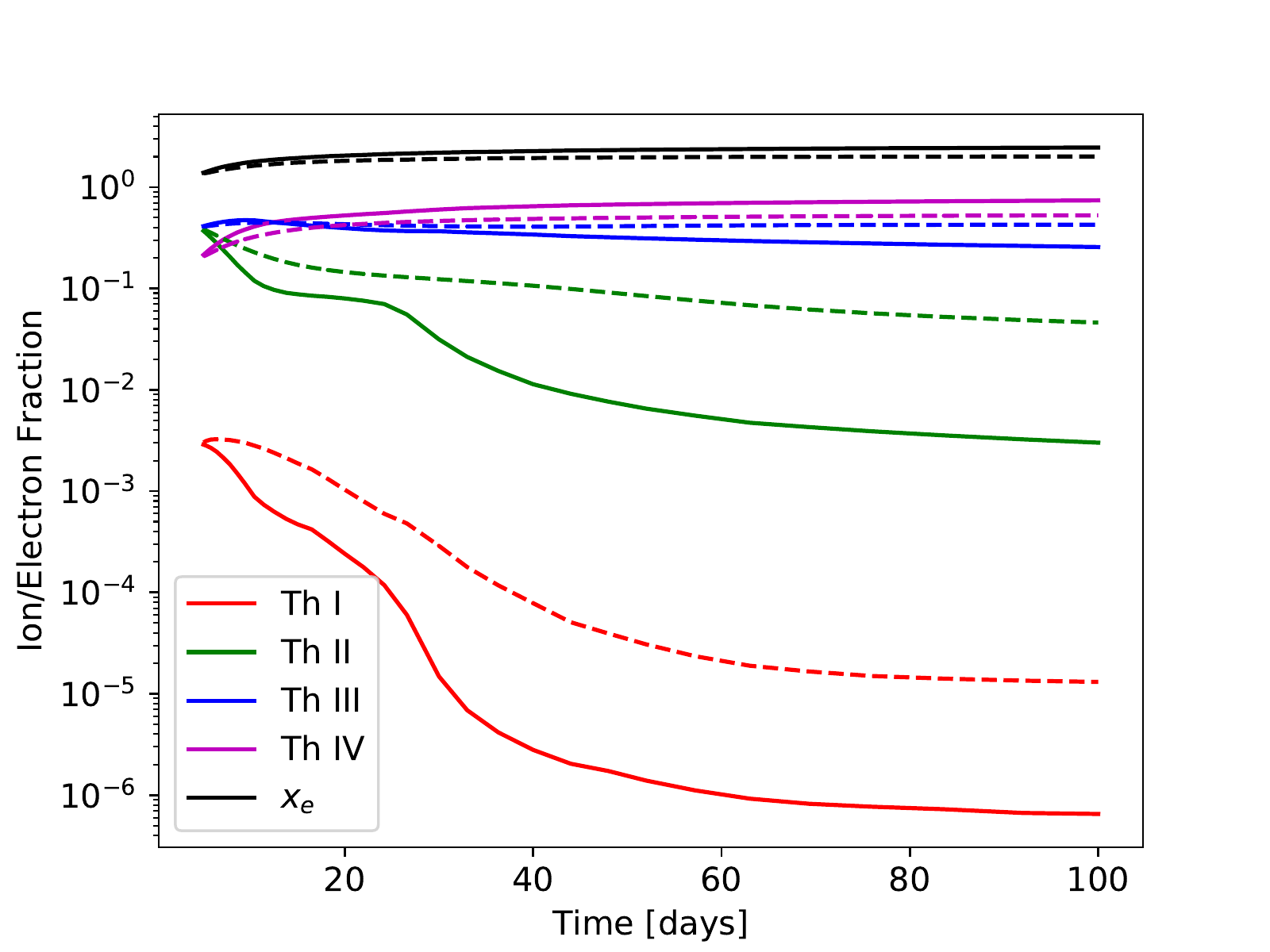}
\caption{Ionisation evolution for the suppressed energy model with day 1 energy deposition of $10^9 \; \mathrm{erg \: s^{-1} \: g^{-1}}$, $M_{\mathrm{ej}} = 0.01\mathrm{\Msol}$, $v_{\mathrm{ej}} = 0.2c$. The solid lines are the steady-state results, while the dashed lines are the time-dependent results. The time-dependent effect on ionisation structure is visible immediately at 5 days, and continues to yield a more neutral fraction for the duration of the evolution until 100 days. The difference in ionisation between steady-state and time-dependent modes appears to stabilise around 60 days, with both modes' ionisation level slowly increasing from then on.}
\label{fig:suppressed_ionfrac}
\end{figure*}

\begin{figure*}
\center
\includegraphics[trim={0.2cm 0.2cm 1.6cm 0.8cm},clip,width = 0.49\textwidth]{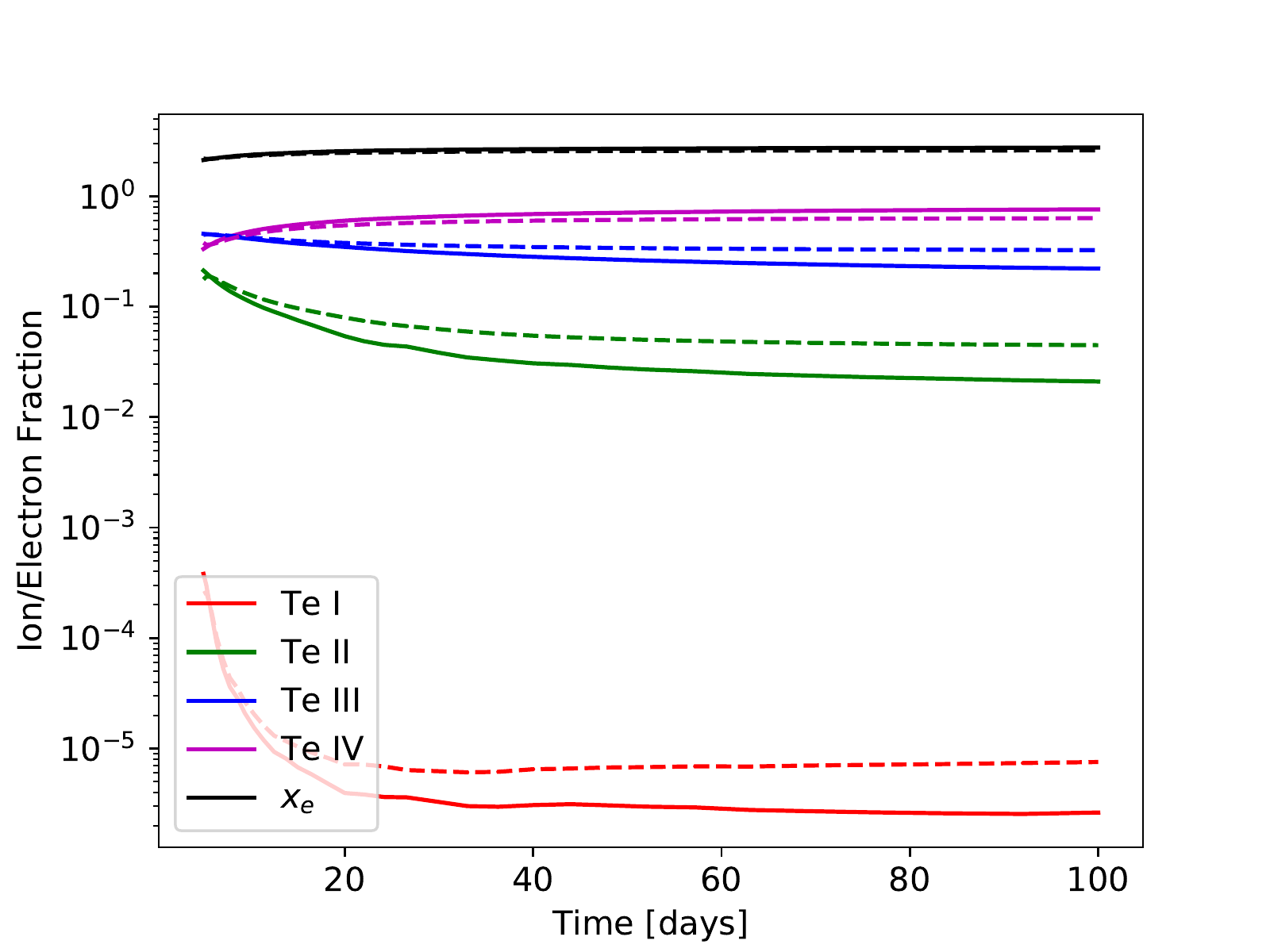}
\includegraphics[trim={0.2cm 0.2cm 1.6cm 0.8cm},clip,width = 0.49\textwidth]{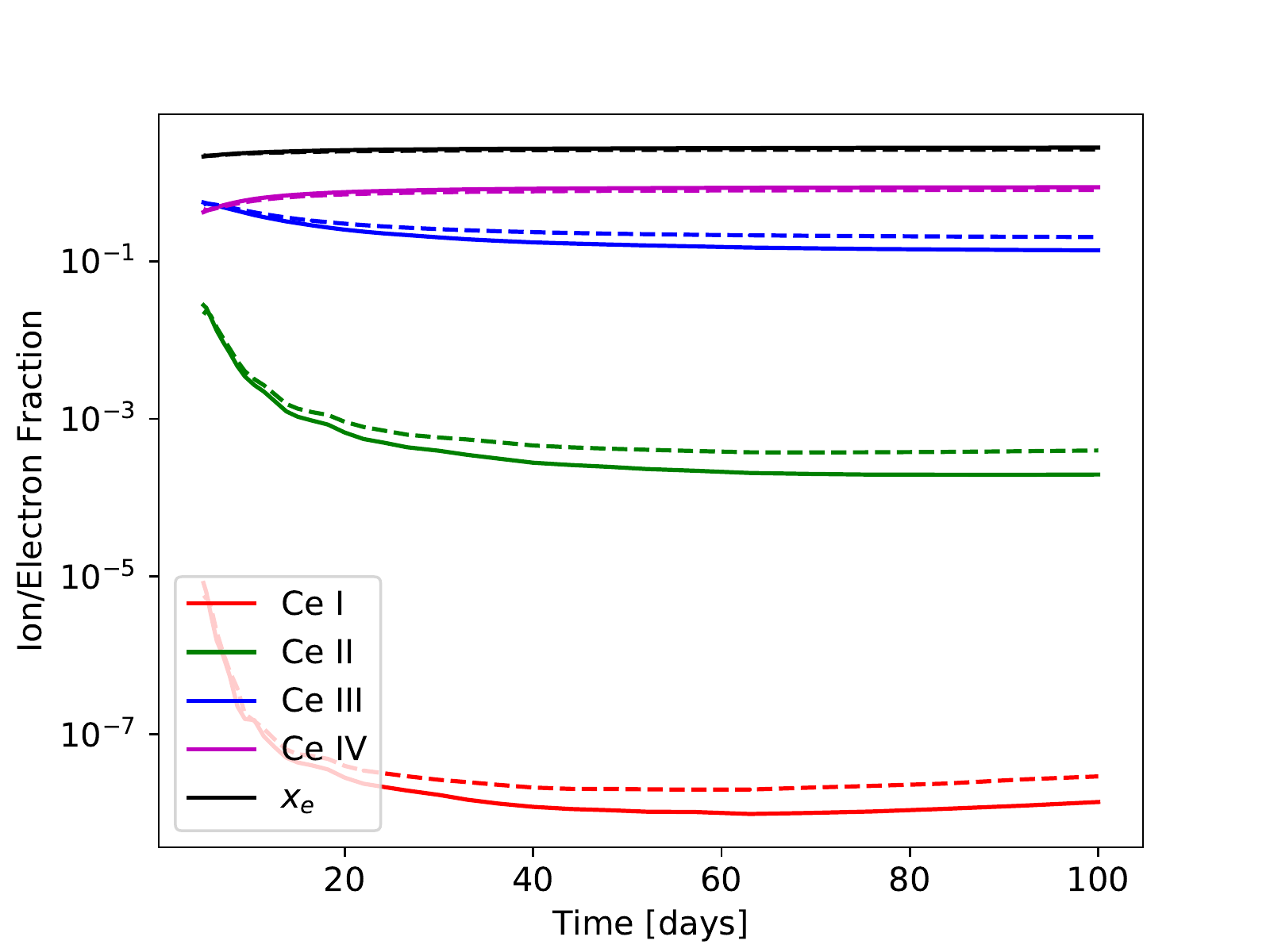}
\includegraphics[trim={0.2cm 0.2cm 1.6cm 0.8cm},clip,width = 0.49\textwidth]{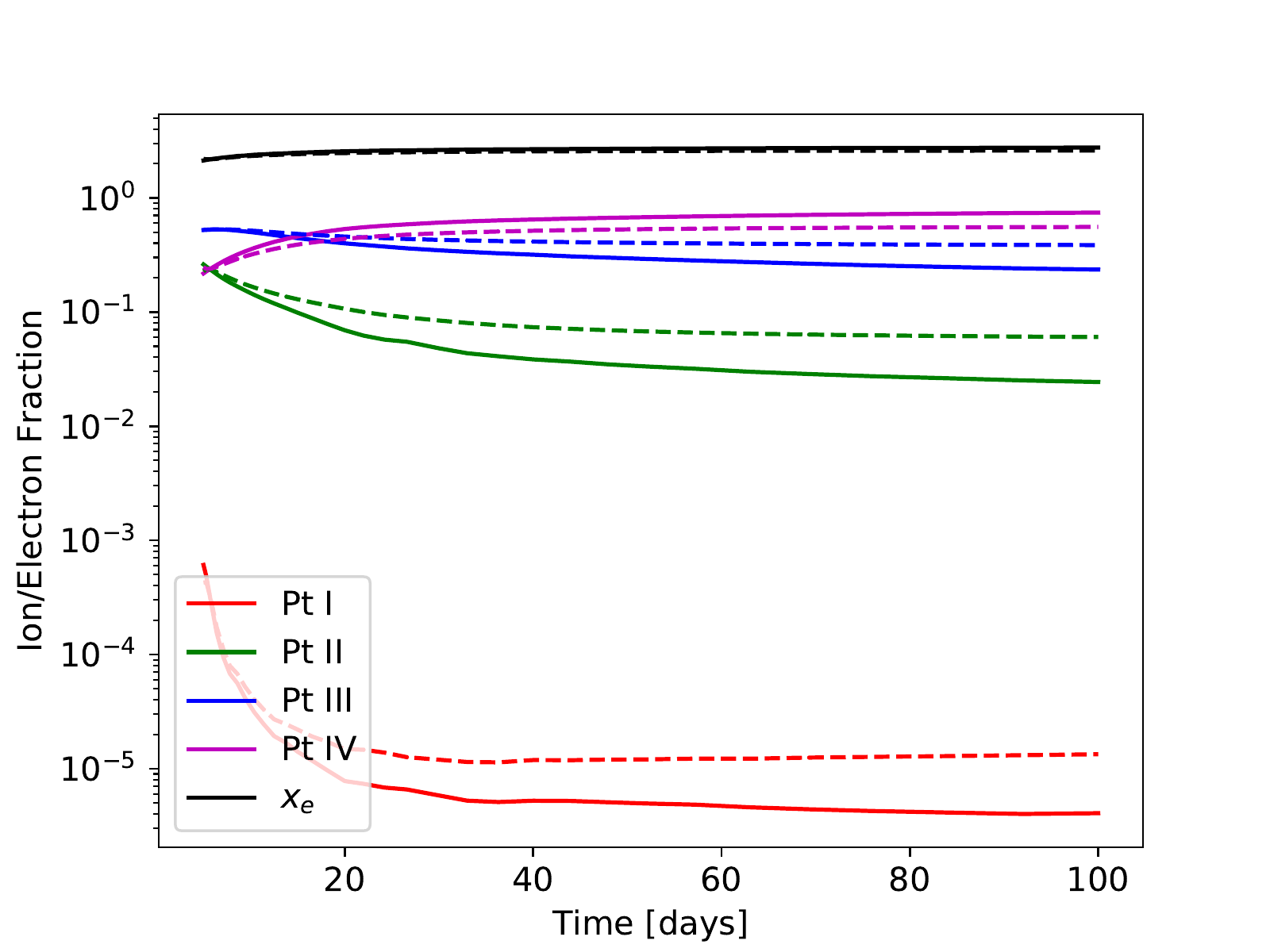}
\includegraphics[trim={0.2cm 0.2cm 1.6cm 0.8cm},clip,width = 0.49\textwidth]{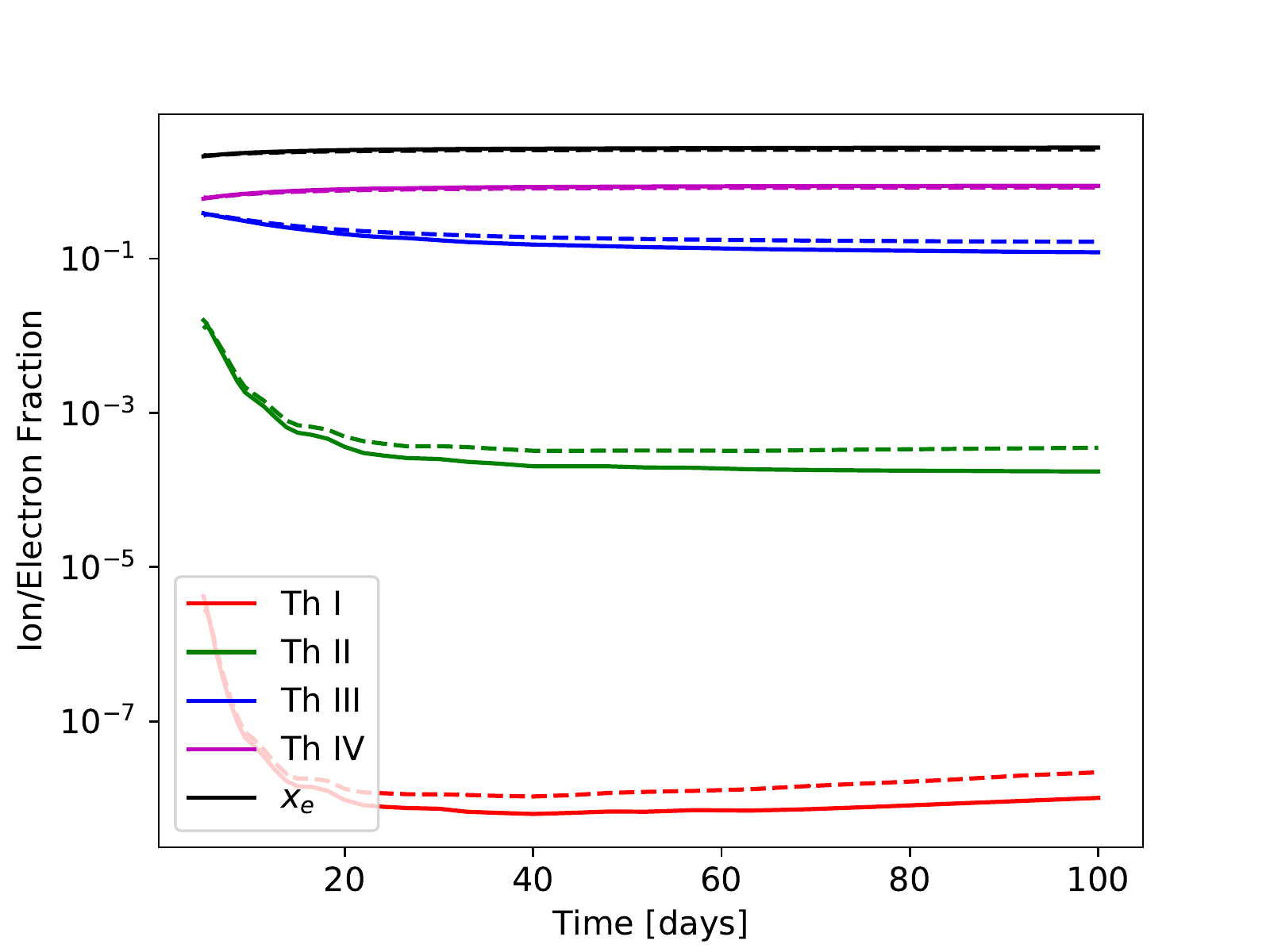}
\caption{Ionisation evolution for the model with steeper $\beta$-decay electron thermalisation, $M_{\mathrm{ej}} = 0.01\mathrm{\Msol}$, $v_{\mathrm{ej}} = 0.2$c. The solid lines are the steady-state results, while the dashed lines are the time-dependent results. The deviation of the time-dependent solution from that of steady-state is visible immediately, especially for the neutral ion fractions. Compared to the suppressed energy model in Figure \ref{fig:suppressed_linecool}, the ejecta is more ionised, and the ionisation structure appears to stabilise earlier around 30 days.}
\label{fig:steep_ionfrac}
\end{figure*}

A significant adiabatic cooling component implies that the bolometric luminosity does not track the exact radioactive energy deposition, as seen in Figure \ref{fig:luminosity}. Although this deviation is quite small for the timespan considered in this study (11 per cent dimmer at 100 days in time-dependent mode), the trend shows diverging luminosities as time goes on. As such, a steeper luminosity decline than predicted by steady-state models could impact future observations which plan to observe nebular phase KN emission. 

It is interesting to note that the bolometric luminosity appears relatively unaffected by the deviation of time-dependent ionisation structure from the steady-state solution. This is due to the fact that at $x_e \gtrsim 1$, as here, the vast majority of non-thermal energy deposition goes into heating rather than ionisation or excitation. Therefore, as long as the energy equation for thermal particles is in steady-state, the bulk of the energy flow is also in steady-state. Time-dependent ionisation can give significant changes to temperature and emergent spectra, but to be able to affect the bolometric light curve on its own, much of the thermal emission would need to be reprocessed by photoionisation. Our simulations indicate this is not the case, and thus a change in ionisation structure has relatively little effect on luminosity.

\subsubsection{Ionisation Evolution}
\label{subsec:suppressed_ionisation}

Alongside the change in temperature evolution, the ionisation structure also deviates significantly from the steady-state solution, as can be seen in Figures \ref{fig:suppressed_ionfrac} and \ref{fig:steep_ionfrac}. As for the models presented in Section \ref{subsec:standard_ionisation}, the time-dependent solutions are overall less ionised than the steady-state solutions. The qualitative evolution of the ionisation is similar to the models with $10^{10} \: \mathrm{erg \: s^{-1} \: g^{-1}}$ day 1 energy deposition, appearing to stabilise around 60 days or so for the suppressed energy model, and slightly earlier around 30 - 40 days for the steep energy model, consistent with an earlier thermalisation break. In both cases, the time-dependent solution yields an ionisation structure with much more neutral atoms and singly ionised ions, as early as 10 - 20 days after merge. However, the overall neutral abundance remains much smaller than the other ions at all epochs $t_d \gtrsim 20$ days. The change in singly ionised abundances is more significant, with solutions ranging from $\sim 1 -  50 $ per cent over the timespan, and depending on the ion. The doubly ionised abundances for every element are also higher in the time-dependent solution, which leads to a lower abundance of triply ionised ions in every case. For Te and Pt, the time-dependent solutions change relative abundance of ions with respect to each other i.e. the triply ionised ion for each element is no longer systematically the most abundant but rather the doubly ionised for $t \gtrsim $40d. As such, ejecta in the time-dependent nebular phase may realistically have a different ionisation structure from that of the steady-state phase, which will have important implications for the emergent spectra.

\begin{figure*}
\center
\includegraphics[trim={0.2cm 0 0.4cm 0.3cm},clip,width = 0.49\textwidth]{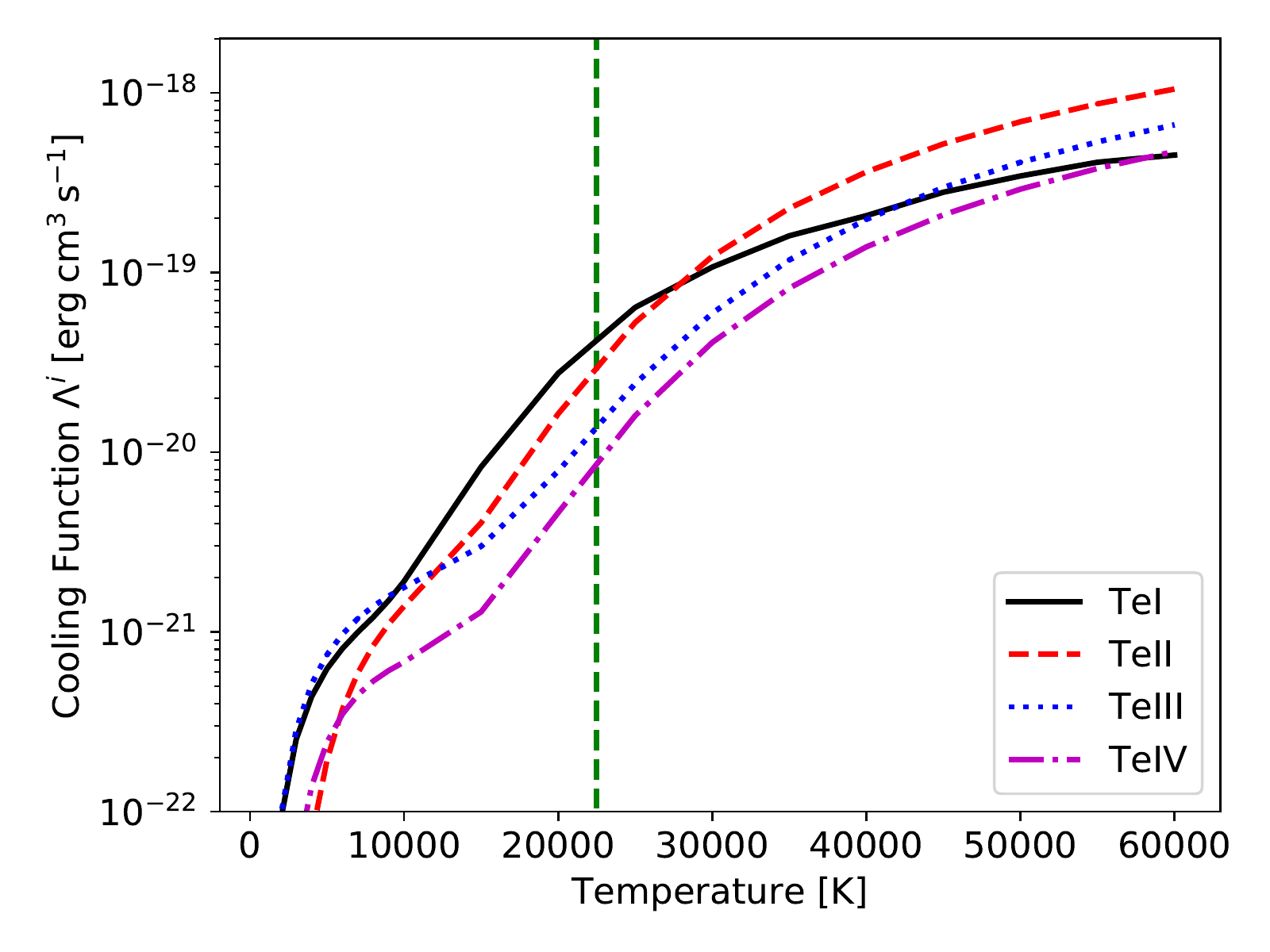}
\includegraphics[trim={0.2cm 0 0.4cm 0.3cm},clip,width = 0.49\textwidth]{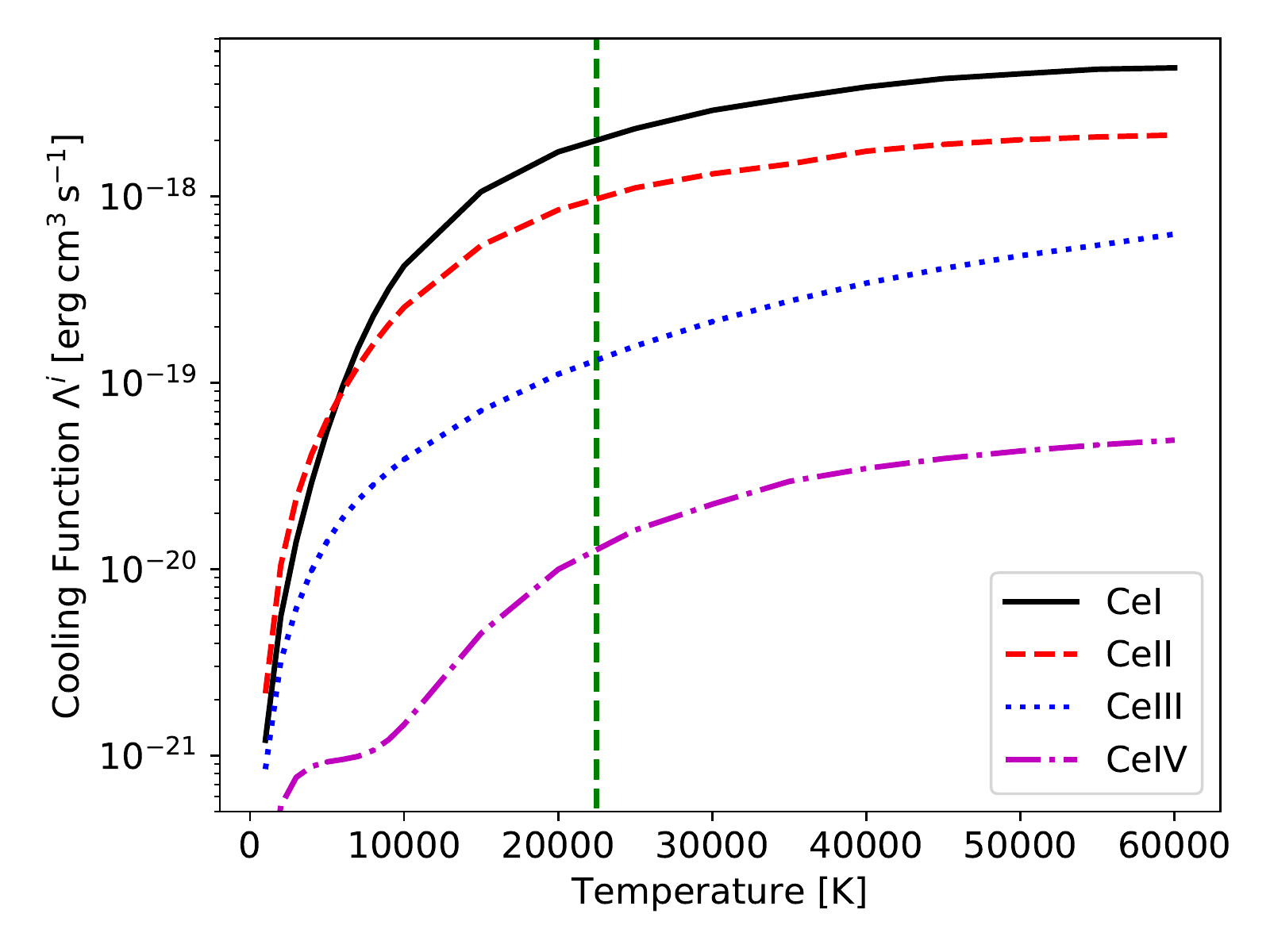}
\includegraphics[trim={0.2cm 0 0.4cm 0.3cm},clip,width = 0.49\textwidth]{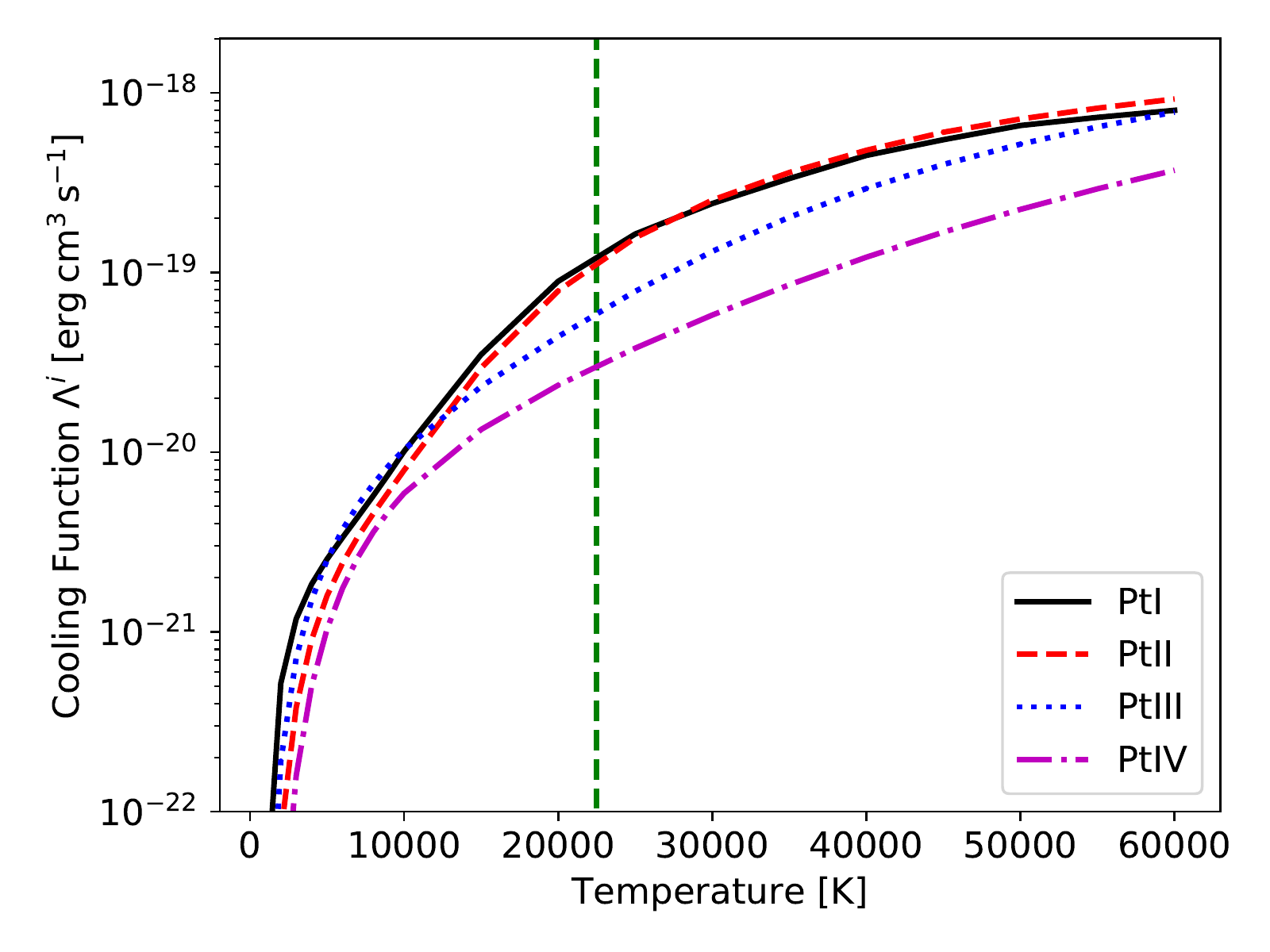}
\includegraphics[trim={0.2cm 0 0.4cm 0.3cm},clip,width = 0.49\textwidth]{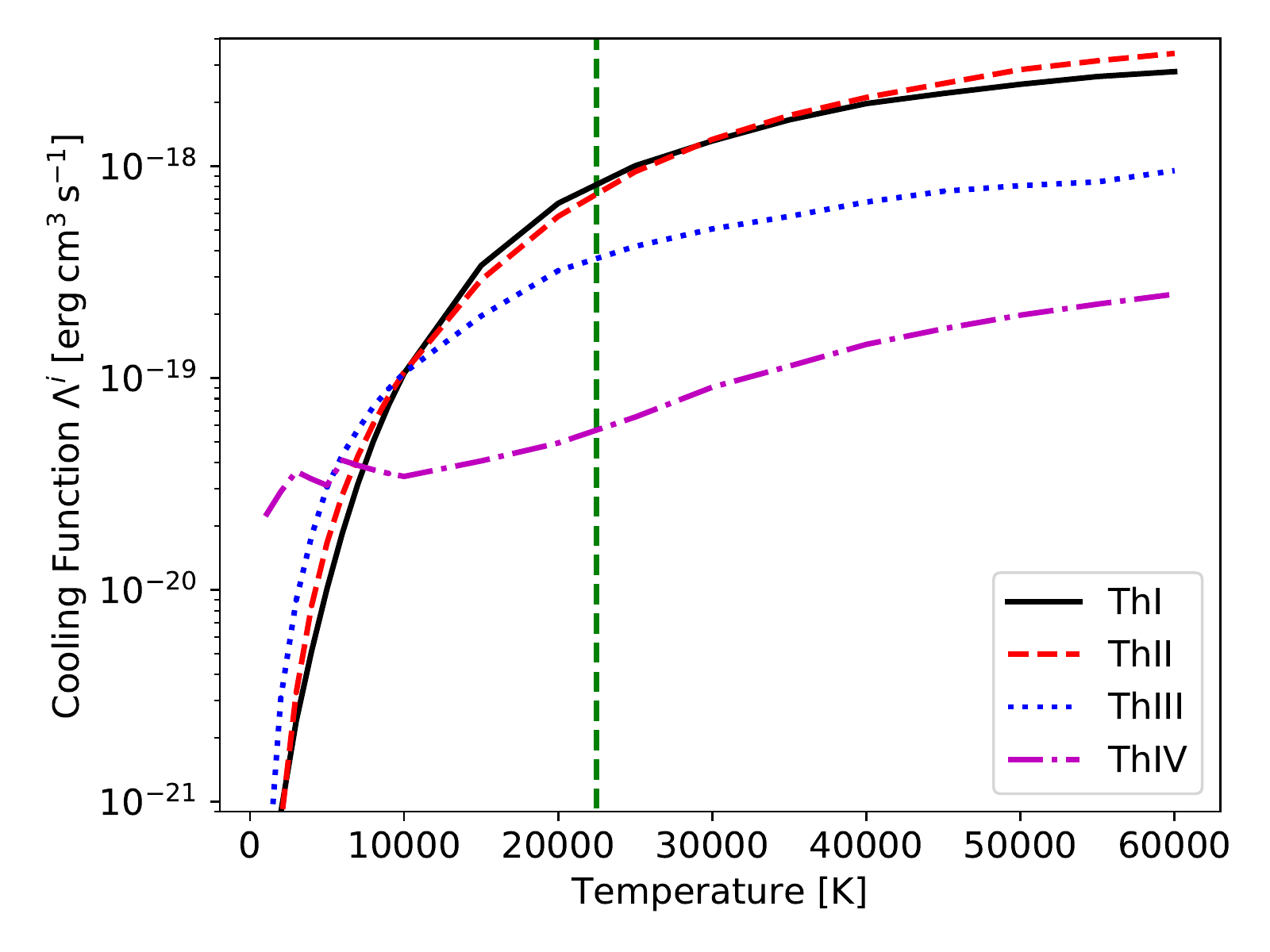}
\caption{Line cooling functions $\Lambda^i$ for the $\left\{n_e,n_i\right\}$ values of the lowest density, suppressed energy model at 100 days after merge. The number density at this epoch was $n=\sim 1.5 \times 10^{2} \mathrm{cm^{-3}}$, the electron fraction $x_e = 2.0$, and the time-dependent temperature solution $T = 22469 \mathrm{K}$ (marked by green dashed line). Note that these curves are not scaled to the respective abundances of each ion, i.e. we plot $\Lambda^i$ and not $x_i \Lambda^i$. For this particular solution, the ions' cooling capacity decreases as ionisation stage increase, with the triply ionised states for every element having the lowest cooling capacity.}
\label{fig:suppressed_linecool}
\end{figure*}

It may seem surprising to have a more neutral time-dependent solution as recombination time-scales become long, suggesting that the gas remains more ionised. However, Equation \ref{eq:ion_balance}, for a two-stage system with $\Psi=\alpha n_+$ can be written as:

\begin{equation}
    x_+ = -\frac{\Gamma}{2\alpha n} + \sqrt{ \left(\frac{\Gamma}{2\alpha n}\right)^2 + \frac{\Gamma}{2\alpha n} - \dot{x}_+}
    \label{eq:twostage_ion}
\end{equation}

\noindent where $\alpha = 10^{-11} \; \mathrm{ cm^3 s^{-1}} $ is the recombination rate as described in Section \ref{subsec:ionandrec}, $n$ is the total number density, $\Gamma$ is the ionisation rate, and $x_+$ is the ion abundance. From this equation, is is clear that growing ionisation (positive $\dot{x}_+$) will yield a less ionised time-dependent solution, whereas declining ionisation (negative $\dot{x}_+$) will yield a more ionised solution. For SNe, time-dependent effects turn on once the ionisation state is well into its decline - thus giving higher ionisation. For these KN models, however, the time-dependent effects become important when ionisation is still growing.

The long term evolution of the ionisation structure is uncertain. The evolution of the radioactive power term,  $t^{-2.8}$ in the standard models, which drives non-thermal ionisation, can be compared to the evolution of recombination per particle, $\alpha n x_e \propto t^{-3}x_e(t)$. As we have solutions with $x_e(t)$ slowly increasing over time, it is appears that ionisation may overcome recombination. However, a higher $x_e$ means that a smaller fraction of radioactive energy is used for ionisation, which also combines with higher ionisation potentials for more ionised states, thus leading to a reduction of overall ionisation rate. With time-scales evolving as $t^3$, the system will 'freeze out' when \emph{all} time-scales become long compared to the evolutionary time. Thus, the long term solution is expected to be that of the ionisation structure at this freeze out time. Though this does not appear to have occurred in the standard model grid within 100 days, the steep energy model with an earlier thermalisation break time shows a very flat evolution beyond $\sim$40 days. Thus, we believe this is a prediction for the final state of the KN.

We focus now on the suppressed energy model, for the following comparison of temperature and ionisation structure. From Figures \ref{fig:reduced_temperature} (left panel) and \ref{fig:suppressed_ionfrac}, we see that these appear to change in relatively similar manners. Initially, there is a small deviation in both solutions, with a noticeable steady-state temperature and ionisation structure increase centred around 30 days. This jump appears missing from the time-dependent temperature solution, though a much shallower ionisation jump can be seen in the time-dependent neutral abundances. This is followed by a stabilisation of both solutions, with a roughly constant difference between steady-state and time-dependent solutions for both temperature and ionisation. 

From these considerations, it is clear that the ionisation structure and temperature affect each other and the resultant overall solution. A higher temperature may lead to an overall higher level of ionisation due to more effective photoionisations (more important for a higher-energy radiation field). In principle, temperature affects ionisation also by thermal collisional ionisations and by its influence on recombination coefficients - however neither of these are treated here. Conversely, the ionisation structure in turn affects the temperature solution by the ions' differing line cooling functions, a detail which is investigated further in the next section.

\subsection{Line Cooling Functions}
\label{subsec:suppressed_linecool}

The effect of ionisation structure on the temperature can be readily explained by considering the various ions' line cooling capacities as functions of density and temperature. As introduced in Section \ref{subsec:temperature}, the line cooling per volume is given by $C_{\mathrm{line}}^i = \Lambda_i(T,n_e,n_i)n_e n_i$, where $\Lambda_i(T,n_e,n_i)$ is the line cooling function. In order to investigate the line cooling function's behaviour, we pick the solution at 100 days for our suppressed energy model with $M_{\mathrm{ej}} = 0.01~\mathrm{\Msol}$, $v_{\mathrm{ej}} = 0.2$c, and map out the line cooling functions for those values of $n_e$ and $n_i$, as shown in Figure \ref{fig:suppressed_linecool}. Taking the lowest-density model at the last epoch means these curves should be close to the low-density limit cooling functions when $\Lambda(T,n_e,n_i) \rightarrow \Lambda(T)$.

It should be noted that the curves shown are not scaled to the ions' respective abundances (i.e. we plot $\Lambda_i$ and not $x_i \Lambda_i$) at that solution. As an example, though the cooling functions for the neutral atoms at the specific temperature solution are larger than for the ions, the neutral fractions at 100 days are all low ($\sim 10^{-5}- 10^{-3}$), and thus cooling from the neutral atoms is in fact negligible ($\ll 1$per cent for all neutrals). 

For this particular model in time-dependent mode, Pt II, Te II and Ce III are the ions doing the most cooling, accounting together for approximately 51 per cent of the total cooling (i.e. $(c_{\mathrm{line}}^{\mathrm{PtII}} + c_{\mathrm{line}}^{\mathrm{TeII}} + c_{\mathrm{line}}^{\mathrm{CeIII}})/c_{\rm total} = 0.51$) . However, the ionisation structure is different in steady-state mode, notably more ionised, such that in this case Pt III, and Te III and IV dominate cooling with $\sim 49$ per cent contribution. Thus, even though the intrinsic cooling capacities of Te III and Te IV are the lowest of all ions included (see Figure \ref{fig:suppressed_linecool}), the large Te and Pt mass fractions amount to 90 per cent of ejecta composition, and the significant abundance of these ions at 100 days allows them to dominate cooling for that solution. Therefore, the temperature change due to ionisation structure variation comes rather from the relative change of these ions, as opposed to the higher abundance of neutral atoms.

Focussing now on intrinsic line cooling capacity (i.e. $\Lambda_i(T,n_e,n_i)$), Figure \ref{fig:suppressed_linecool} shows that neutral and singly ionised ions are usually better coolants at $T \geq 10^4 \mathrm{K}$ than their doubly or triply ionised counterparts. For this particular solution at 100 days with a time-dependent temperature of $T \approx 22500 \mathrm{K}$, the triply ionised states are consistently the worst coolers. This is especially striking for Ce, where Ce III has a line cooling function an order of magnitude greater than Ce IV across almost the entire temperature range. As such, having less triply ionised species will lower the temperature, because the overall cooling capacity is increased. Since the heating rate comes directly from radioactive energy deposition, and the temperature solution is found by solving Equation \ref{eq:temperature_equation}, enhanced line cooling will generally lead to a lower temperature. 

This result illustrates the non-linear and complex nature of time-dependent ionisation and temperature, where one can affect the other in various ways. Notably, it is not always true that a more neutral ionisation structure will lead to lower temperatures. Should the time-dependent ionisation solution favour a higher level of ionisation for the same temperature, the opposite result will naturally occur. Furthermore, the effect on temperature of ionisation structure will also depend on the temperature, since the line cooling functions can show diverse behaviour with respect to temperature, as well as ion and electron density. It is important to note that the cooling curves in Figure \ref{fig:suppressed_linecool} correspond to the specific solution of our chosen model at 100 days, with fixed ion and electron densities, and thus do not cover the entire model parameter space and epochs in a replete manner. In general, the line cooling functions can have strong dependencies on $n_e$ and $n_i$.

\section{Summary and Conclusions}
\label{sec:conclusions}

We describe the first steps in adapting the NLTE spectral synthesis code \texttt{SUMO} to model kilonovae, and present initial results for the temperature and ionisation evolution for 5-100 days after the merger, corresponding roughly to the onset of the nebular phase until the upper limit of current KN observations. In particular, we investigate the effect of using the full time-dependent temperature and ionisation equations on the evolution of gas state quantities compared to using the steady-state approximation. Understanding the validity, and possible break-down, of the steady-state approximation will be useful for future more detailed modelling efforts.

Our ejecta model is a single-zone, homologously expanding ejecta with a solar abundance composition consisting of the four elements Te, Ce, Pt and Th, each with the first four ionisation stages. In this work we have calculated new energy levels and A-values for these ions using the \texttt{FAC} code \citep{Gu:08}. We test our model over a range of ejecta parameters (decay power, ejecta mass, and ejecta velocity), selected to broadly cover those of various types of KN ejecta. We apply a radioactive energy deposition scheme to this ejecta generally following the works of \citet{Kasen.Barnes:19} and \citet{Waxman.etal:18}, allowing for both $\beta$ and $\alpha$ decay.

Analytic considerations point to possible time-dependent effects after a month or so for certain ejecta parameters, in particular for the ionisation structure. Numeric solutions are needed, however, to get a more accurate picture. Our simulations indicate that the analytic estimates give conservative steady-state time-scales for most of parameter space, with only minor changes seen for $t \leq $100 days. The strongest effect is observed for our lowest-density model with low decay power: here the time-dependent ionisation solution is significantly more neutral across all epochs already from 5 days. This modifies the effective line cooling function, resulting a temperature solution cooler by $\sim 2500$ K. In addition to this, adiabatic cooling reaches 30 per cent at 100 days for the same model. This gives a lower temperature than in steady-state, and a bolometric light curve declining faster than the instantaneous radioactive heating - a $\sim$10 per cent effect at 100 days. A model in which electrons are assumed not to be magnetically confined also shows a similar effect. 

A caveat to this result is that late time energy deposition from effectively thermalising fission products may continue to provide higher power to the ejecta regardless of $\alpha$ or $\beta$ decay contributions. As such, the effects seen in the low power ejecta may be suppressed, should heavy nuclei decaying by fission be present in significant amounts. On the other hand, ejecta tending to lower atomic mass ranges, and thus entirely dominated by $\alpha$ and $\beta$ decay, would be susceptible to the effects described here.

The emerging picture is that time-dependent terms typically give only minor effects for $\lesssim 100$ days, but these can become significant for certain types of ejecta - the crucial characteristics we identify here are low power and/or low density. Notably, should the energy generation rate be low, dominated by $\beta$ decay, or the electrons avoid magnetic trapping, time-dependent ionisation effects, which in turn affect the temperature, can be highly significant already from $\sim$10 days onwards. When adiabatic cooling becomes important, the bolometric luminosity will also cease to track the instantaneous energy deposition. Both of these could have possible consequences for interpreting the late evolution of AT2017gfo. A complete understanding of to which degree steady-state holds in kilonovae, as well as improved accuracy in future models, will however require systematic calculations of r-process recombinations rates, and collisional cross sections, as these impact the relevant time-scales.

\section*{Acknowledgements}

We acknowledge funding from the European Research Council (ERC) under the European Union's Horizon 2020 Research and Innovation Program (ERC Starting Grant 803189  -- SUPERSPEC, PI Jerkstrand).

The computations were enabled by resources provided by the Swedish National Infrastructure for Computing (SNIC), partially funded by the Swedish Research Council through grant agreement no. 2018-05973. 

We thank the anonymous referee for the constructive feedback, leading to a more thorough discussion of the results presented in this work.

%%%%%%%%%%%%%%%%%%%%%%%%%%%%%%%%%%%%%%%%%%%%%%%%%%
\section*{Data Availability}

The data underlying this article will be shared on reasonable request to the corresponding author.



\bibliographystyle{mnras}
\bibliography{biblio} % if your bibtex file is called example.bib

%\begin{thebibliography}{99}
%\end{thebibliography}

%%%%%%%%%%%%%%%%%%%%%%%%%%%%%%%%%%%%%%%%%%%%%%%%%%

%%%%%%%%%%%%%%%%% APPENDICES %%%%%%%%%%%%%%%%%%%%%

\appendix

\newpage

\section{Steady-State Considerations for Excitation Structure}
\label{app:ssexc}

Consider a two-level atom. The rate equation for the i=1 state in the neutral species is:

\begin{equation}
 \frac{\rm{d}x_{n1}}{\rm{d}t} = -C_{up}n_e x_{n1} + x_{n2}A \beta + \alpha_{n1}n_e   
\end{equation}

\noindent With $x_{n1} + x_{n2}=x_n$,

\begin{equation}
 \frac{\rm{d}x_{n1}}{\rm{d}t} = -C_{up}n_e x_{n1} + A \beta (x_n-x_{n1}) + \alpha_{n1}n_e 
\end{equation}

\noindent or
 
\begin{equation}
 \frac{\rm{d}x_{n1}}{\rm{d}t} = (-C_{up}n_e - A\beta)x_{n1} + A \beta x_n + \alpha_{n1}n_e
\end{equation}
 
\noindent The equation can be rewritten as
 
\begin{equation}
 x_{n1} = \frac{A \beta x_n + \alpha_{n1}n_e - \dot{x}_{n1}}{C_{up} n_e + A \beta}
\end{equation}
 
\noindent The steady-state solution is then accurate when

\begin{equation}
    \dot{x}_{n1} \ll  A\beta x_n + \alpha_{n1} n_e
\end{equation} 

\noindent or, when the time-scale for changing the relative population in $i=1$ ($x_{n1}/x_n$) is longer than $1/(A \beta + \alpha n_e)$. The radiative transition rate \textit{A} is rarely smaller than $\sim 10^{-3}~\mathrm{s^{-1}}$, so $1/A$ is of order $\gtrsim 10^3$s. Optically thick lines ($\beta < 1$) typically give larger values than this for $A\beta$. The level populations cannot change on time-scales significantly shorter than the expansion or radioactive time-scales, $\sim t \gg 10^3 $s, so the fast rates of spontaneous decays guarantees steady-state for the excitation structure at all times.

\section{Atomic Data - Comparison to the NIST ASD}
\label{app:atomicdata}

\begin{table*}
    \caption{Number of calculated levels and lines (allowed and forbidden with an upper level below the first experimental ionisation limit) for the ions included in this work. Levels$_\mathrm{raw}$ are the total number of fine-structure levels produced from the configurations of Tab. \ref{tab:configs}, while Levels$_\mathrm{red}$ and Lines$_\mathrm{red}$ correspond to the reduced number of levels and lines following a cut at the the first experimental ionisation limit. For Ce I this leads to 1236 levels which are cut further to the practical upper limit of 999 levels to be treated per ion in the \texttt{SUMO} code at present. The final two columns show the literature sources for the experimental levels and ionisation limit from the NIST ASD as shown in Figure \ref{fig:levels_all}.}
    \centering
    \begin{tabular}{c|c|c|c|c|c}
    Ion & Levels$_\mathrm{raw}$ & Levels$_\mathrm{red}$ & Lines$_\mathrm{red}$ & NIST Energy Level Source & NIST Ionisation Limit Source  \\
    \hline \\
    Te I   &  245 &   99 & 4077 & \citet{Morillon.Verges:75a,Morillon.Verges:75b,Cantu.etal:83} & \citet{Kieck.etal:19} \\
    Te II  &  165 &  158 & 10651 & \citet{Handrup.Mack:64,Tauheed.etal:09} & \citet{Handrup.Mack:64}\\ 
    Te III &   57 &   57 & 1432 & \citet{Moore:71} & \citet{Tauheed.etal:11} \\ 
    Te IV  &   18 &   18 & 153 & \citet{Moore:71} & \citet{Crooker.etal:64}\\ 
    Ce I   & 1920 &  999 & 314269 & \citet{Martin:78} & \citet{Worden.etal:78}\\ 
    Ce II  &  459 &  459 & 69999 & \citet{Martin:78} & \citet{Johnson.Nelson:17}\\
    Ce III &  237 &  235 & 18710 & \citet{Martin:78} & \citet{Martin:78}\\ 
    Ce IV  &   10 &   10 & 42 &  \citet{Martin:78} & \citet{Reader.Wyart:09}\\
    Pt I   &  152 &  110 & 4726 & \citet{Blaise.etal:92a} & \citet{Jakubek.Simard:00}\\
    Pt II  &  248 &  232 & 20285 & \citet{Reader.etal:88,Blaise.etal:92b} & \citet{Moore:71}\\
    Pt III &  555 &  551 & 105890 & \\
    Pt IV  &  781 &  780 & 211123 & \citet{Azarov.Gayasov:2016} & \citet{Rodrigues.etal:04}\\
    Th I   &  822 &  590 & 114716 & \citet{Redman.etal:14} & \citet{Kohler.etal:97}\\
    Th II  &  343 &  343 & 41678 & \citet{Redman.etal:14} & \citet{Herrera-Sancho.etal:13}\\
    Th III &   79 &   79 & 2269 & \citet{Redman.etal:14} & \citet{Blaise.etal:92c}\\
    Th IV  &   15 &   15 & 97 & \\
    \end{tabular}
    \label{tab:data_sources}
\end{table*}

\begin{table*}
\caption{
Electron configurations treated in the atomic structure calculations. In the notation employed here, the closed shells, the electron core, are common for all ions of a given element. The bold-faced configurations are the ones used to optimise the radial orbitals. This is limited to the ground configuration in all cases except for Pt I (see section \ref{sec:atomic_data} for details on the atomic structure method). The list of target configurations is compiled from comparisons with the NIST ASD \citep{NIST_ASD} and earlier works \citep[e.g.][]{tanaka:opacities:2020, fontes:2020} }
\label{tab:configs}
\resizebox{\textwidth}{!}{
\begin{tabular}{llllllllllll}
\textbf{Ion} &
  \textbf{Core} &
  \textbf{Ground} &
  \multicolumn{9}{l}{\textbf{Excited Configurations}} \\ \hline 
Te I &
  {[}Kr{]}4d$^{10}$ &
  \textbf{5s$^2$ 5p$^4$} &
  5s$^2$ 5p$^3$ 6s &
  5s$^2$ 5p$^3$ 6p &
  5s$^2$ 5p$^3$ 5d &
  5s$^2$ 5p$^3$ 7s &
  5s$^2$ 5p$^3$ 7p &
  5s$^2$ 5p$^3$ 6d &
  5s$^2$ 5p$^3$ 4f &
  5s$^2$ 5p$^3$ 8s &
  5s$^2$ 5p$^3$ 7d \\
Te II &
   &
  \textbf{5s$^2$ 5p$^3$} &
  5s 5p$^4$ &
  5s$^2$ 5p$^2$ 6s &
  5s$^2$ 5p$^2$ 5d &
  5s$^2$ 5p$^2$ 6p &
  5s$^2$ 5p$^2$ 7s &
  5s$^2$ 5p$^2$ 6d &
  5s$^2$ 5p$^2$ 4f &
  5s$^2$ 5p$^2$ 7p &
  5s$^2$ 5p$^2$ 8s \\
Te III &
   &
  \textbf{5s$^2$ 5p$^2$} &
  5s 5p$^3$ &
  5s$^2$ 5p 5d &
  5s$^2$ 5p 6s &
  5s$^2$ 5p 6p &
  5s$^2$ 5p 6d &
  5s$^2$ 5p 7s &
   &
   &
   \\
Te IV &
   &
  \textbf{5s$^2$ 5p} &
  5s 5p$^2$ &
  5s$^2$ 5d &
  5s$^2$ 6s &
  5s$^2$ 6p &
  5s$^2$ 6d &
  5s$^2$ 7s &
   &
   &
   \\
Ce I &
  {[}Cd{]} &
  \textbf{5p$^6$ 4f 5d 6s$^2$} &
  5p$^6$ 4f 5d$^2$ 6s &
  5p$^6$ 4f$^2$ 6s$^2$ &
  5p$^6$ 4f$^2$ 5d 6s &
  5p$^6$ 4f 5d 6s 6p &
  5p$^6$ 4f 5d$^3$ &
  5p$^6$ 4f 6s$^2$ 6p &
  5p$^6$ 4f$^2$ 6s 6p &
  5p$^6$ 4f 5d$^2$ 6p &
  5p$^6$ 4f$^2$ 5d$^2$ \\
Ce II &
   &
  \textbf{5p$^6$ 5d$^2$ 4f} &
  5p$^6$ 4f 5d 6s &
  5p$^6$ 4f$^2$ 6s &
  5p$^6$ 4f$^2$ 5d &
  5p$^6$ 4f 6s$^2$ &
  5p$^6$ 4f 5d 6p &
  5p$^6$ 4f$^2$ 6p &
  5p$^6$ 4f 6s 6p &
   &
   \\
Ce III &
   &
  \textbf{5p$^6$ 4f$^2$} &
  5p$^6$ 4f 5d &
  5p$^6$ 4f 6s &
  5p$^6$ 5d$^2$ &
  5p$^6$ 4f 6p &
  5p$^6$ 5d 6s &
  5p$^6$ 4f 6d &
  5p$^6$ 4f 7s &
  5p$^6$ 5d 6p &
  5p$^6$ 4f 5f \\
 &
   &
   &
  5p$^6$ 4f 7p &
  5p$^6$ 4f 8s &
  5p$^6$ 4f 7d &
  5p$^6$ 4f 6f &
  5p$^6$ 4f 5g &
  5p$^6$ 6p$^2$ &
  5p$^6$ 5d 6d &
   &
   \\
Ce IV &
   &
  \textbf{5p$^6$ 4f} &
  5p$^6$ 5d &
  5p$^6$ 6s &
  5p$^6$ 6p &
  5p$^6$ 6d &
  5p$^6$ 7s &
   &
   &
   &
   \\
Pt I &
  {[}Xe{]}4f$^{14}$ &
  \textbf{5d$^9$ 6s} &
  \textbf{5d$^{10}$} &
  \textbf{5d$^9$ 6p} &
  5d$^9$ 7s &
  5d$^8$ 6s$^2$ &
  5d$^8$ 6s 6p &
  5d$^8$ 6s 7s &
   &
   &
   \\
Pt II &
   &
  \textbf{5d$^9$} &
  5d$^8$ 6s &
  5d$^7$ 6s$^2$ &
  5d$^8$ 6p &
  5d$^8$ 7s &
  5d$^8$ 6d &
  5d$^8$ 8s &
  5d$^8$ 7d &
   &
   \\
Pt III &
   &
  \textbf{5d$^8$} &
  5d$^7$ 6s &
  5d$^7$ 6p &
  5d$^7$ 7s &
  5d$^6$ 6s 6p &
   &
   &
   &
   &
   \\
Pt IV &
   &
  \textbf{5d$^7$} &
  5d$^6$ 6s &
  5d$^6$ 6p &
  5d$^6$ 6d &
  5d$^6$ 7s &
  5d$^6$ 7p &
   &
   &
   &
   \\
Th I &
  {[}Rn{]} &
  \textbf{6d$^2$ 7s$^2$} &
  6d$^3$ 7s &
  5f 6d$^2$ 7s &
  5f 6d 7s$^2$ &
  6d 7s$^2$ 7p &
  6d$^2$ 7s 7p &
  5f 7s$^2$ 7p &
  6d$^4$ &
  5f 6d 7s 7p &
  5f$^2$ 7s$^2$ \\
Th II &
   &
  \textbf{6d 7s$^2$} &
  6d$^2$ 7s &
  5f 7s$^2$ &
  5f 6d 7s &
  6d$^3$ &
  5f 6d$^2$ &
  6d 7s 7p &
  5f$^2$ 7s &
  5f 7s 7p &
  5f 6d 7p \\
Th III &
   &
  \textbf{5f 6d} &
  6d$^2$ &
  6d 7s &
  7s$^2$ &
  5f$^2$ &
  5f 7p &
  6d 7p &
  7s 7p &
  5f 8s &
   \\
Th IV &
   &
  \textbf{5f} &
  6d &
  7s &
  7p &
  7d &
  8s &
  6f &
  8d &
  9s &
\end{tabular}%
}
\end{table*}

\begin{figure*}
\center
\includegraphics[trim={0.3cm 0.3cm 0.3cm 0cm},width = 0.32\textwidth]{Atomic_Data/fig1.pdf}
\includegraphics[trim={0.3cm 0.3cm 0.3cm 0cm},width = 0.32\textwidth]{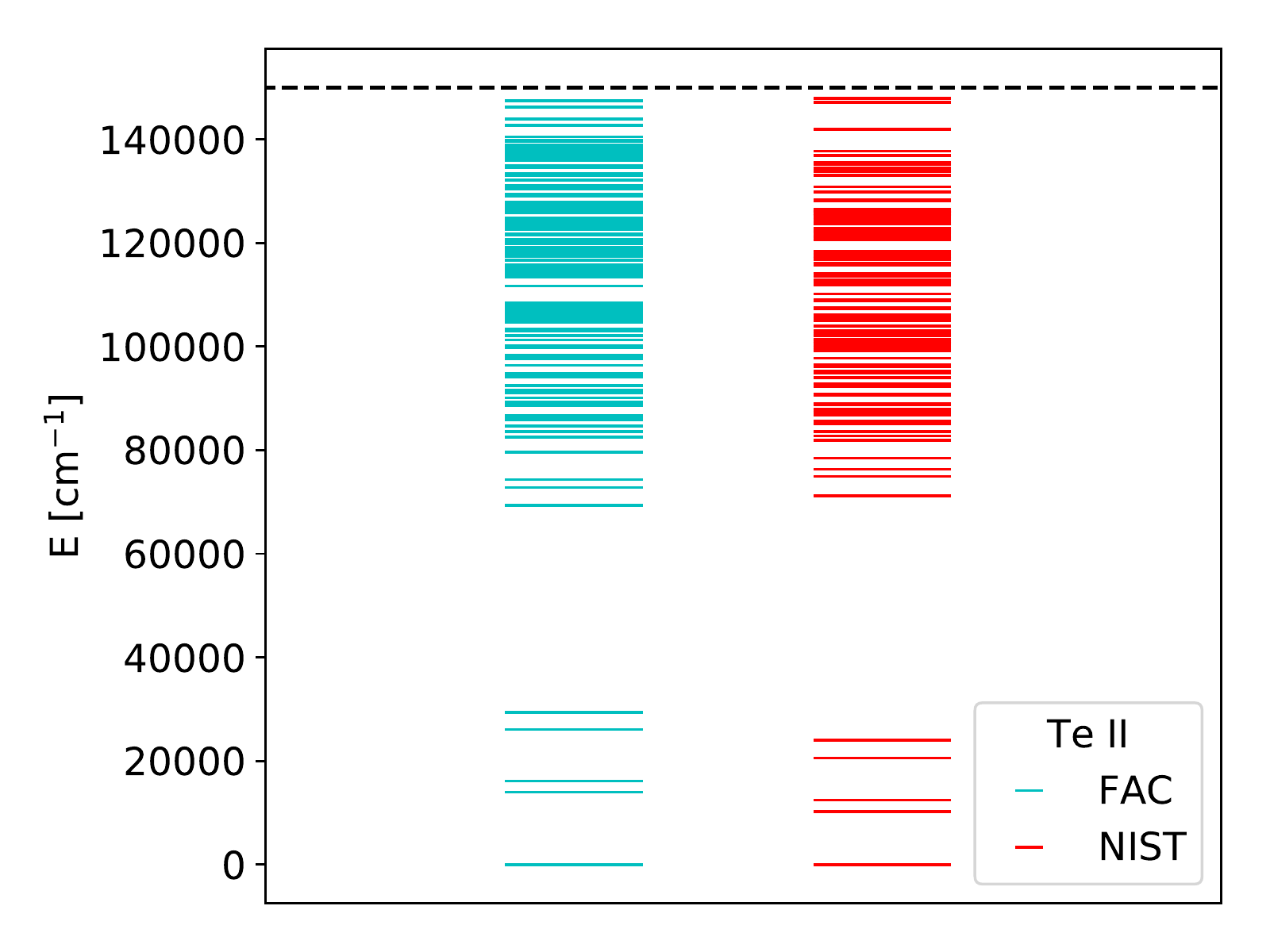}
\includegraphics[trim={0.3cm 0.3cm 0.3cm 0cm},width = 0.32\textwidth]{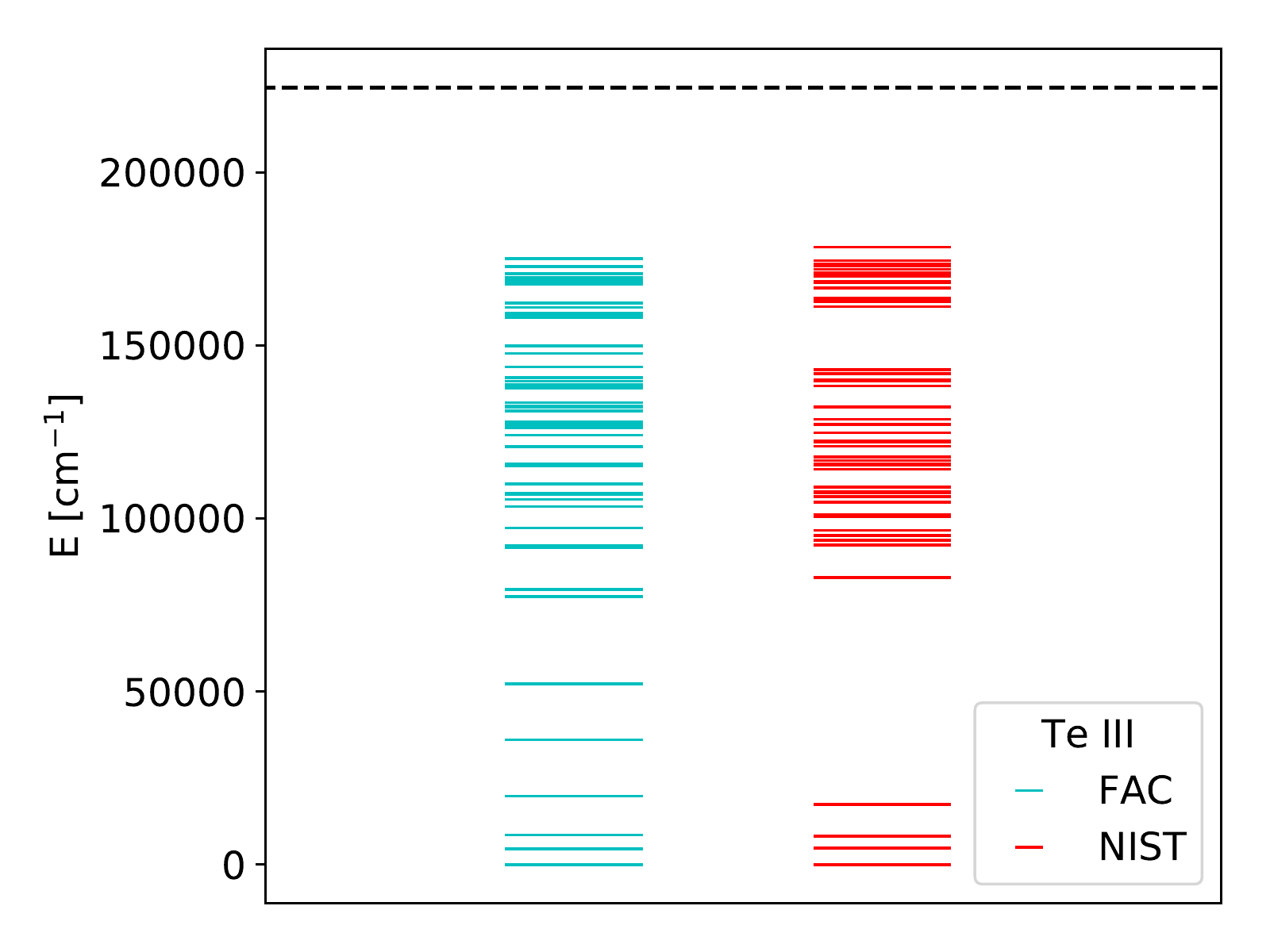}
\includegraphics[trim={0.3cm 0.3cm 0.3cm 0cm},width = 0.32\textwidth]{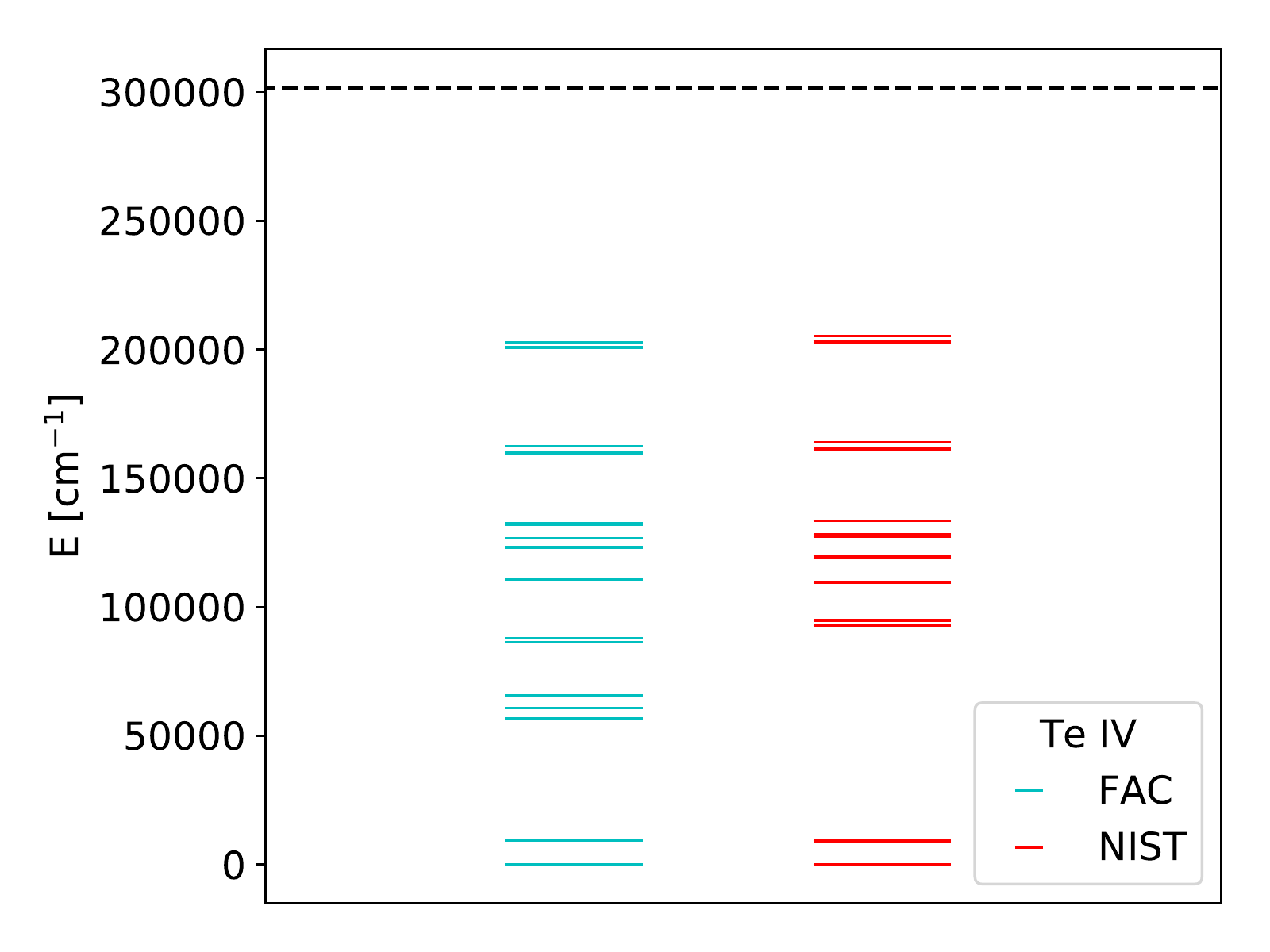}
\includegraphics[trim={0.3cm 0.3cm 0.3cm 0cm},width = 0.32\textwidth]{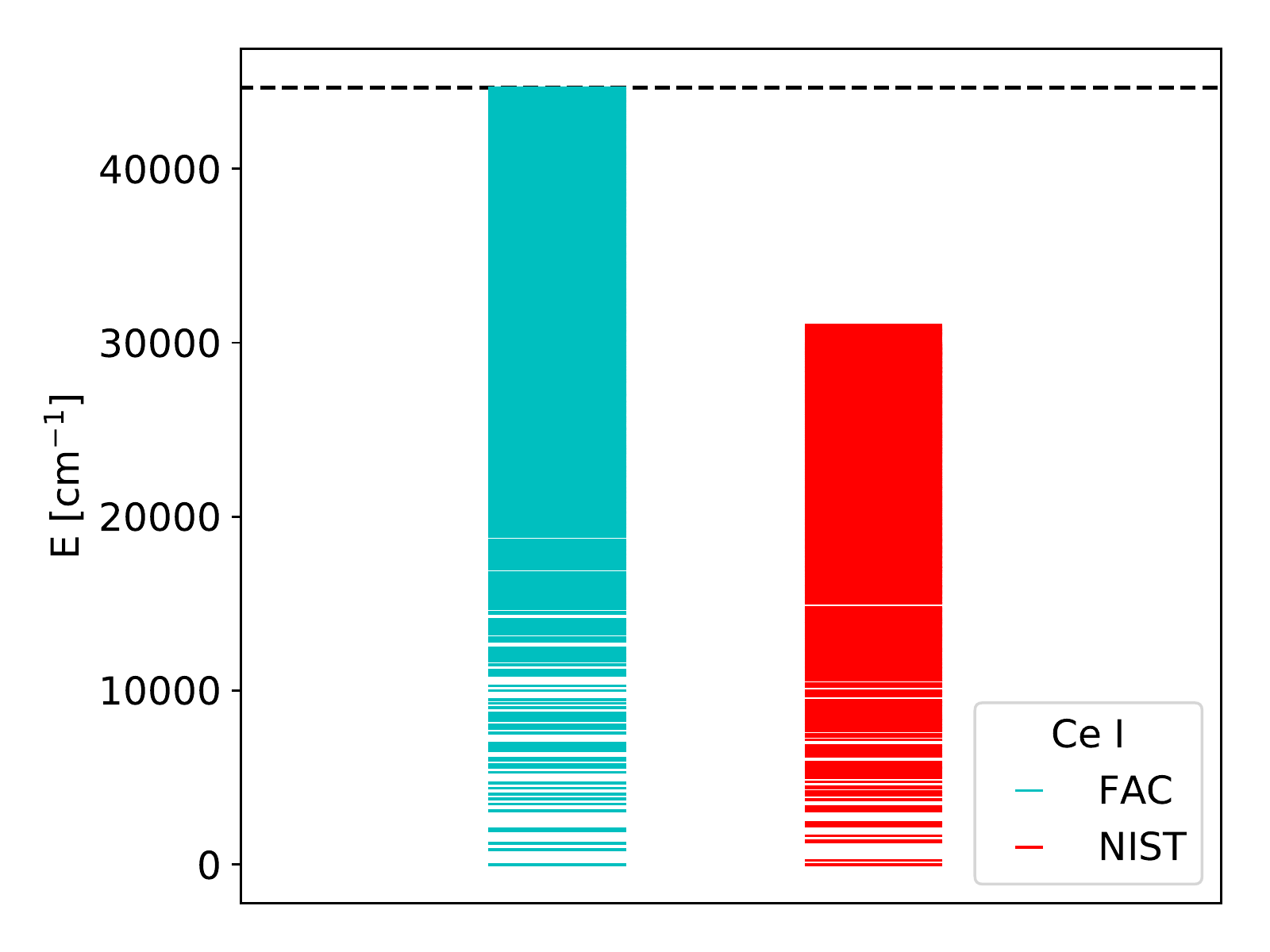}
\includegraphics[trim={0.3cm 0.3cm 0.3cm 0cm},width = 0.32\textwidth]{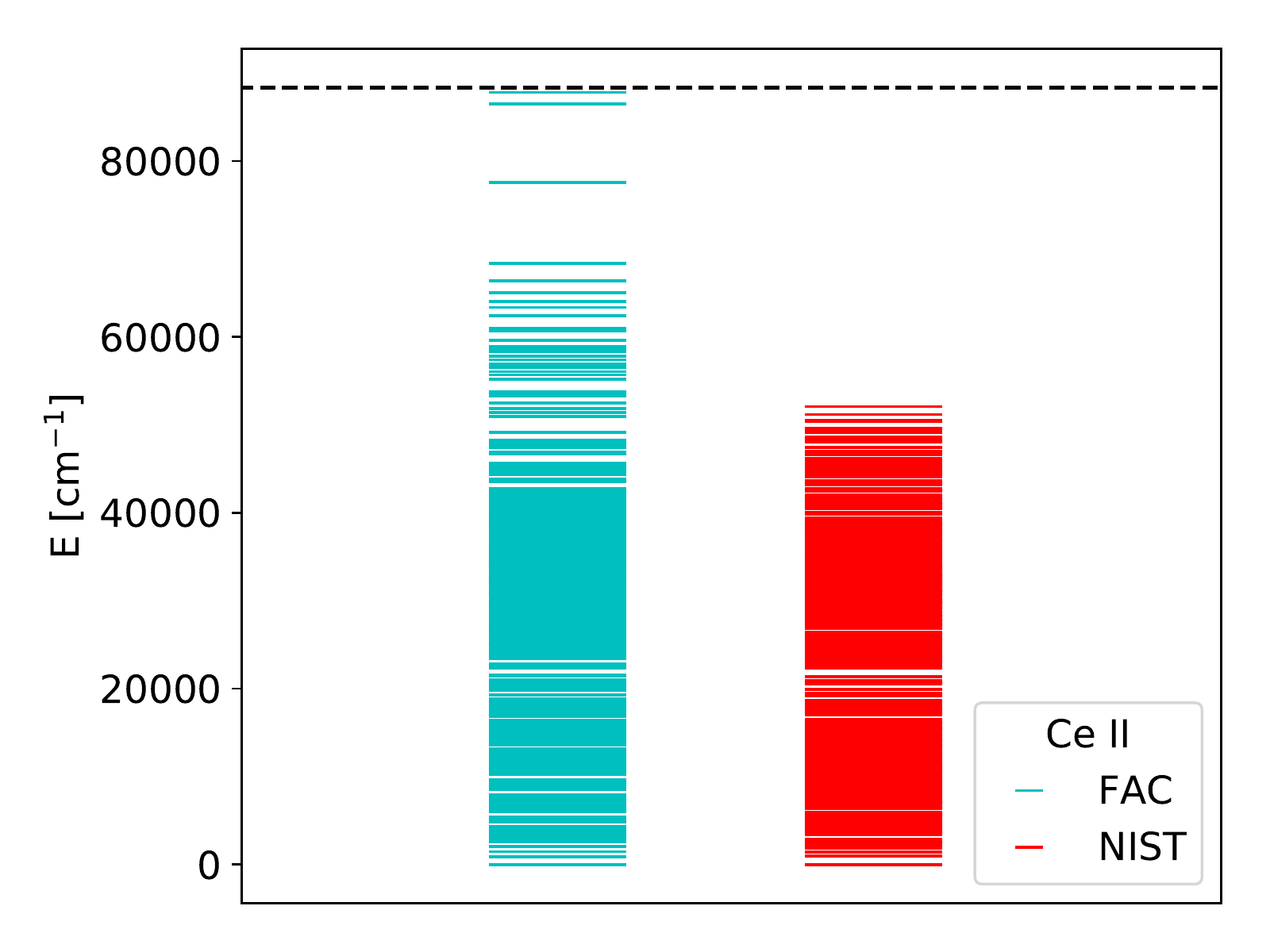}
\includegraphics[trim={0.3cm 0.3cm 0.3cm 0cm},width = 0.32\textwidth]{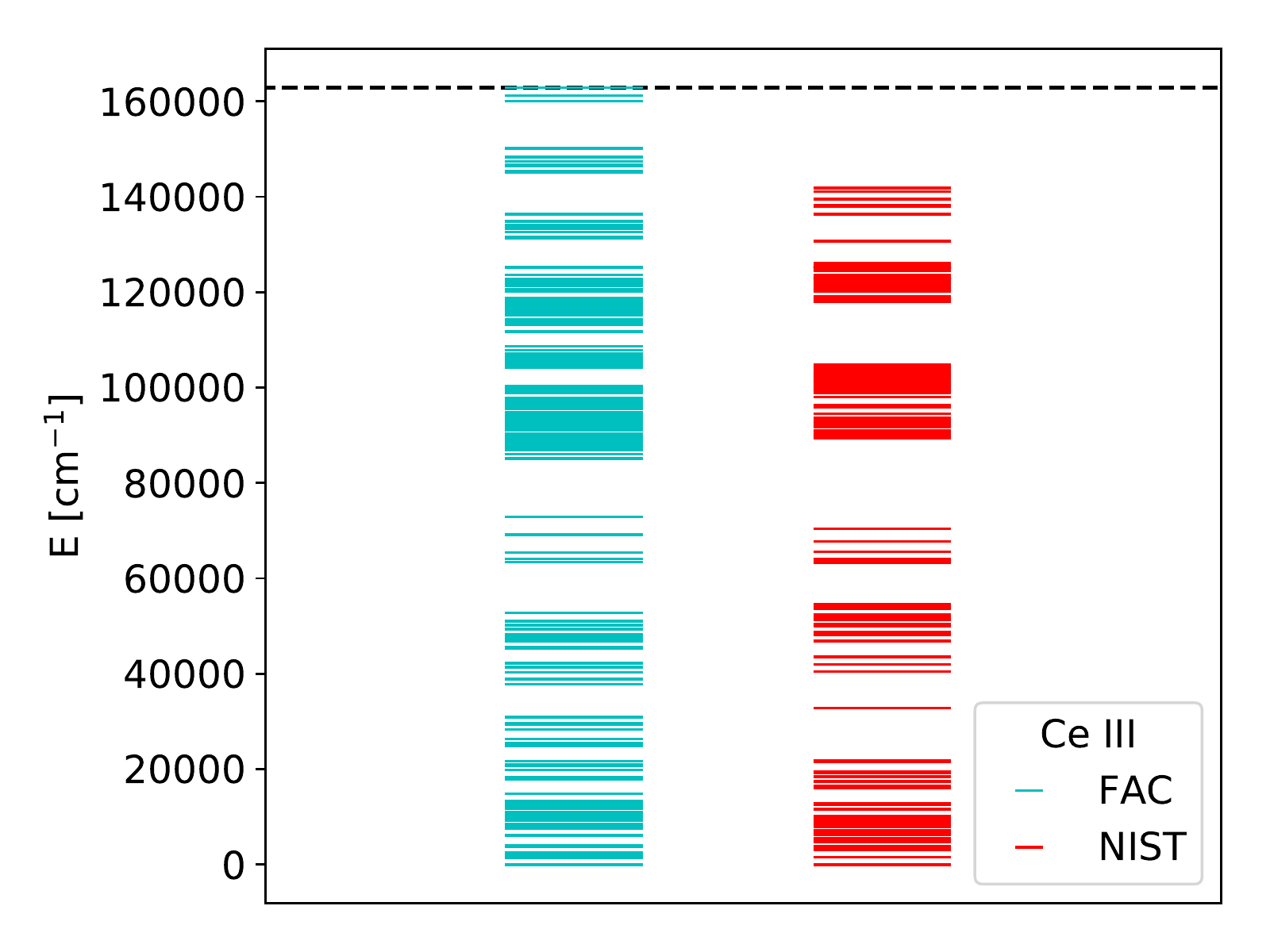}
\includegraphics[trim={0.3cm 0.3cm 0.3cm 0cm},width = 0.32\textwidth]{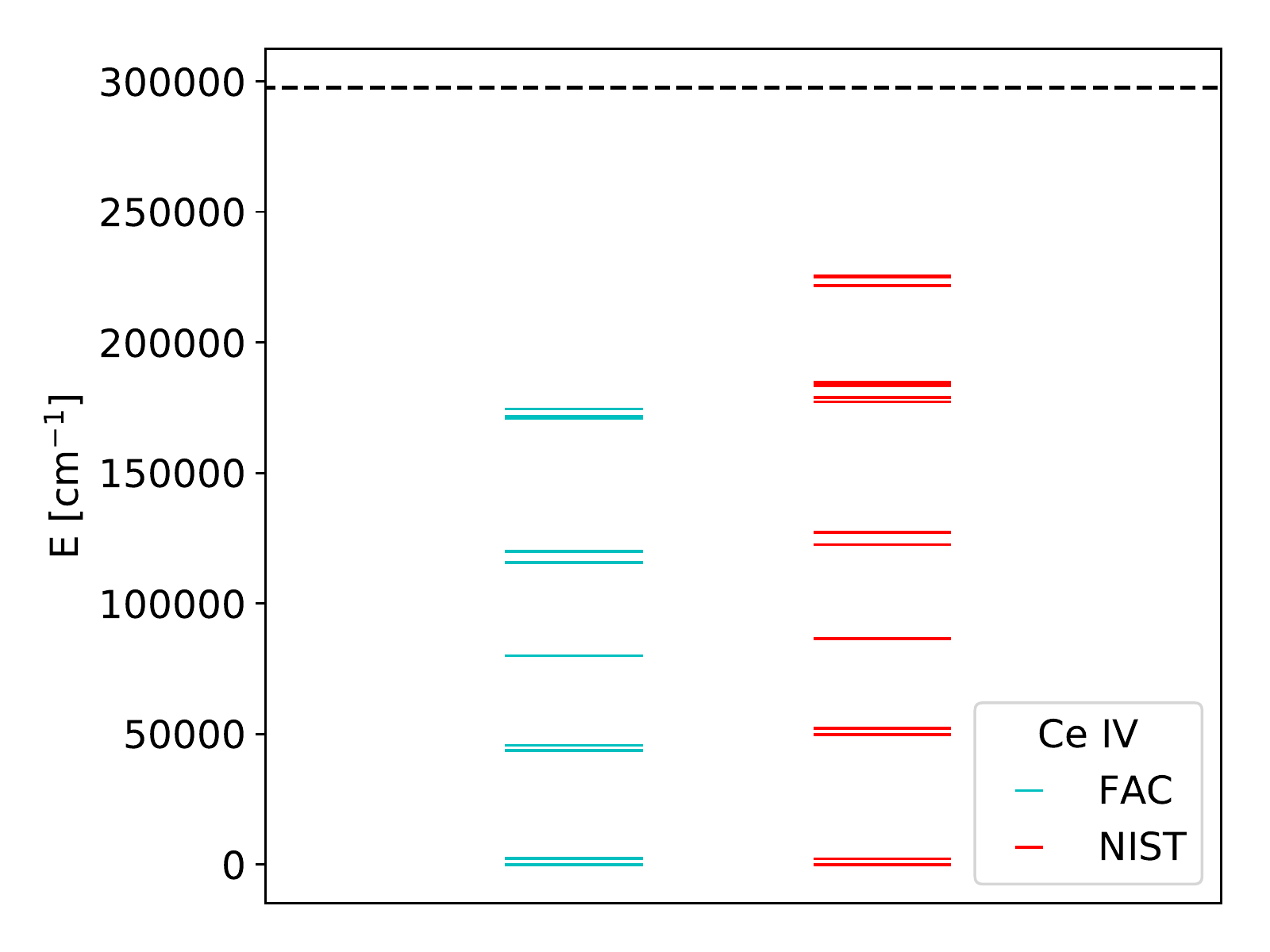}
\includegraphics[trim={0.3cm 0.3cm 0.3cm 0cm},width = 0.32\textwidth]{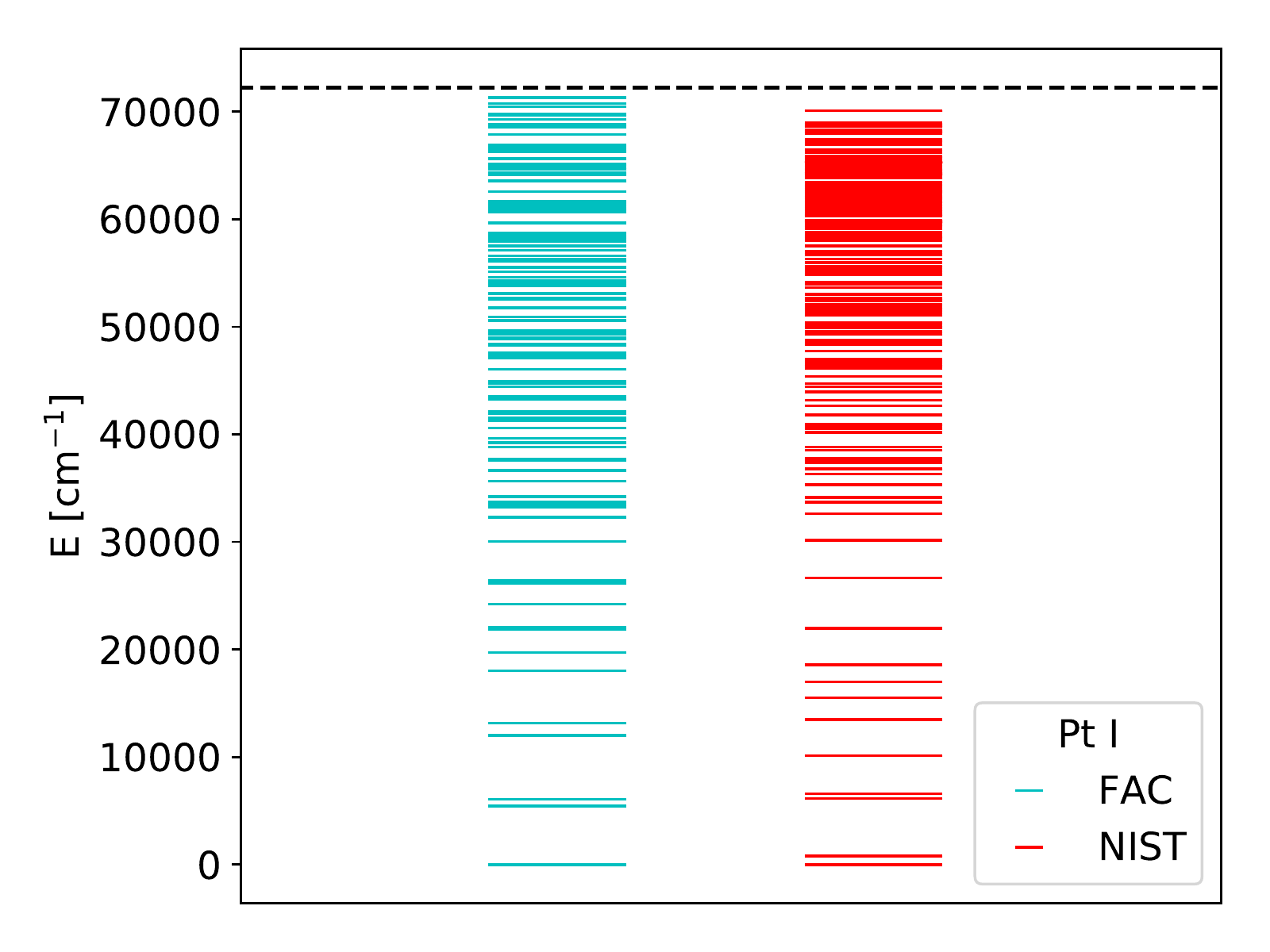}
\includegraphics[trim={0.3cm 0.3cm 0.3cm 0cm},width = 0.32\textwidth]{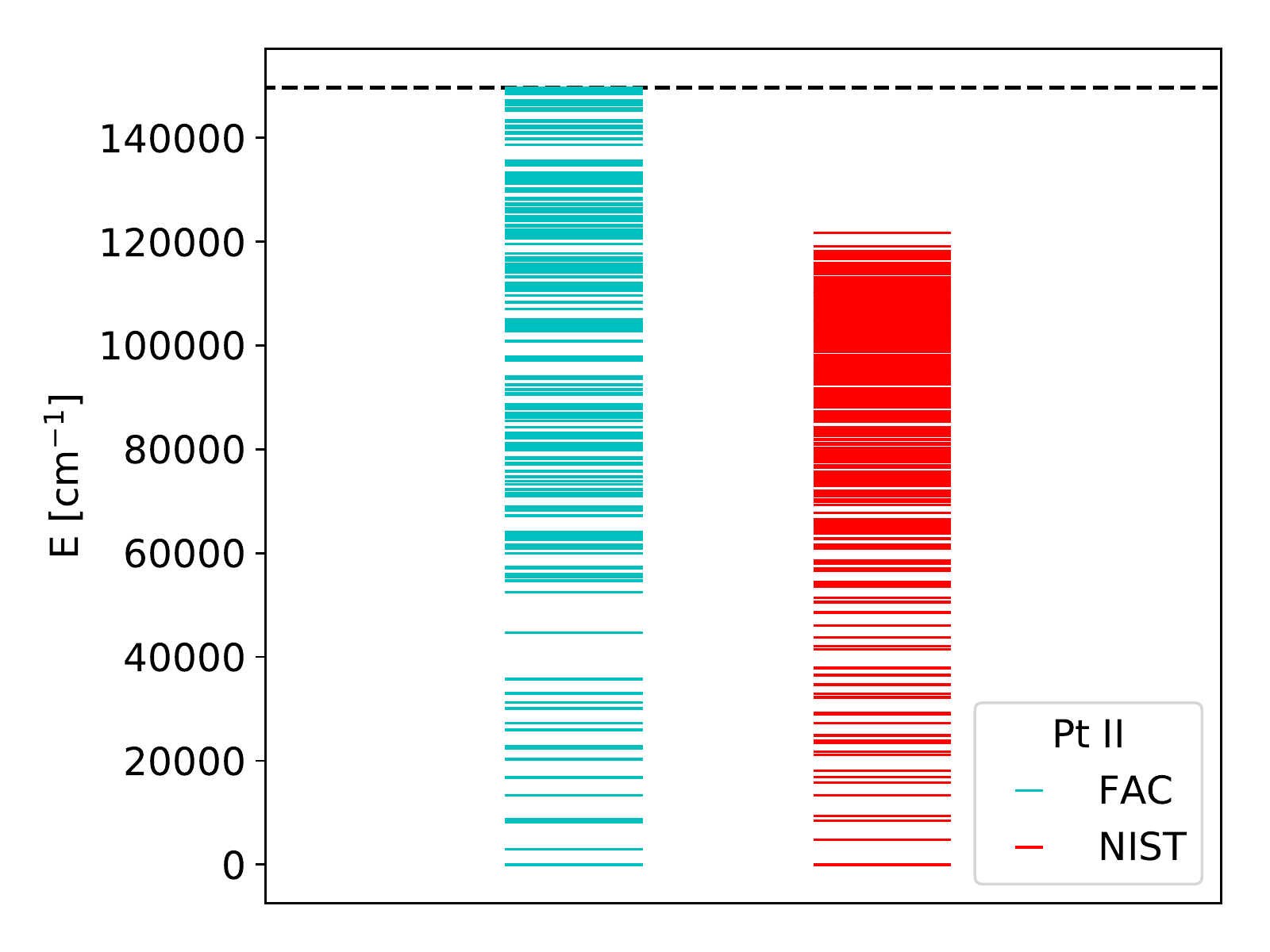}
\includegraphics[trim={0.3cm 0.3cm 0.3cm 0cm},width = 0.32\textwidth]{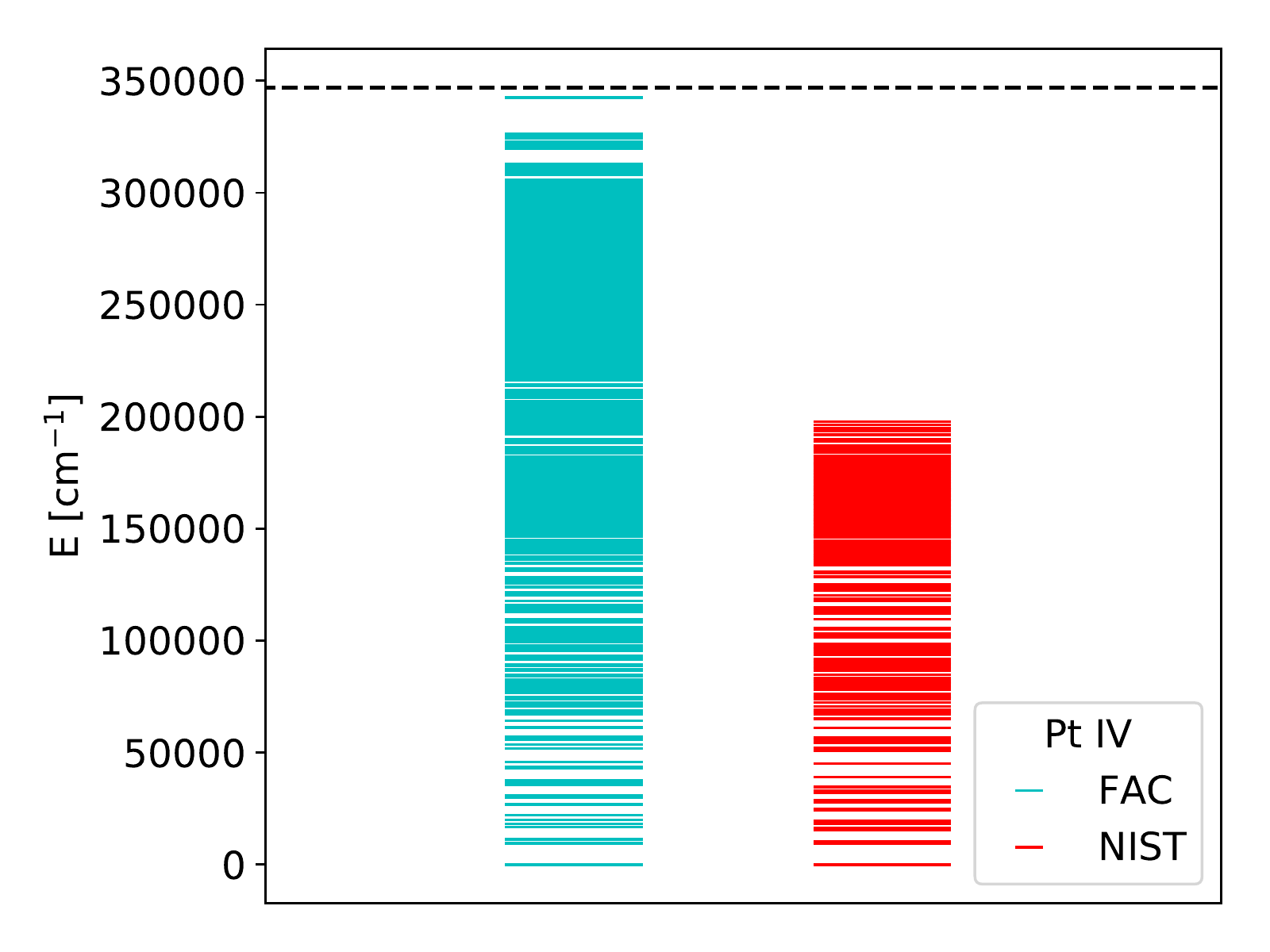}
\includegraphics[trim={0.3cm 0.3cm 0.3cm 0cm},width = 0.32\textwidth]{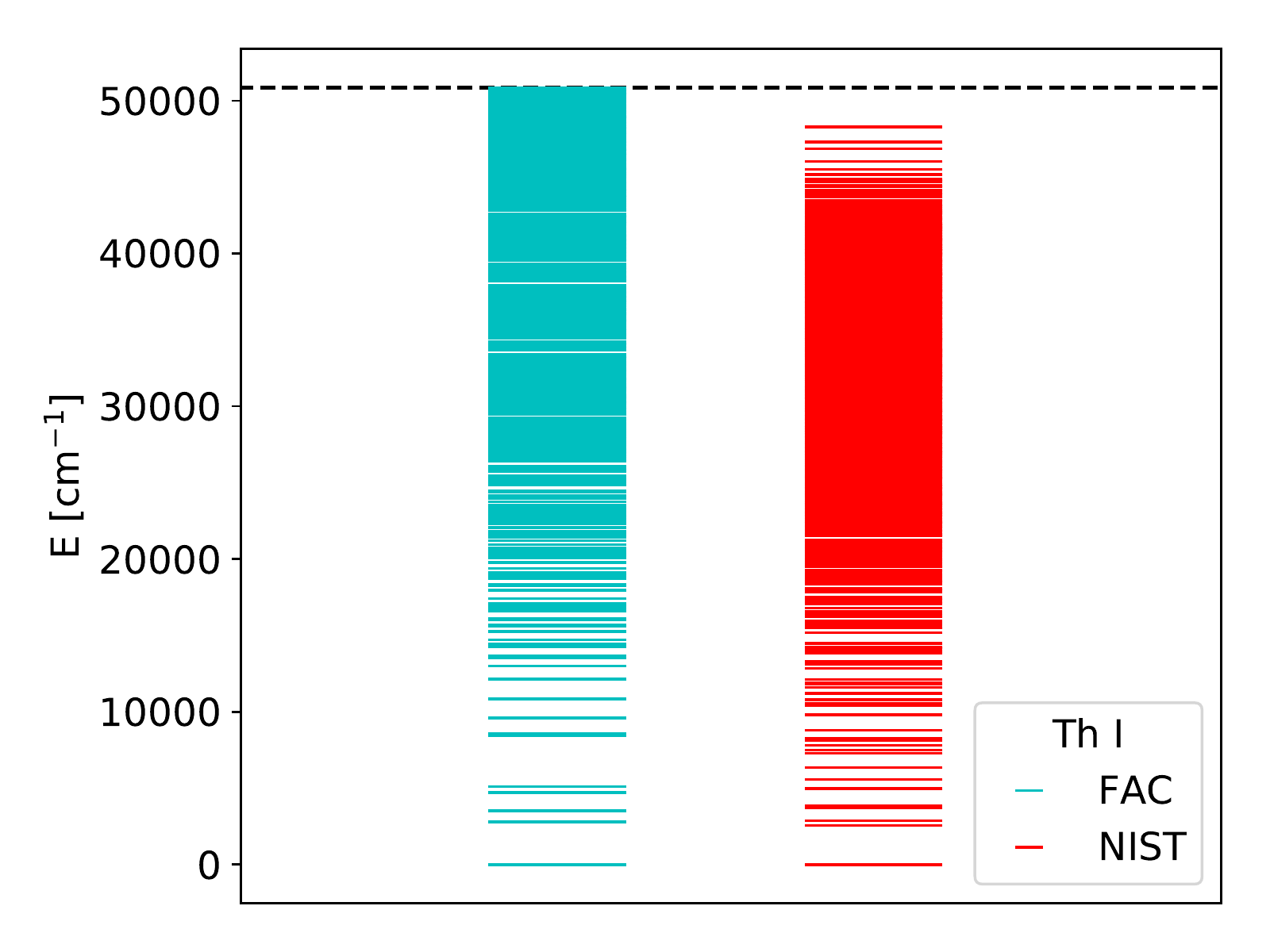}
\includegraphics[trim={0.3cm 0.3cm 0.3cm 0cm},width = 0.32\textwidth]{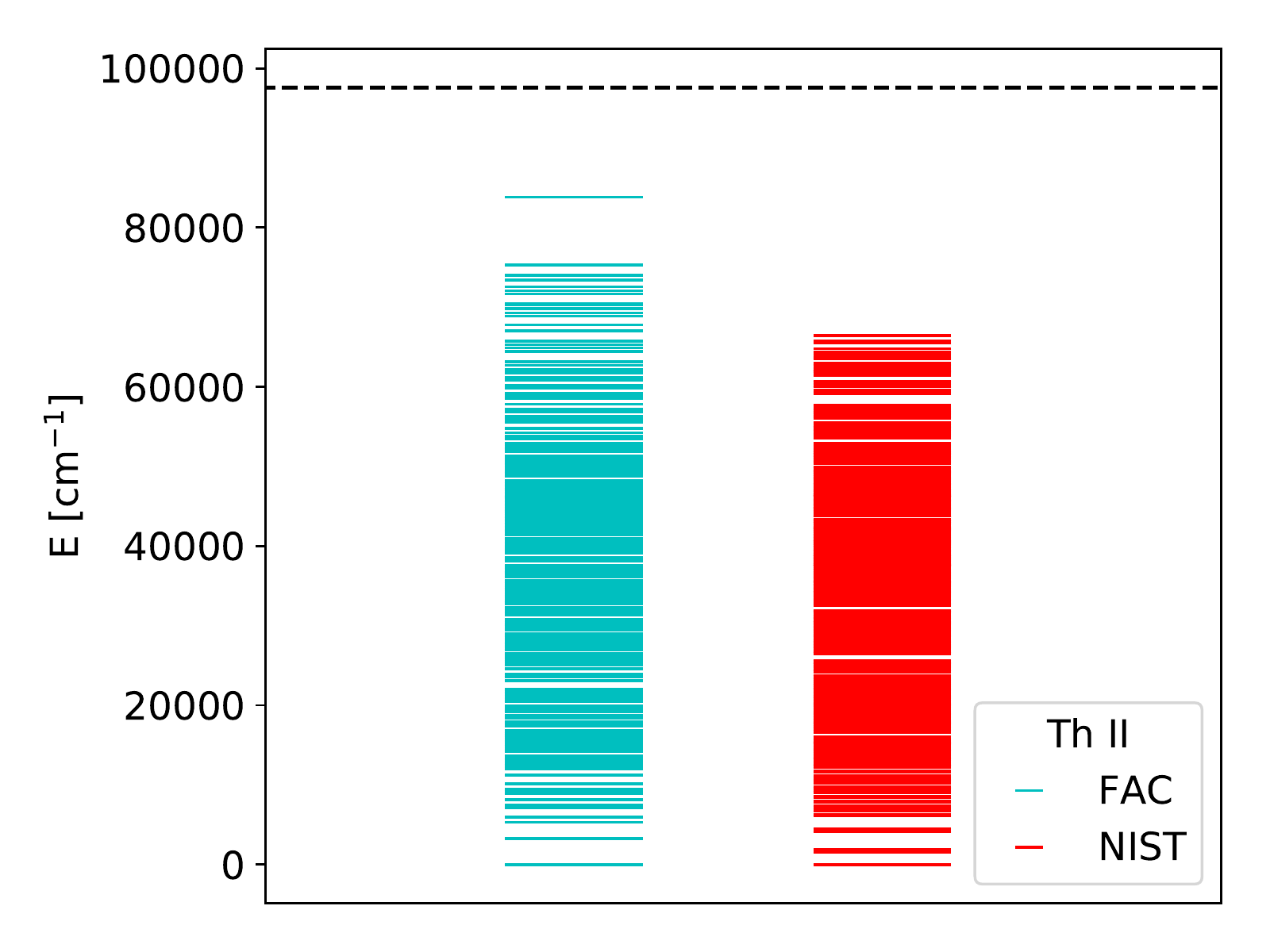}
\includegraphics[trim={0.3cm 0.3cm 0.3cm 0cm},width = 0.32\textwidth]{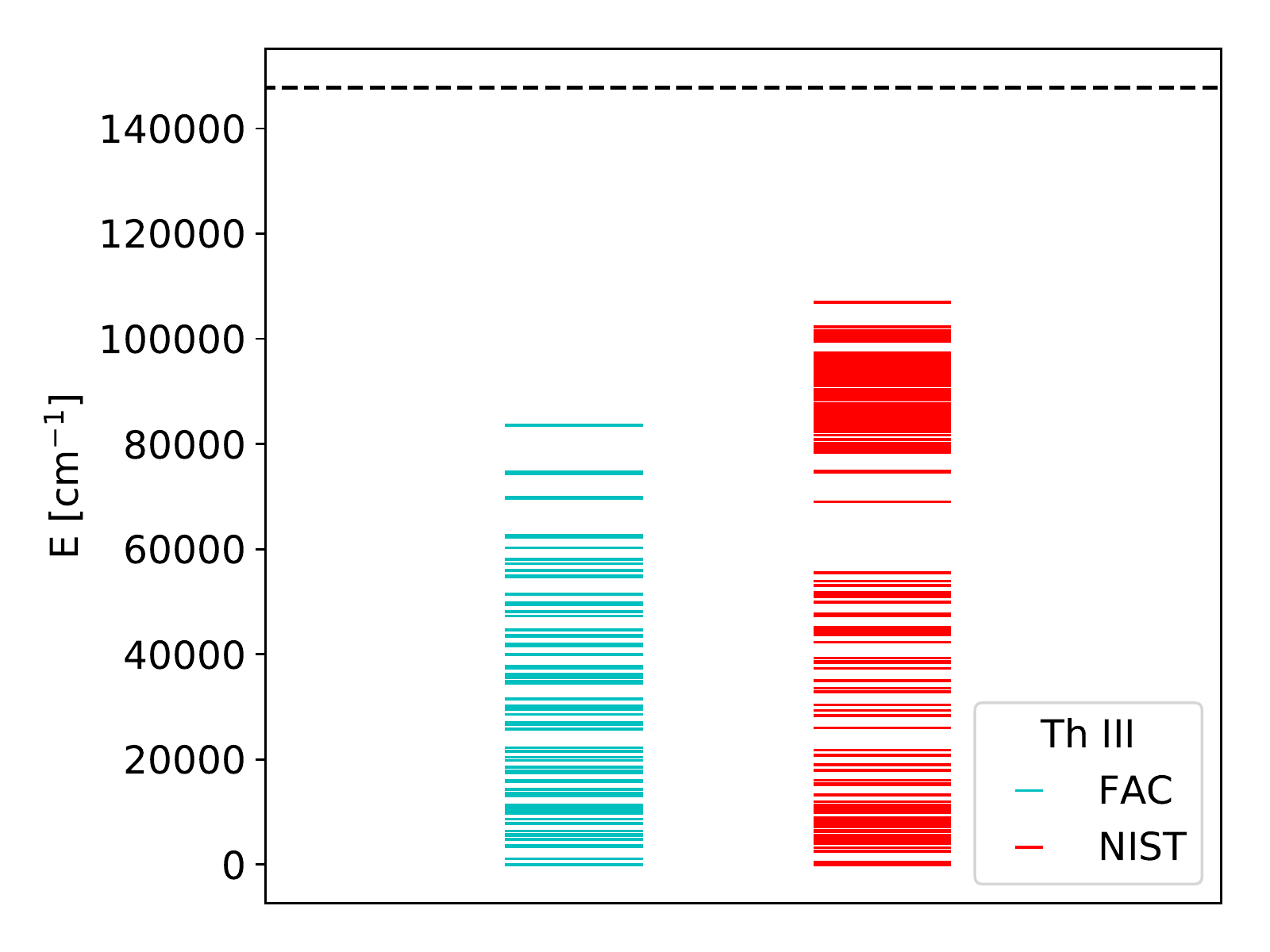}
\caption{Calculated energy levels in cm$^{-1}$ for the ions included in this study (left columns) compared to the data from the NIST ASD (right columns) \citep{NIST_ASD}. The dashed line represents the ionisation limit as given by NIST. In order of descending rows: Te, Ce, Th and Pt, while ionisation level increases from left to right. Note that NIST has no data for Pt III and Th IV, and thus these plots are not included here. The source of the NIST data when available for each ion is presented in Table \ref{tab:data_sources} and the list of included configurations is given in Tab. \ref{tab:configs}}.
\label{fig:levels_all}
\end{figure*}

Energy level plots for all the ions included in this study can be found in Figure \ref{fig:levels_all}. The theoretical energy levels calculated in this work are plotted in the left columns in blue, while the energy levels from the NIST atomic spectra database (ASD) \citep{NIST_ASD} are plotted in the right columns in red, when available. The included theoretical level data are cut at the ionisation limit, also taken from the NIST ASD. We note that energy level data for Pt III and Th IV are missing from the database, and as such are not plotted here. The NIST data come from a variety of sources, which are presented in Table \ref{tab:data_sources}. The electron configurations used for the atomic data calculations with \texttt{FAC} are presented in Table \ref{tab:configs}.

\section{Recombination Coefficients of Light Elements}
\label{app:recombination}

As mentioned in the main body of the paper in Section \ref{subsec:ionandrec}, atomic data for r-process elements and their ions are often crucially missing from experimental databases. Notably, interaction cross sections relating to ionisation and recombination are not readily available for heavy elements, and as such must either be calculated from atomic physics codes, or parametrised in some other way. Here we present the evolution of radiative and dielectronic recombination coefficients for C, N, O, Ne, Mg, Si, S, Ar, Ca, Fe and Ni as compiled by \citet{Arnaud.Rothenflug:85} for a temperature range of 1000 - 60 000 K, covering broadly the range of temperatures relevant to KN as found in this study. These recombination coefficients are fits that have been found from calculations by \citet{Aldrovandi.Pequignot:73,Nussbaumer.Story:83}, with some individual calculations and corrections made by \citet{Arnaud.Rothenflug:85} themselves.

\begin{figure*}
\center
\includegraphics[trim={0.3cm 0.0cm 0.3cm 0cm},width = 0.32\textwidth]{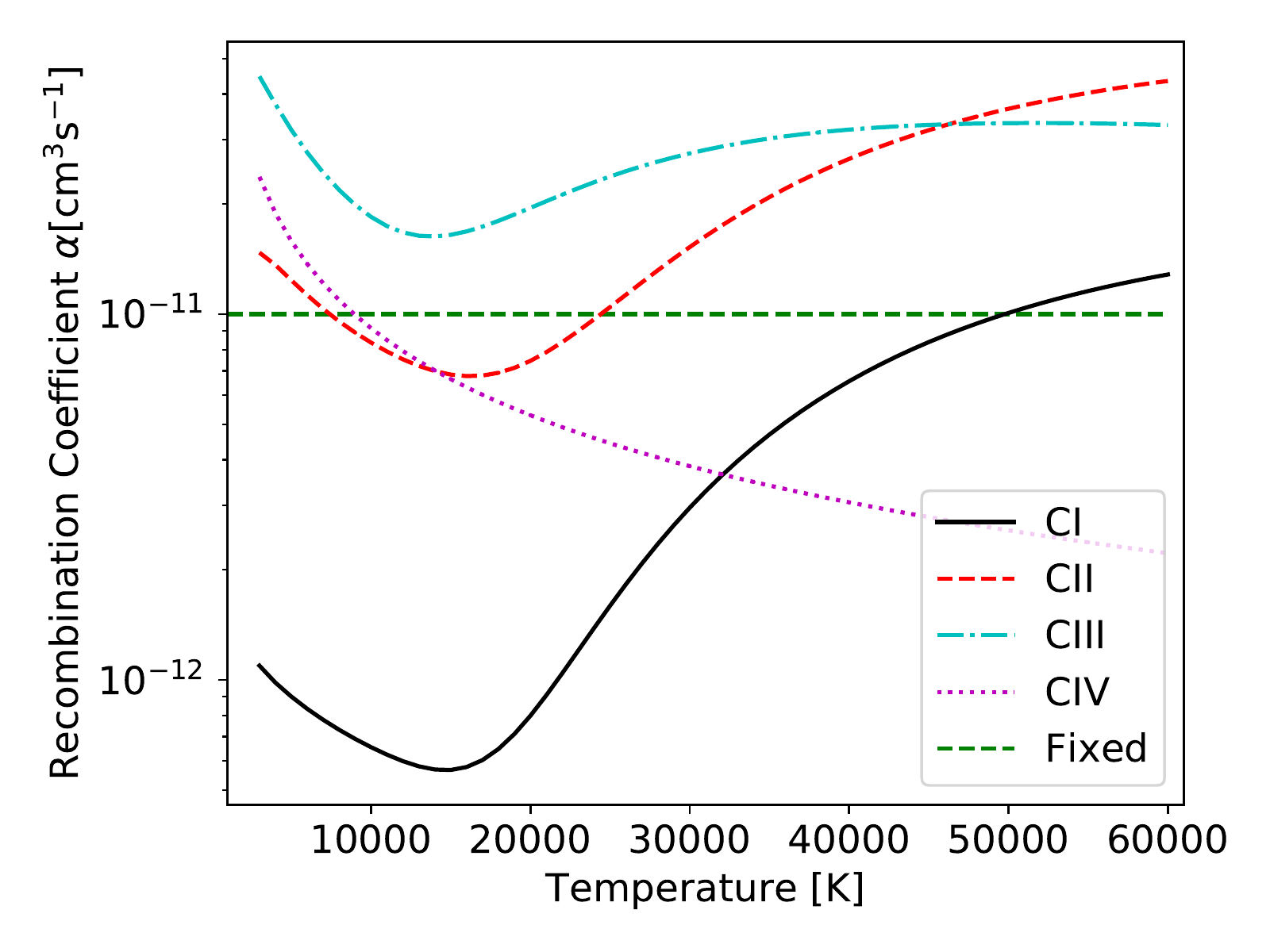}
\includegraphics[trim={0.3cm 0.0cm 0.3cm 0cm},width = 0.32\textwidth]{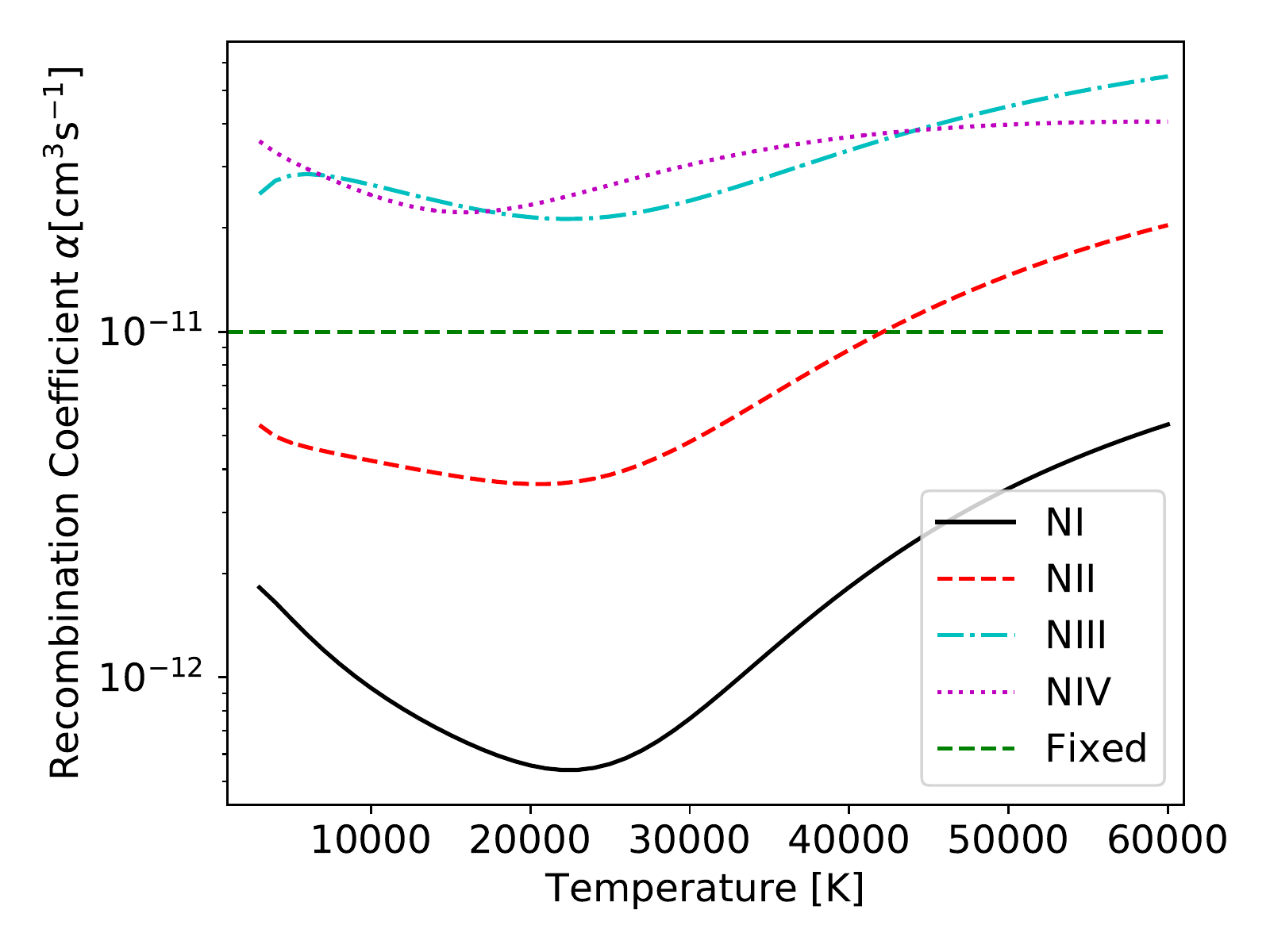}
\includegraphics[trim={0.3cm 0.0cm 0.3cm 0cm},width = 0.32\textwidth]{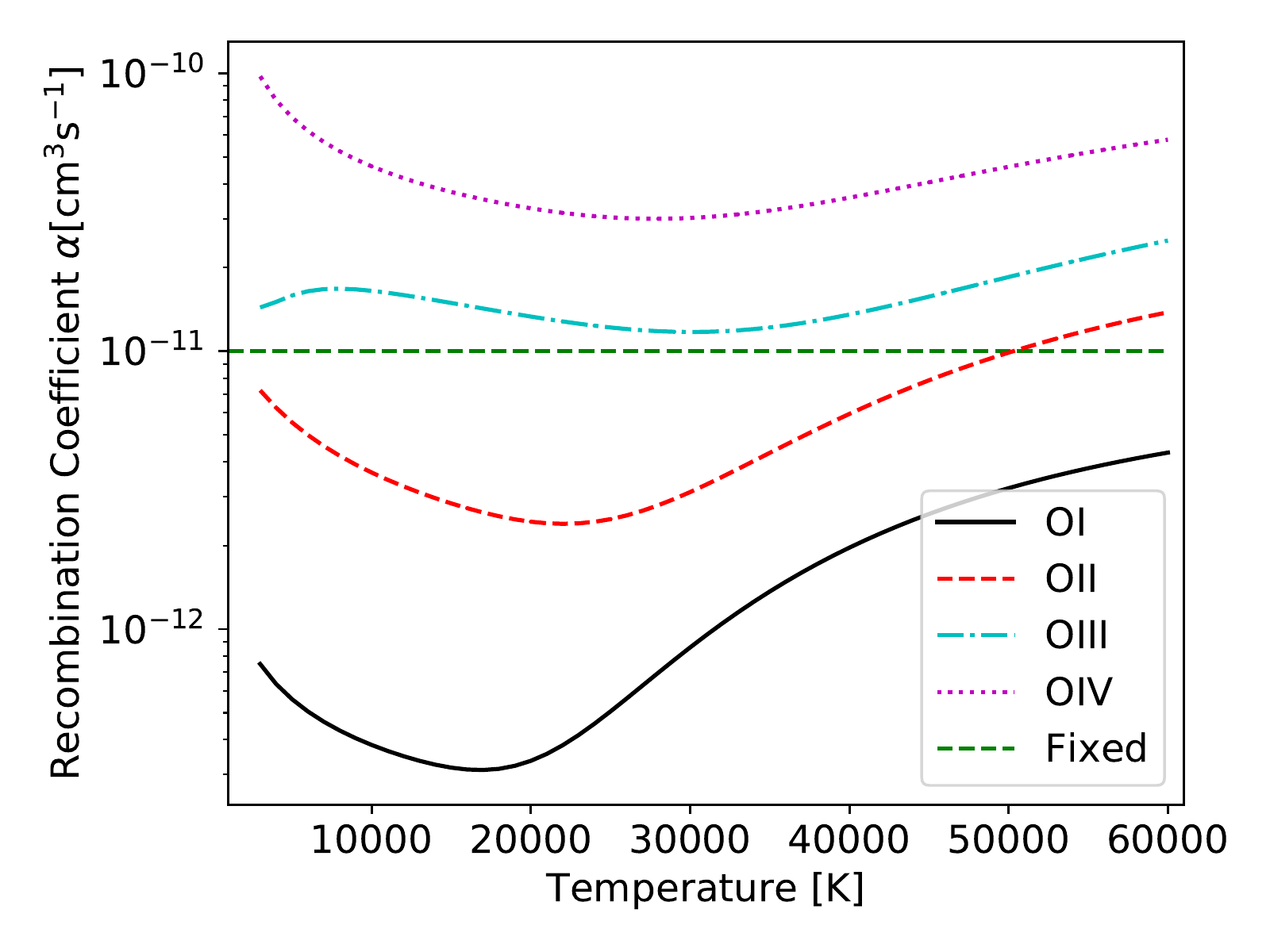}
\includegraphics[trim={0.3cm 0.0cm 0.3cm 0cm},width = 0.32\textwidth]{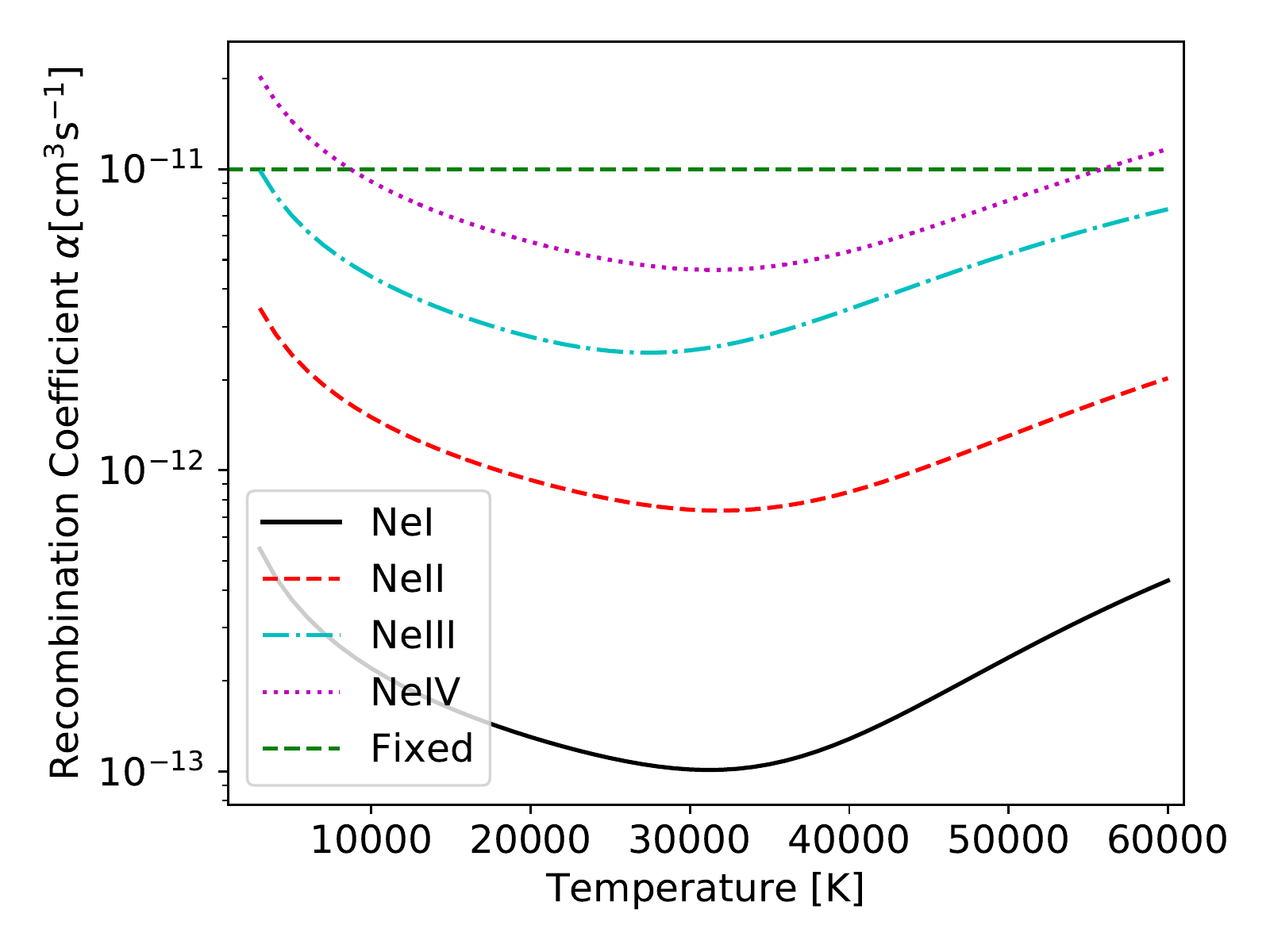}
\includegraphics[trim={0.3cm 0.0cm 0.3cm 0cm},width = 0.32\textwidth]{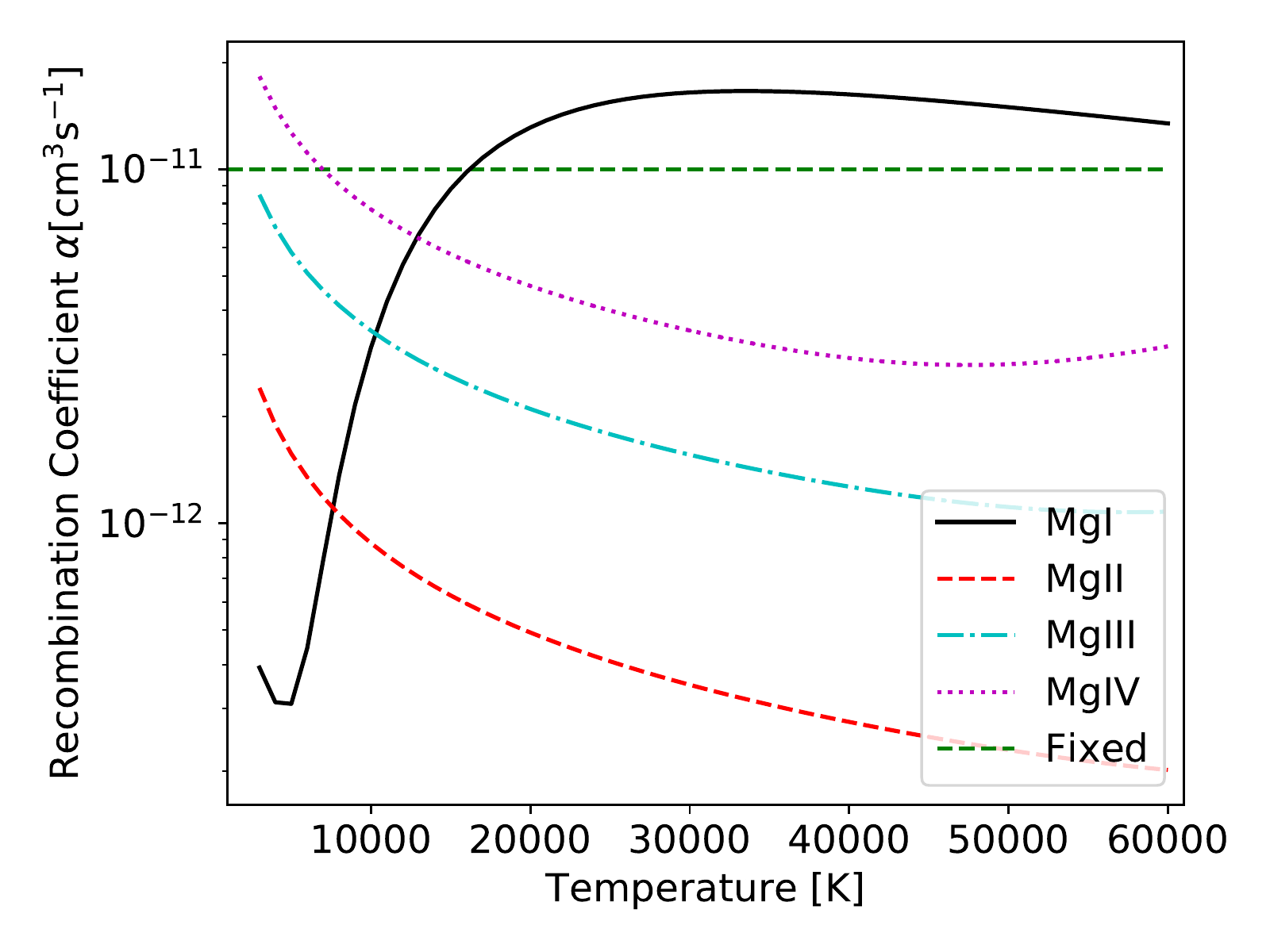}
\includegraphics[trim={0.3cm 0.0cm 0.3cm 0cm},width = 0.32\textwidth]{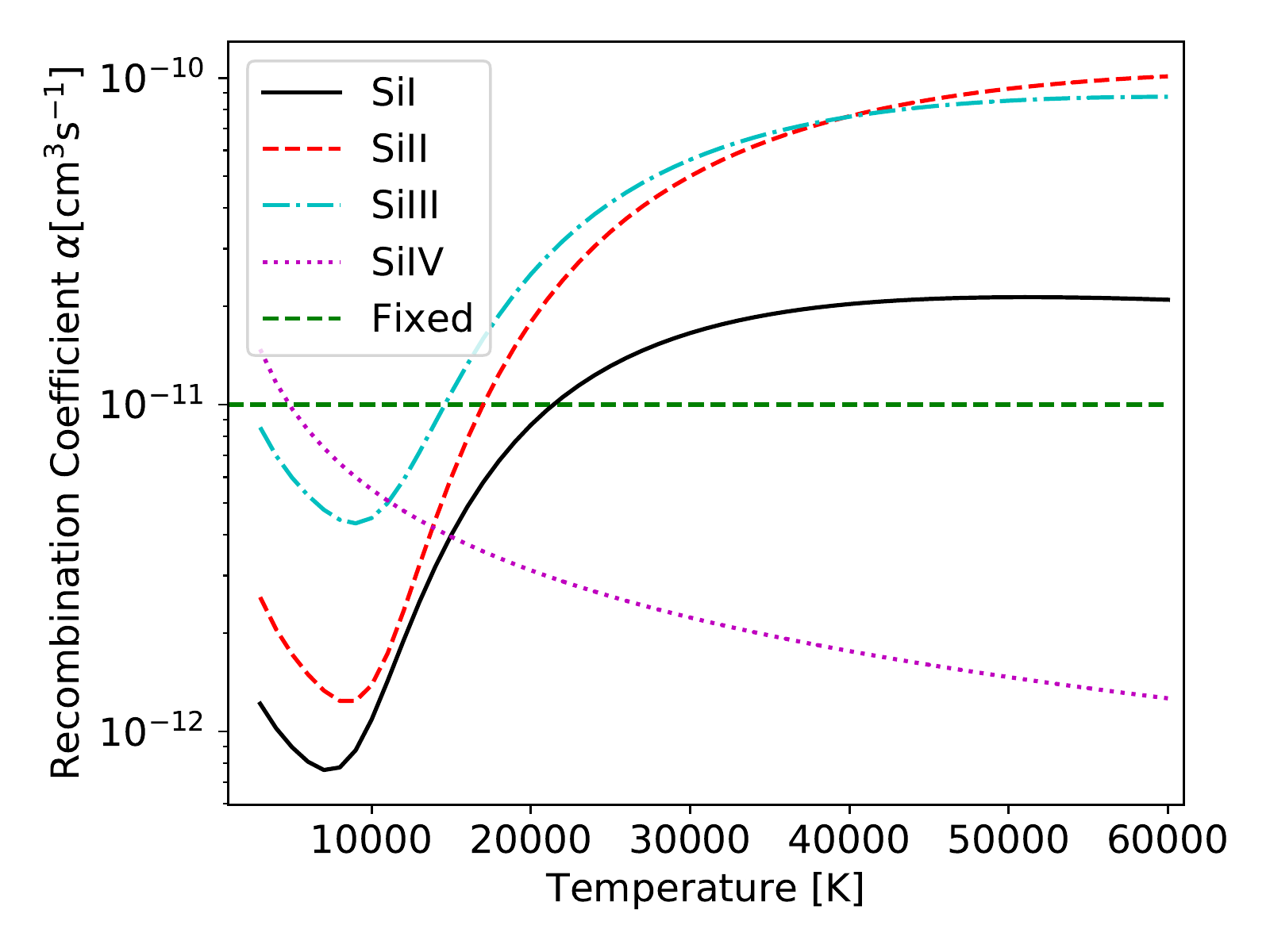}
\includegraphics[trim={0.3cm 0.0cm 0.3cm 0cm},width = 0.32\textwidth]{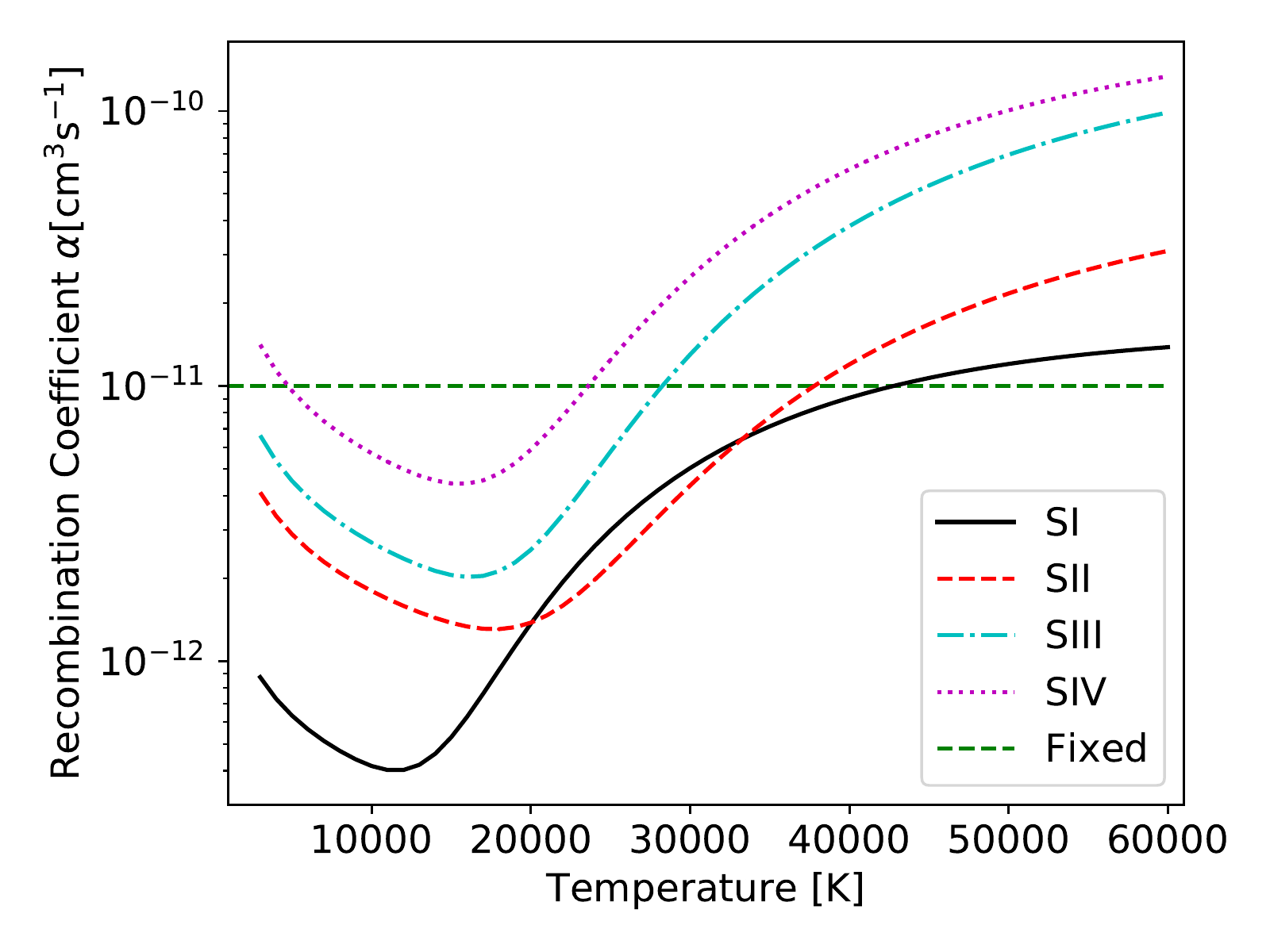}
\includegraphics[trim={0.3cm 0.0cm 0.3cm 0cm},width = 0.32\textwidth]{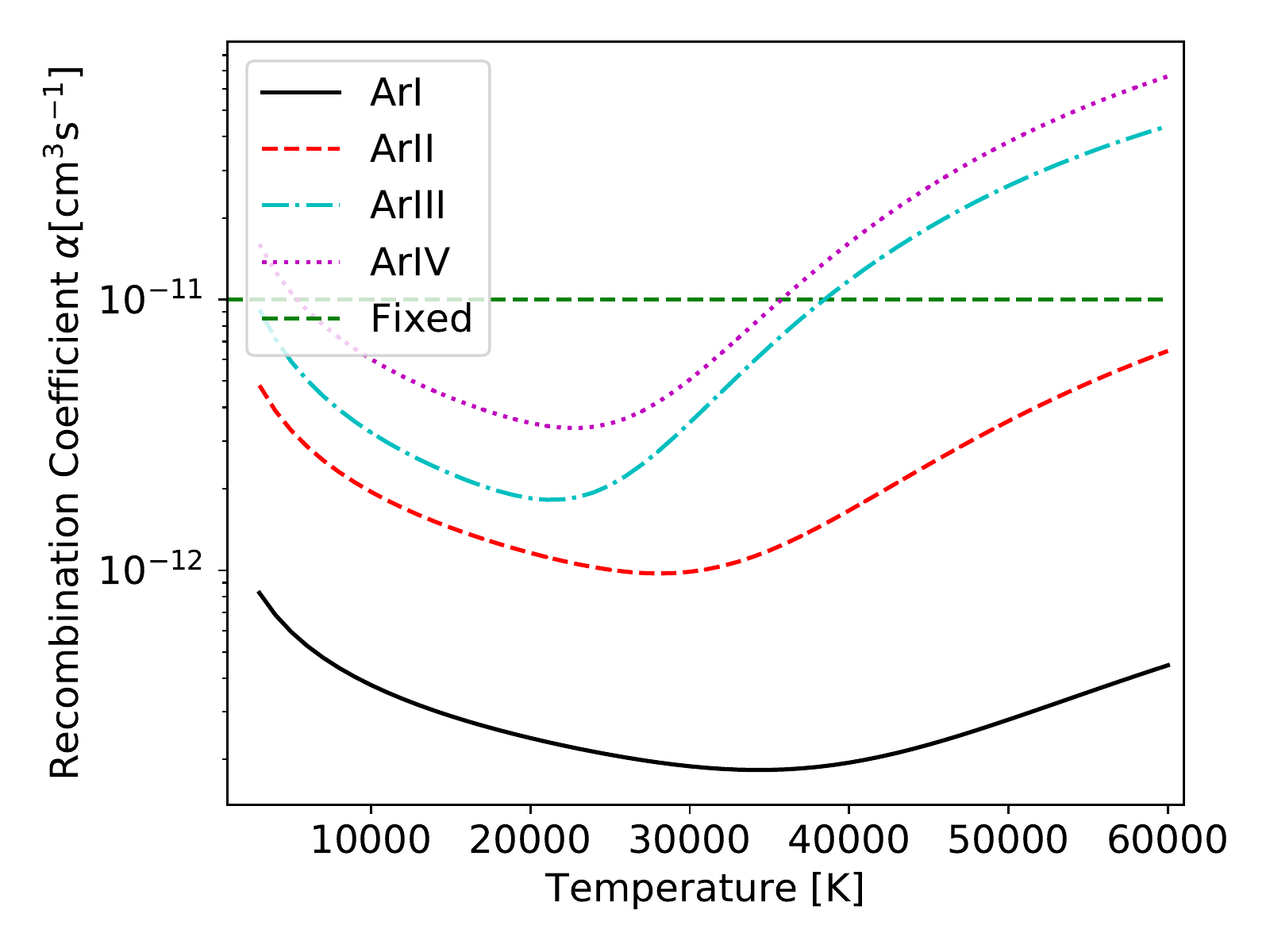}
\includegraphics[trim={0.3cm 0.0cm 0.3cm 0cm},width = 0.32\textwidth]{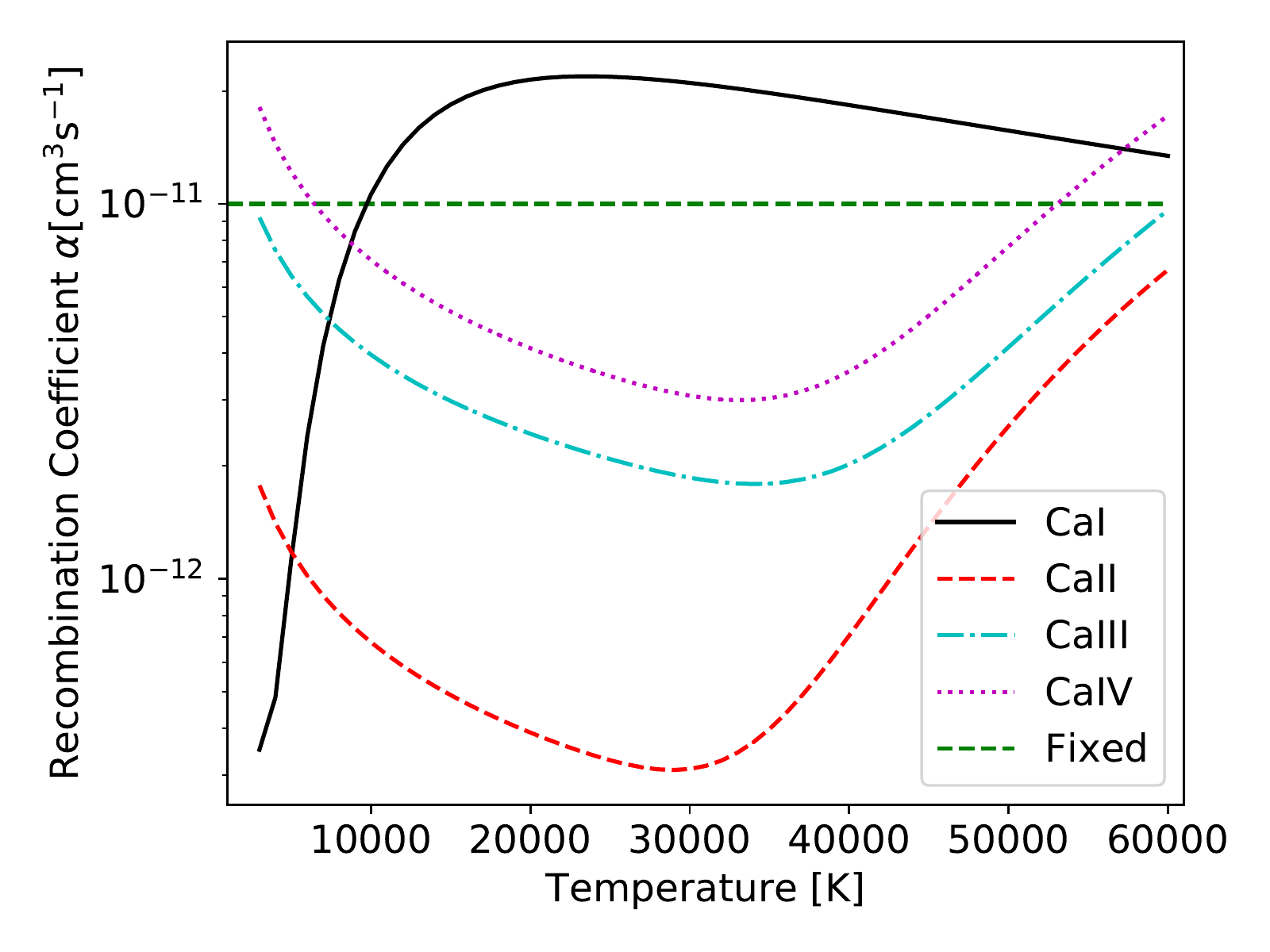}
\includegraphics[trim={0.3cm 0.0cm 0.3cm 0cm},width = 0.32\textwidth]{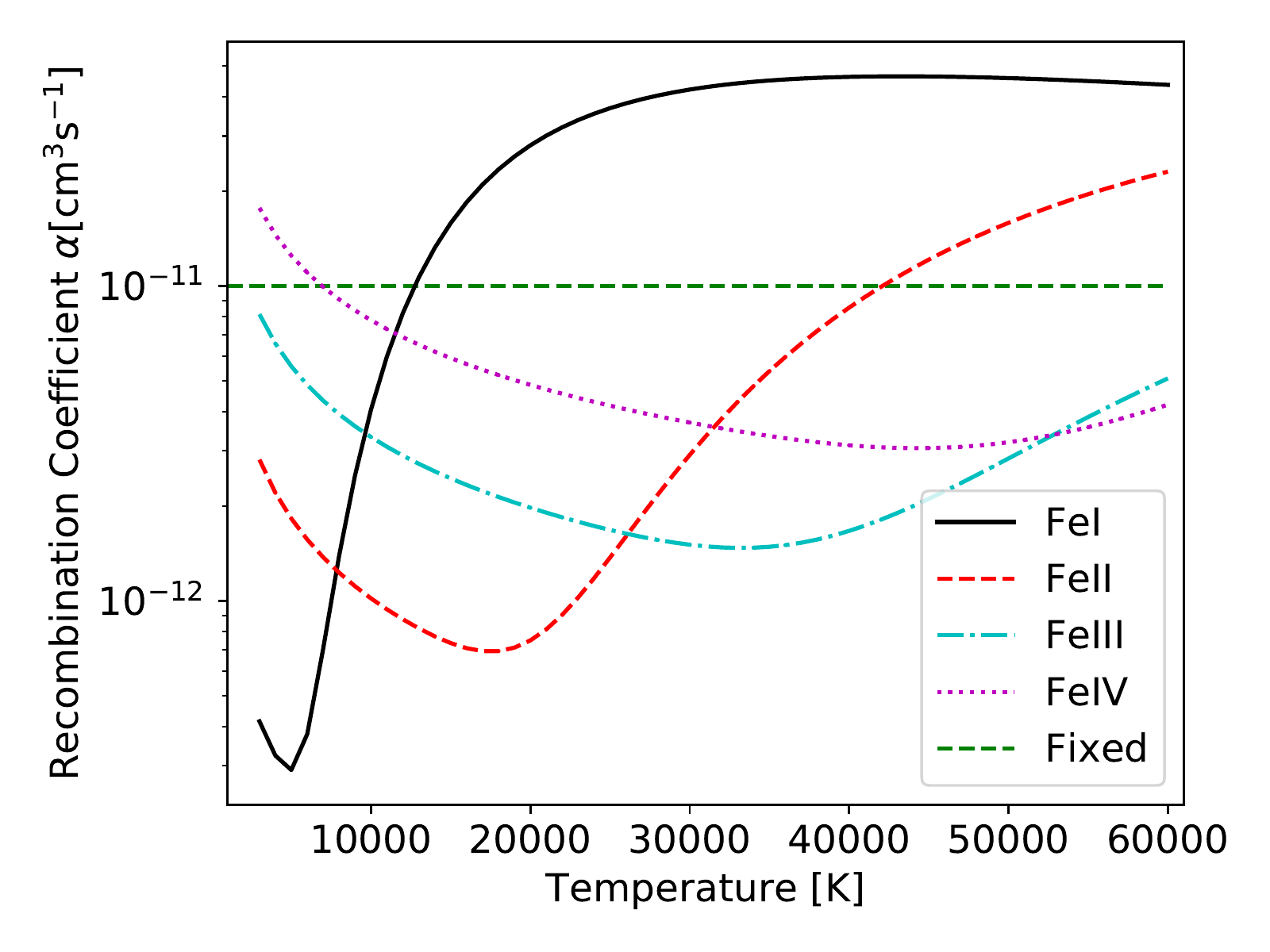}
\includegraphics[trim={0.3cm 0.0cm 0.3cm 0cm},width = 0.32\textwidth]{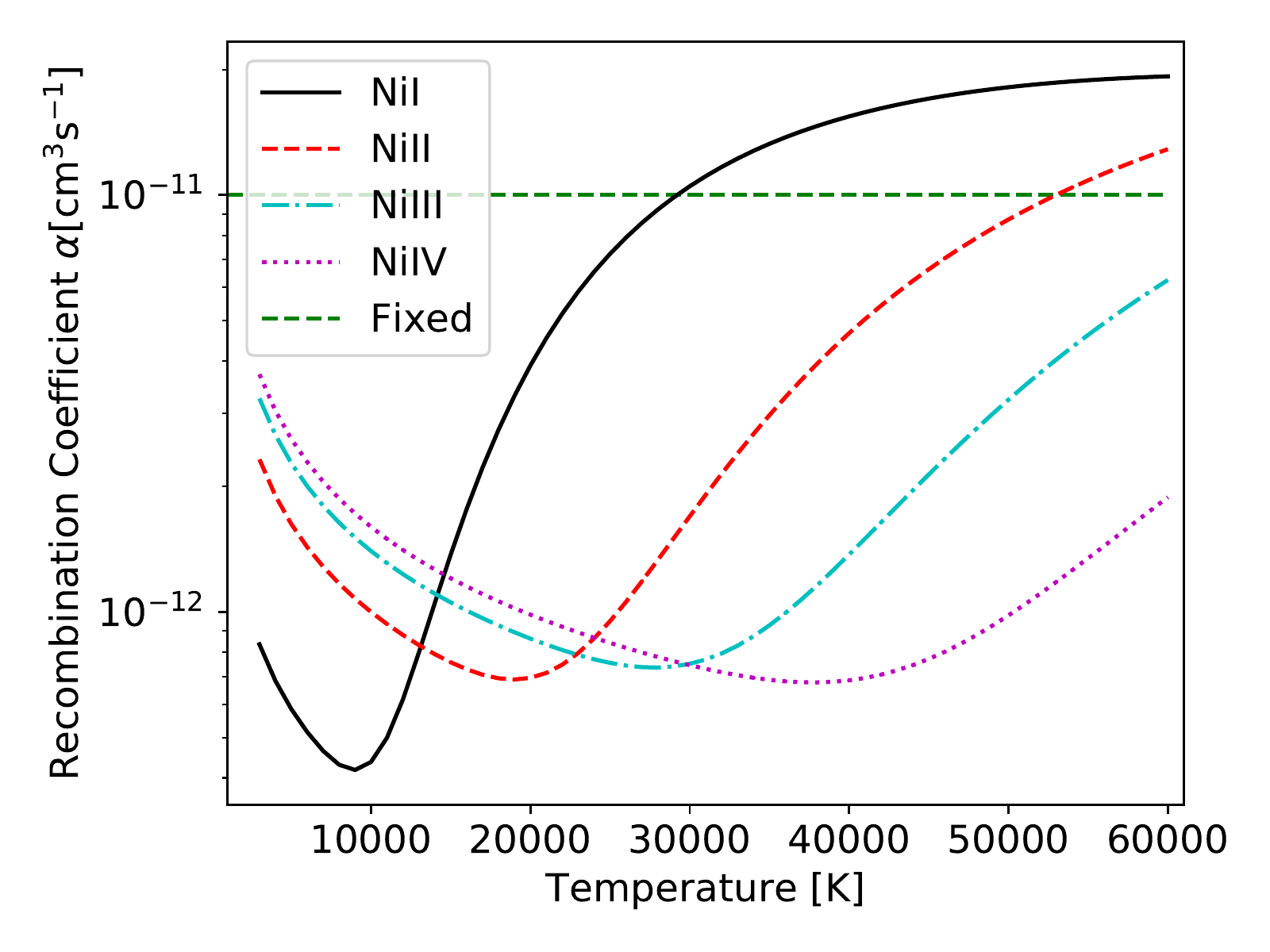}
\caption{Recombination coefficients for light elements considered when determining the treatment of recombination for the r-process elements as compiled by \citet{Arnaud.Rothenflug:85}. The dashed green horizontal line represents our fiducial choice of a constant $\alpha = 10^{-11} \; \mathrm{cm^3 \: s^{-1}}$. }
\label{fig:recrates}
\end{figure*}

These plots are presented in Figure \ref{fig:recrates}, and highlight the importance of shell properties on recombination coefficients. Following the complex evolution of combination coefficients for lighter elements, we decide to adopt a fixed value of $\alpha = 10^{-11} \; \mathrm{cm^3 \: s^{-1}}$. Although this value may over or underestimate recombination for certain ions and temperature ranges, it does appear to provide a decent overall average for the temperatures under consideration.

\section{Time-Step Testing}
\label{app:time-step}

\begin{figure}
    \centering
    \includegraphics[trim={0.5cm 0.5cm 1.0cm 0.5cm},width=\linewidth]{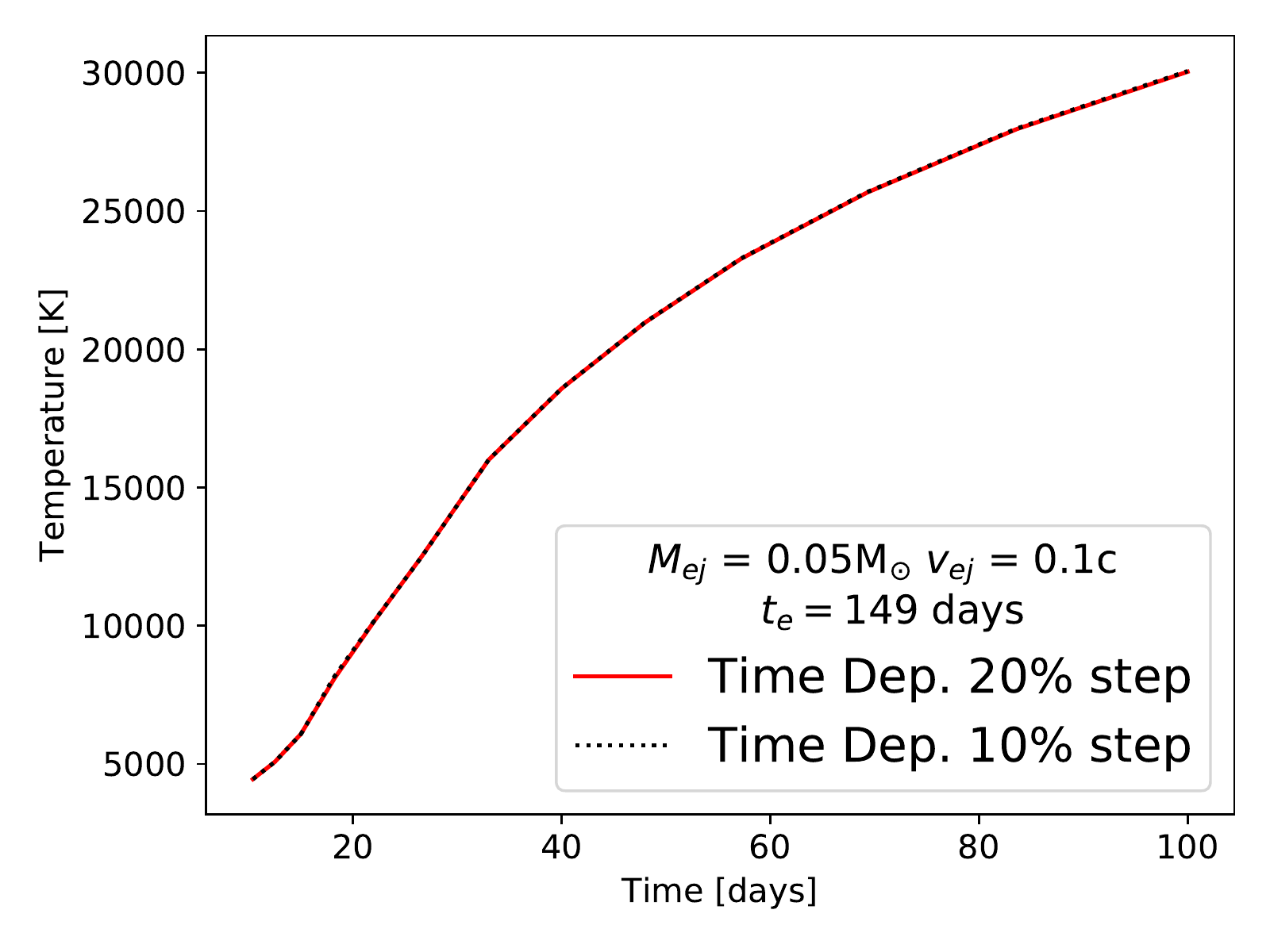}
    \caption{Time-step testing: Temperature evolution for $M_{\mathrm{ej}} = 0.05\mathrm{\Msol}$, $v_{\mathrm{ej}} = 0.1$c when run with 10 and 20 per cent time-steps in time-dependent mode. We note that the time-dependent solutions are almost identical regardless of time-step. As such, since the temperature solutions with $\Delta t = 0.2t_d$ converge accurately to the same solution as the smaller time-step, the solution of the smaller 10 per cent time-step can be taken to be accurate.}
    \label{fig:temperature_bigstep}
\end{figure}

In order to verify the stability and accuracy of the time-steps used in time-dependent mode, we ran our fiducial 'standard' model with $M_{\mathrm{ej}} = 0.05~\mathrm{\Msol}$, $v_{\mathrm{ej}} = 0.1$c in time-dependent mode using both 10 and 20 per cent time-steps, i.e. $\Delta t = 0.1, 0.2 t_d$. The results for these runs are plotted in Figures \ref{fig:temperature_bigstep} and \ref{fig:ionfrac_bigstep} below. We note that the solutions are identical to within less than 1 per cent for both temperature and ionisation, and so are overlapping almost exactly when plotted. Since the larger time-step reproduces the same results as the smaller 10 per cent time-step, the value of $\Delta t = 0.1t_d$ is acceptably accurate and is used throughout this study.

\begin{figure*}
\center
\includegraphics[trim={0.2cm 0.2cm 1.6cm 0.8cm},clip,width = 0.49\textwidth]{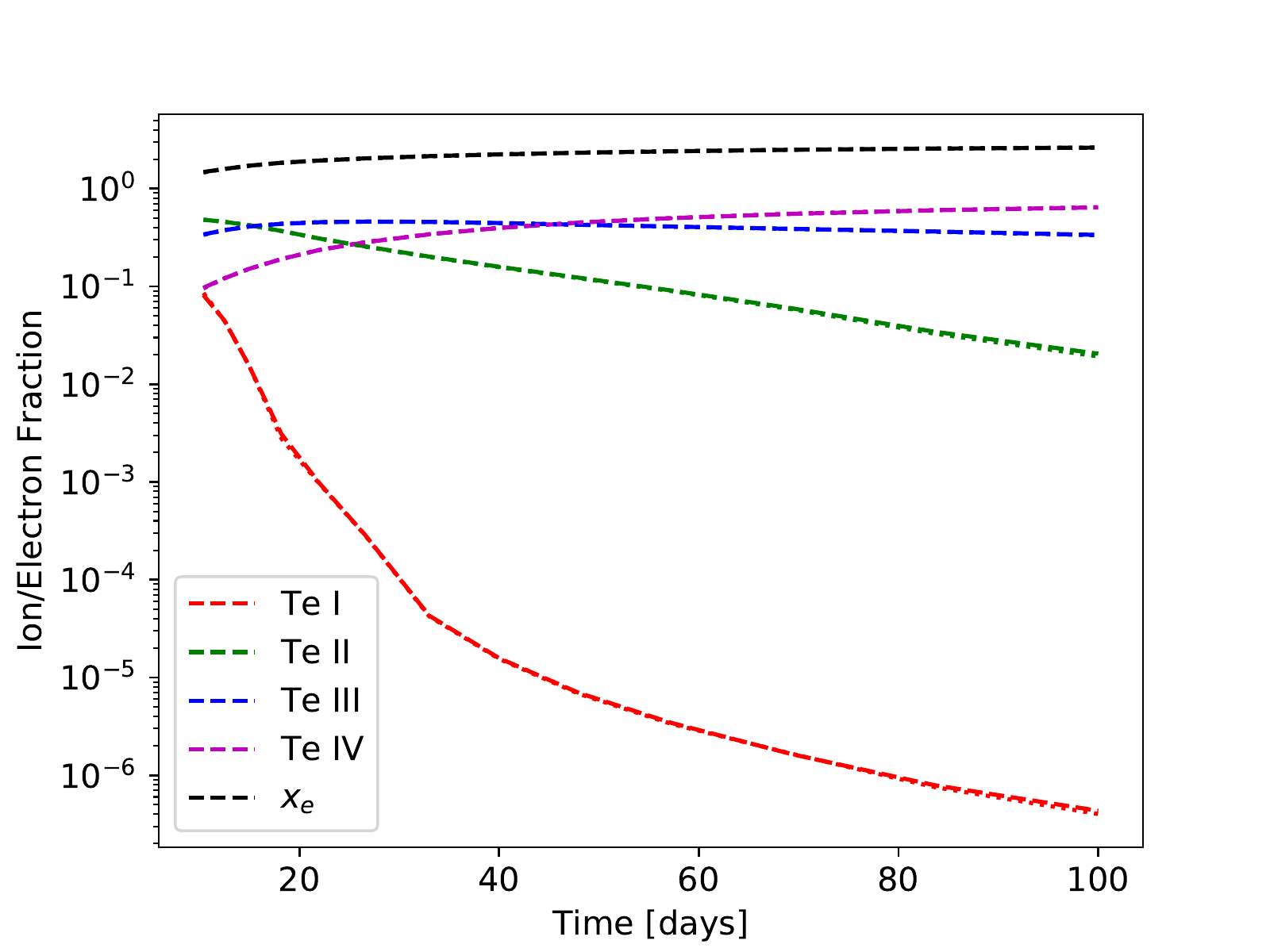}
\includegraphics[trim={0.2cm 0.2cm 1.6cm 0.8cm},clip,width = 0.49\textwidth]{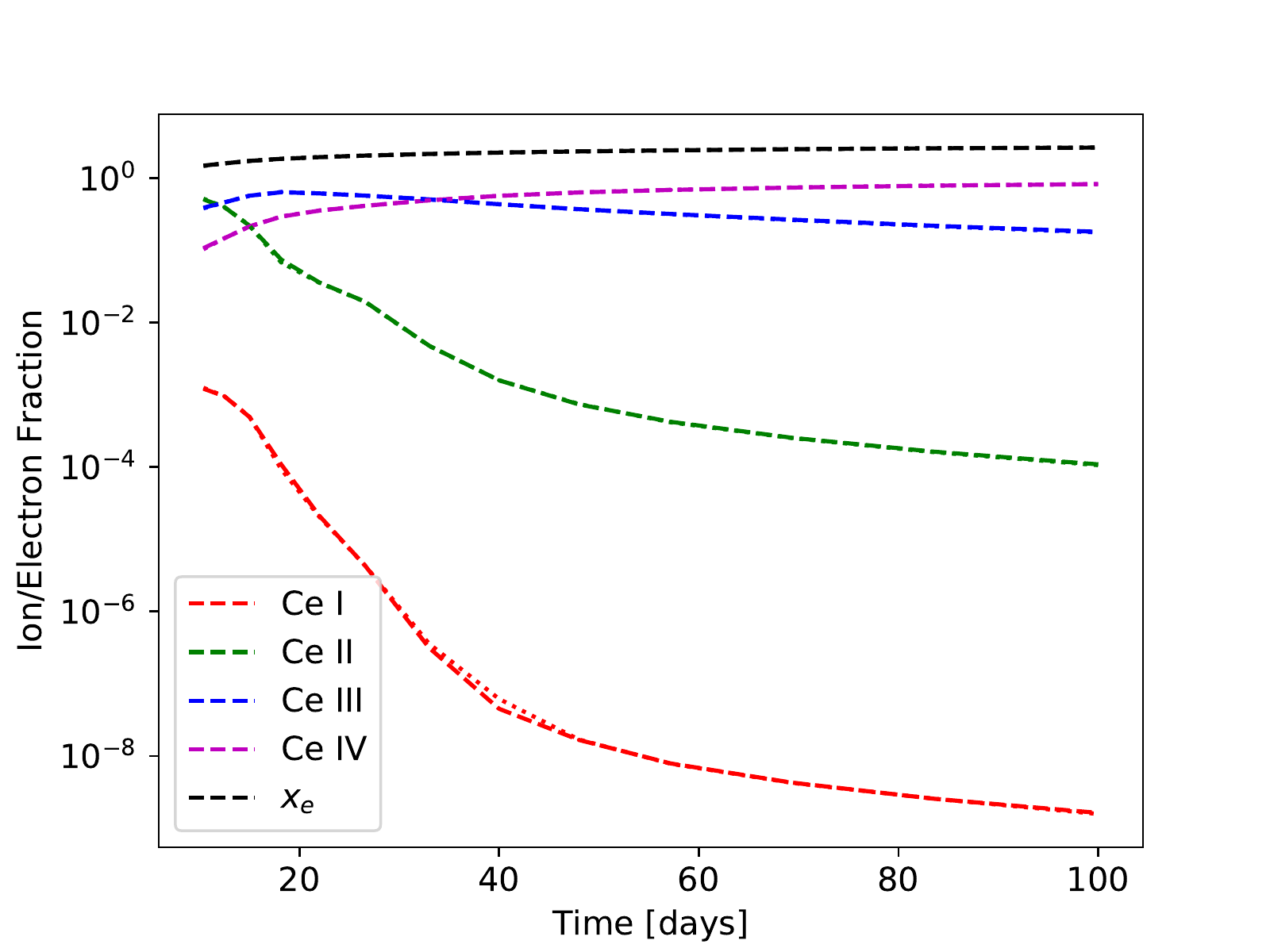}
\includegraphics[trim={0.2cm 0.2cm 1.6cm 0.8cm},clip,width = 0.49\textwidth]{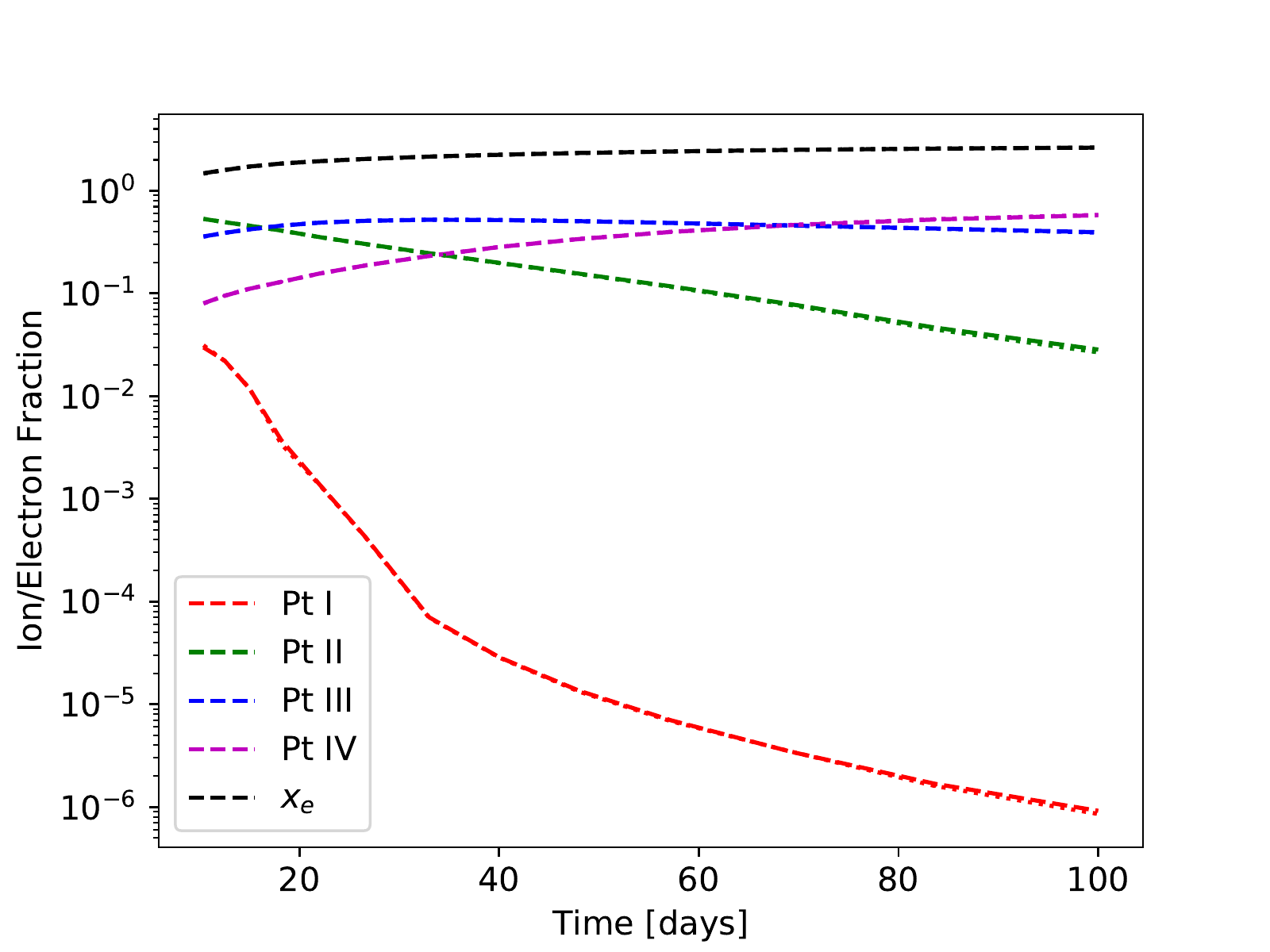}
\includegraphics[trim={0.2cm 0.2cm 1.6cm 0.8cm},clip,width = 0.49\textwidth]{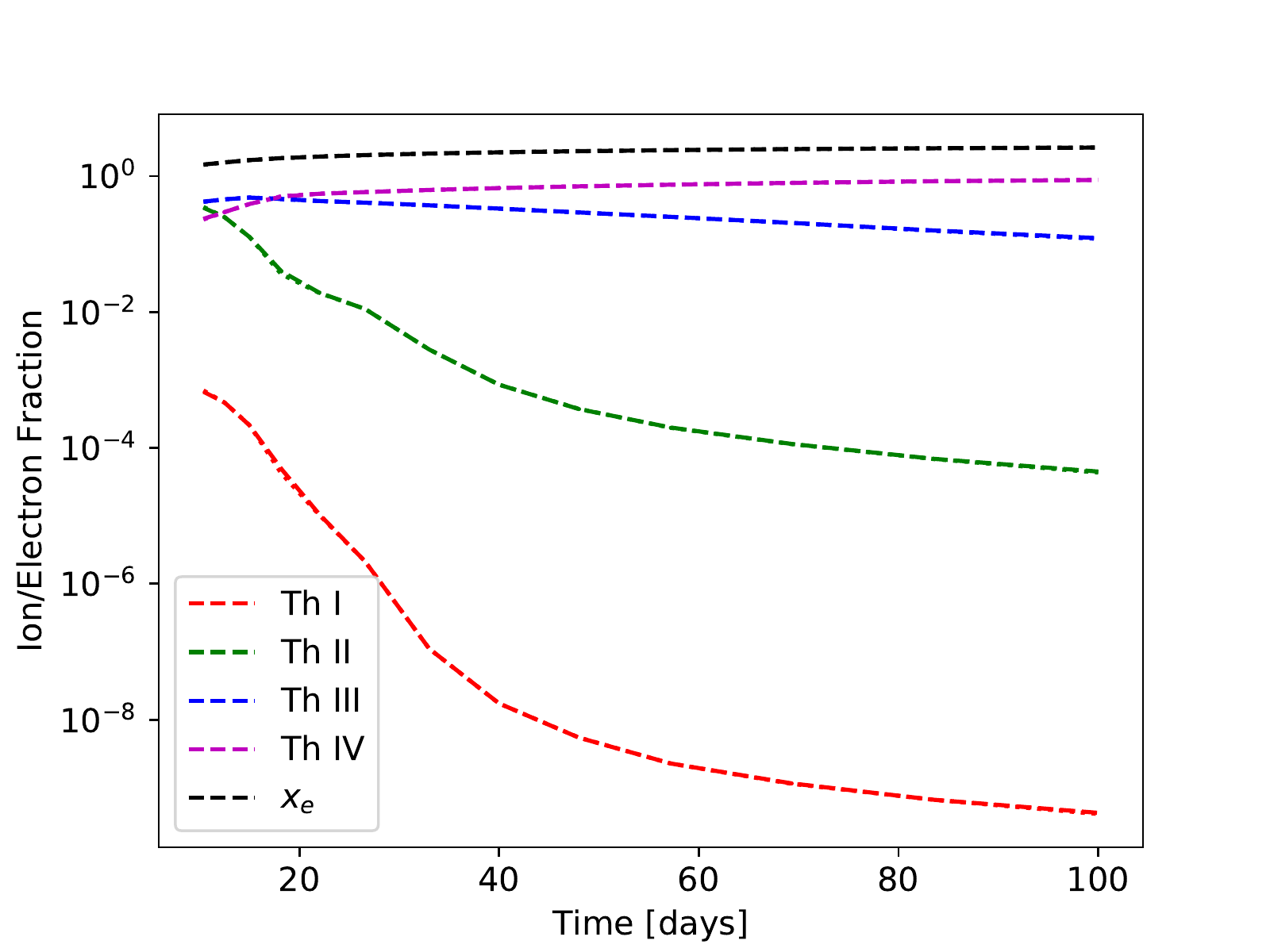}
\caption{Time-step testing: ionisation structure evolution for $M_{\mathrm{ej}} = 0.05~\mathrm{\Msol}$, $v_{\mathrm{ej}} = 0.1$c when run with 10 and 20 per cent time-steps in time-dependent mode. The dashed lines are the solutions with $\Delta t = 0.2 t_d$, while the dotted lines are with the smaller 10 per cent time-step. We note that the time-dependent solutions are almost identical and the curves overlap. As such, since the ionisation structure solutions with $\Delta t = 0.2t_d$ converge accurately to the same solution as the smaller time-step, the solution of the smaller 10 per cent time-step can be taken to be accurate.}
\label{fig:ionfrac_bigstep}
\end{figure*}

\section{Supporting material}
\label{app:results}

\begin{figure*}
    \centering
    \includegraphics[trim={0.2cm 0.2cm 0.4cm 0.3cm},clip,width = 0.49\textwidth]{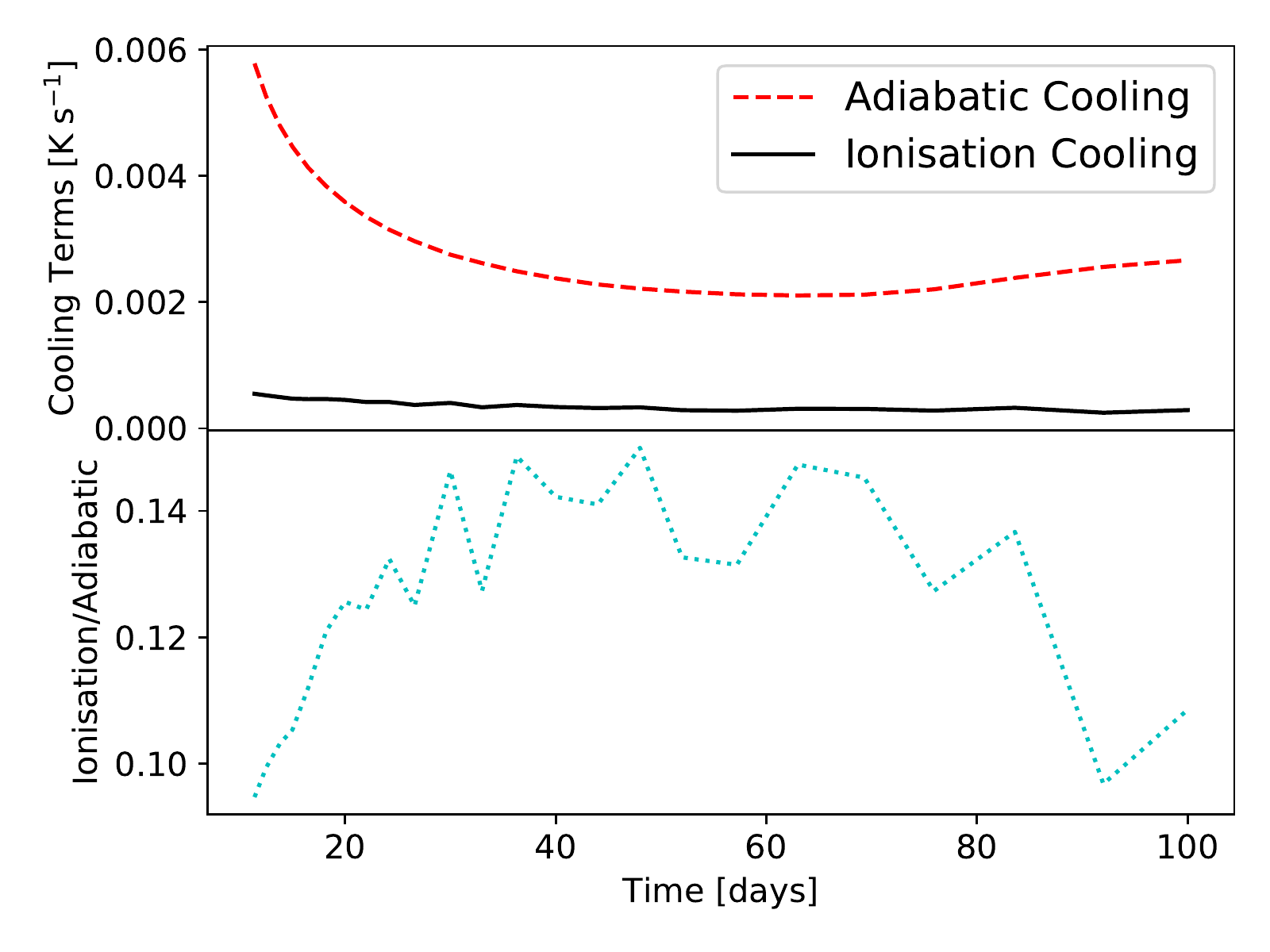}
    \includegraphics[trim={0.2cm 0.2cm 0.4cm 0.3cm},clip,width = 0.49\textwidth]{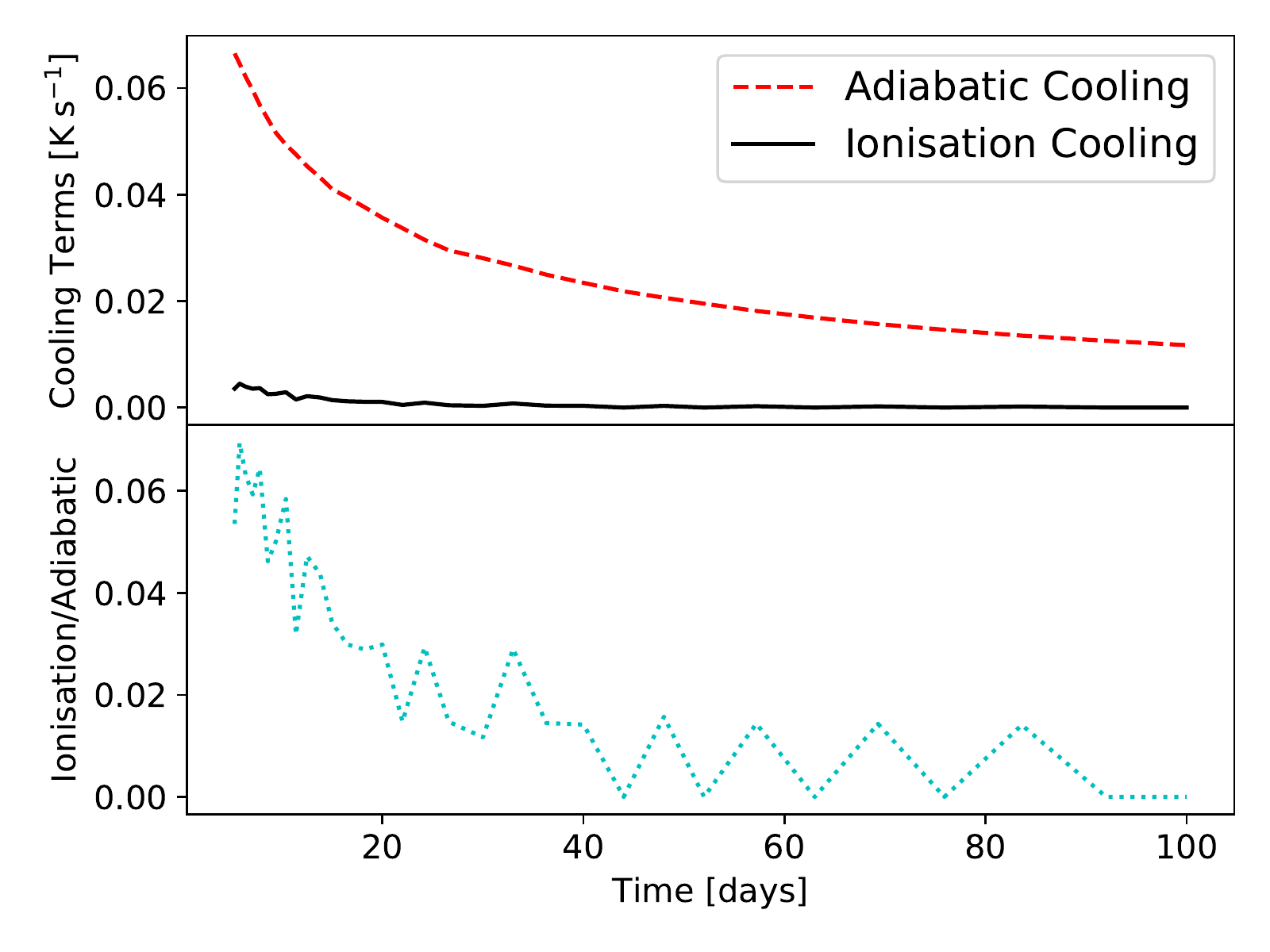}
    \caption{The importance of the ionisation cooling term vs. the adiabatic cooling for the highest (left panel, $M_{\mathrm{ej}} = 0.1\mathrm{\Msol}$, $v_{\mathrm{ej}} = 0.05$c) and lowest (right panel, $M_{\mathrm{ej}} = 0.01\mathrm{\Msol}$, $v_{\mathrm{ej}} = 0.2$c) density models respectively. The ionisation cooling term is always smaller than the adiabatic cooling term, often significantly. In the case of the densest model, the ionisation cooling appears to be non-negligible compared to adiabatic cooling, however since adiabatic cooling was at most $\sim 0.1$ per cent of total cooling in that model, the ionisation cooling is even less significant with respect to total cooling.}
    \label{fig:ioncool}
\end{figure*}

\begin{figure*}
\center
\includegraphics[trim={0.2cm 0.2cm 1.6cm 0.8cm},clip,width = 0.49\textwidth]{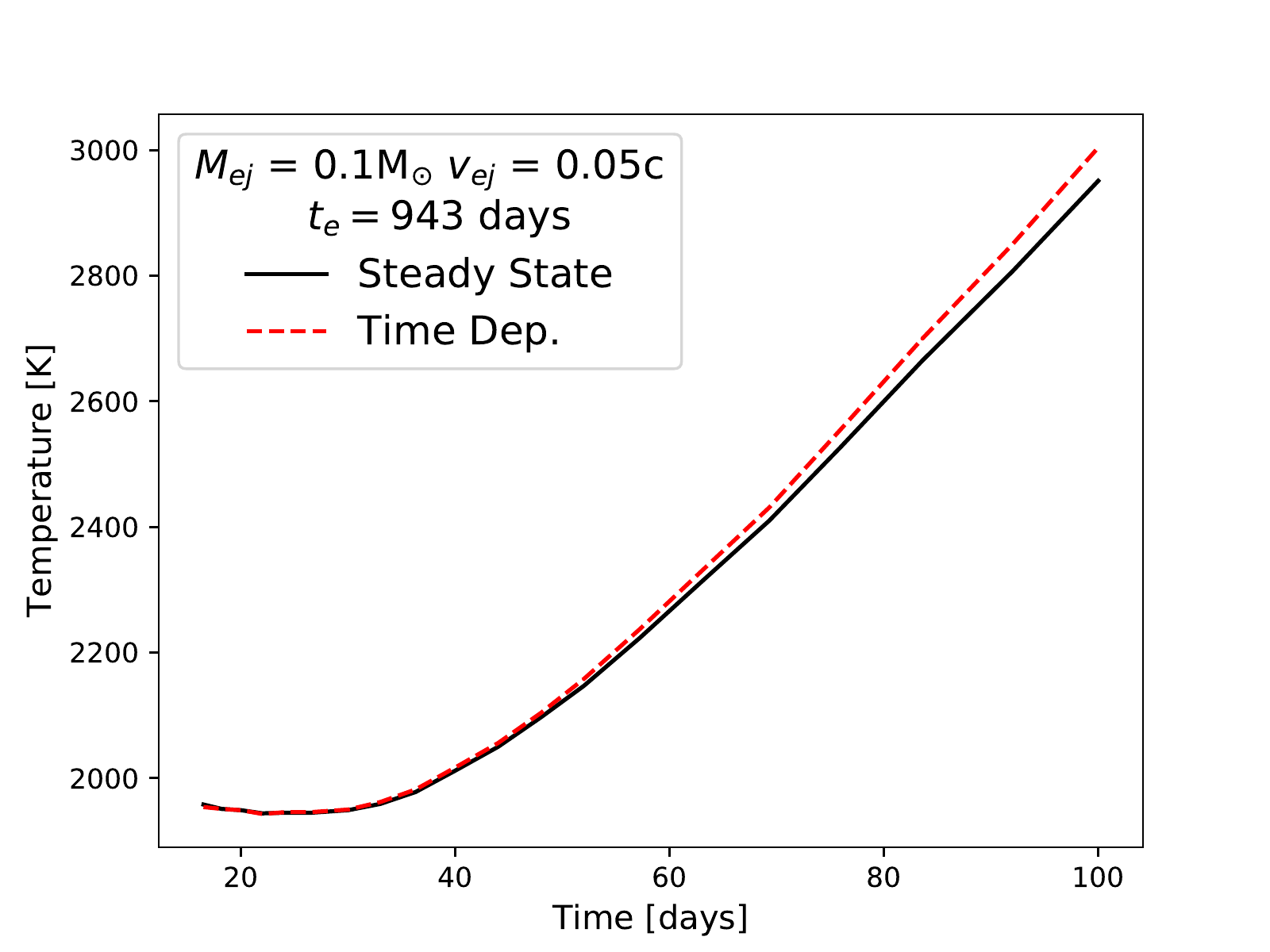} \\
\includegraphics[trim={0.2cm 0.2cm 1.6cm 0.8cm},clip,width = 0.49\textwidth]{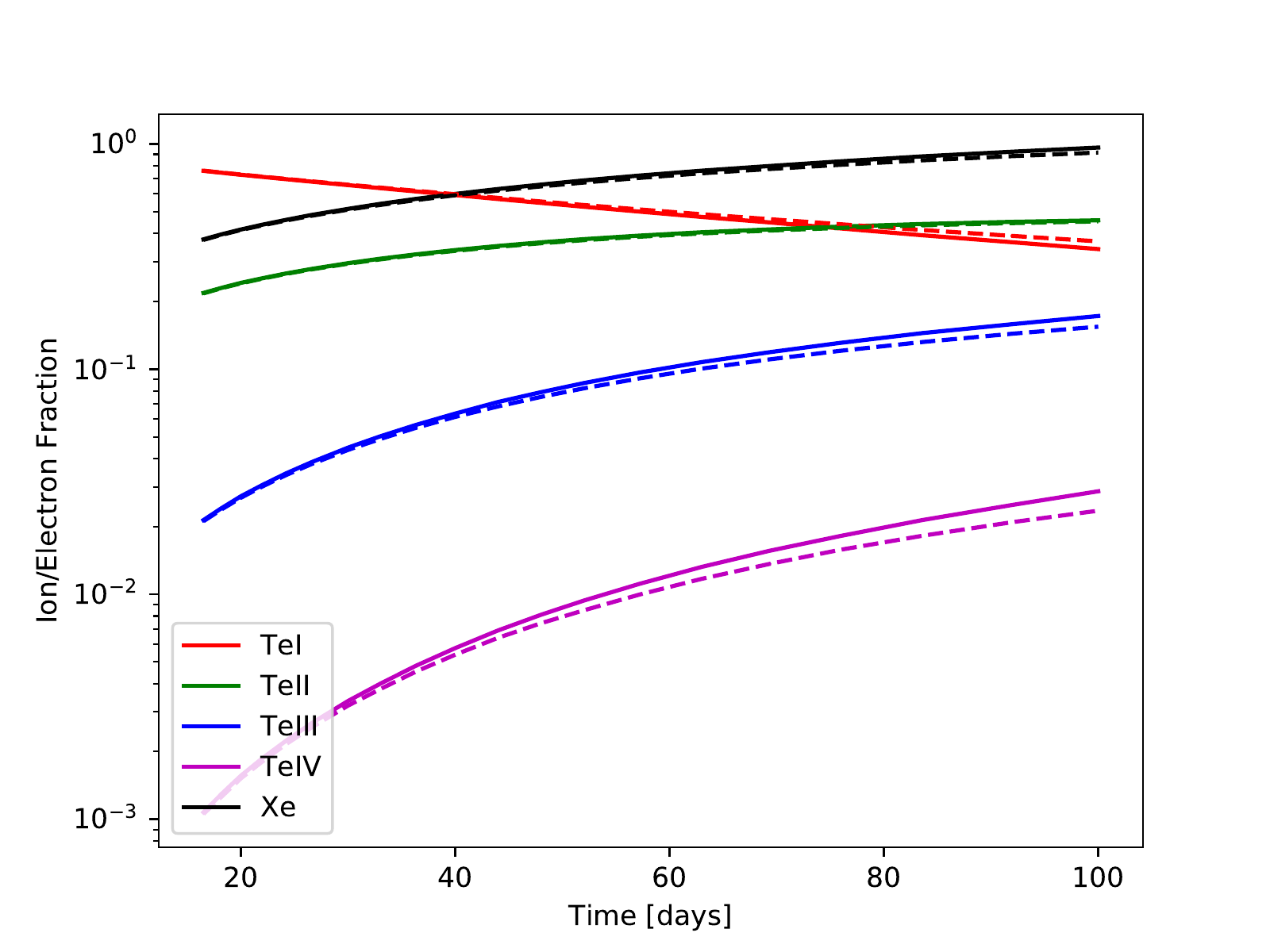}
\includegraphics[trim={0.2cm 0.2cm 1.6cm 0.8cm},clip,width = 0.49\textwidth]{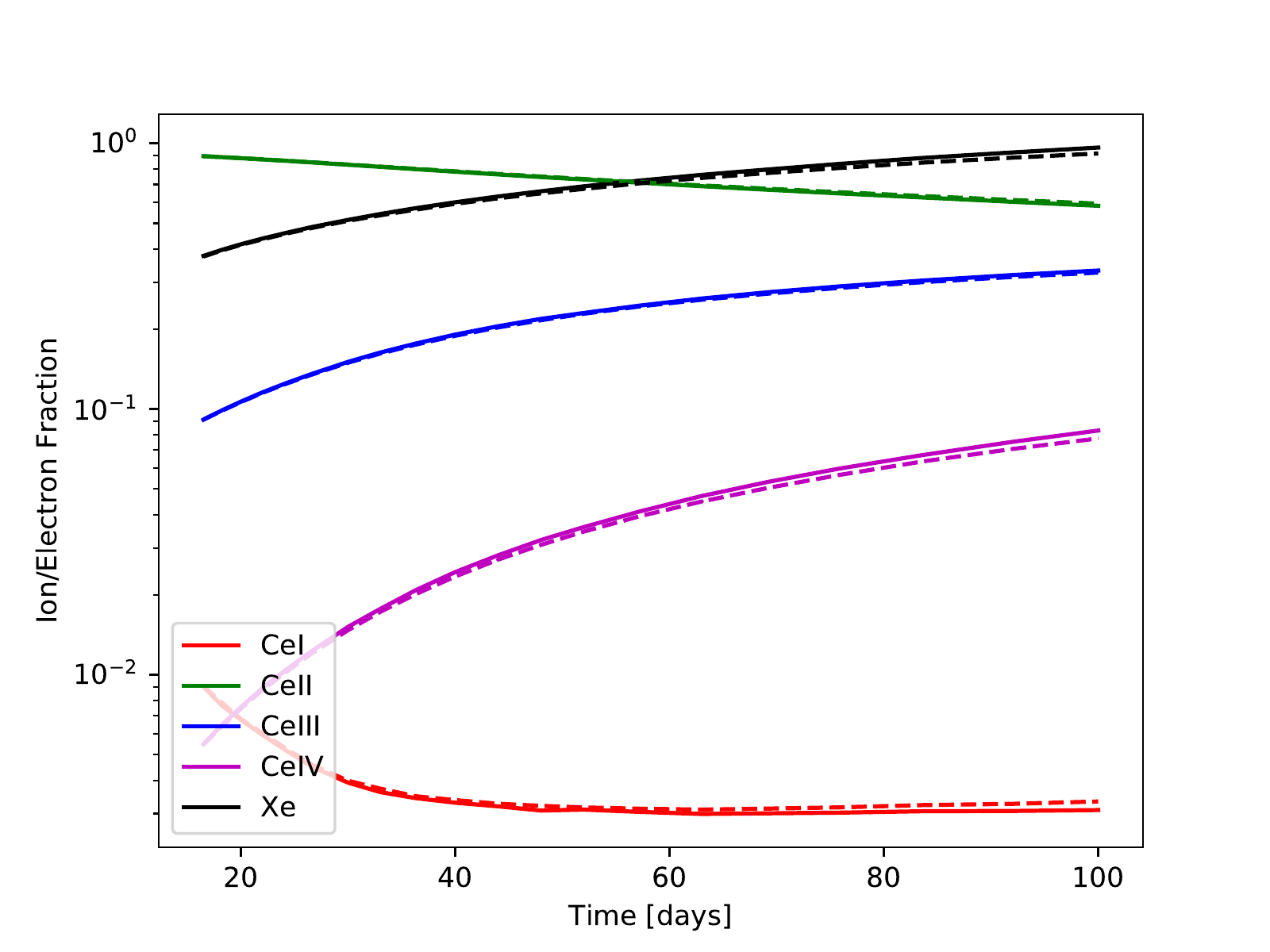}
\includegraphics[trim={0.2cm 0.2cm 1.6cm 0.8cm},clip,width = 0.49\textwidth]{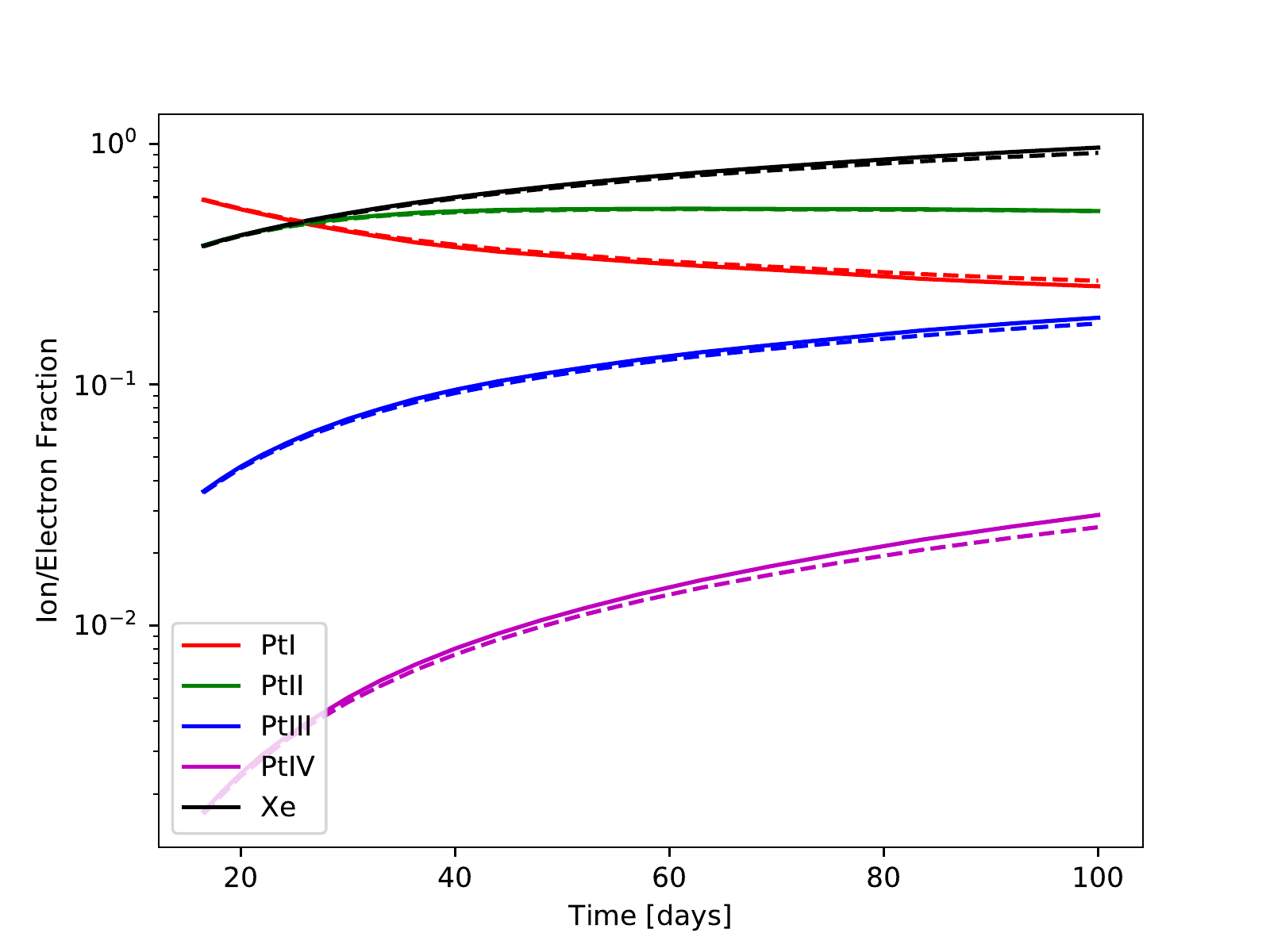}
\includegraphics[trim={0.2cm 0.2cm 1.6cm 0.8cm},clip,width = 0.49\textwidth]{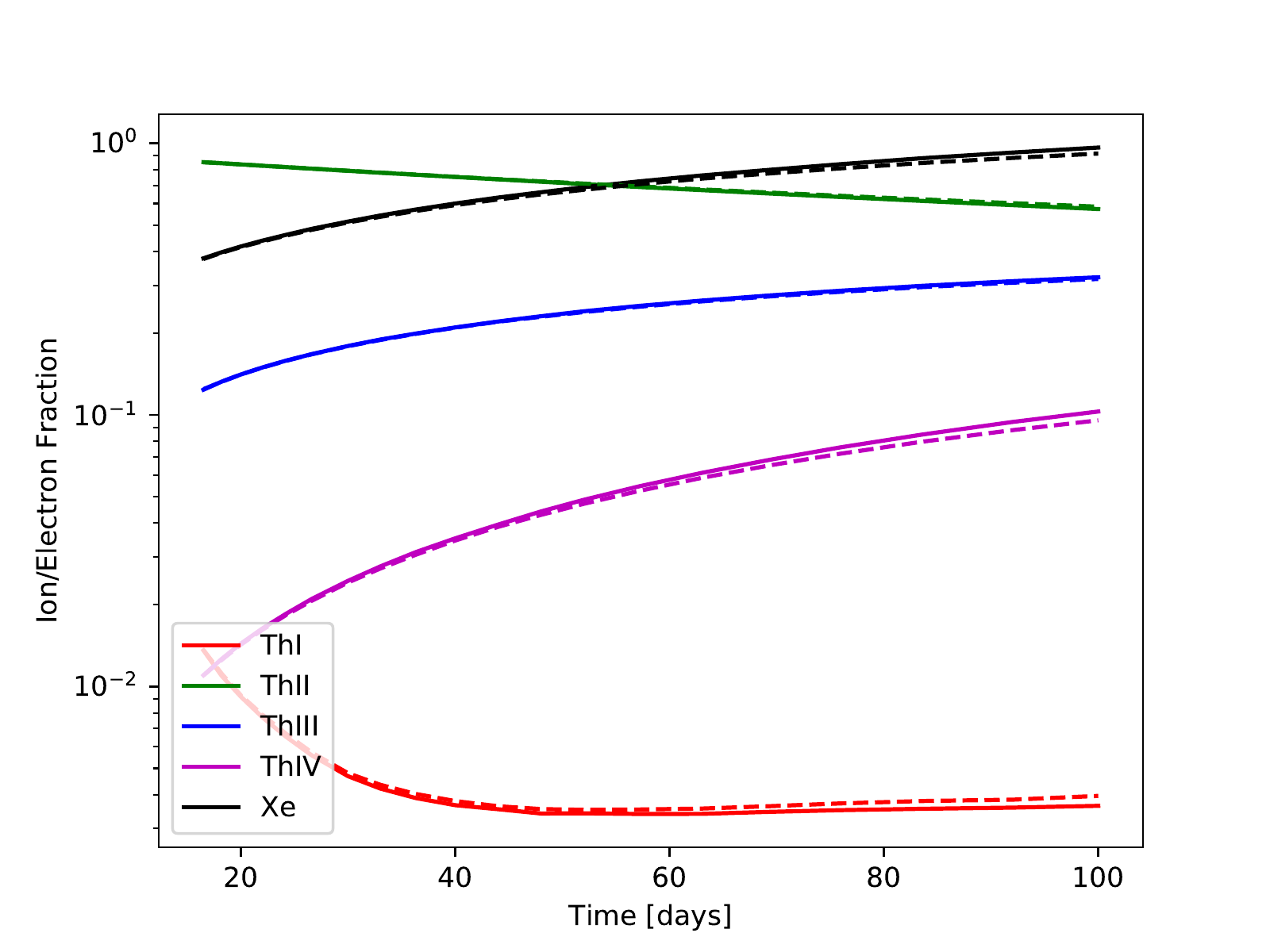}
\caption{Temperature evolution and ionisation structure evolution for the model with $M_{\mathrm{ej}} = 0.1\mathrm{\Msol}$, $v_{\mathrm{ej}} = 0.05$c and suppressed initial energy deposition. The solid lines are the steady-state results, while the dashed lines are the time-dependent results.}
\label{fig:01M_005v_suppressed}
\end{figure*}

\begin{figure*}
\center
\includegraphics[trim={0.2cm 0.2cm 1.6cm 0.8cm},clip,width = 0.49\textwidth]{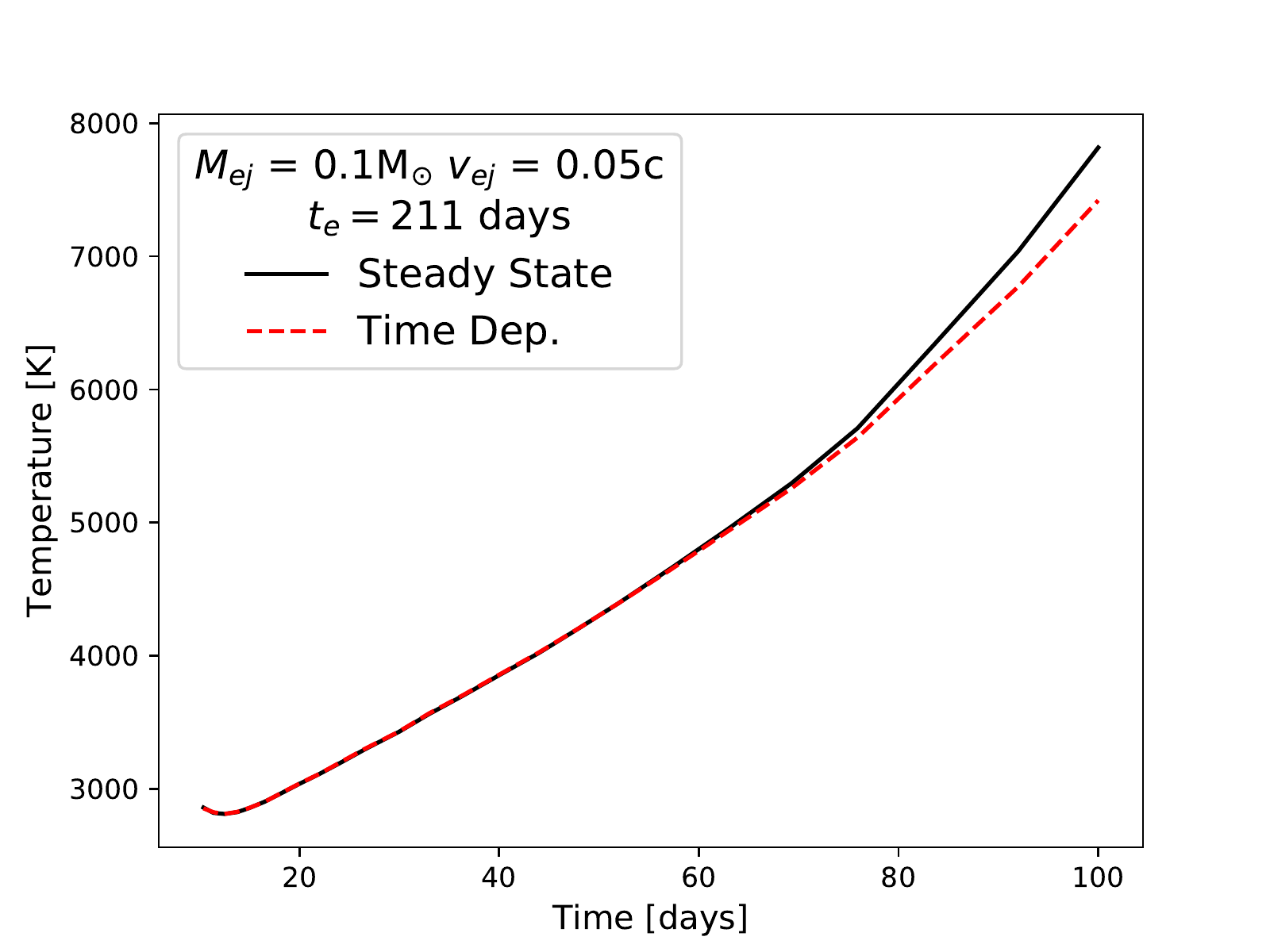} \\
\includegraphics[trim={0.2cm 0.2cm 1.6cm 0.8cm},clip,width = 0.49\textwidth]{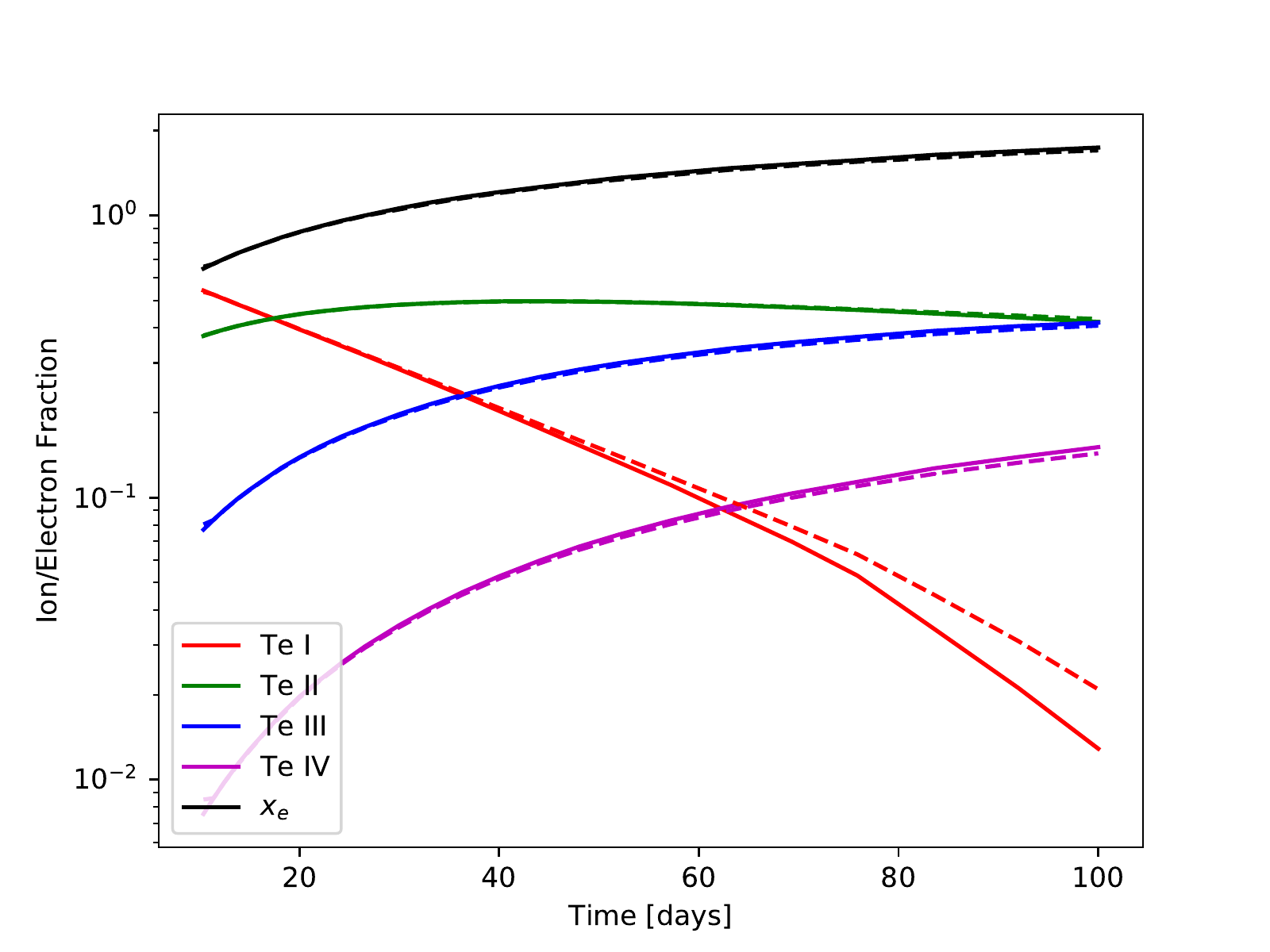}
\includegraphics[trim={0.2cm 0.2cm 1.6cm 0.8cm},clip,width = 0.49\textwidth]{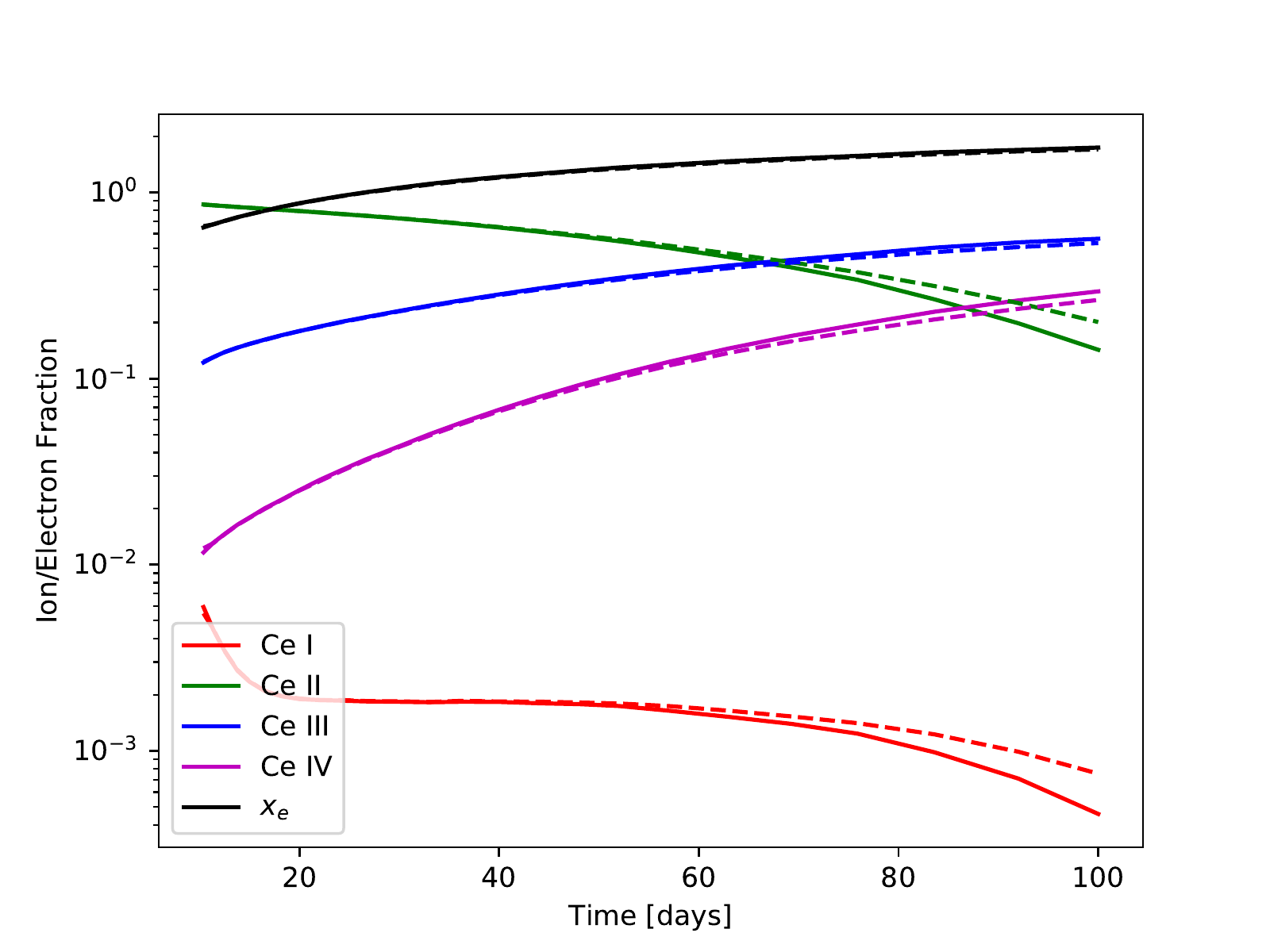}
\includegraphics[trim={0.2cm 0.2cm 1.6cm 0.8cm},clip,width = 0.49\textwidth]{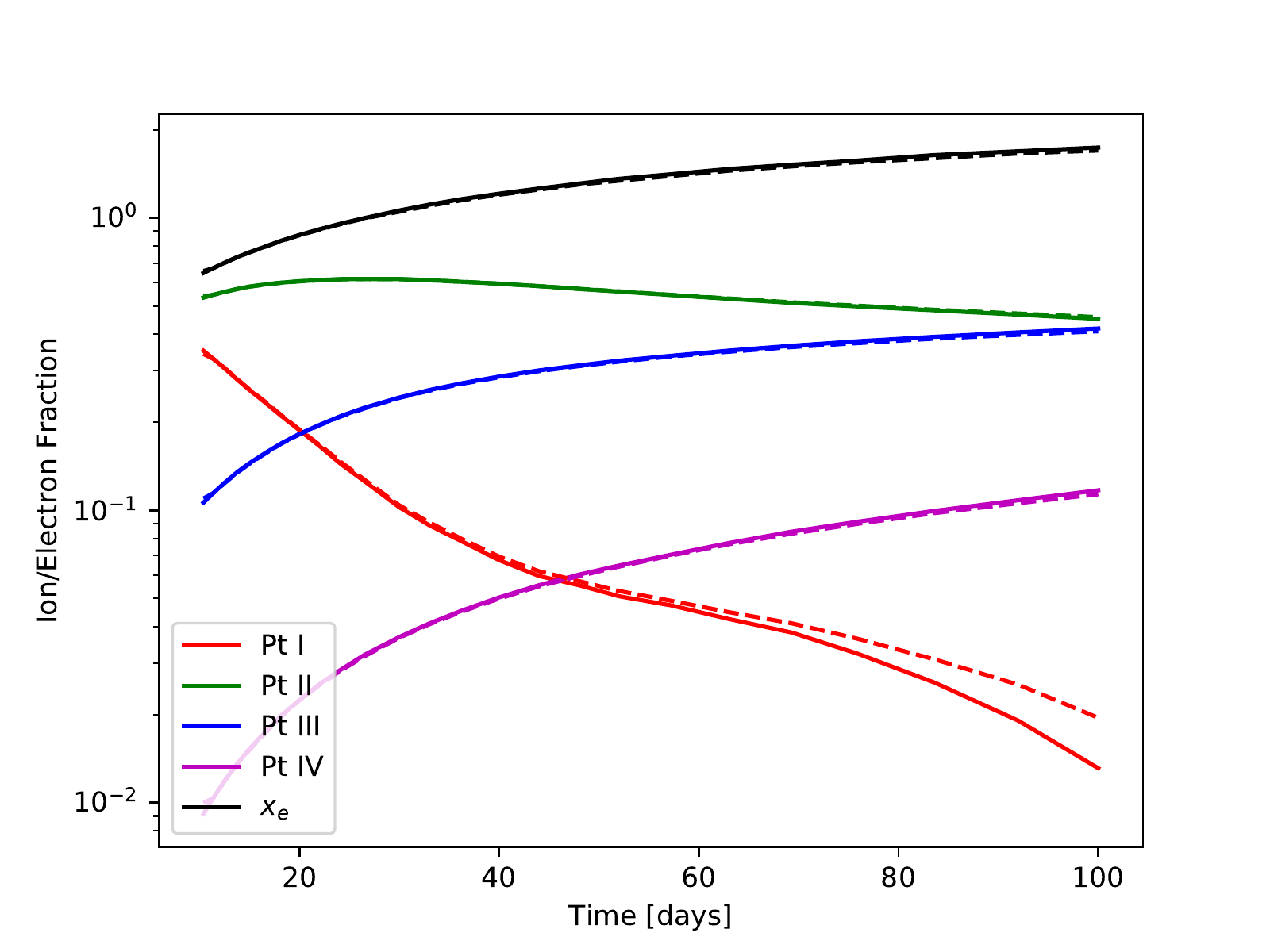}
\includegraphics[trim={0.2cm 0.2cm 1.6cm 0.8cm},clip,width = 0.49\textwidth]{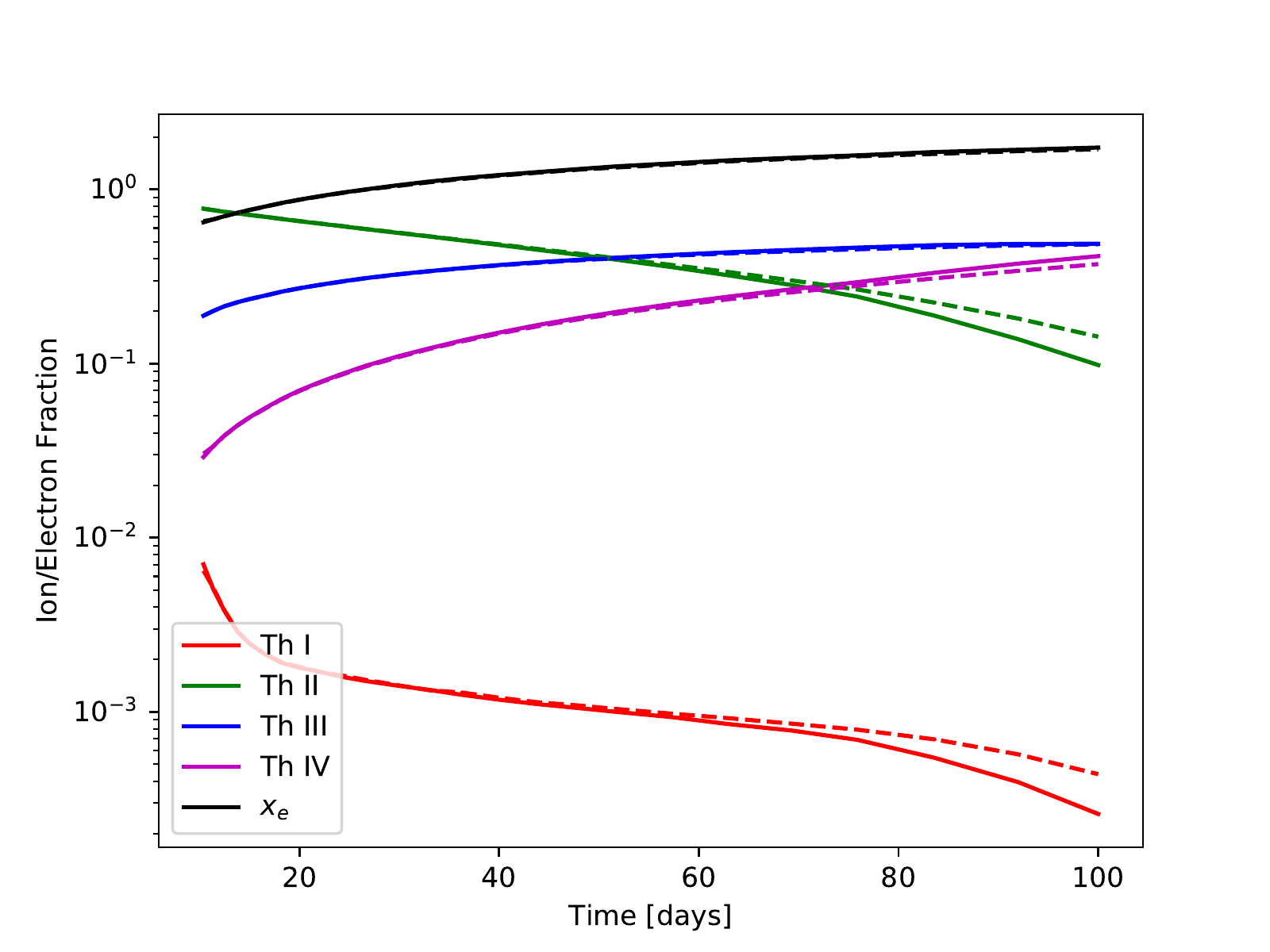}
\caption{Temperature evolution and ionisation structure evolution for the model with $M_{\mathrm{ej}} = 0.1\mathrm{\Msol}$, $v_{\mathrm{ej}} = 0.05$c and steeper thermalisation. The solid lines are the steady-state results, while the dashed lines are the time-dependent results.}
\label{fig:01M_005v_steep}
\end{figure*}

Figure \ref{fig:ioncool} shows the ratio of the 'ionisation cooling' term to the adiabatic term for the least and most dense models, in order to visually demonstrate that the ionisation cooling term remains small. The rest of the ionisation structure evolution results for the main model grid are shown below in Figures \ref{fig:001M_005v_ionfrac} - \ref{fig:01M_02v_ionfrac}. These are identical in format to Figures \ref{fig:005M_01v_ionfrac} and \ref{fig:01M_005v_ionfrac} in Section \ref{sec:results}, and these are therefore not included here.  Finally, Figures \ref{fig:01M_005v_suppressed} and \ref{fig:01M_005v_steep} show the results from the densest model with $M_{\mathrm{ej}} = 0.1\mathrm{\Msol}$ and $v_{\mathrm{ej}} = 0.05$c, with suppressed energy deposition, and steeper thermalisation, analogous to Figures \ref{fig:reduced_temperature} through \ref{fig:steep_ionfrac} for the low density model with $M_{\mathrm{ej}} = 0.01\mathrm{\Msol}$ and $v_{\mathrm{ej}} = 0.2$c in the main body.

%Ionisation cooling term plot

%Main model grid ionisation structure plots
\begin{figure*}
\center
\includegraphics[trim={0.2cm 0.2cm 1.6cm 0.8cm},clip,width = 0.49\textwidth]{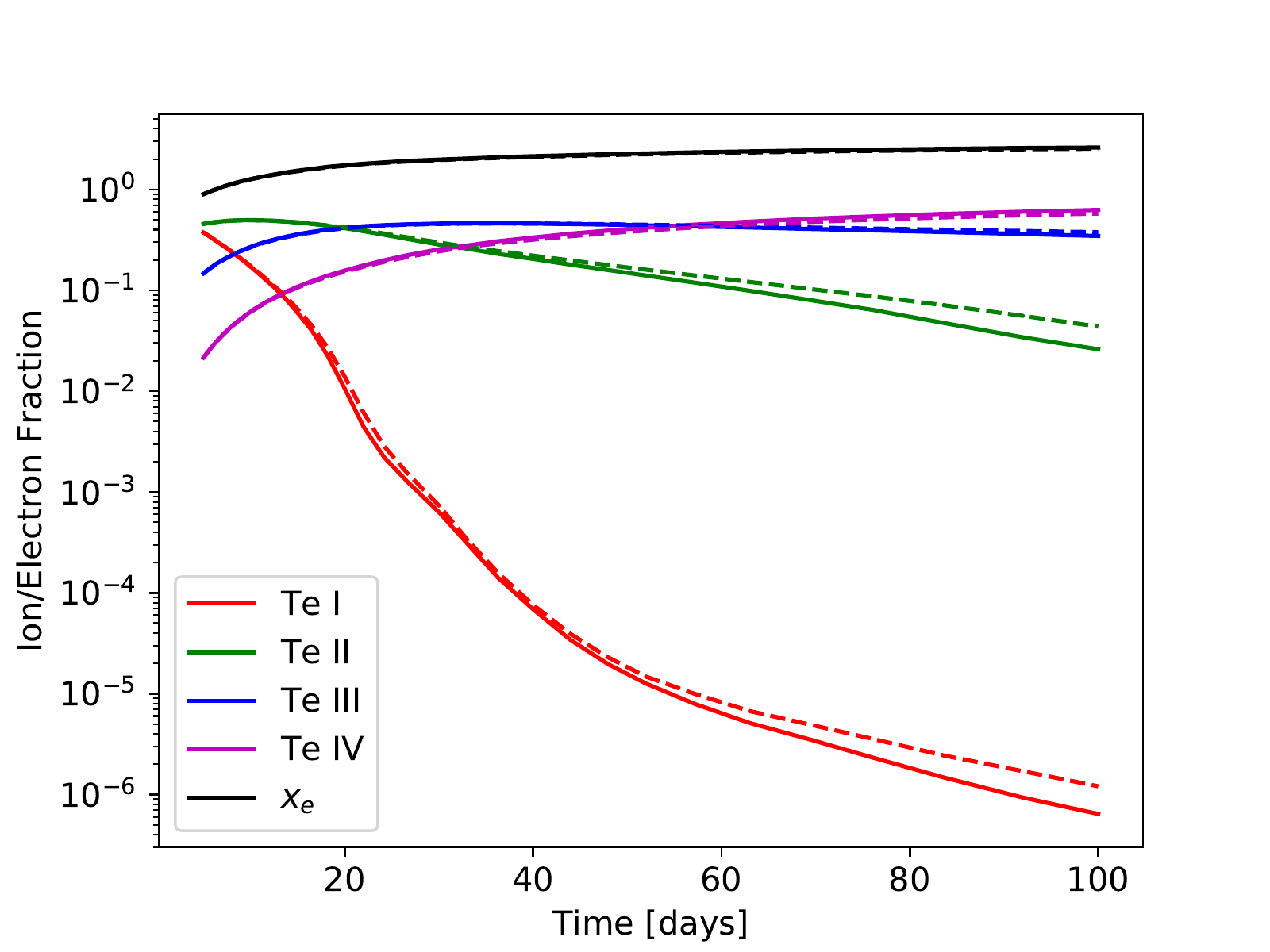}
\includegraphics[trim={0.2cm 0.2cm 1.6cm 0.8cm},clip,width = 0.49\textwidth]{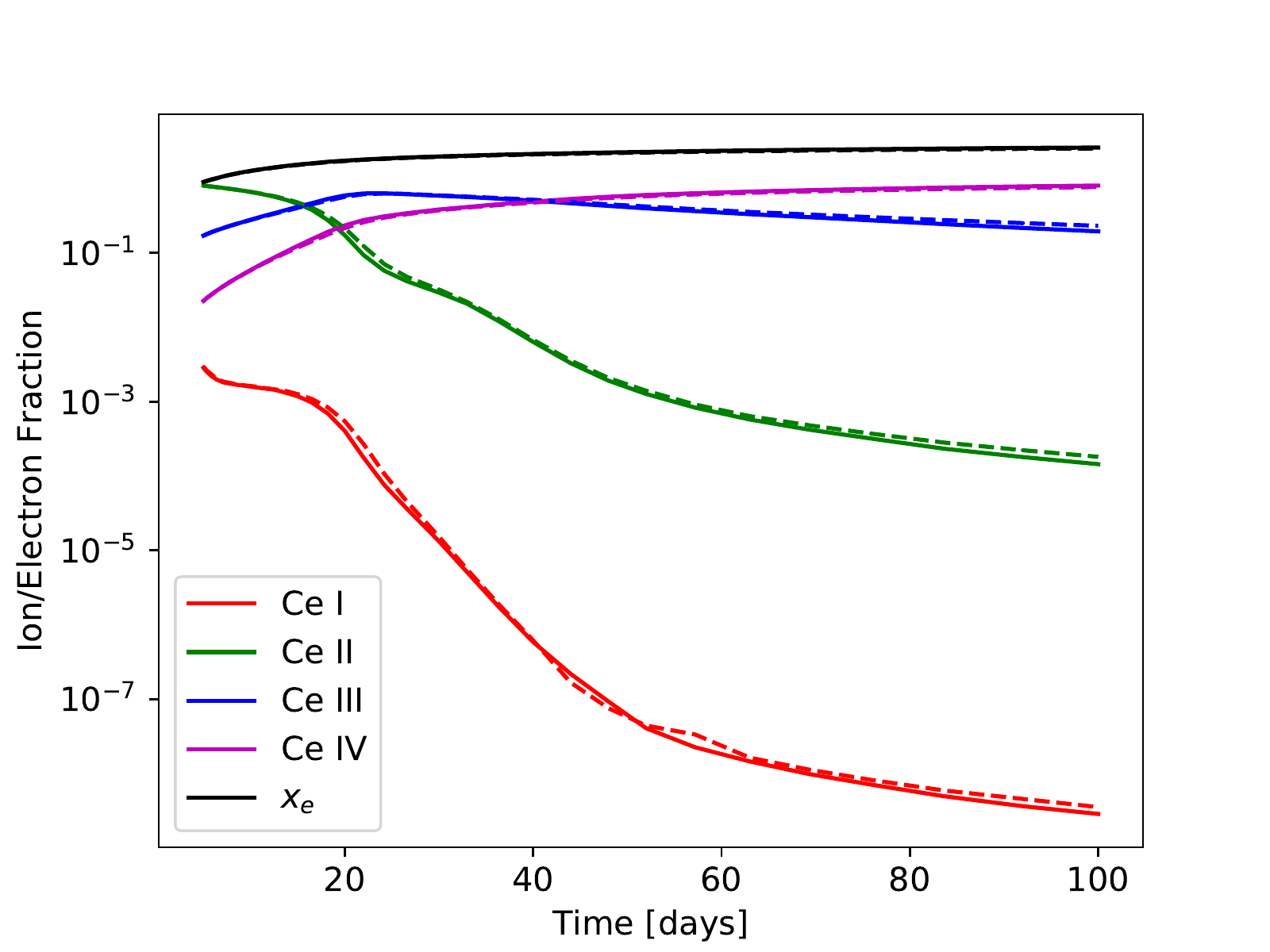}
\includegraphics[trim={0.2cm 0.2cm 1.6cm 0.8cm},clip,width = 0.49\textwidth]{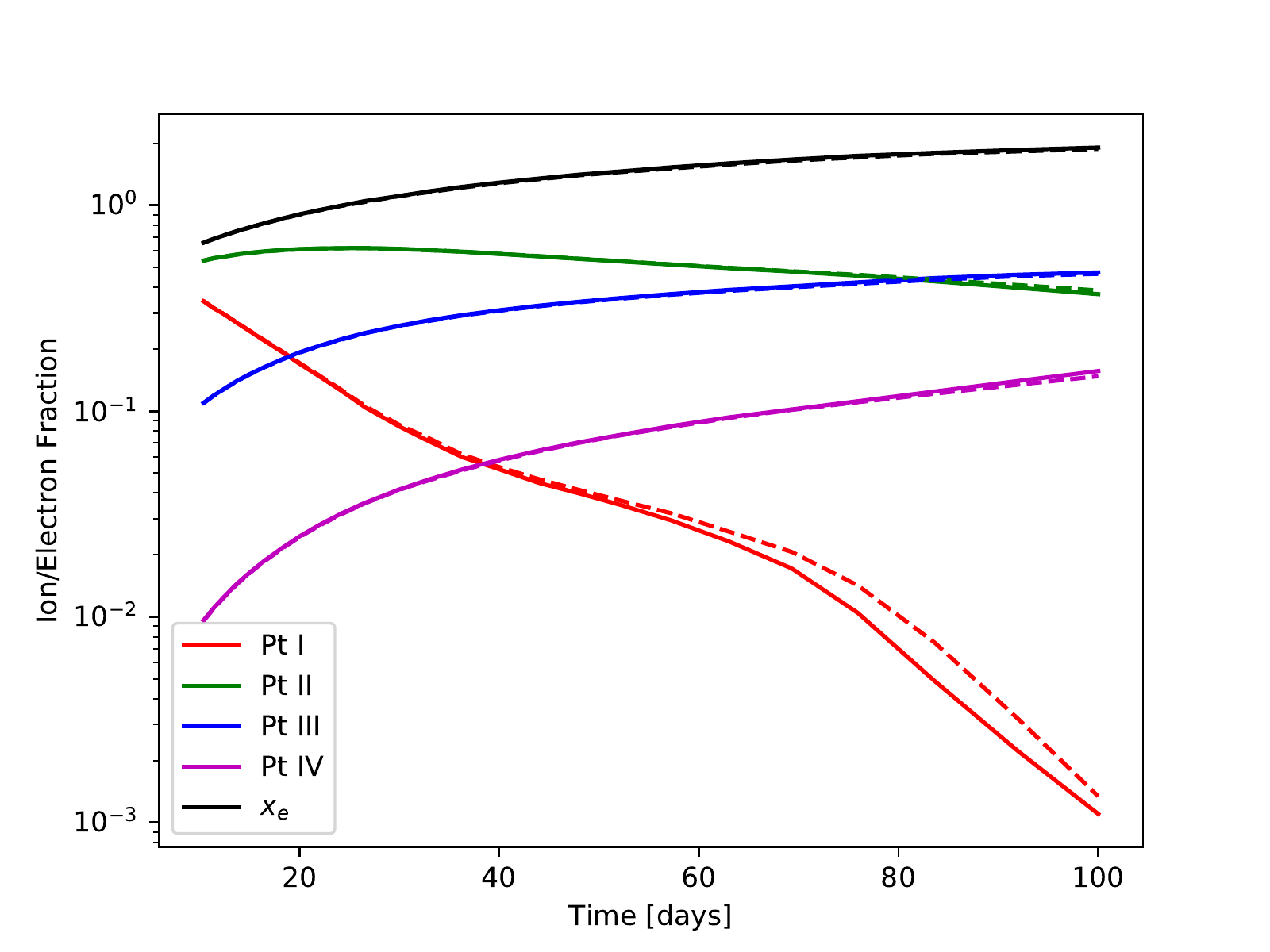}
\includegraphics[trim={0.2cm 0.2cm 1.6cm 0.8cm},clip,width = 0.49\textwidth]{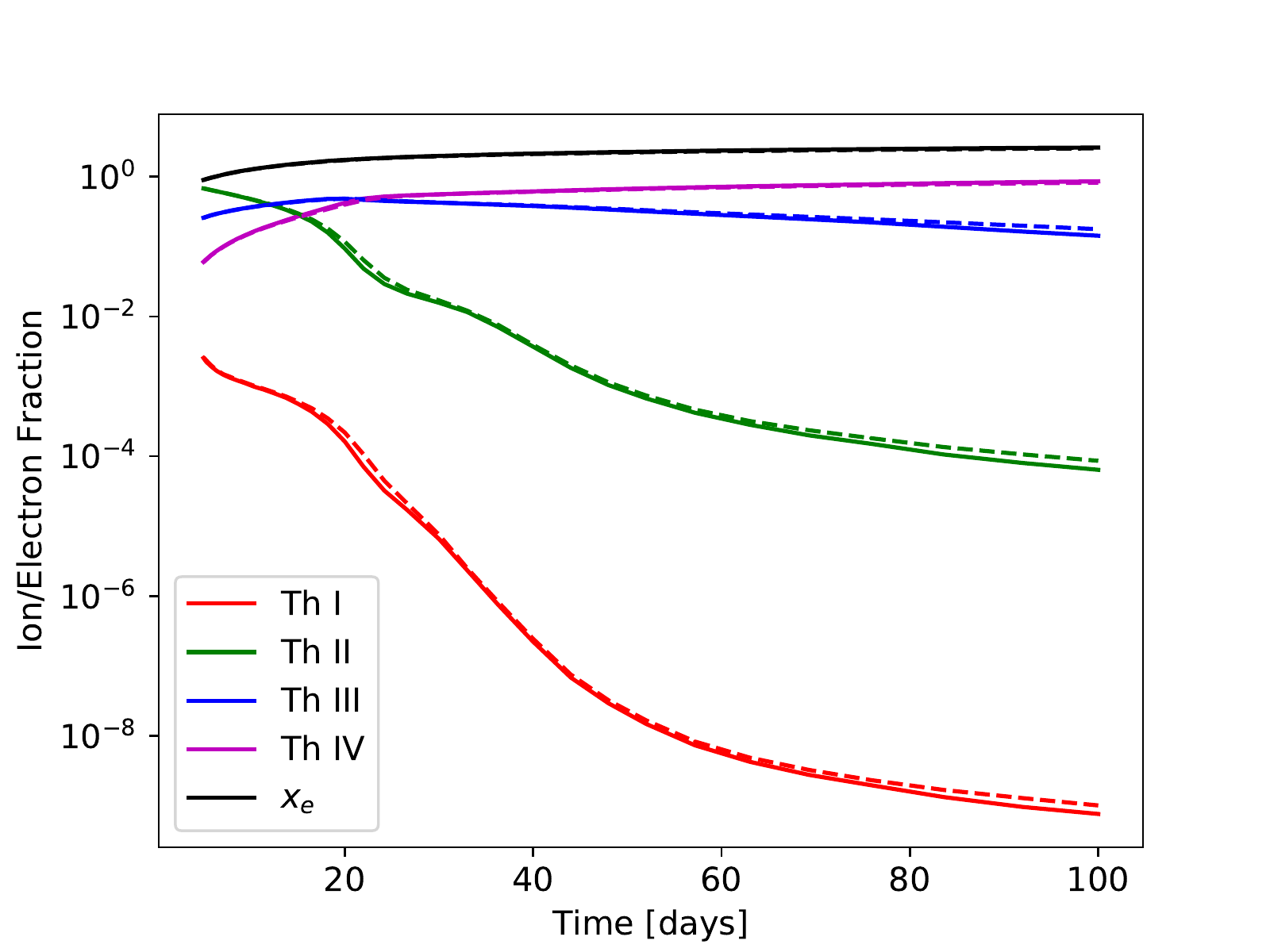}
\caption{Ionisation structure evolution for the model with $M_{\mathrm{ej}} = 0.01\mathrm{\Msol}$, $v_{\mathrm{ej}} = 0.05$c. The solid lines are the steady-state results, while the dashed lines are the time-dependent results.}
\label{fig:001M_005v_ionfrac}
\end{figure*}

\begin{figure*}
\center
\includegraphics[trim={0.2cm 0.2cm 1.6cm 0.8cm},clip,width = 0.49\textwidth]{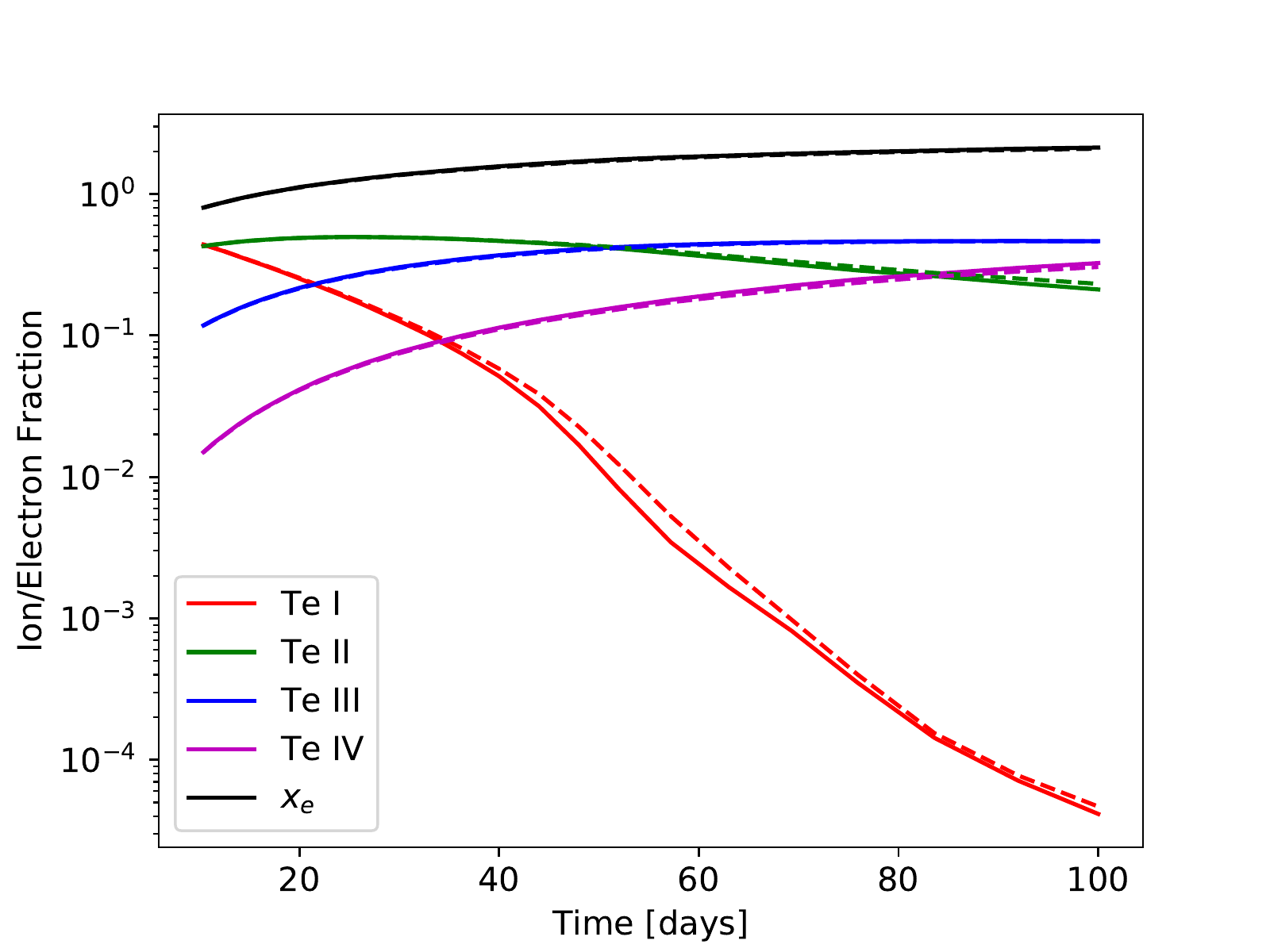}
\includegraphics[trim={0.2cm 0.2cm 1.6cm 0.8cm},clip,width = 0.49\textwidth]{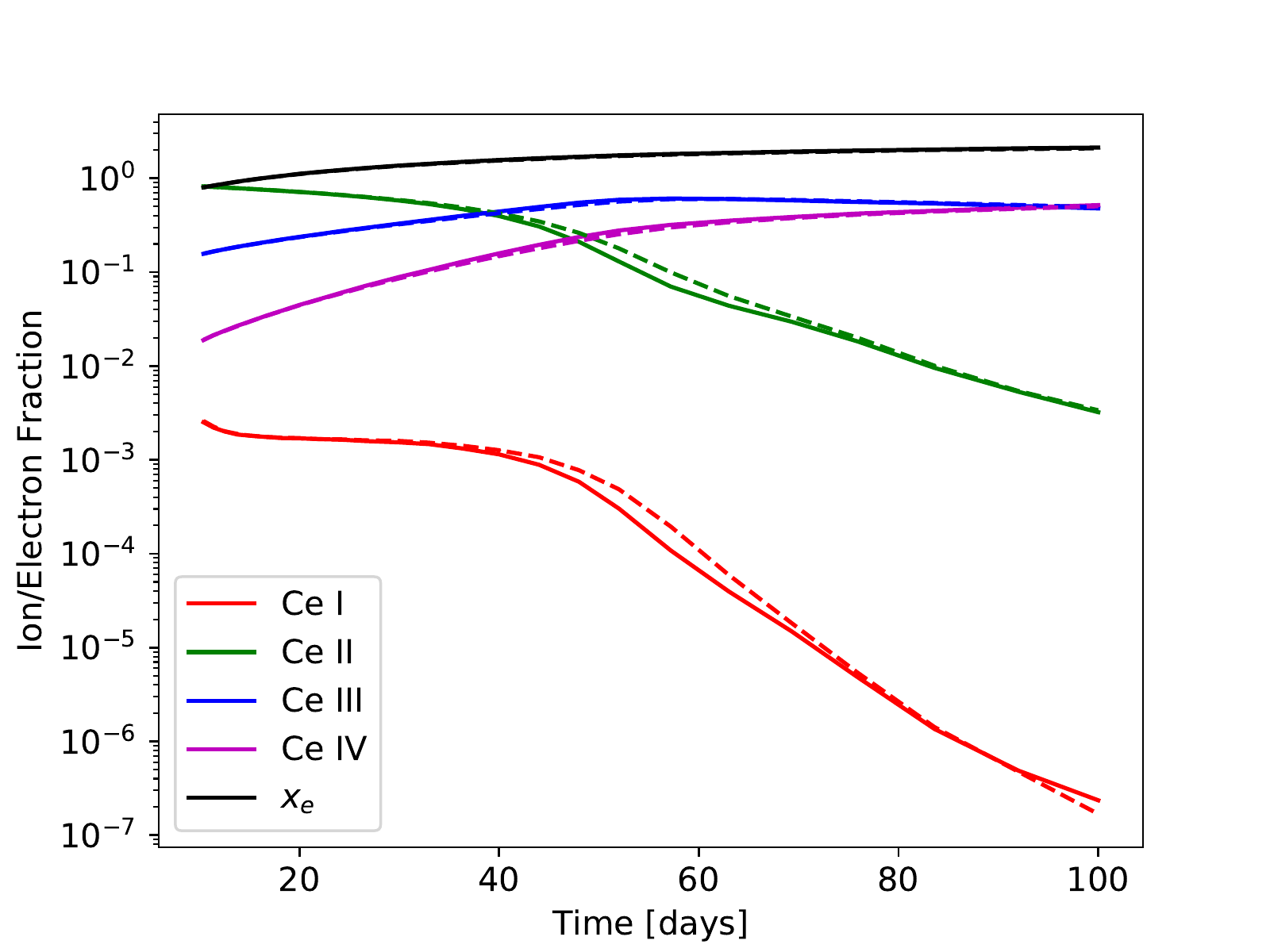}
\includegraphics[trim={0.2cm 0.2cm 1.6cm 0.8cm},clip,width = 0.49\textwidth]{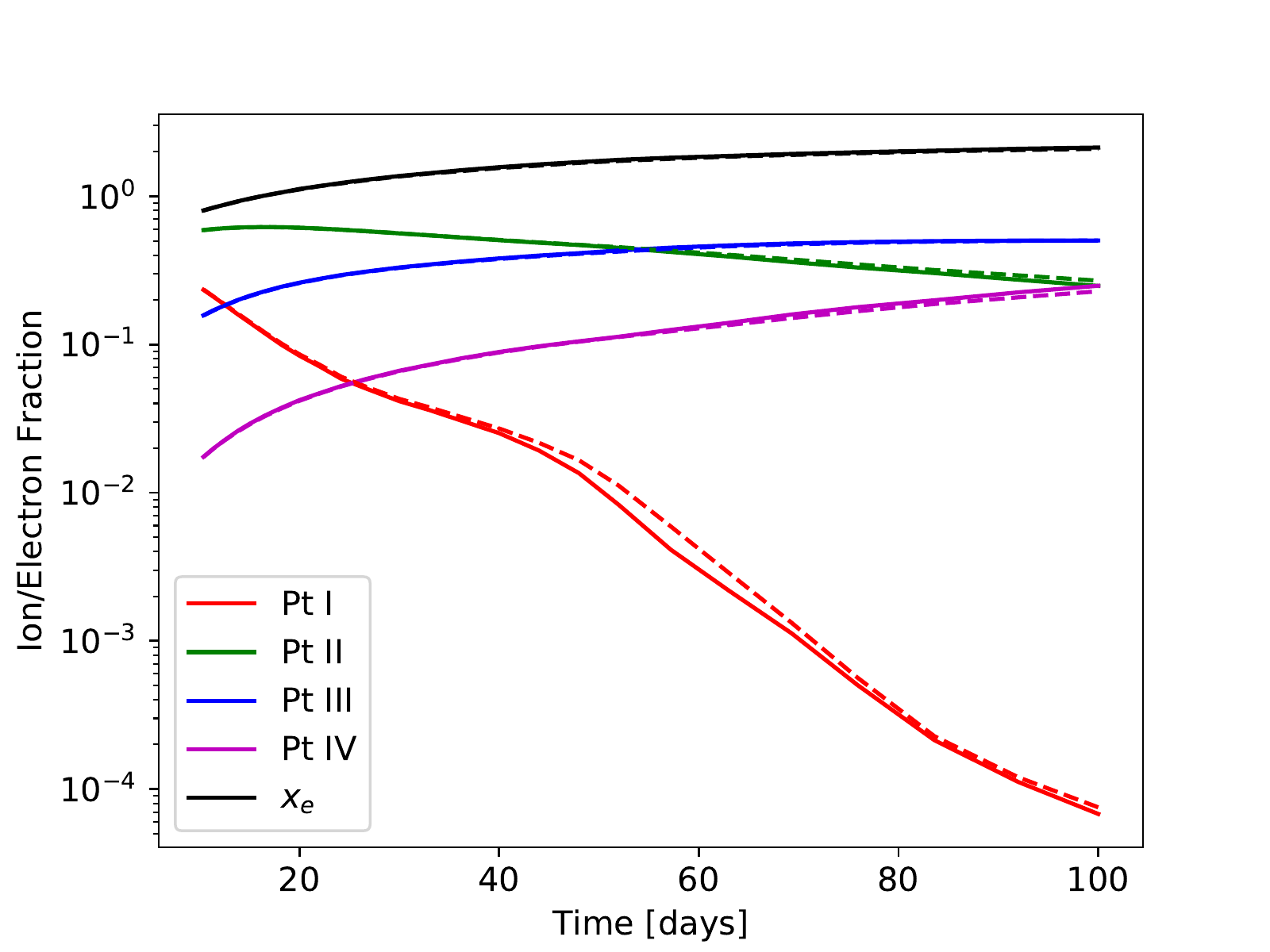}
\includegraphics[trim={0.2cm 0.2cm 1.6cm 0.8cm},clip,width = 0.49\textwidth]{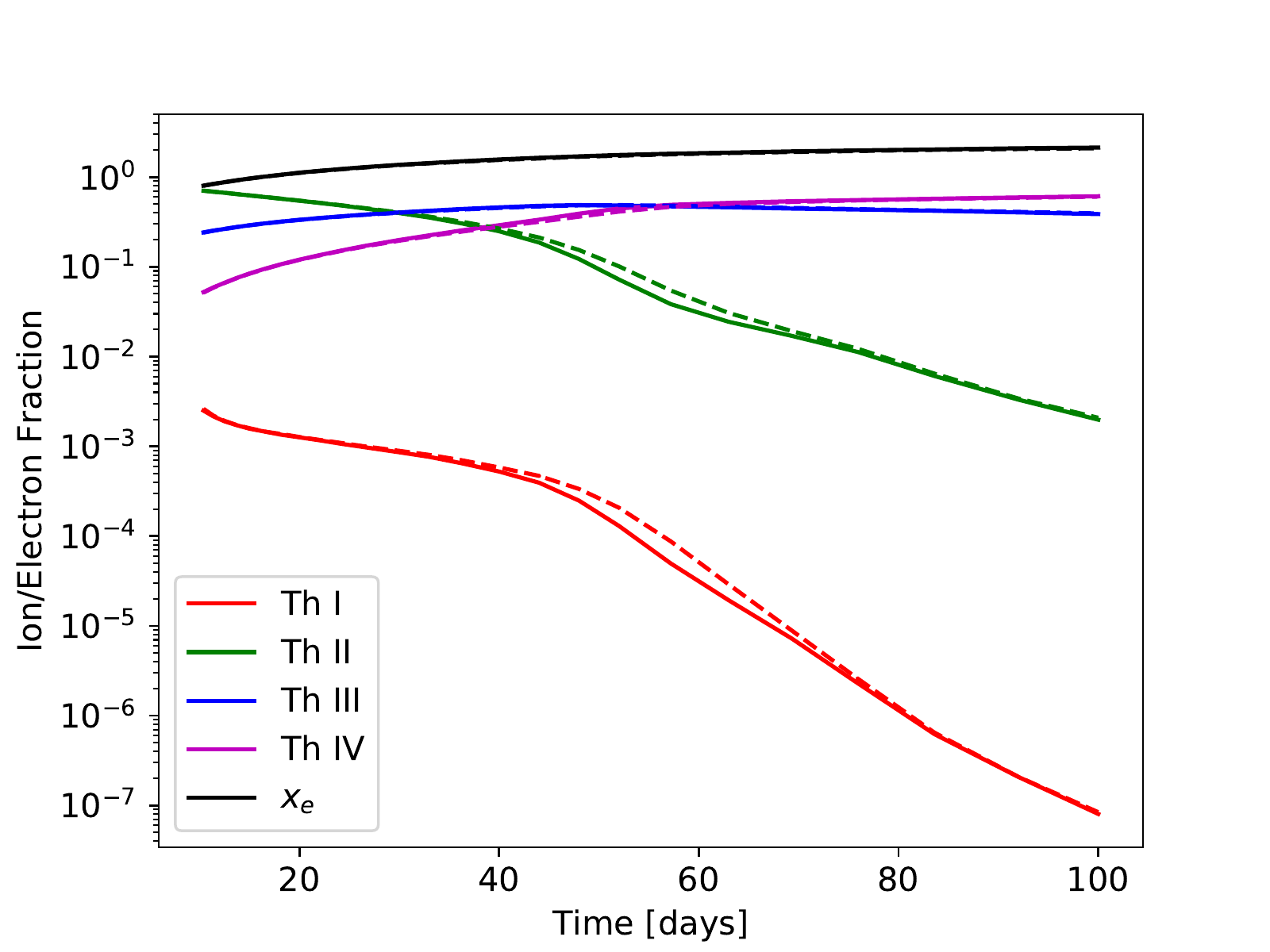}
\caption{Ionisation structure evolution for the model with $M_{\mathrm{ej}} = 0.05\mathrm{\Msol}$, $v_{\mathrm{ej}} = 0.05$c. The solid lines are the steady-state results, while the dashed lines are the time-dependent results.}
\label{fig:005M_005v_ionfrac}
\end{figure*}

\begin{figure*}
\center
\includegraphics[trim={0.2cm 0.2cm 1.6cm 0.8cm},clip,width = 0.49\textwidth]{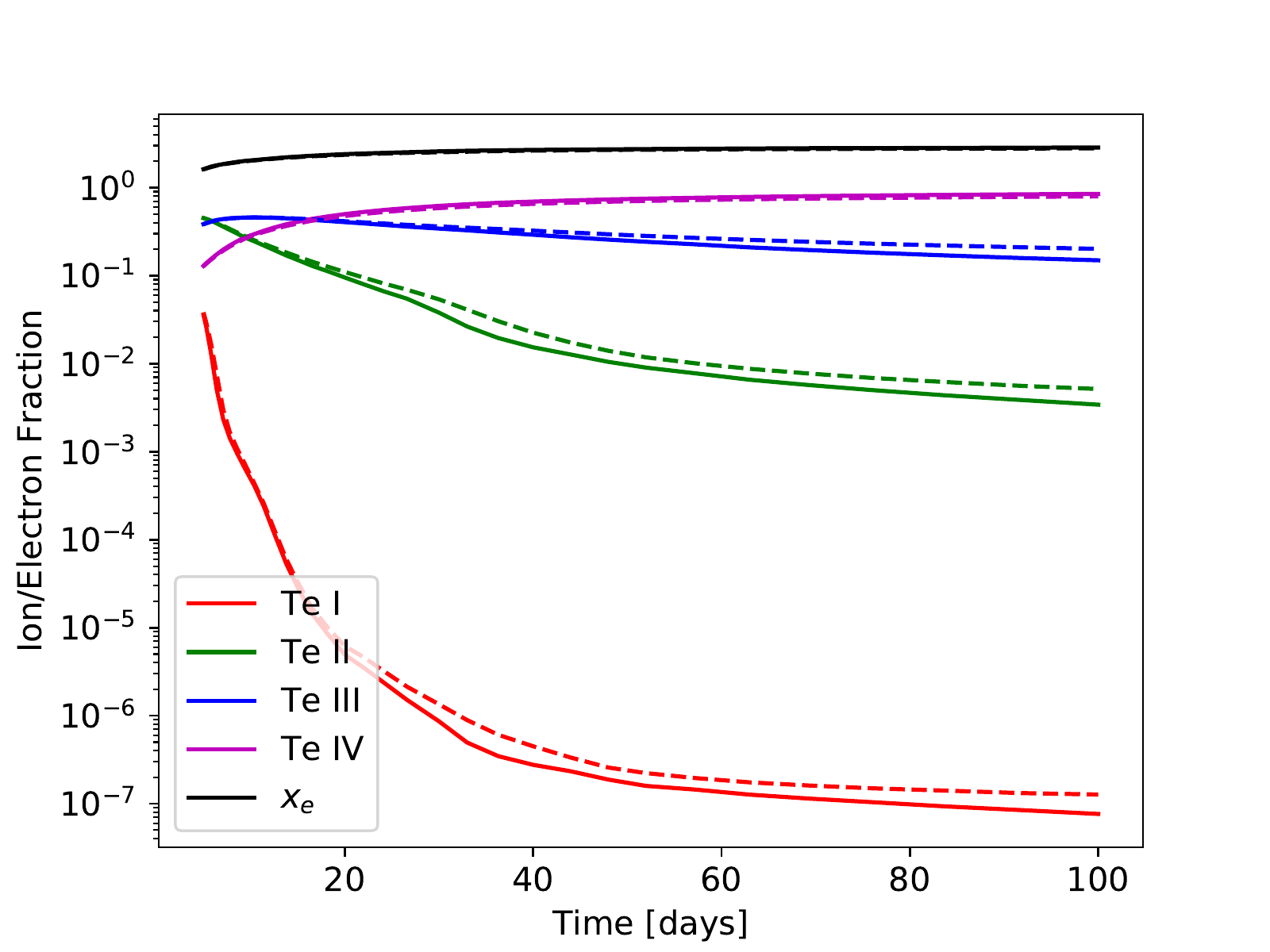}
\includegraphics[trim={0.2cm 0.2cm 1.6cm 0.8cm},clip,width = 0.49\textwidth]{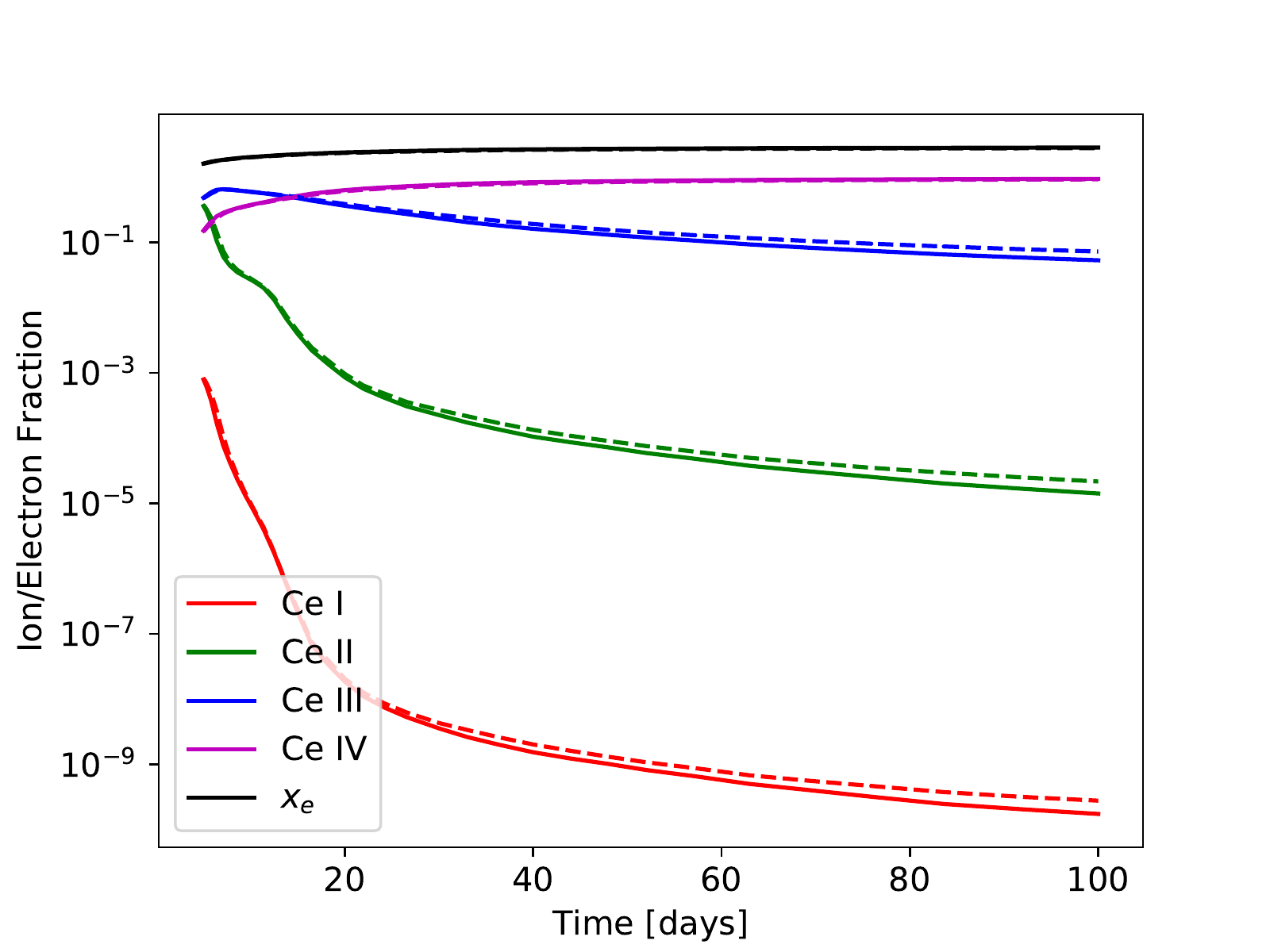}
\includegraphics[trim={0.2cm 0.2cm 1.6cm 0.8cm},clip,width = 0.49\textwidth]{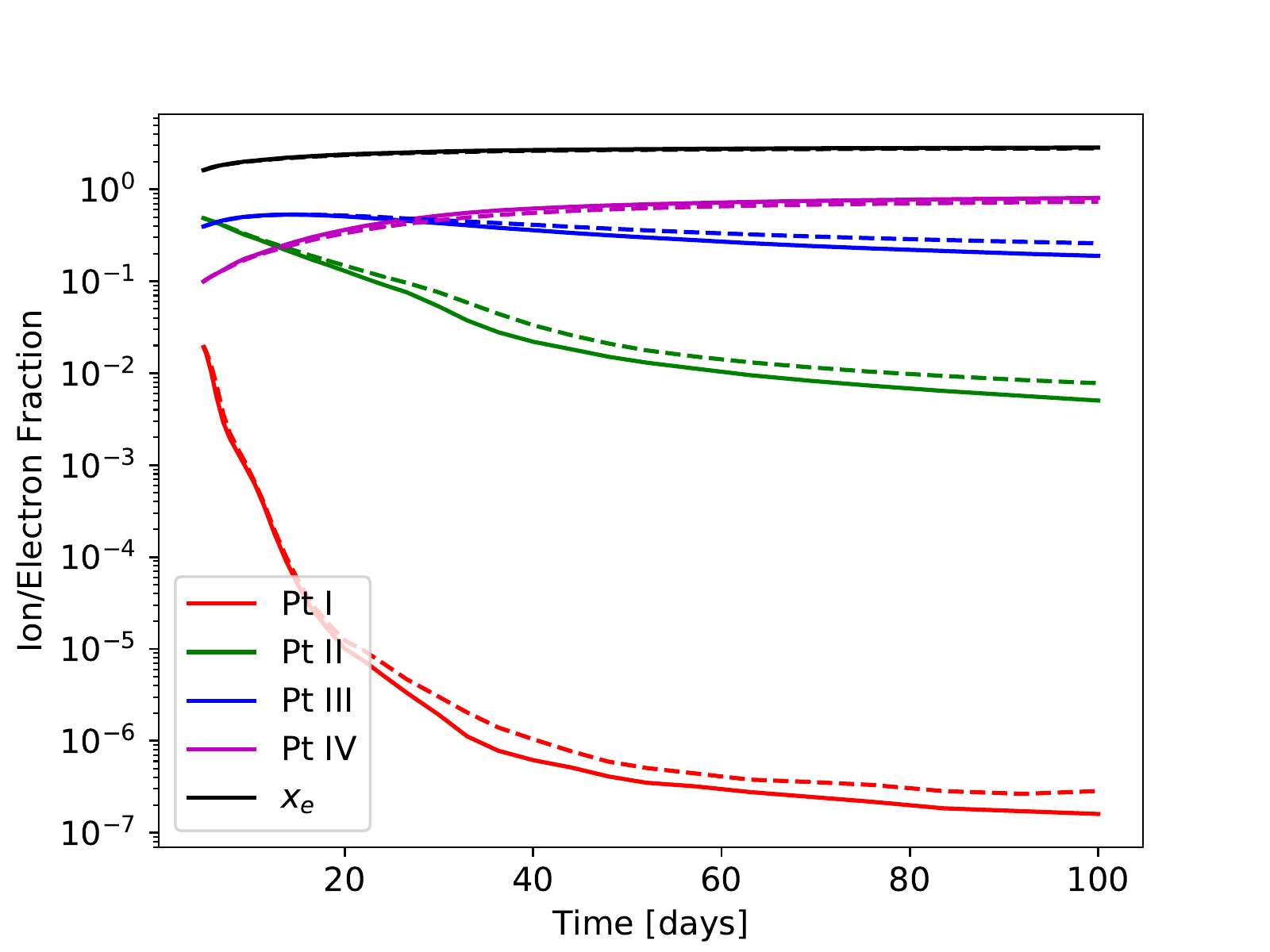}
\includegraphics[trim={0.2cm 0.2cm 1.6cm 0.8cm},clip,width = 0.49\textwidth]{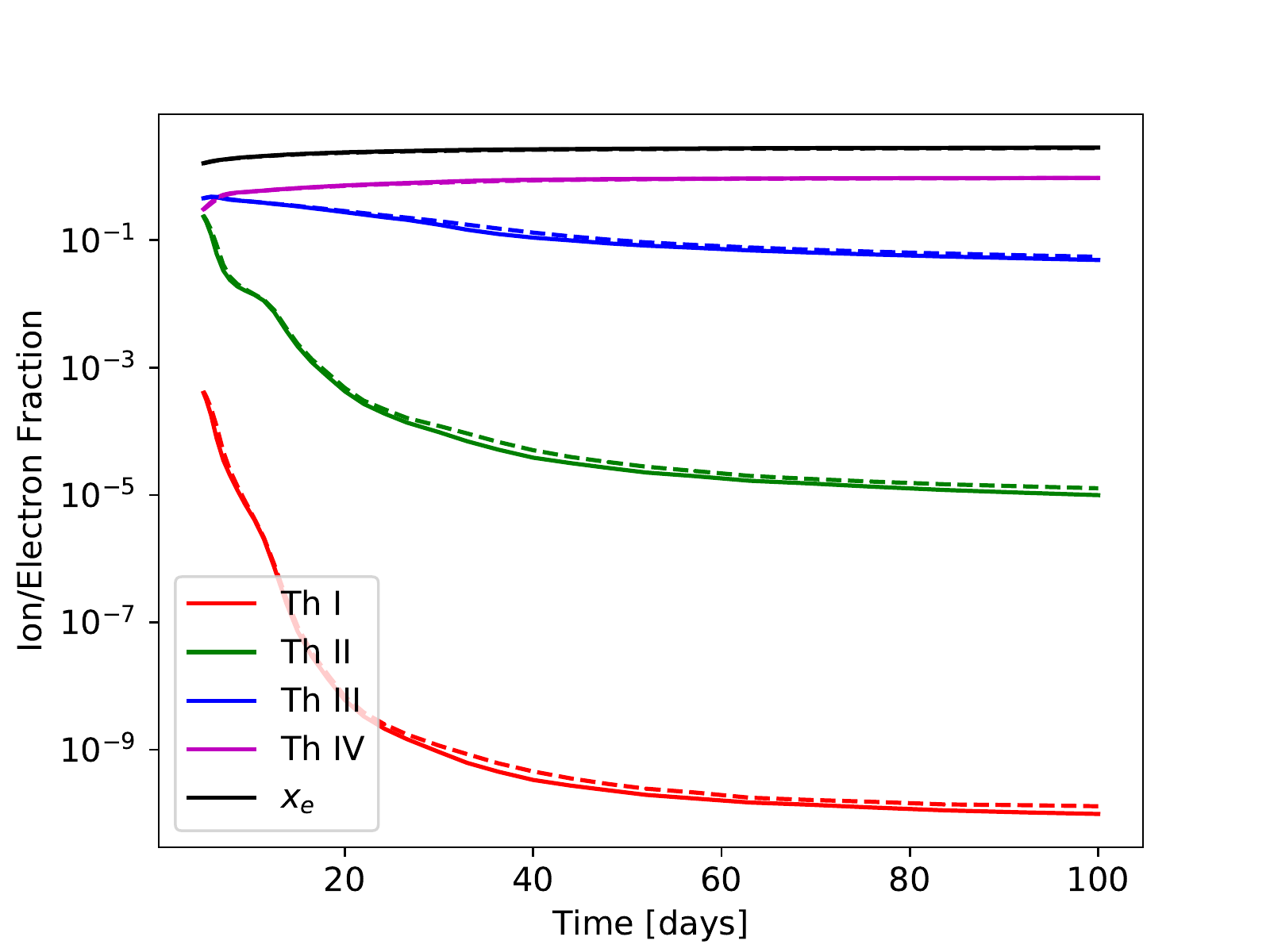}
\caption{Ionisation structure evolution for the model with $M_{\mathrm{ej}} = 0.01\mathrm{\Msol}$, $v_{\mathrm{ej}} = 0.1$c. The solid lines are the steady-state results, while the dashed lines are the time-dependent results.}
\label{fig:001M_01v_ionfrac}
\end{figure*} 

\begin{figure*}
\center
\includegraphics[trim={0.2cm 0.2cm 1.6cm 0.8cm},clip,width = 0.49\textwidth]{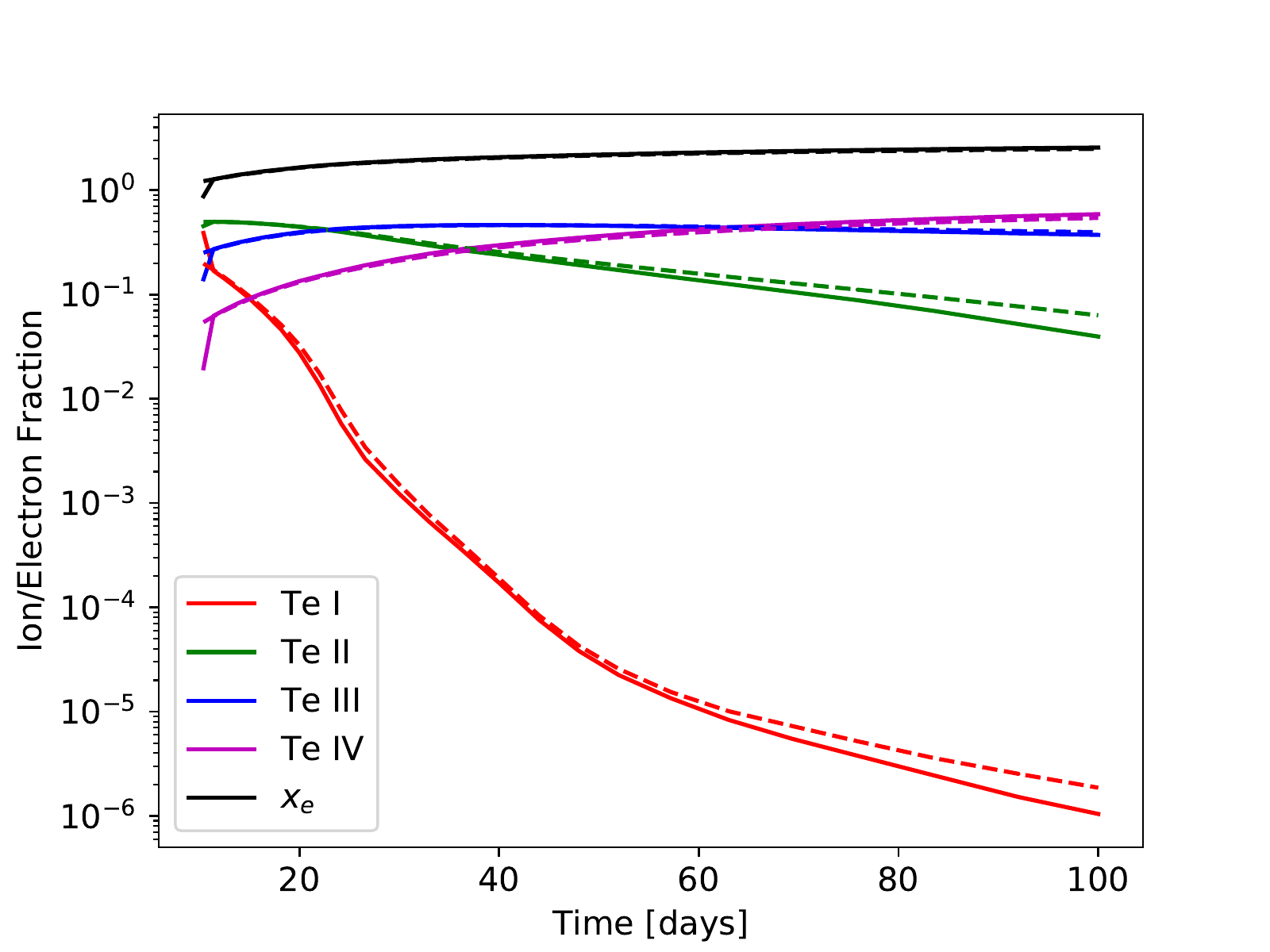}
\includegraphics[trim={0.2cm 0.2cm 1.6cm 0.8cm},clip,width = 0.49\textwidth]{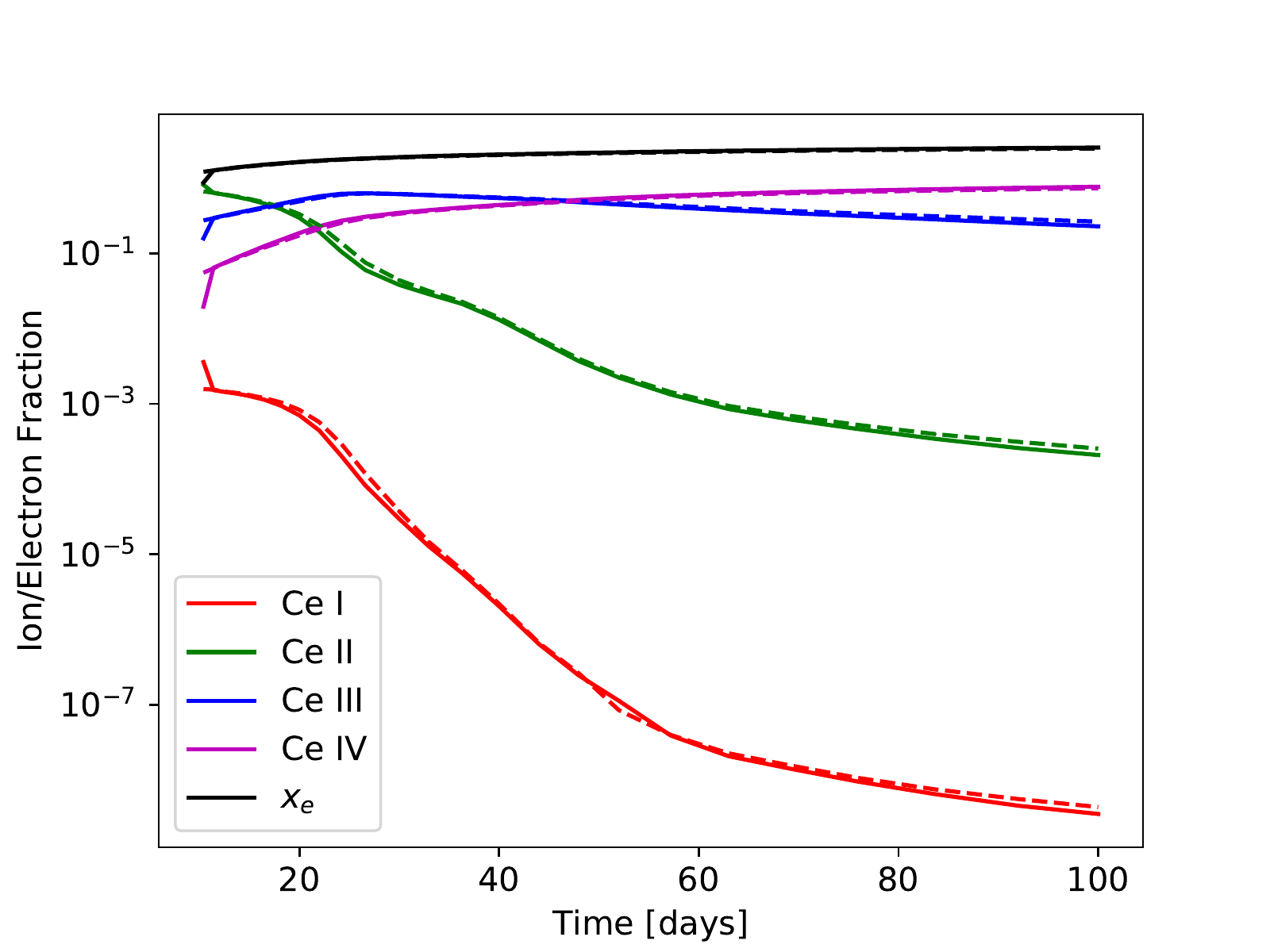}
\includegraphics[trim={0.2cm 0.2cm 1.6cm 0.8cm},clip,width = 0.49\textwidth]{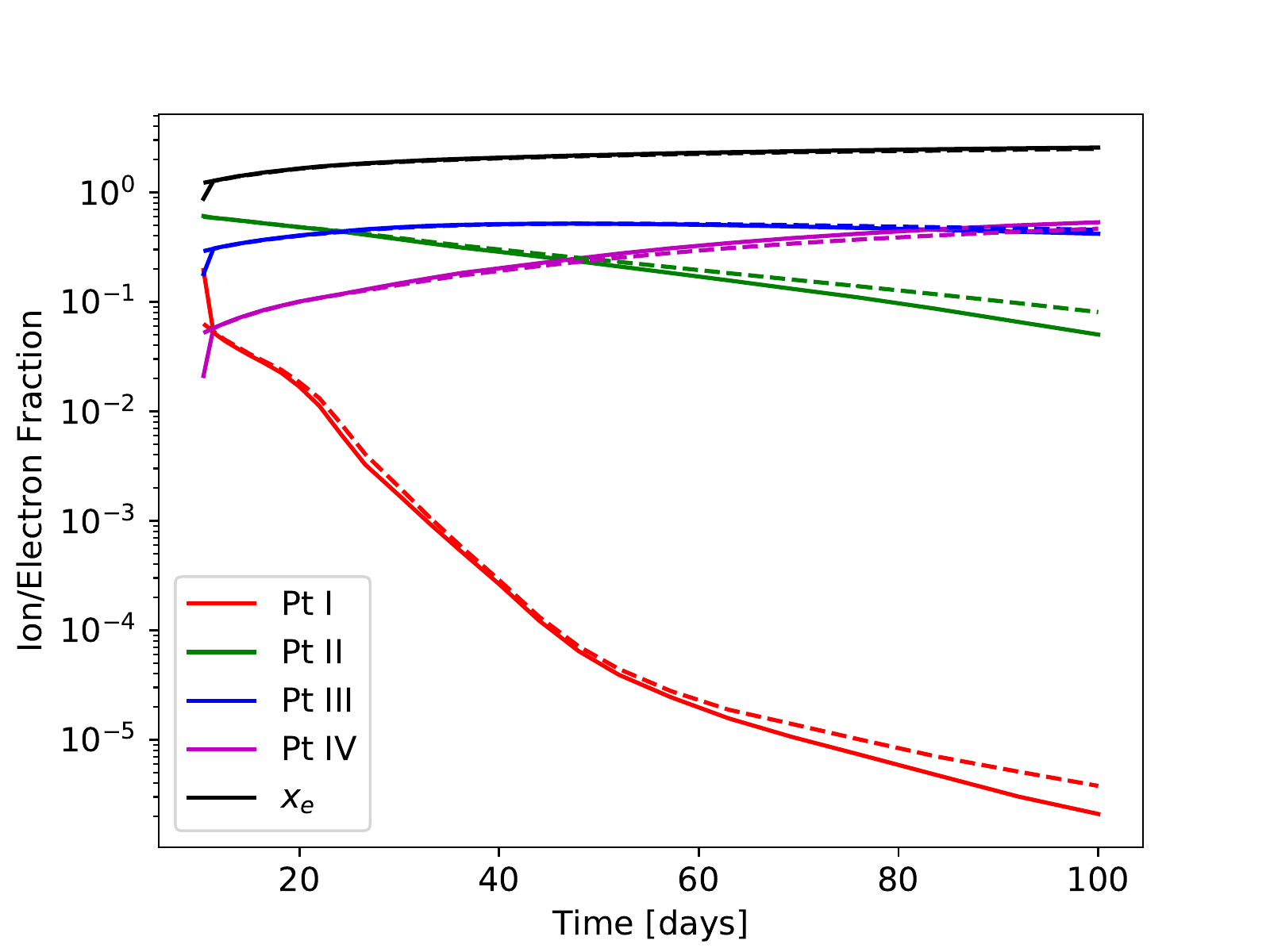}
\includegraphics[trim={0.2cm 0.2cm 1.6cm 0.8cm},clip,width = 0.49\textwidth]{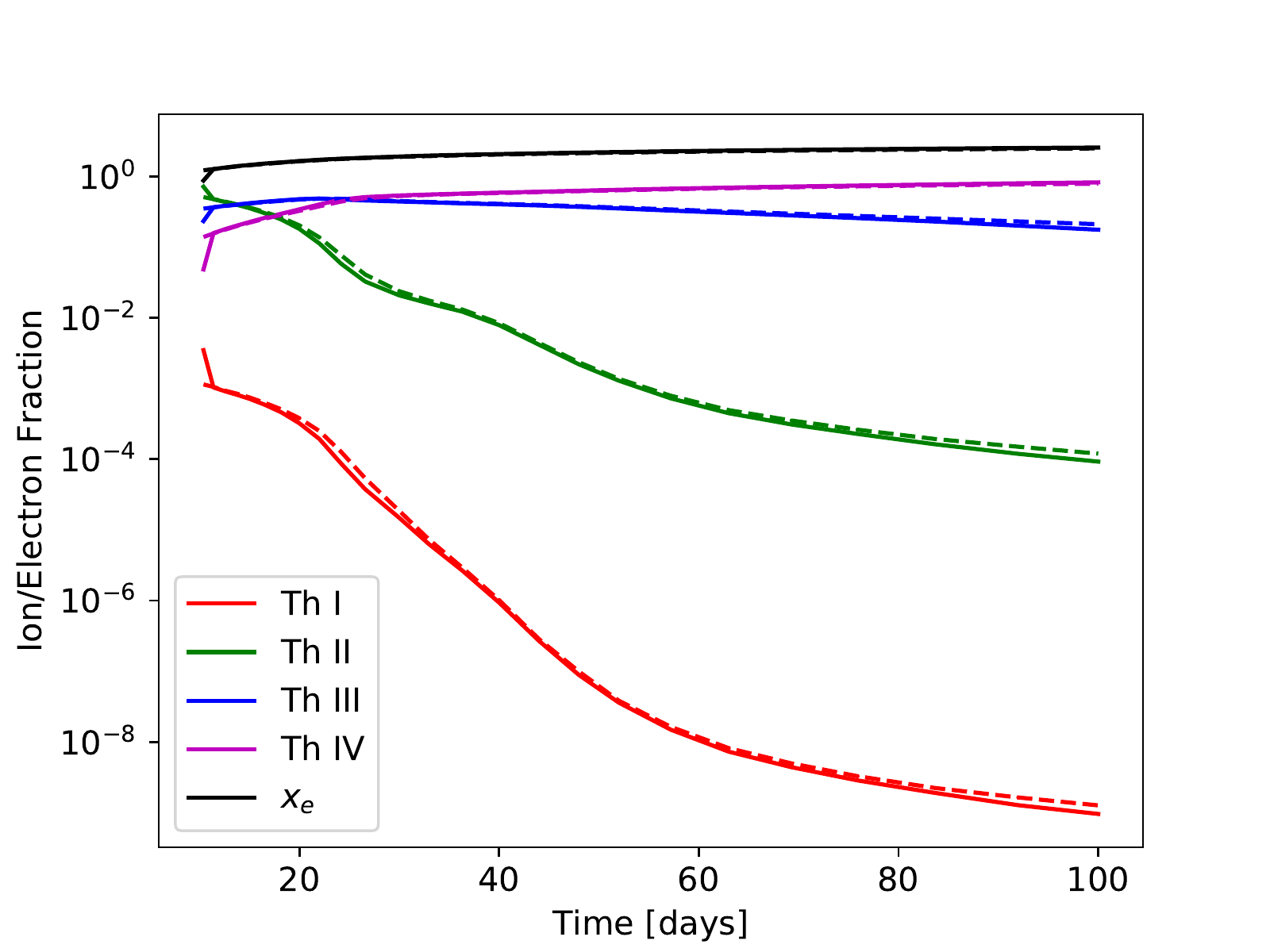}
\caption{Ionisation structure evolution for the model with $M_{\mathrm{ej}} = 0.1\mathrm{\Msol}$, $v_{\mathrm{ej}} = 0.1$c. The solid lines are the steady-state results, while the dashed lines are the time-dependent results. The sharp drop in steady-state fractions at the first epoch is a numeric artefact.}
\label{fig:01M_001v_ionfrac}
\end{figure*} 

\begin{figure*}
\center
\includegraphics[trim={0.2cm 0.2cm 1.6cm 0.8cm},clip,width = 0.49\textwidth]{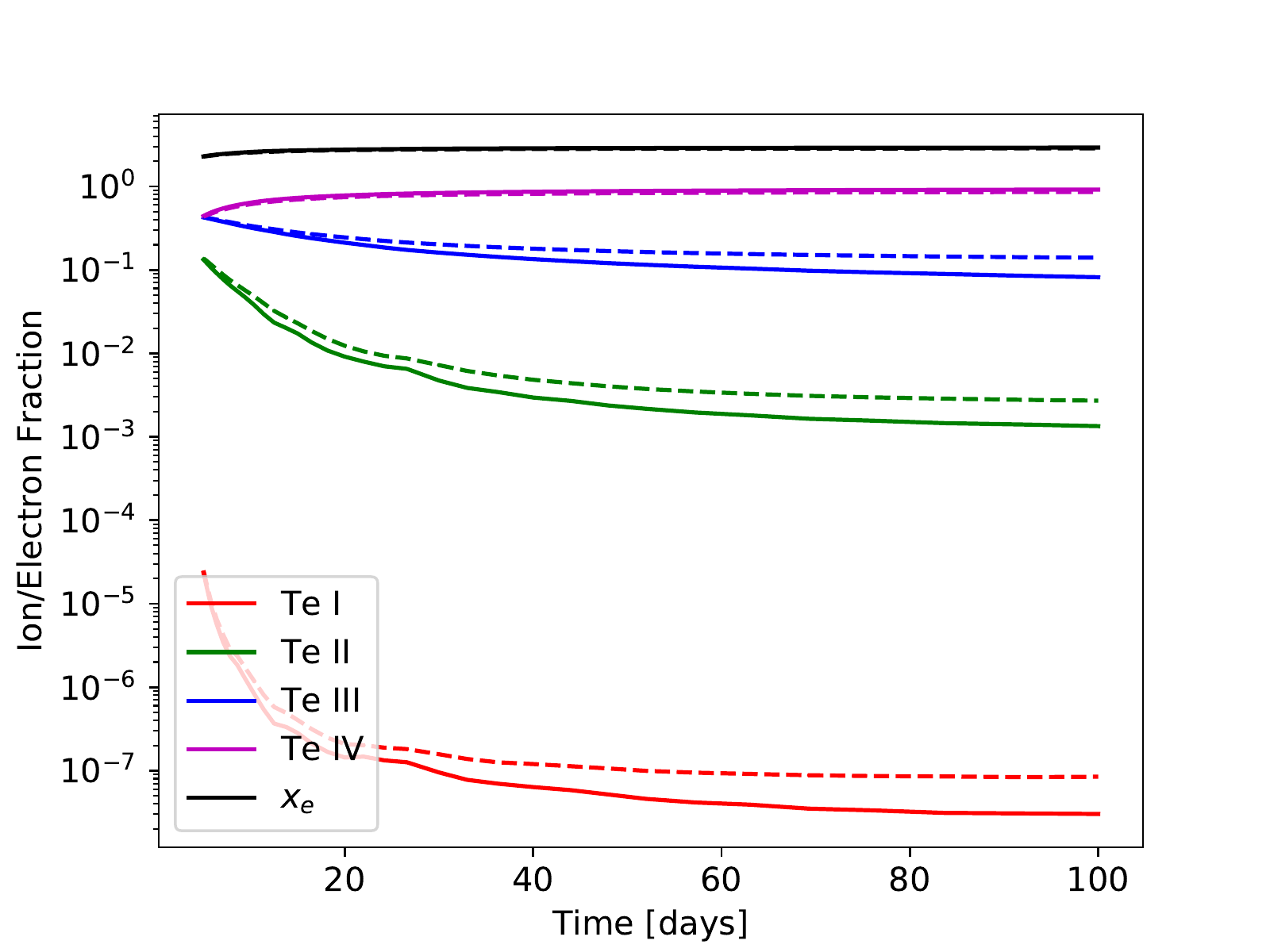}
\includegraphics[trim={0.2cm 0.2cm 1.6cm 0.8cm},clip,width = 0.49\textwidth]{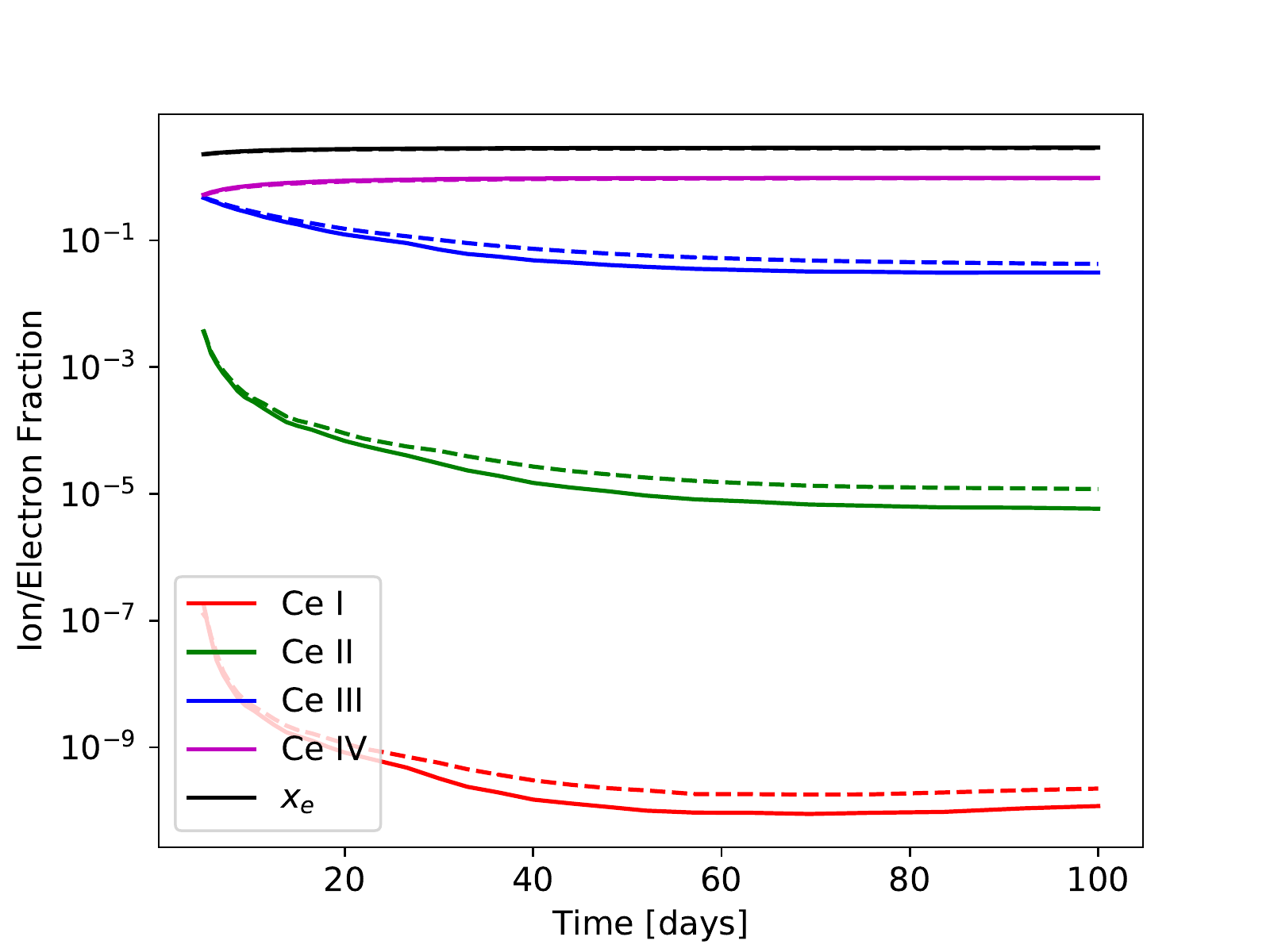}
\includegraphics[trim={0.2cm 0.2cm 1.6cm 0.8cm},clip,width = 0.49\textwidth]{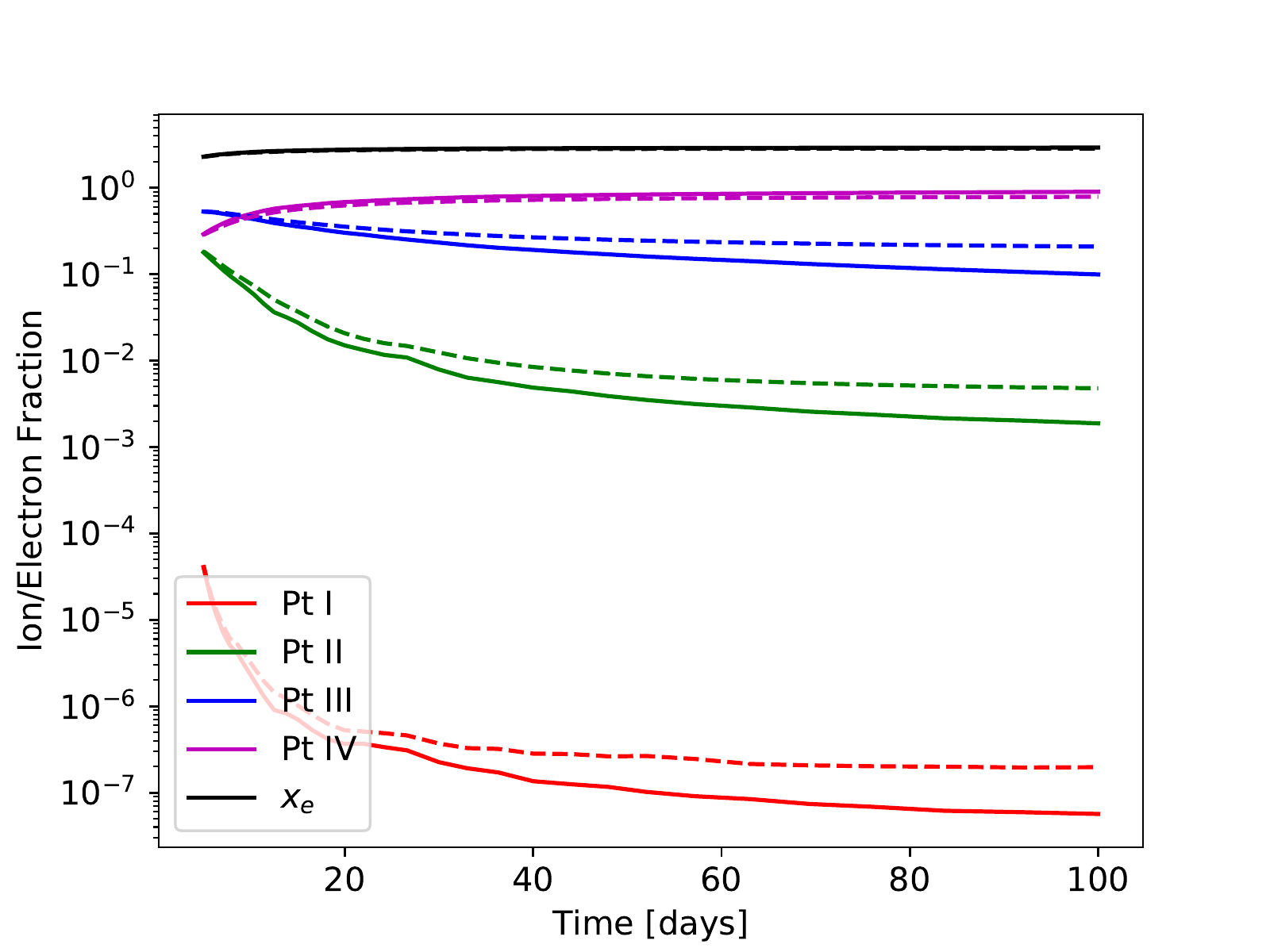}
\includegraphics[trim={0.2cm 0.2cm 1.6cm 0.8cm},clip,width = 0.49\textwidth]{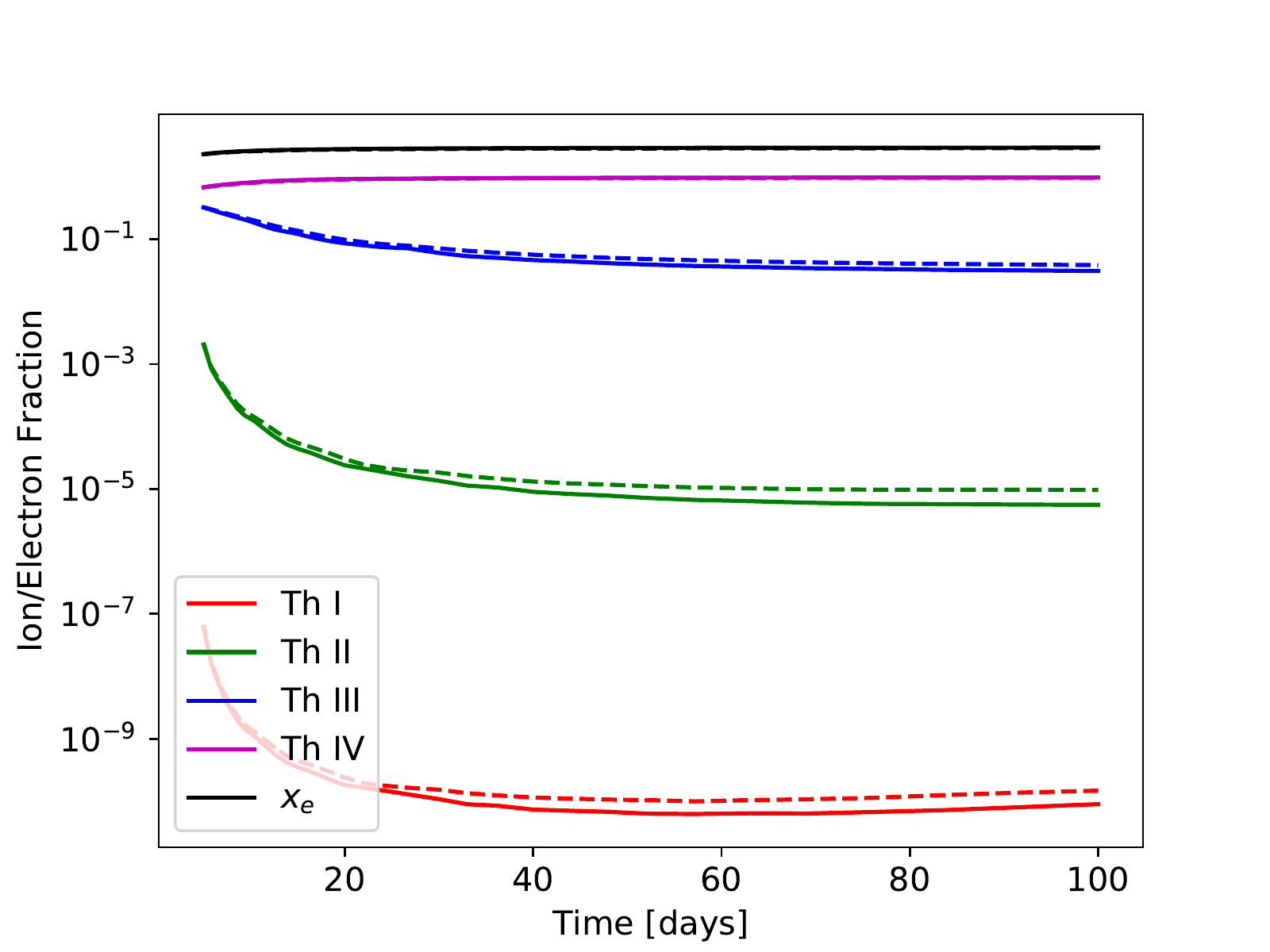}
\caption{Ionisation structure evolution for the model with $M_{\mathrm{ej}} = 0.01\mathrm{\Msol}$, $v_{\mathrm{ej}} = 0.2$c. The solid lines are the steady-state results, while the dashed lines are the time-dependent results.}
\label{fig:001M_02v_ionfrac}
\end{figure*} 

\begin{figure*}
\center
\includegraphics[trim={0.2cm 0.2cm 1.6cm 0.8cm},clip,width = 0.49\textwidth]{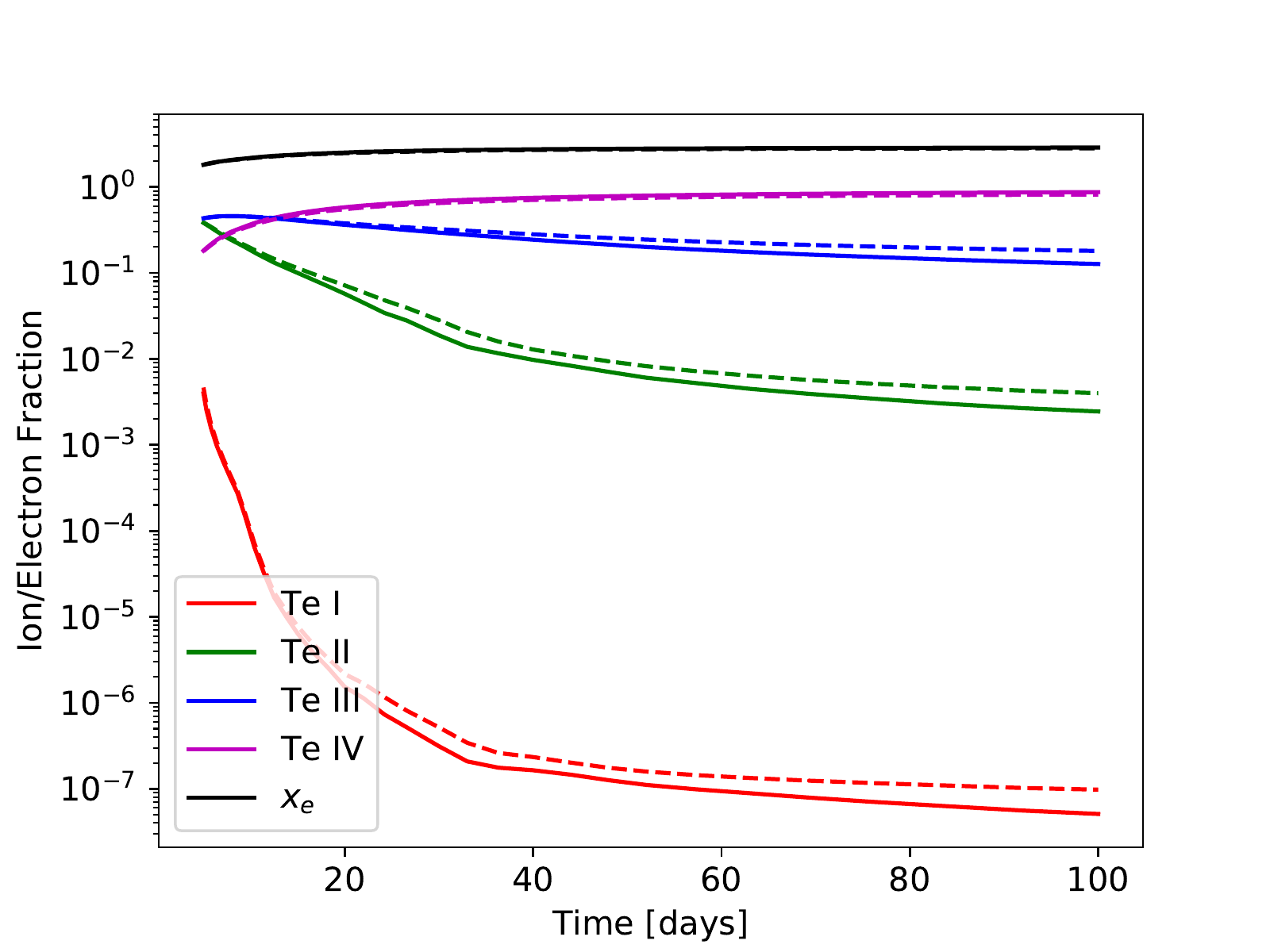}
\includegraphics[trim={0.2cm 0.2cm 1.6cm 0.8cm},clip,width = 0.49\textwidth]{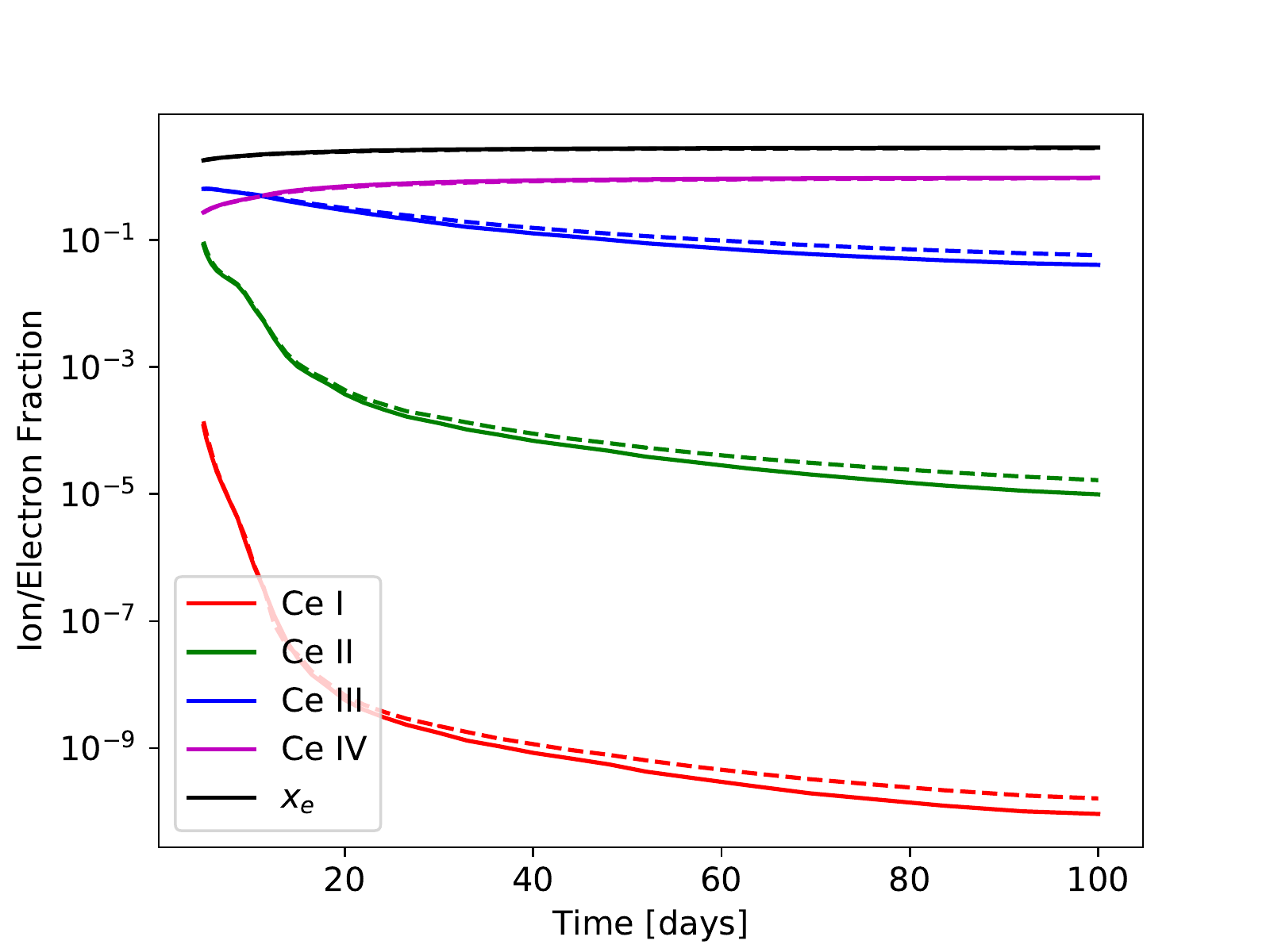}
\includegraphics[trim={0.2cm 0.2cm 1.6cm 0.8cm},clip,width = 0.49\textwidth]{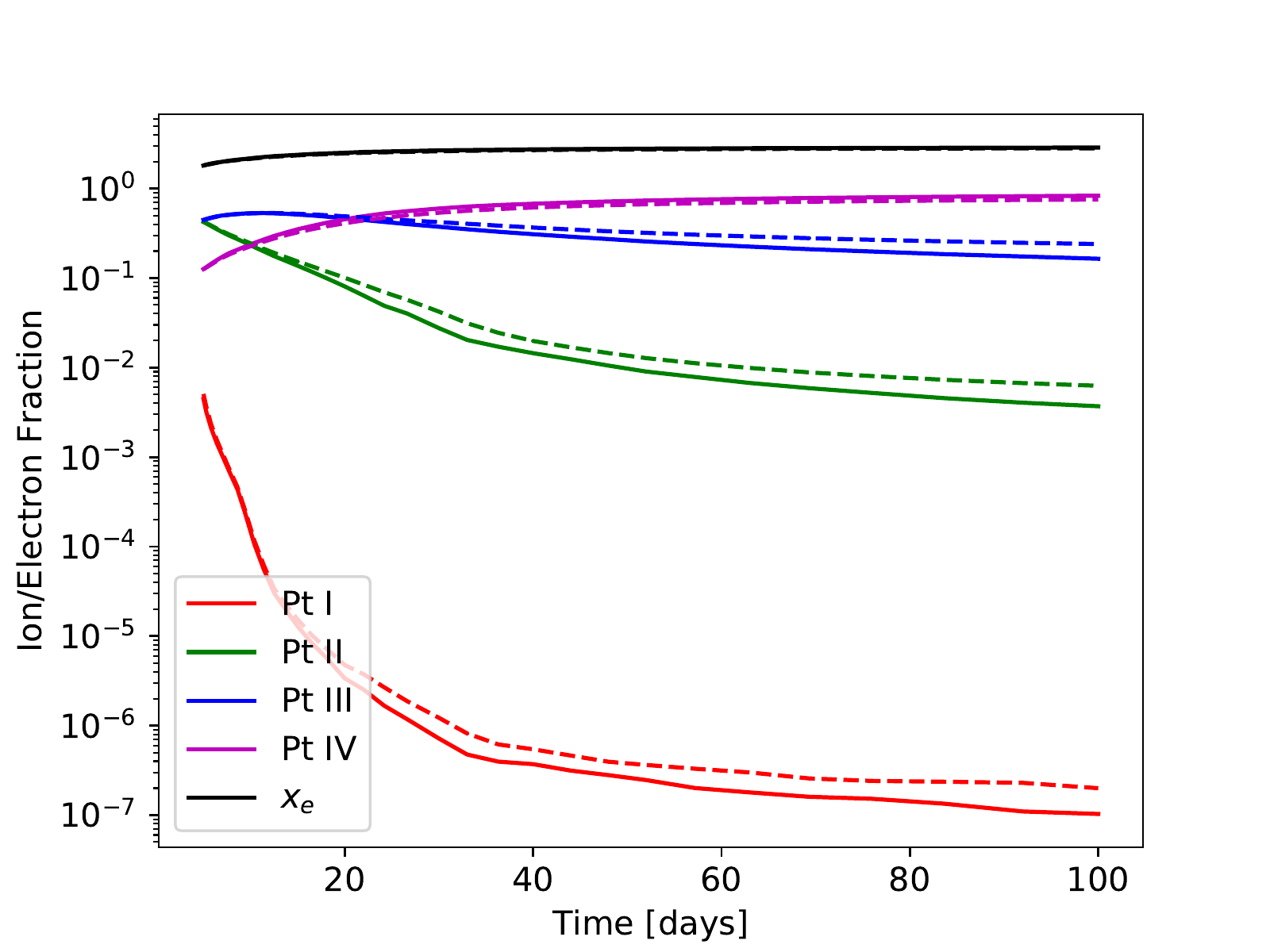}
\includegraphics[trim={0.2cm 0.2cm 1.6cm 0.8cm},clip,width = 0.49\textwidth]{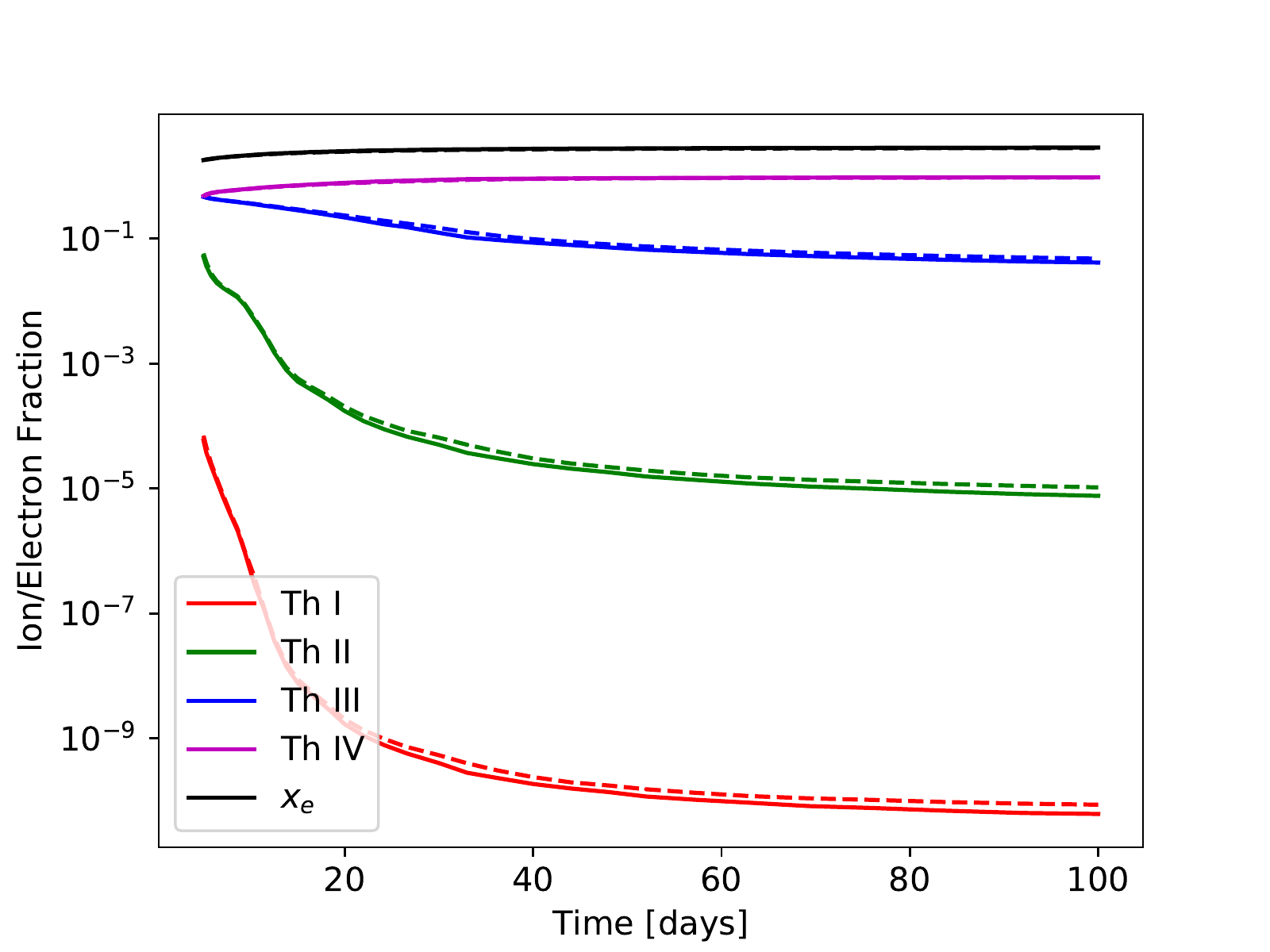}
\caption{Ionisation structure evolution for the model with $M_{\mathrm{ej}} = 0.05\mathrm{\Msol}$, $v_{\mathrm{ej}} = 0.2$c. The solid lines are the steady-state results, while the dashed lines are the time-dependent results.}
\label{fig:005M_02v_ionfrac}
\end{figure*} 

\begin{figure*}
\center
\includegraphics[trim={0.2cm 0.2cm 1.6cm 0.8cm},clip,width = 0.49\textwidth]{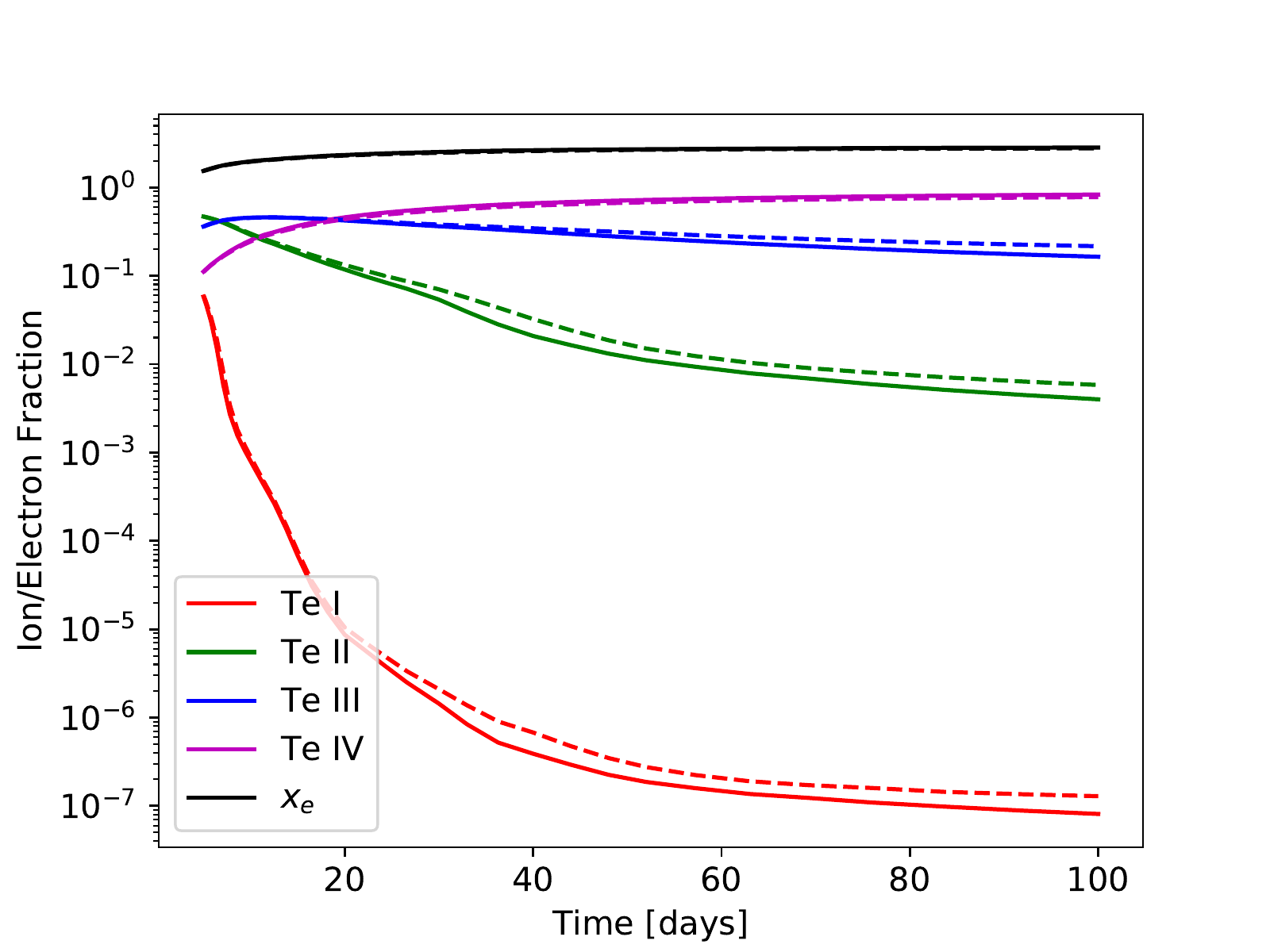}
\includegraphics[trim={0.2cm 0.2cm 1.6cm 0.8cm},clip,width = 0.49\textwidth]{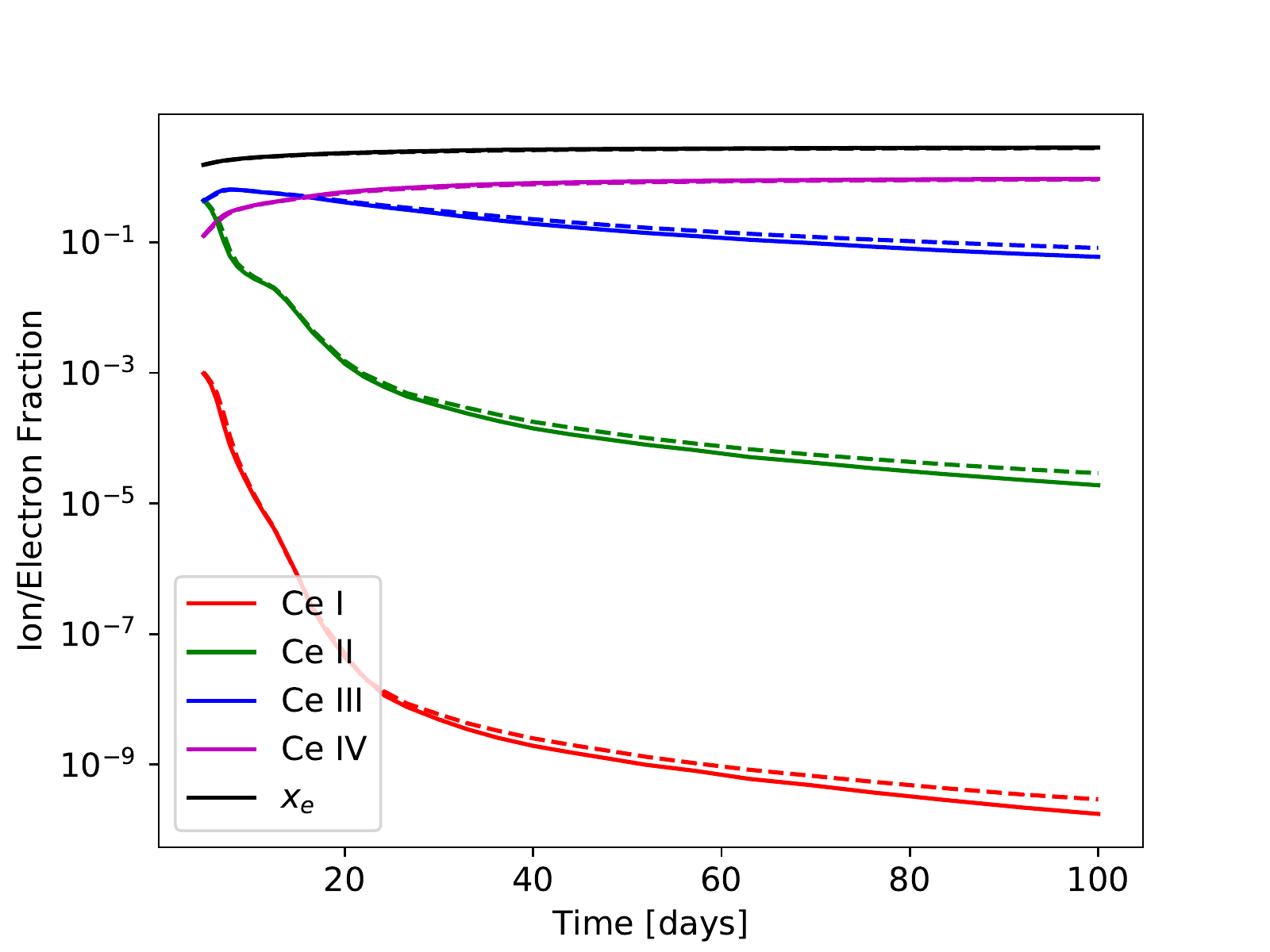}
\includegraphics[trim={0.2cm 0.2cm 1.6cm 0.8cm},clip,width = 0.49\textwidth]{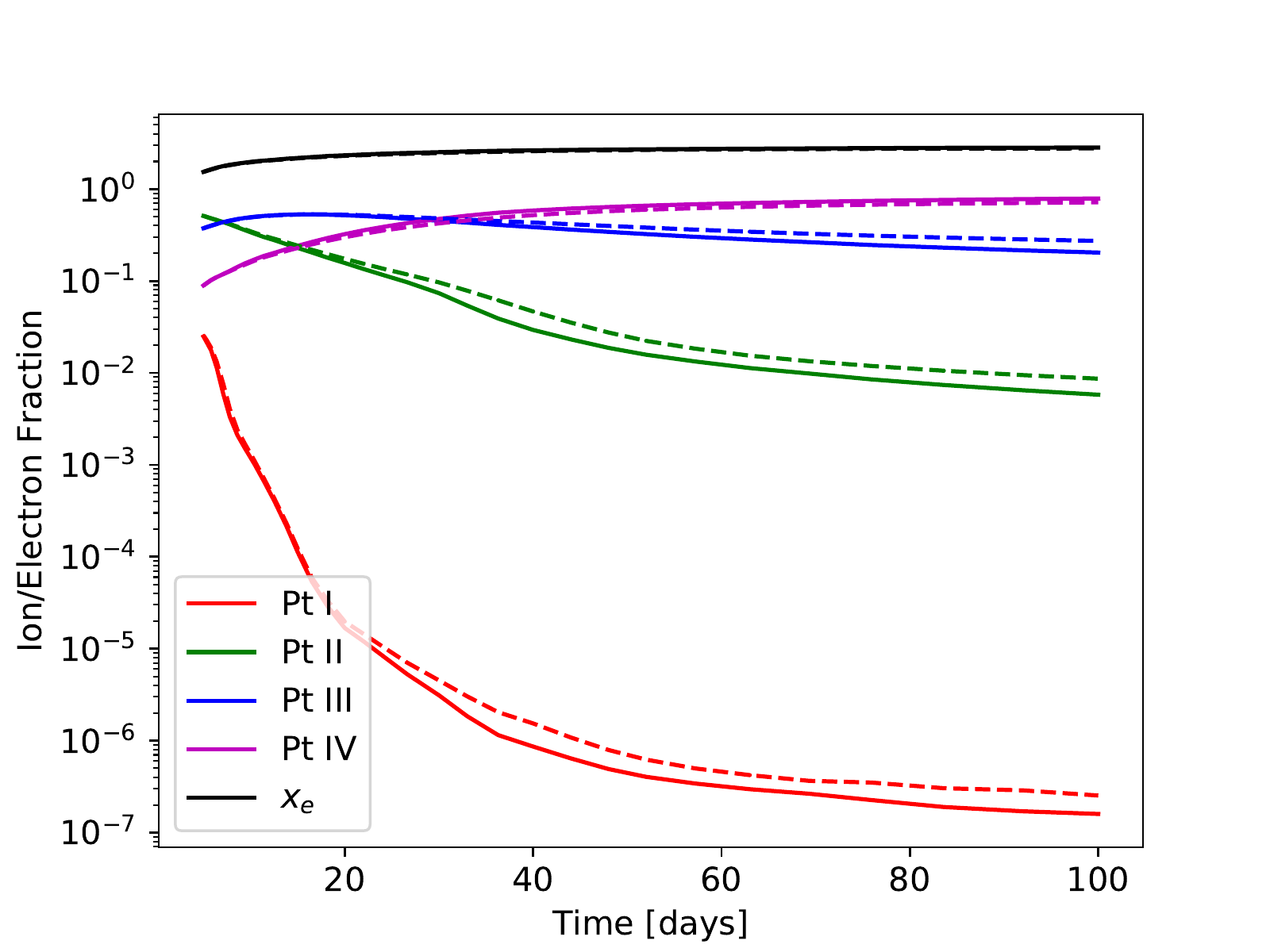}
\includegraphics[trim={0.2cm 0.2cm 1.6cm 0.8cm},clip,width = 0.49\textwidth]{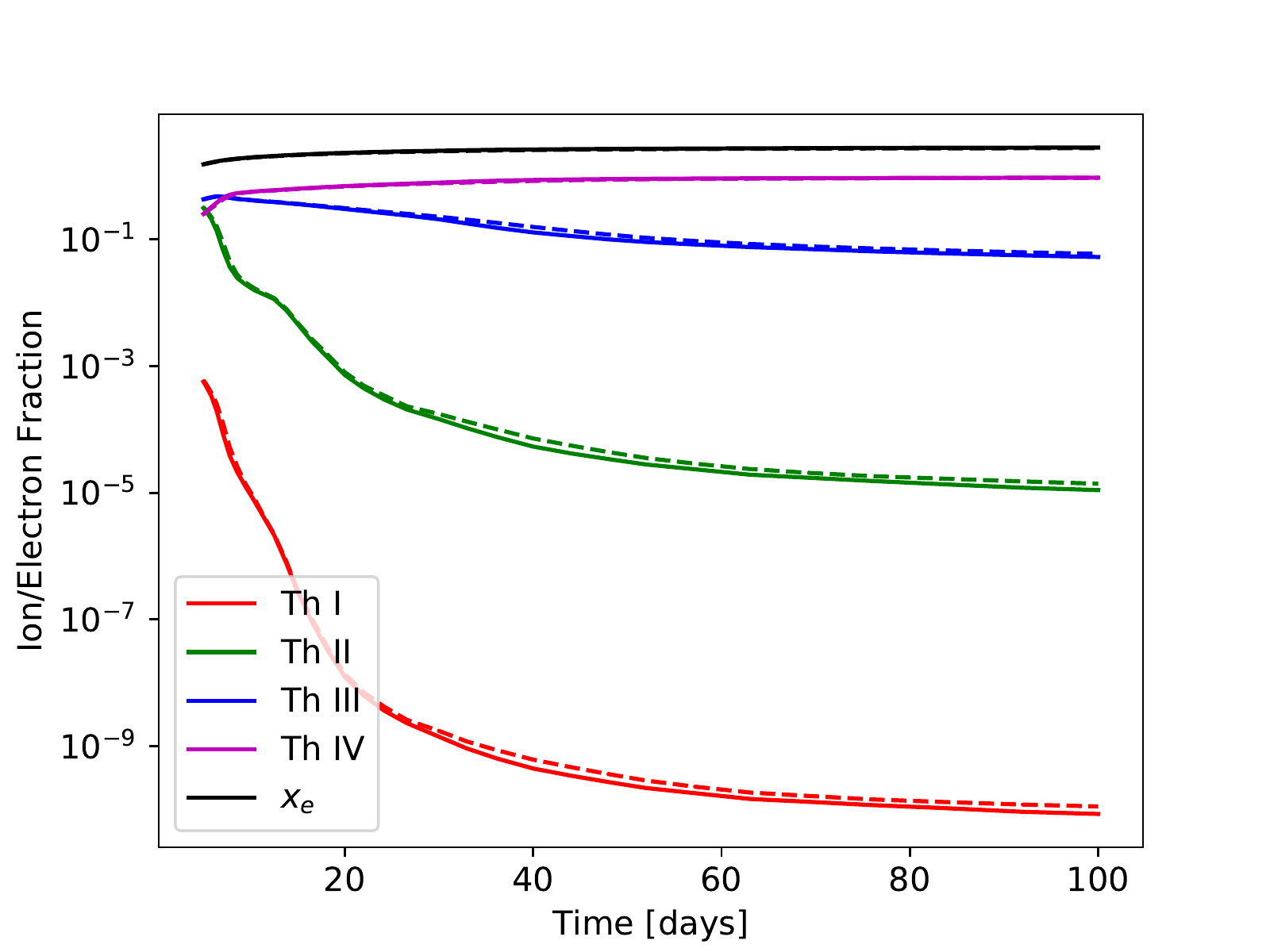}
\caption{Ionisation structure evolution for the model with $M_{\mathrm{ej}} = 0.1\mathrm{\Msol}$, $v_{\mathrm{ej}} = 0.2$c. The solid lines are the steady-state results, while the dashed lines are the time-dependent results.}
\label{fig:01M_02v_ionfrac}
\end{figure*} 

%% Dense model reduced energy plots

\section{Non-Thermal Electron Escape and Magnetic Trapping}
\label{app:electron_freestream}

The relativistic formula for electron velocity is
\begin{equation}
    v_e/c = \sqrt{1 - 1/\left(\hat{E} + 1\right)^2},
\end{equation}
where $\hat{E}$ is the energy in units of $m_ec^2=0.511$ MeV. Newborn $\sim$ MeV electrons emitted by $\beta$ decay thus have $v_e/c=0.94,$ and stay at over half the speed of light for $E>0.1$ MeV, at which point they have lost almost all their energy. As such, electrons emitted by $\beta$ decay in the ejecta will be firmly in the relativistic regime. These electrons will lose energy both by collisional ionisation of the plasma as well as by adiabatic losses, and have different thermalisation times depending on the conditions assumed within the ejecta, especially with regards to the degree of magnetic field trapping. 

For the magnetically confined case, the thermalisation time is the time-scale at which the collisional energy loss rate, $\dot{E}_{\mathrm{coll}}=\mbox{const} \times v_e(E)^{-1} t^{-3} \ln(E/\chi)$, 
equals the adiabatic loss rate, $\dot{E}_{\mathrm{ad}}=xE/t$, where $x=1$ in the relativistic limit and $x=2$ in the non-relativistic one. The constant for collisional losses encompasses the constant ejecta mass and velocity, as well as constants in front of the ionisation cross section evolving as $ln(E/\chi)$. Equivalently, this is when the collisional loss time $E/\dot{E}_{\mathrm{coll}}$ exceeds the adiabatic loss time $E/\dot{E}_{\mathrm{ad}}=t/x$. Solving for the thermalisation time-scale in the confined case, we find:

\begin{equation}
    t_{\rm{trap}}(E) = \left(\mbox{const}^{-1} \; v_e(E) \; \ln{(E/\chi)} \; E^{-1} \right)^{1/2}
    \label{eq:t_trap}
\end{equation}

In the case of electron escape, the key value is the escape time-scale, $\left(v_{\mathrm{ej}}/c\right)t$, which is typically smaller than the adiabatic time-scale since $v_{\mathrm{ej}}/c\sim 0.1$, and so $t_{esc} \sim \frac{v_{\mathrm{ej}}}{c} t_{\mathrm{ad}}$ for $x \approx 1$ in the relativistic regime. Solving for the thermalisation time in this case yields:

\begin{equation}
\begin{split}
    t_{\rm{free}}(E) &= \left(\mbox{const}^{-1} \; v_e(E) \; ln(E/\chi) \; E^{-1} \right)^{1/2} \; \sqrt{\frac{v_{\mathrm{ej}}}{c}} \\
    &= t_{\rm{trap}}(E) \; \sqrt{\frac{v_{\mathrm{ej}}}{c}}
\end{split}
\end{equation}

which gives a rough indication that the thermalisation break will start a factor $\sim$3 earlier in the free-streaming case for $v_{\mathrm{ej}} \sim 0.1$c.

The asymptotic power law is also different. We have

\begin{equation}
    \dot{q}(t) = \int_{t-\left(v_{\mathrm{ej}}/c\right) t}^t \dot{n}({t'}) \dot{E}_{\rm coll}(t,t') dt' 
\end{equation}

Since $v_{\mathrm{ej}}/c \ll 1$ and $\dot{n}$ and $\dot{E}$ change on longer time-scale $t$, we can move them outside the integral giving

\begin{equation}
   \dot{q}(t) \approx \dot{n}(t) \times E_d(t) \times \frac{\mbox{const} \times v_e(E_d(t))^{-1} t^{-3} \ln(E_d(t)/\chi) \frac{v_{\mathrm{ej}}}{c}t}{E_d(t)} 
\end{equation}
\noindent where $E_d(t)$ is the decay energy of particles. The first two terms represent the decay power at $t$. If $E_d(t)$ is constant the asymptotic thermalization factor in the free-streaming case goes as $t^{-2}$, as originally derived by \citet{Waxman.etal:19}. If $E_d(t) \propto t^{-0.3}$, as argued by \citet{Kasen.Barnes:19,Hotokezaka.etal:20}, this is flattened somewhat.

%%%%%%%%%%%%%%%%%%%%%%%%%%%%%%%%%%%%%%%%%%%%%%%%%%

% Don't change these lines
\bsp	% typesetting comment
\label{lastpage}
\end{document}